\documentclass{egpubl}
\usepackage{sca2026}

\SpecialIssueSubmission

\CGFccby

\usepackage[T1]{fontenc}
\usepackage{dfadobe}

\usepackage{cite}
\BibtexOrBiblatex
\electronicVersion
\PrintedOrElectronic
\usepackage{graphicx}

\usepackage{egweblnk}
\hypersetup{pdfsubject={},pdfkeywords={}}

\usepackage[table,dvipsnames,x11names]{xcolor}
\usepackage{graphicx}
\usepackage{subcaption}
\usepackage{float}
\usepackage{tabularx}
\usepackage{amsmath}

\usepackage{amssymb}
\usepackage{wrapfig}
\usepackage{boldline}
\usepackage{arydshln}
\usepackage{tikz}
\usepackage{tabularx}
\usepackage{rotating}
\usepackage{multirow}
\usepackage{makecell}
\usepackage[percent]{overpic}

\usepackage{booktabs}
\usepackage[normalem]{ulem}

\renewcommand{\vec}[1]{\mathbf{#1}}
\newcommand{\mat}[1]{\mathbf{#1}}

\newcommand{\todo}[1]{}

\newcommand\hide[1]{}

\definecolor{vcacolor}{RGB}{123,50,210}

\definecolor{jtcolor}{RGB}{153,51,255}

\newcommand\eg{e.g.,~}
\newcommand\Fig[1]{Figure~\ref{fig:#1}}

\newcommand\Sec[1]{Section~\ref{sec:#1}}
\newcommand\App[1]{Appendix~\ref{app:#1}}

\newcommand\Tab[1]{Table~\ref{tab:#1}}

\definecolor{gold}{RGB}{221, 196, 65}
\definecolor{silver}{RGB}{215, 215, 215}
\definecolor{bronze}{RGB}{126, 66, 5}
\definecolor{grayband}{RGB}{245, 245, 245}
\definecolor{lightgreen}{RGB}{195, 255, 195}
\definecolor{none}{RGB}{0, 0, 0}

\newcommand{\dotmark}[1]{%
  \,\tikz[baseline=-0.675ex]\fill[#1] (0,0) circle (0.7ex);%
}

\definecolor{mygray}{HTML}{666666}
\newcommand{\overlaylabel}{\fontsize{7}{8}\selectfont}

\newcommand{\mg}[1]{\text{#1}\dotmark{gold}}
\newcommand{\ms}[1]{\text{#1}\dotmark{silver}}
\newcommand{\mb}[1]{#1\phantom{\dotmark{gold}}}

\newcommand{\mn}[1]{#1\phantom{\dotmark{gold}}}

\newcolumntype{g}{>{\columncolor[HTML]{F0F0F0}}c}

\DeclareMathOperator*{\argmin}{arg\,min}

\newcommand{\gar}{\vec{x}_{G}}

\newcommand{\bind}{\vec{x}_{R}}
\newcommand{\deformationDelta}{\Delta\vec{x}}

\newcommand{\latent}{\vec{z}}
\newcommand{\skinning}{\mathcal{W}}
\newcommand{\unskinning}{\mathcal{W}^{-1}}

\newcommand{\pose}{\vec{\theta}}
\usepackage{xspace}
\newcommand{\dataset}[1]{\textsc{#1}\xspace}
\newcommand{\jenShirt}{\dataset{Jen's Shirt}}
\newcommand{\jenPants}{\dataset{Jen's Pants}}
\newcommand{\vto}{\dataset{VTO}}

\title[Pose-based Neural Clothing for Animated Characters]%
      {Pose-based Neural Clothing for Animated Characters}

\author[Heidenreich et al.]
{\parbox{\textwidth}{\centering
        Julian N. Heidenreich$^{1,2}$\orcid{0009-0004-2418-6193},
        Vinicius C. Azevedo$^{1}$\orcid{0009-0002-4133-4309},
        Jakob Buhmann$^{1}$\orcid{0009-0008-3038-4881},
        Lento Manickathan$^{1}$\orcid{0000-0002-0425-4679}, \\
        Arnold Moon$^{3}$,
        Paul Kanyuk$^{3}$\orcid{0009-0004-6022-6789},
        Amit H. Bermano$^{1}$\orcid{0000-0002-3592-1112},
        Jingwei Tang$^{1}$\orcid{0009-0000-6005-7808}
        }
        \\
{\parbox{\textwidth}{\centering
         $^1$DisneyResearch|Studios, Switzerland \qquad
         $^2$ETH Z\"{u}rich, Switzerland \qquad
         $^3$Pixar Animation Studios, United States of America
       }
}
}

\begin{document}

\newsavebox{\rightblockbox}

\teaser{
    \centering
    \vspace{-4mm}
    \makebox[\linewidth][c]{%
        \begin{minipage}{0.975\linewidth}
            \centering
            \setlength{\tabcolsep}{0.0015\linewidth}

            \sbox{\rightblockbox}{%
                \begin{minipage}[t]{0.605\linewidth}
                    \vspace{0pt}
                    \centering
                    \begin{tabular}{cccc}
                        \includegraphics[width=0.245\linewidth]{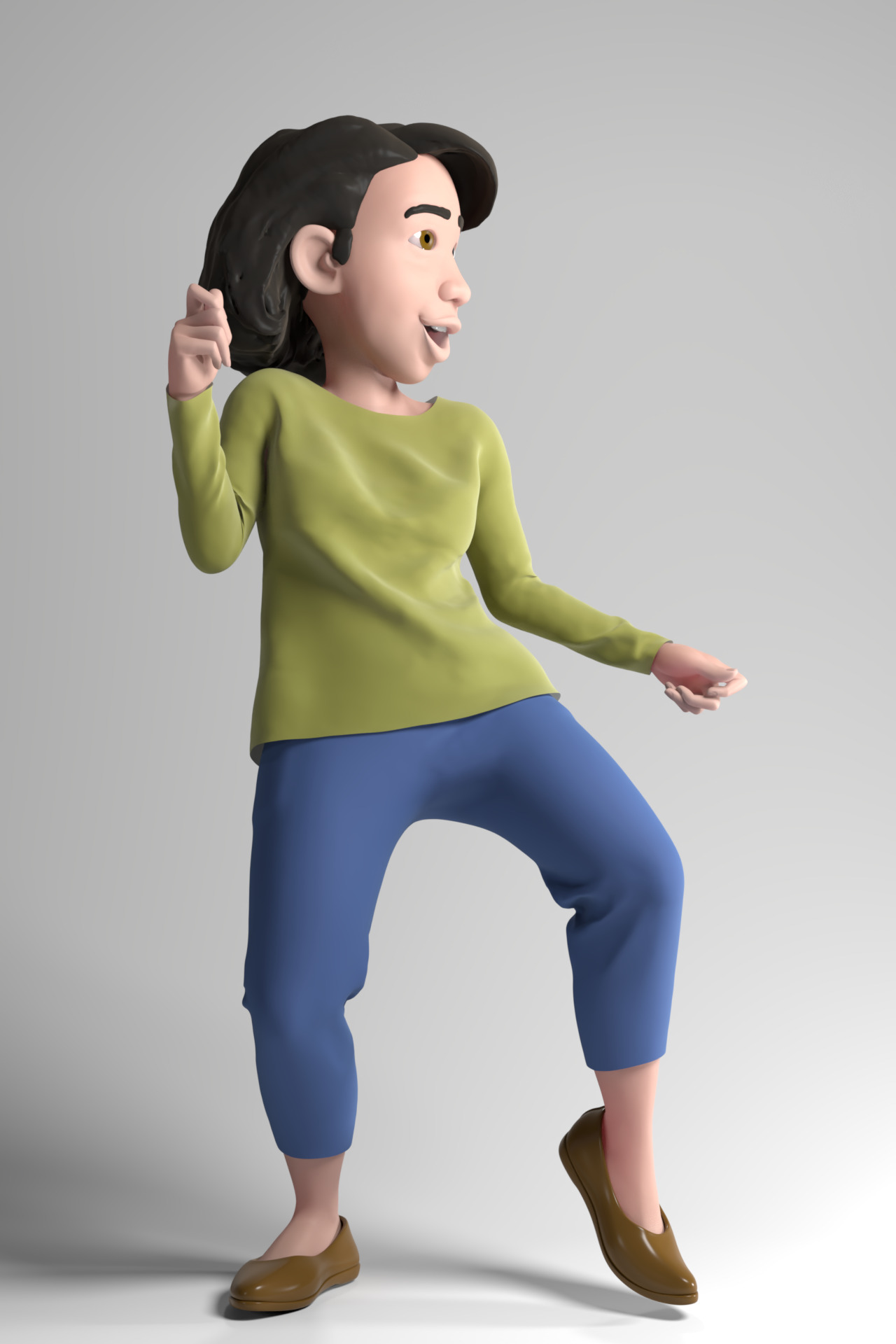} &
                        \includegraphics[width=0.245\linewidth]{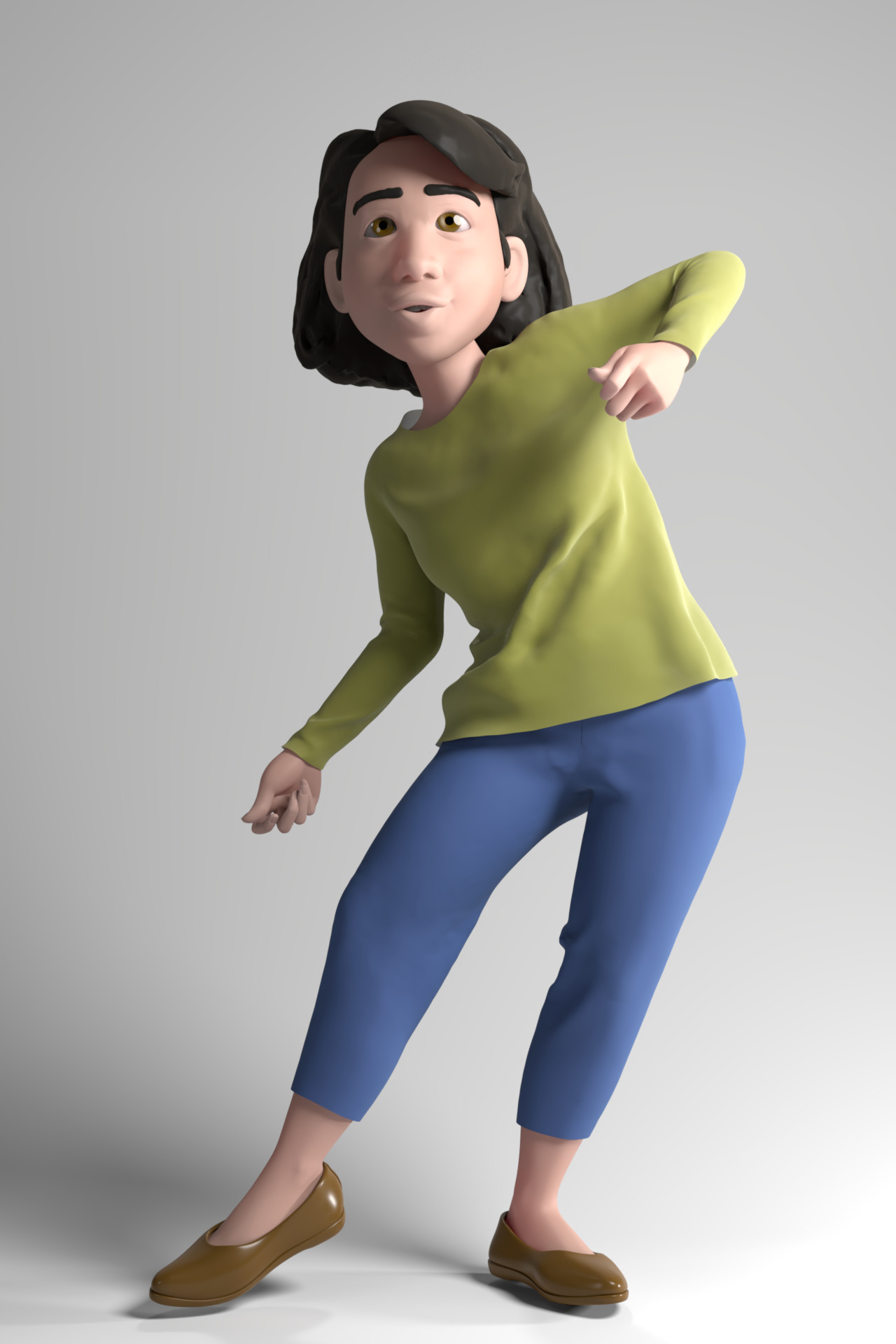} &
                        \includegraphics[width=0.245\linewidth]{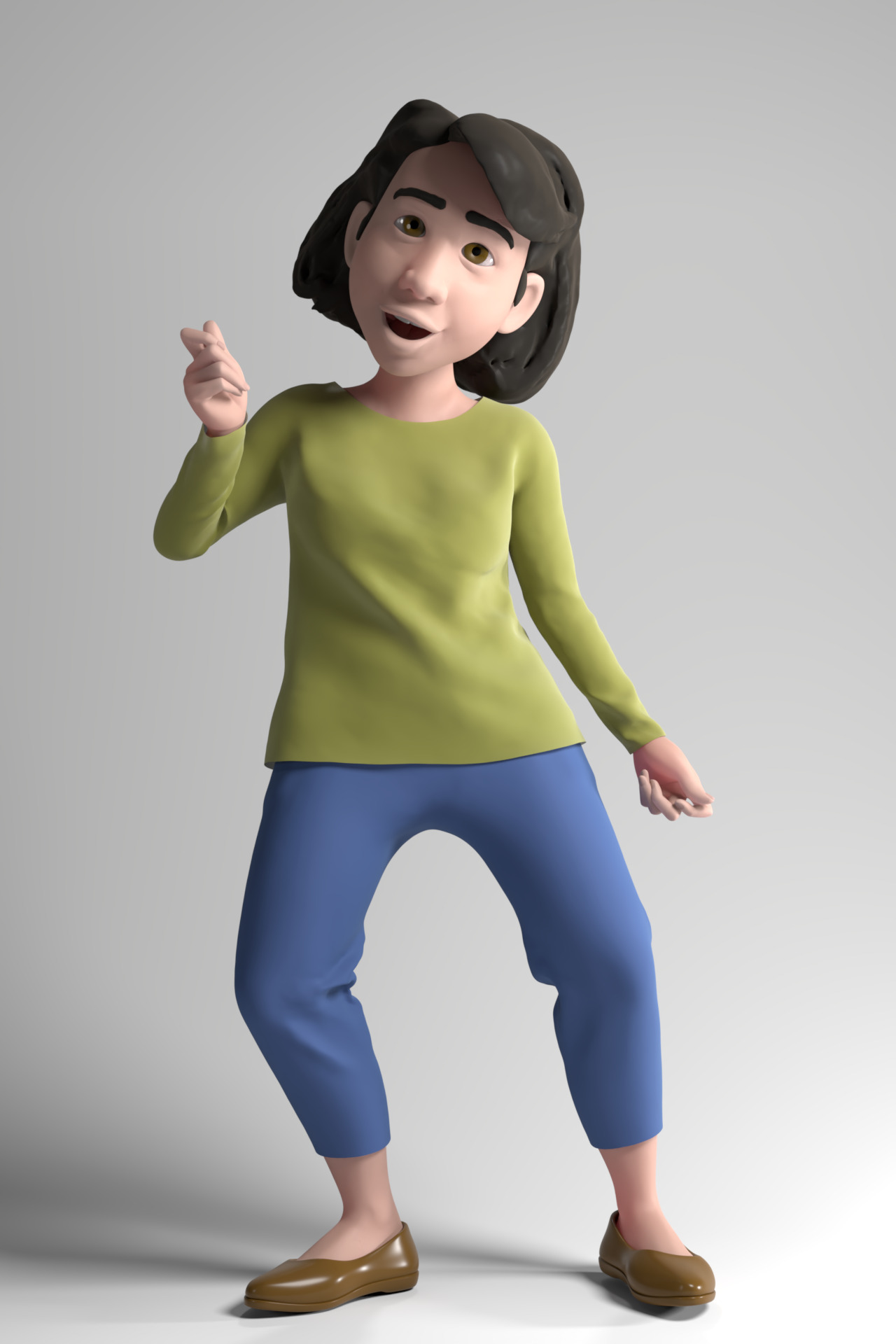} &
                        \includegraphics[width=0.245\linewidth]{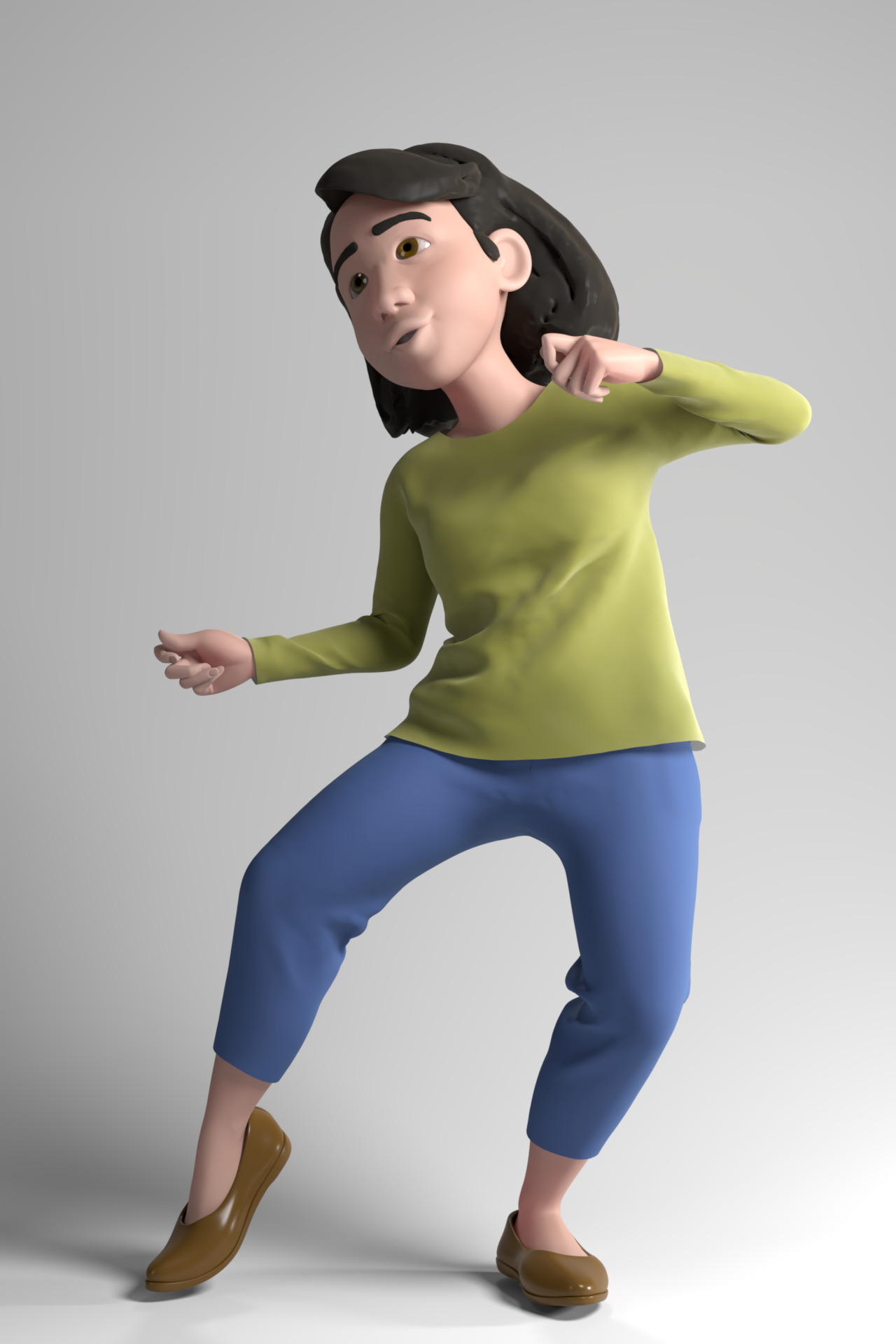} \\
                        \includegraphics[width=0.245\linewidth]{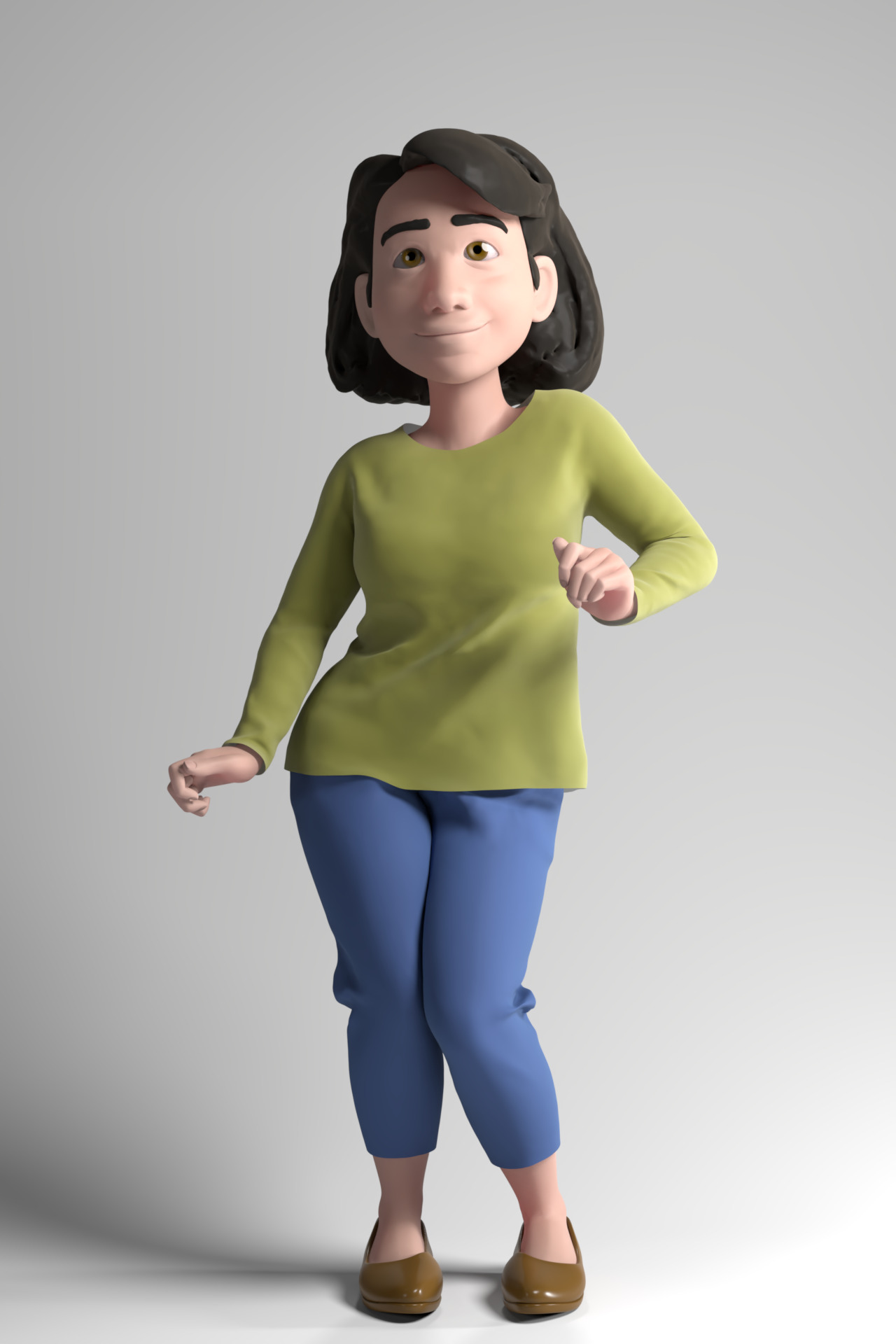} &
                        \includegraphics[width=0.245\linewidth]{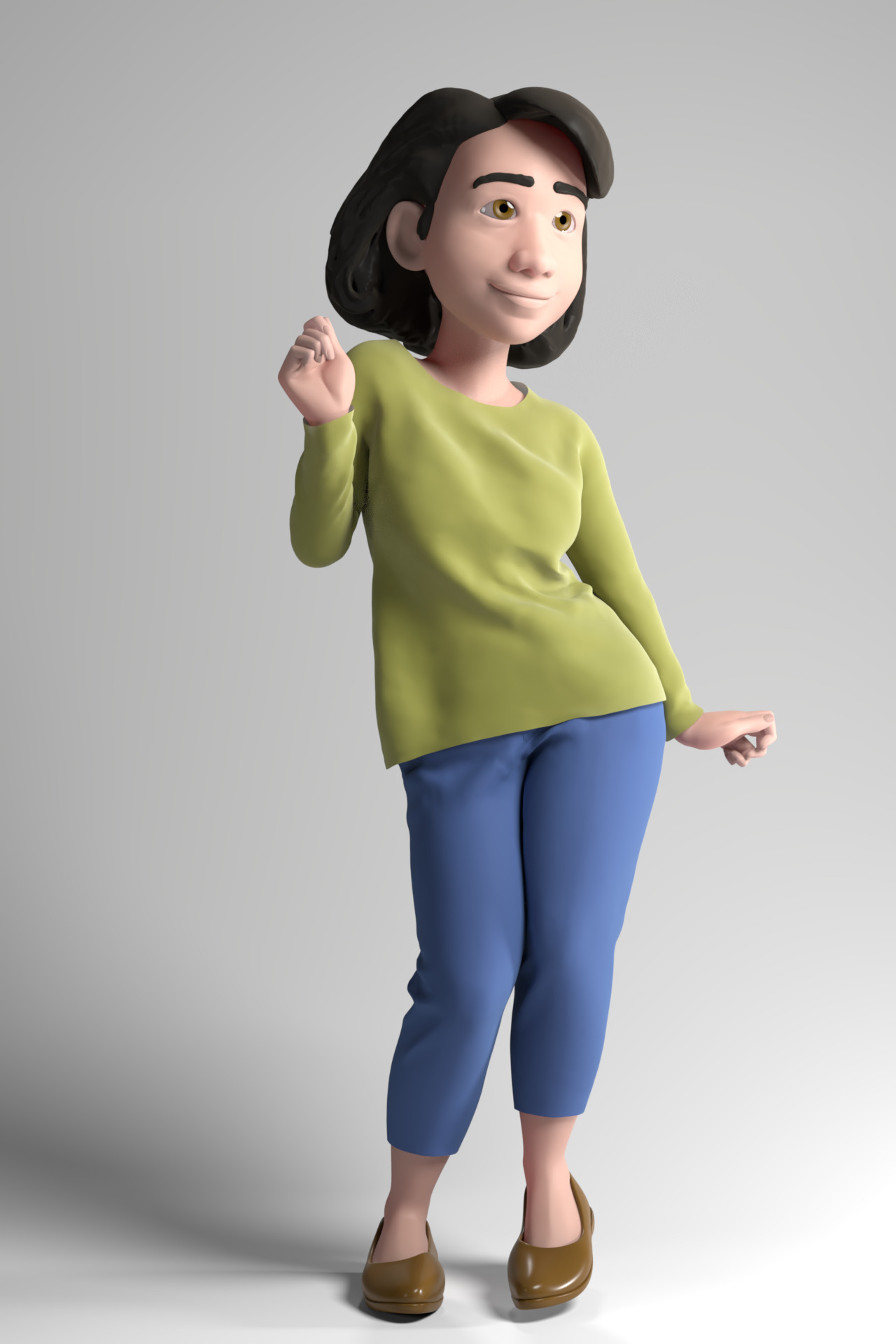} &
                        \includegraphics[width=0.245\linewidth]{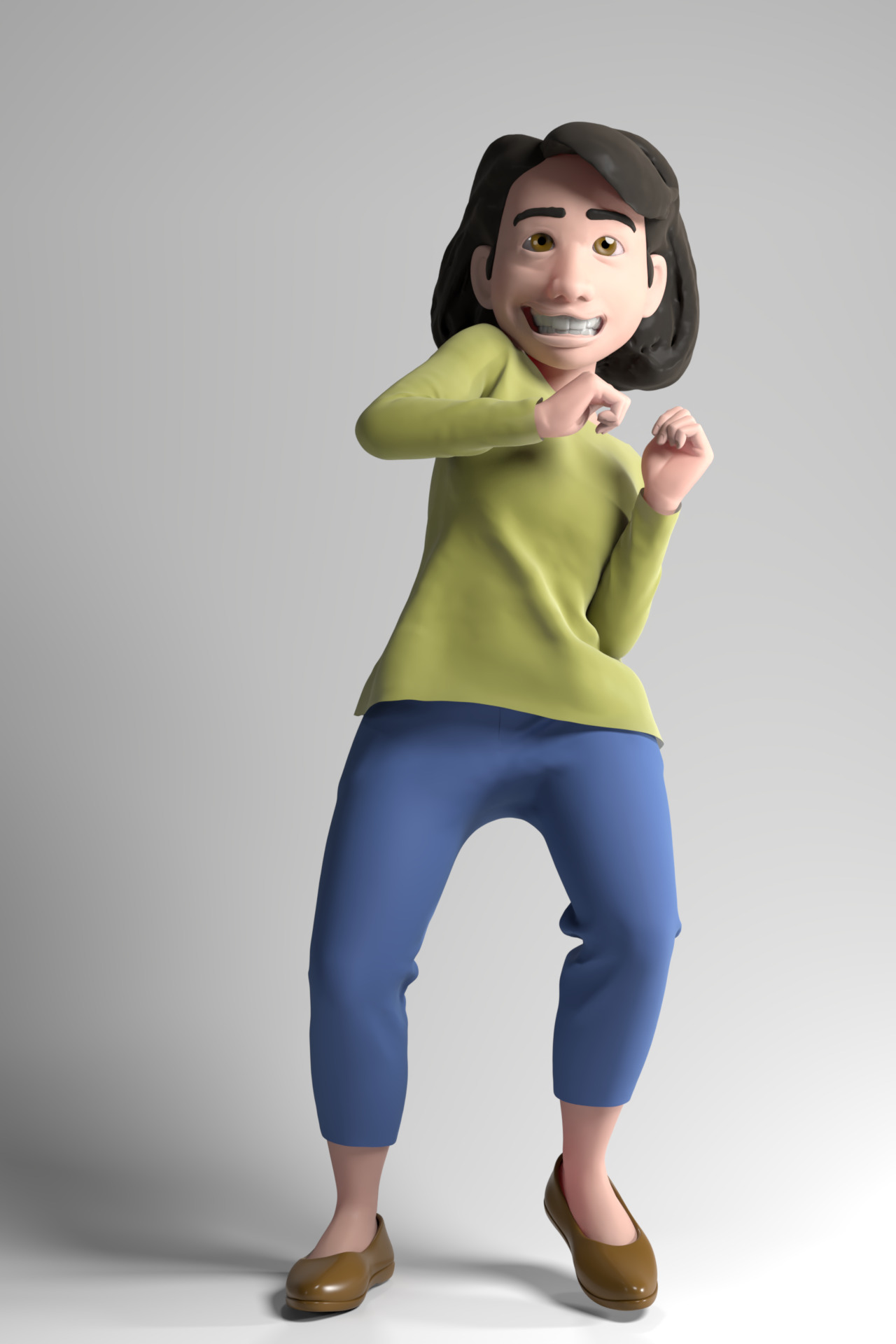} &
                        \includegraphics[width=0.245\linewidth]{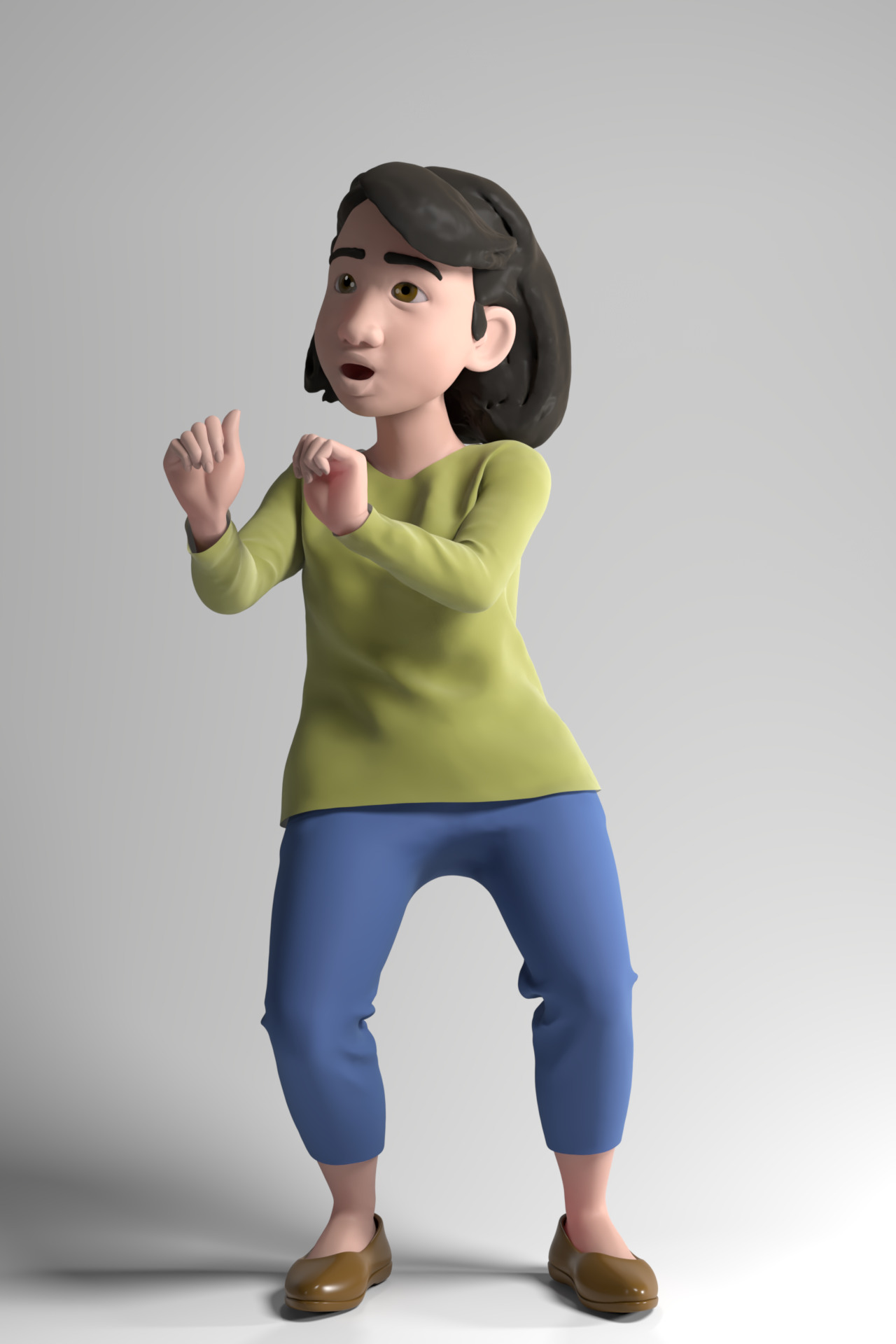}
                    \end{tabular}
                \end{minipage}%
            }

            \noindent
            \begin{minipage}[t]{0.39\linewidth}
                \vspace{0pt}
                \centering
                \includegraphics[
                    width=\linewidth,
                    trim={14.5cm 0.48cm 2.5cm 1.72cm},clip,
                    height=\dimexpr\ht\rightblockbox+\dp\rightblockbox\relax,
                    keepaspectratio
                ]{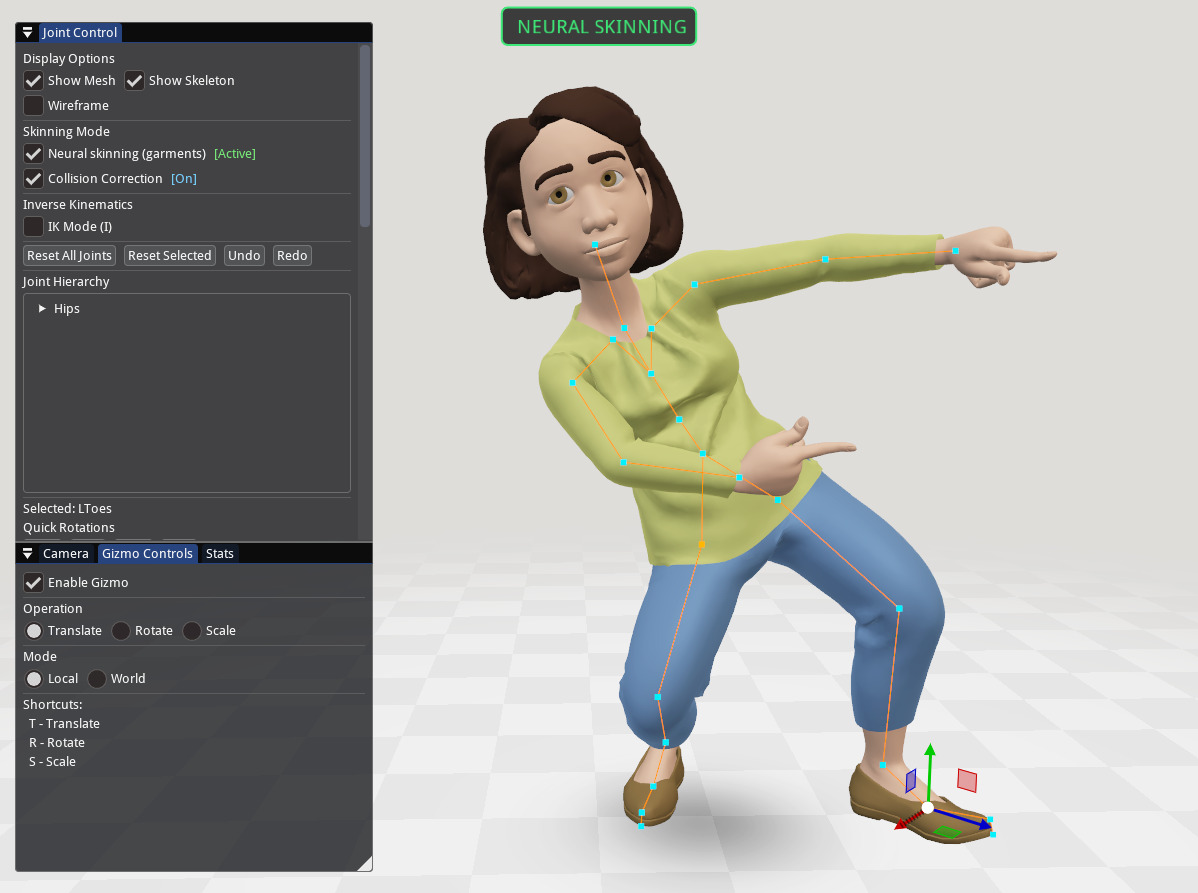}
            \end{minipage}
            \hfill
            \usebox{\rightblockbox}
        \end{minipage}%

    }

    \caption{Using a slim, progressively trained, and mesh-aware Variational Auto-Encoder (VAE) architecture, our method deforms garments during character posing (left), generating realistic wrinkles at interactive rates. We demonstrate this through example frames showing real-time computed deformations for both the shirt and pant (right).}
    \vspace{2mm}
    \label{fig:TeaserImage}
}

\maketitle

\thispagestyle{empty}
\makeatletter
\renewcommand{\@oddhead}{}
\renewcommand{\@evenhead}{}
\renewcommand{\@oddfoot}{\hfil\thepage\hfil}
\renewcommand{\@evenfoot}{\hfil\thepage\hfil}
\makeatother

\begin{abstract}
    Traditional garment simulation for animated characters is a time-consuming bottleneck for artists due to slow simulations and tedious parameter tunings.
    Recent data-driven methods offer a powerful alternative, capable of learning from a dataset of simulated data to generate high-quality deformations.
    A lightweight network design allows data-driven systems to achieve a real-time inference speed. This significantly improves the artist workflow, as they can interactively pose a character and see high-quality folds and wrinkles at animation time, thereby allowing to make creative decisions ahead of time.
    However, existing neural clothing methods have presented appealing results on datasets based on realistic human models (\eg SMPL), but when trained on animated character datasets, they struggle to generalize to unseen poses.
    In this paper, we aim at designing neural networks and training strategies to generalize better to unseen poses in real-time to enable the interactive posing applications. We focus on a production-oriented setting where the model is trained for a single garment in a quasi-static setup.
    Technically, we leverage a smooth, expressive latent space by first training a Variational Auto-Encoder (VAE). We then progressively train the VAE encoder to transition from reconstruction to pose-driven prediction. To guide this prediction, we condition the network on both low-dimensional pose parameters and character body surface displacements. Finally, we introduce a novel local gradient histogram loss that mitigates the over-smoothing effect of reconstruction losses.
    Extensive evaluations on production-level assets and public datasets demonstrate that our method produces sharp, realistic wrinkles, smooth large-scale drape, and temporal coherence in real-time, offering a robust and easily integrable solution for interactive posing workflows.


\end{abstract}

\section{Introduction}
Cloth deformation plays a central role in conveying both realism and style in digital characters. In production settings such as film, interactive media, virtual try-on, and web-based character customization, \emph{interactive posing} is essential in artistic decisions: artists expect immediate visual feedback when posing a character, including how garments fold and wrinkle. Standard physically-based cloth simulation is therefore often unsuitable for interactive use, due to its dynamic formulation and computational cost, even though it produces highly realistic results.
Moreover, production cloth is rarely ``off-the-shelf'': studios rely on highly specialized solvers and carefully tuned pipelines refined over the years of show-specific expertise. It is not uncommon for artists to perform manual interventions on certain frames to resolve difficult cases: persistent body inter-penetrations, unstable contacts causing popping or jitter, loss of volume in layered folds, or solver artifacts such as cloth snagging and sticking in extreme poses are some examples. As a result, even though deploying a faster GPU-based simulator \cite{wu_safe_2020, wu_gpu_2022, guo_fast_2025} can potentially reach interactive rates,
they are often insufficient to reproduce the final production-approved look. These constraints motivate a data-driven alternative: a supervised neural deformer trained directly on production-generated targets, capturing both the solver’s tuned behavior and the artist-approved outcomes.

A vast literature \cite{patel_tailornet_2020, santesteban_snug_2022, santesteban_self-supervised_2021, bertiche_neural_2022, grigorev_hood_2023, zhang_neural_2024, bertiche_pbns_2021} has explored learning-based approaches to improve the performance--realism trade-off of cloth simulation, typically by approximating full simulations using neural networks that operate over time or leverage simplified dynamics.
While these learning-based methods present appealing results on datasets featuring realistic human models (\eg SMPL \cite{loper_smpl_2015}), our experiments in \Sec{results} reveal that they struggle when trained on animated characters, often producing inaccurate details and visual artifacts.  Furthermore, these approaches often assume access to simulation states or poses in previous frames, which are typically unavailable during interactive character posing.
Additionally, while many existing methods train their models to generalize across multiple body shapes and garment materials \cite{patel_tailornet_2020, grigorev_hood_2023}, we intentionally focus on a single-mesh paradigm. This targeted approach allows us to train a lightweight network that faithfully preserves the unique style avoiding the loss of detail often caused by generic priors and guarantees real-time inference.
Finally, although some works incorporate physically-based unsupervised \cite{bertiche_pbns_2021, bertiche_neural_2022, santesteban_snug_2022, lee_clothcombo_2023, grigorev_hood_2023, grigorev_contourcraft_2024} terms to improve visual quality, production-level garment deformations are hard to reproduce with simple energy terms, compelling us to adhere to a strictly supervised setting to ensure our results faithfully replicate targeted artistic intent.

We aim to train real-time neural deformers that predict cloth deformations from character poses. Additionally, we target a single garment simulated using a quasi-static setup for a single character.
The design space, the input representations, and training strategies of our network are extensively explored in the paper.
We start with a popular architecture, employing spiral convolutions for mesh-aware computations \cite{chentanez_cloth_2020}, which outperforms simpler options such as variants of global and local MLPs \cite{bailey_fast_2018, patel_tailornet_2020}.
Our first insight decouples the task of predicting pose-dependent deformations from the generation of plausible, high-frequency wrinkle patterns: before training the deformer, we first train a Variational Auto-Encoder (VAE) to reconstruct a modest set of deformations to create a smooth, concise, and expressive latent space. Our second insight is to progressively train the VAE encoder to transition from reconstruction to deformation prediction from body poses, thus simplifying the task. We find that this progressive strategy significantly improves the visual quality while maintaining architectural complexity and keeping inference times at interactive rate. In addition to conditioning the re-purposed encoder on low-dimensional pose parameters, we splat body surface displacements onto the garment mesh using linear blend skinning weights to provide deformation guidance on the mesh level. Finally, we leverage a novel local gradient histogram loss over surface neighborhoods to preserve high-frequency wrinkles and mitigate the over-smoothing that is commonly associated with reconstruction-based losses.

We evaluate our approach against state-of-the-art methods from both academia \cite{bailey_fast_2018, patel_tailornet_2020, chentanez_cloth_2020, santesteban_self-supervised_2021} and production pipelines \cite{autodesk_autodesk_2025, epic_games_unreal_2025} on production-level shirt and pants assets, as well as previously published virtual try-on datasets. Across datasets, our method produces sharp, realistic wrinkles while maintaining smooth large-scale drape at interactive rates, and exhibits temporally coherent behavior as a byproduct of the structured latent space despite operating per-frame.
Extensive ablations further justify each stage of our pipeline and its design choices. Overall, our experiments demonstrate a production-ready approach that is easy to integrate into existing workflows, giving artists a robust and high-quality preview in real-time.

\section{Related Work}

\subsection{Numerical Simulation of Cloth}
Cloth simulation is a long-standing and widely researched topic in computer animation. Since the debut of the seminal Baraff--Witkin model \cite{baraff_large_1998}, many improvements have been proposed to simulate fabrics with higher fidelity \cite{bridson_simulation_2005, choi_stable_2005, thomaszewski_continuum-based_2009, li_stable_2015, kim_finite_2020}. For a detailed analysis regarding production implementation practicalities, we refer to Kim et al. \cite{kim_dynamic_2020}. Recent efforts have greatly boosted the efficiency of cloth solvers on both CPUs and GPUs \cite{wu_safe_2020, wu_gpu_2022, guo_fast_2025, lan_efficient_2024, lu_high-performance_2025, chen_offset_2025}. However, implementing these efficient solvers in production settings often poses practical challenges. Production solvers must handle broken animations, non-physical setups, self-intersections, solver-specific heuristics, and artist-authored corrections. Neural networks, such as our method, are not only efficient, but also easily portable to different digital content creation software.

One well-researched area within numerical simulation is sub-space solvers \cite{kim_skipping_2009, james_precomputing_2003, treuille_model_2006, huang_subspace_2006, pan_subspace_2015, yang_expediting_2015, mukherjee_incremental_2016, brandt_hyper-reduced_2018, luo_nnwarp_2018, weidner_conjac_2020}, thus reducing the excessive degrees of freedom for a given application. Many learning-based cloth deformation methods \cite{santesteban_self-supervised_2021, su2022deepcloth, de2023drapenet} reflect this idea by training Auto-Encoders to reduce the garments into smooth and expressive latent space. We similarly reduce the dimensionality of the problem by training a VAE on deformation deltas.

Collision handling \cite{bridson_simulation_2005,stam_nucleus_2009, fratarcangeli_vivace_2016,tang_i-cloth_2018, wu_safe_2020} also plays an important role in cloth simulations. Most neural clothing methods either perform simple repulsion-based post-processing at inference time to resolve collisions \cite{santesteban_learning-based_2019, patel_tailornet_2020, santesteban_snug_2022, chen_deep_2023, li_detail-aware_2023}, or formulate a loss term based on interpenetration distance  \cite{santesteban_self-supervised_2021, gundogdu_garnet_2020, bertiche_pbns_2021, bertiche_neural_2022}. We utilize the post-processing approach in our method as we find it sufficiently robust and efficient.

\begin{figure*}[htbp]
	\centering
    \includegraphics[width=1.0\linewidth]{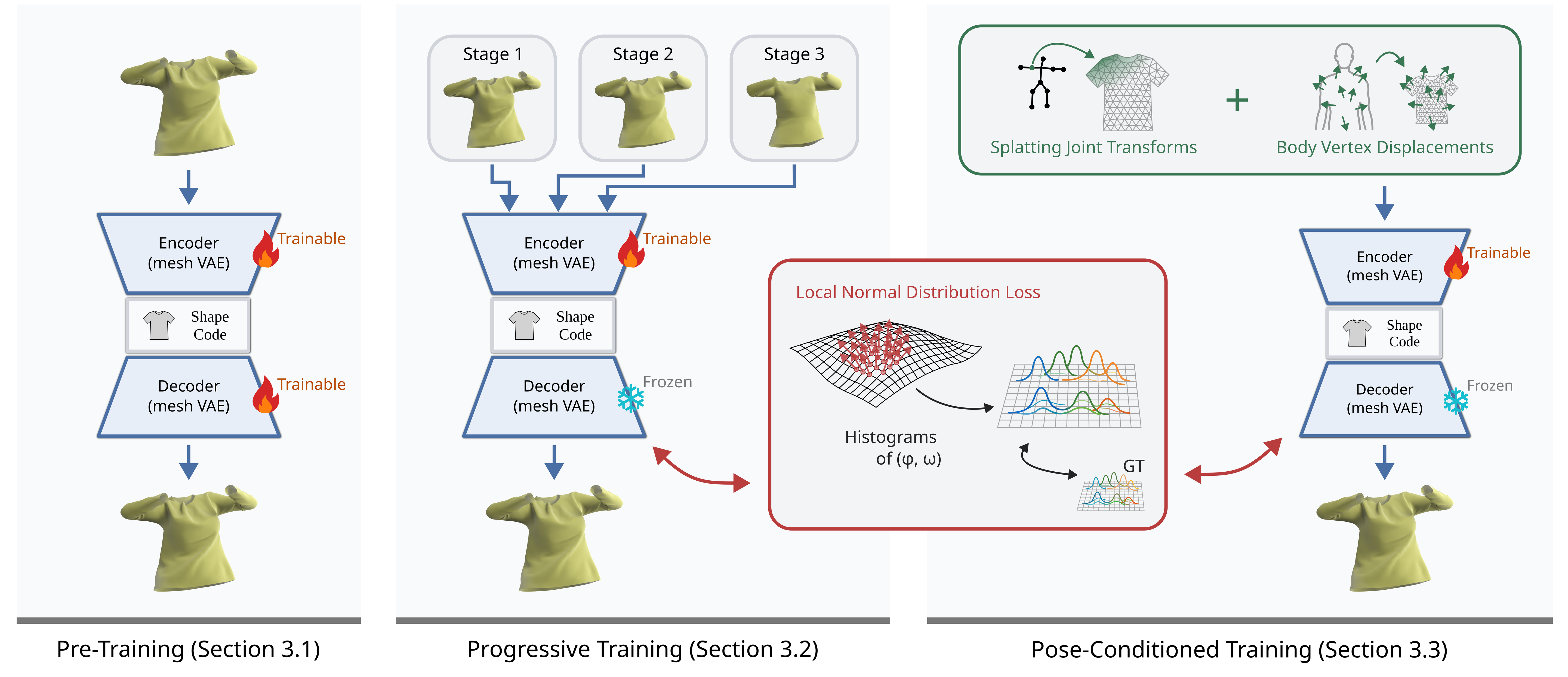}
	\caption{
    Overview of our real-time neural cloth deformer framework. We first pre-train a mesh VAE to reconstruct simulated rest-pose garment deformations, yielding a compact, well-behaved latent space and a garment-valid decoder that synthesizes high-frequency wrinkle detail (left, \Sec{latent_VAE}). We then progressively train the autoencoder by conditioning the encoder on increasingly Laplacian-smoothed versions of the deformation while keeping the full-resolution target fixed, enabling stable detail completion from weak geometric cues (middle, \Sec{progressive_training}). Finally, we replace the geometric encoder input with mesh-aligned pose conditioning, constructed by splatting joint features (and optional body cues) onto garment vertices. A pose-conditioned encoder is then trained to predict latent codes decoded into high-fidelity deformations for unseen poses at interactive rates (right, \Sec{mesh_aware_model}). To further preserve fine-scale wrinkles, we additionally introduce a local normal distribution loss that matches neighborhood-level orientation statistics to make the model less sensitive to small spatial phase shifts (\Sec{lnd}).
    }
	\label{fig:OverviewImage}
\end{figure*}

\subsection{Learning-based Methods}

\paragraph*{Cloth regression.} Many prior works regress garment deformations directly from character poses or refine a coarse simulation to create more realistic wrinkles. Data-driven methods were first explored to predict cloth deformations without relying on neural networks \cite{kim2008drivenshape, feng_deformation_2010, wang_example-based_2010, kavan_physics-inspired_2011, guan_drape_2012, wu_example-based_2021}. TailorNet \cite{patel_tailornet_2020} trains MLPs in two stages to separate low- and high-frequency deformations. \cite{li_ctsn_2024} extends the two-stream idea and trained Graph Transformer to obtain high-frequency wrinkles. Bertiche et al. \cite{bertiche_cloth3d_2020} solve the same problem by training a Conditional VAE. Santesteban et al. \cite{santesteban_learning-based_2019} extend this work with RNNs to capture dynamic cloth motion from time-varying animation. Chentanez et al. \cite{chentanez_cloth_2020} proposed a new spiral convolution operation for regressing garment deformations. Our work is in these lines: we regress deformations directly from character poses with a single feed-forward pass.

The above works all create or reuse a cloth deformation dataset through physically-based simulations and perform supervised trainings. Another option is to employ a set of physically-based unsupervised losses \cite{bertiche_pbns_2021, bertiche_neural_2022, santesteban_snug_2022, lee_clothcombo_2023, grigorev_hood_2023, grigorev_contourcraft_2024}. Using unsupervised losses relieves the burden of dataset creation while providing high-quality wrinkle configurations. However, it is difficult, if not impossible, to replicate production simulation setups in this manner due to their complexities. We thus train our models in fully supervised settings.

Instead of directly regressing 3D deformations, some methods represent deformations via 2D displacement maps and use convolutional neural networks \cite{lahner_deepwrinkles_2018, jin_pixel-based_2020, su2022deepcloth, guo_high-fidelity_2025}. Though leveraging well-studied 2D network architectures, handling seams and distortions in texture maps is often challenging. Other methods train an upsampler of low-resolution cloth to obtain fine details \cite{chen_multi-feature_2021, zhang_deep_2021, chen_deep_2023, zhang_neural_2024}. However, these works usually use coarse simulations as input, which is not always available in production settings.

\paragraph*{Neural simulations.} Another type of approach to learn cloth deformations is to replace the physically-based simulator with a neural network. A Graph Network-based Simulator (GNS) based on message passing  is leveraged to learn particle-based \cite{sanchez-gonzalez_learning_2020} and mesh-based \cite{pfaff_learning_2021} simulations. HOOD \cite{grigorev_hood_2023} proposes a hierarchical message passing scheme to enhance the simulation quality. Subsequent works of HOOD handle multi-layered garments \cite{grigorev_contourcraft_2024} and address self-collisions \cite{liao_senc_2024, ban_sags-gnn_2025}. Transformer-based architectures have also been explored in the context of neural cloth simulation \cite{shao_towards_2023, li_neural_2024}. Although demonstrating generalization capabilities to unseen meshes, GNS and transformers are usually expensive to evaluate due to their iterative or global nature. Most of these works also rely on physically-based unsupervised losses. As our focus is on real-time inference speed for production settings, we do not adopt this category of approach.

\section{Method}

The target of our neural cloth deformer is to efficiently regress a high-quality wrinkled garment mesh given input character joint transforms $\pose$ and a single baseline garment rest shape $\bind$ for quasi-static scenarios. Similar to previous approaches \cite{loper_smpl_2015, bailey_fast_2018, zheng_multi-resolution_2024, patel_tailornet_2020}, our predictor outputs rest-pose deformations that are subsequently skinned to obtain the final garment positions $\gar$ for a single frame as
\begin{equation}
\gar = \skinning\left(\bind + \mathcal{N}_{\Phi}(\pose),\ \pose\right),
\label{eq:lbs_prediction}
\end{equation}
where $\skinning$ is a skinning function, \eg linear blend skinning (LBS), extended for the garment mesh and $\mathcal{N}_{\Phi}(\pose)$ predicts per-vertex rest-space displacements. Assuming the skinning function is invertible, the ground-truth rest-space deformations can be computed as:
\begin{equation}
\deformationDelta = \unskinning (\gar, \theta) - \bind.
\end{equation}
Training on $\deformationDelta$ factors out world transforms and facilitates normalization. A general pose-to-mesh neural network predictor can then be trained with the following objective:
\begin{equation}
\argmin_{\Phi} \sum_i w_i \mathcal{L}_i( \mathcal{F}\left(\mathcal{N}_{\Phi}(\pose)\right), \mathcal{F}(\deformationDelta)),
\label{eq:neural_regression}
\end{equation}
where $\Phi$ are the trainable network weights, $\mathcal{F}$ is a user-specified feature extractor (\eg computing vertex normals), and $\mathcal{L}_i$ is a weighted ($w_i$) loss function that can represent distance metrics such as $L^2$, cosine similarity or a local norm distribution loss (\Sec{lnd}); or even physically-inspired energy-based formulations \cite{bertiche_pbns_2021, bertiche_neural_2022}. Our work focuses on supervised losses, since the goal is to preserve the artistic intent for a particular setting, as captured by carefully tuned cloth simulations and occasional artist-authored frame edits.

Previous methods differ mainly in how they couple pose information to mesh deformations. A straightforward option is a global MLP that maps joint transforms directly to rest-space deformations. However, fully connected architectures scale poorly with mesh resolution; to keep parameter counts tractable, previous approaches predict a reduced set of scalar weights coupled with learned linear basis (e.g., blendshape-style subspaces) ~\cite{santesteban_learning-based_2019, bertiche_deepsd_2021}. One issue is that these learned linear basis are oblivious to the mesh structure, which hinders the capacity of the network to efficiently learn non-linear local features that can improve generalization. A more structured alternative is the per-joint MLP formulation of~\cite{bailey_fast_2018}, which leverages the spatial locality implied by skinning. In this setting, each joint $j$ has its own MLP $\mathcal{N}^j_{\Phi}(\pose_j)$ that takes the joint transform $\pose_j$ as input and predicts rest-space displacements only for vertices influenced by that joint, implemented via a vertex mask $M(\cdot)$ derived from skinning weights. The posed garment is then obtained by aggregating the per-joint contributions:
\begin{equation}
{\gar} =  \skinning\left(\bind + \sum_jM^j\left(\mathcal{N}^j_{\Phi}(\pose_j)\right), \pose\right).
\label{eq:per_joint_mlp_prediction}
\end{equation}

This formulation is highly effective for learning pose-dependent body corrections that bridge the gap between a simple LBS and high-quality rig deformations. Naively extending such an approach to neural garments, however, has its issues.
For character rigs, rest-space corrections are typically smooth and localized around the driving joints; in contrast, cloth wrinkles are inherently non-local, and small changes in a shoulder or elbow can induce folds that propagate across the entire garment surface.
Figure~\ref{fig:ModelImprovements}(a) shows a per-joint MLP baseline. The predicted deformations exhibit over-smoothed wrinkles, indicating that the per-joint locality assumption is insufficient to reproduce the production-quality wrinkling patterns required in our setting.
In the next sections we will discuss how our proposed training pipeline can effectively capture long-range dependencies with a novel mesh-aware architecture that is tailored to efficiently condition pose information.

\begin{figure}
	\centering
	\begin{minipage}[c]{0.52\linewidth}
		\vfill
		\centering
		\includegraphics[width=\linewidth]{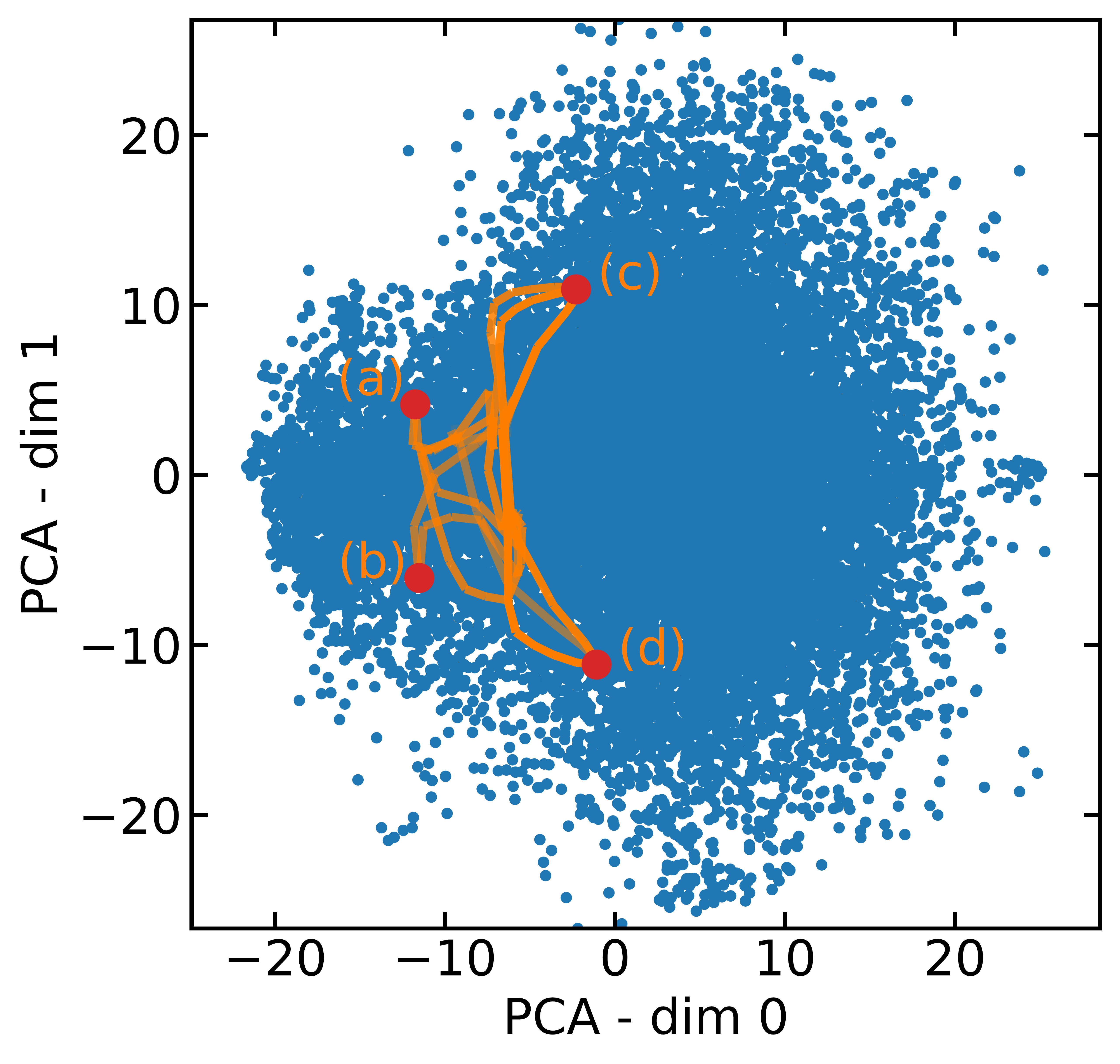}
		\vfill
	\end{minipage}
	\hfill
	\begin{minipage}[c]{0.45\linewidth}
		\centering
		\setlength{\tabcolsep}{0.0050\linewidth}
		\begin{tabular}{cc}
			\includegraphics[width=0.45\linewidth]{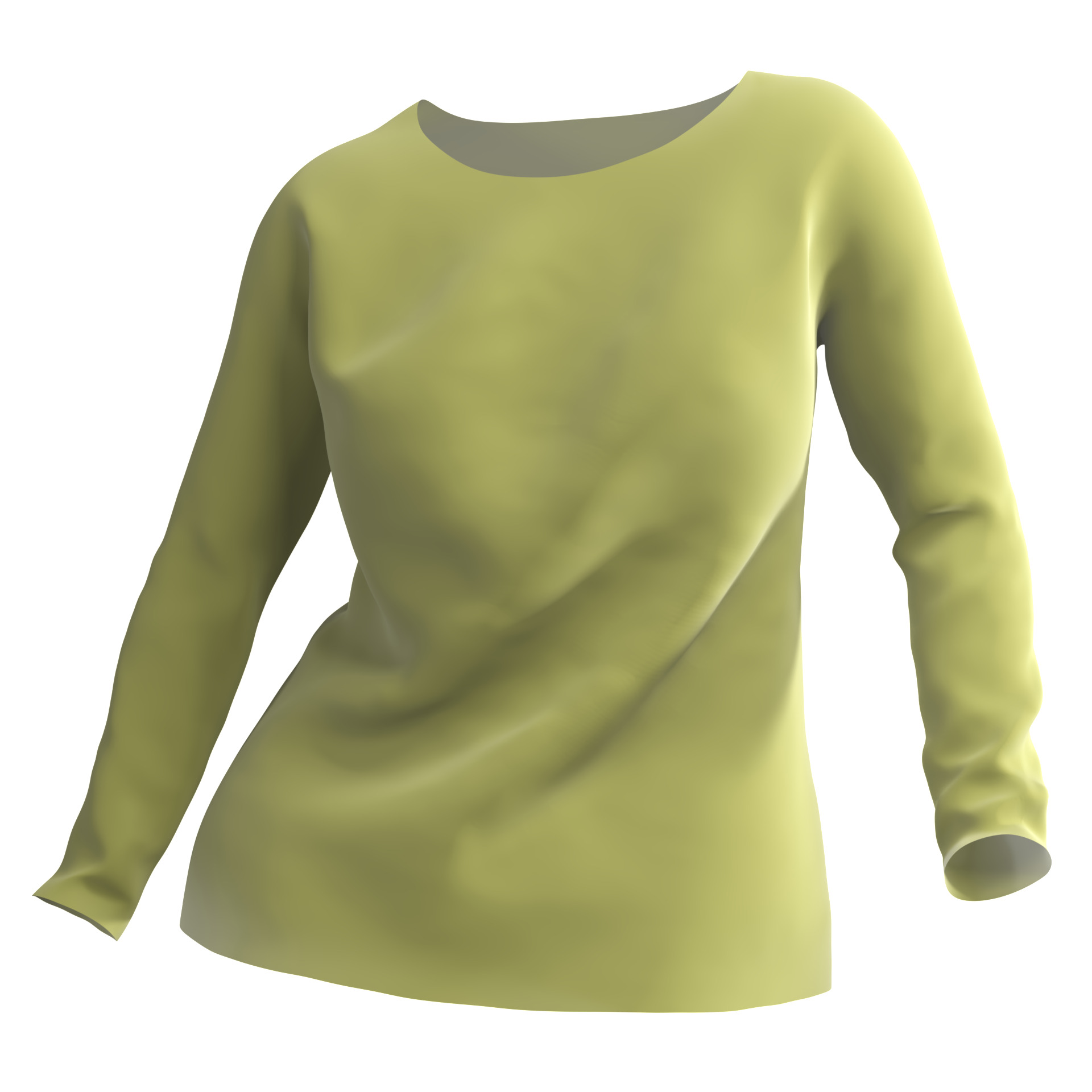} &
			\includegraphics[width=0.45\linewidth]{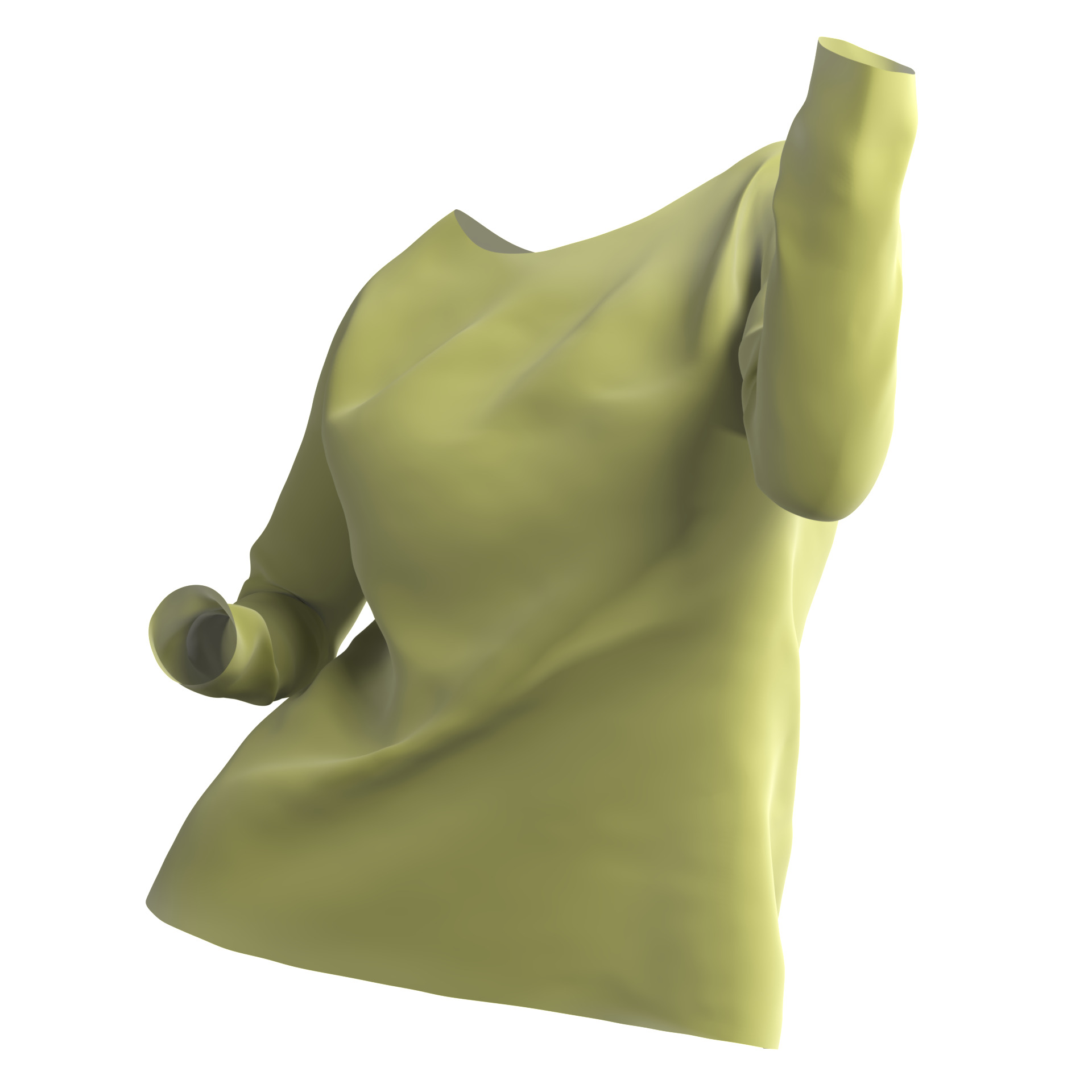} \\[-1.5ex]
			\footnotesize{(a)} & \footnotesize{(c)} \\
			\includegraphics[width=0.45\linewidth]{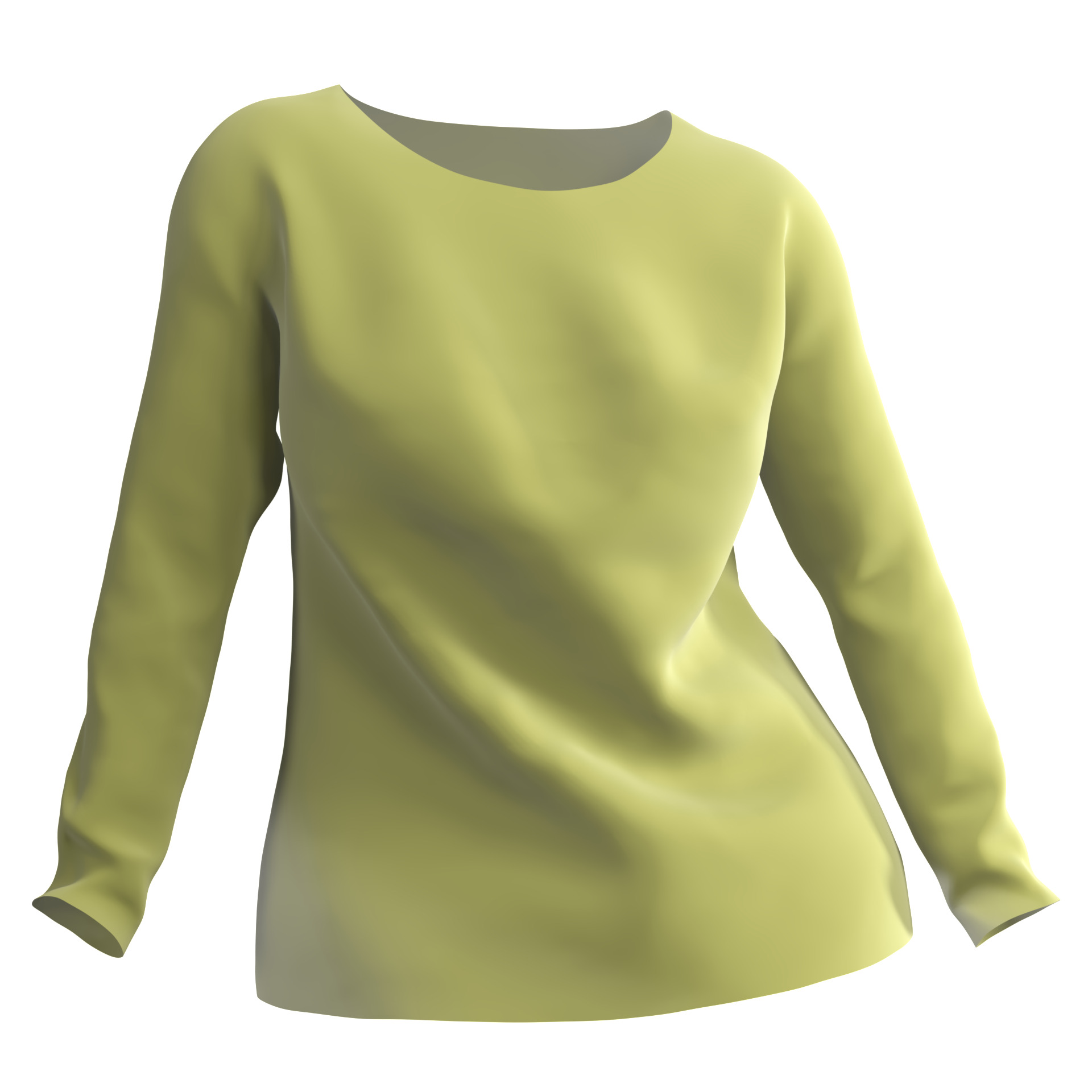} &
			\includegraphics[width=0.45\linewidth]{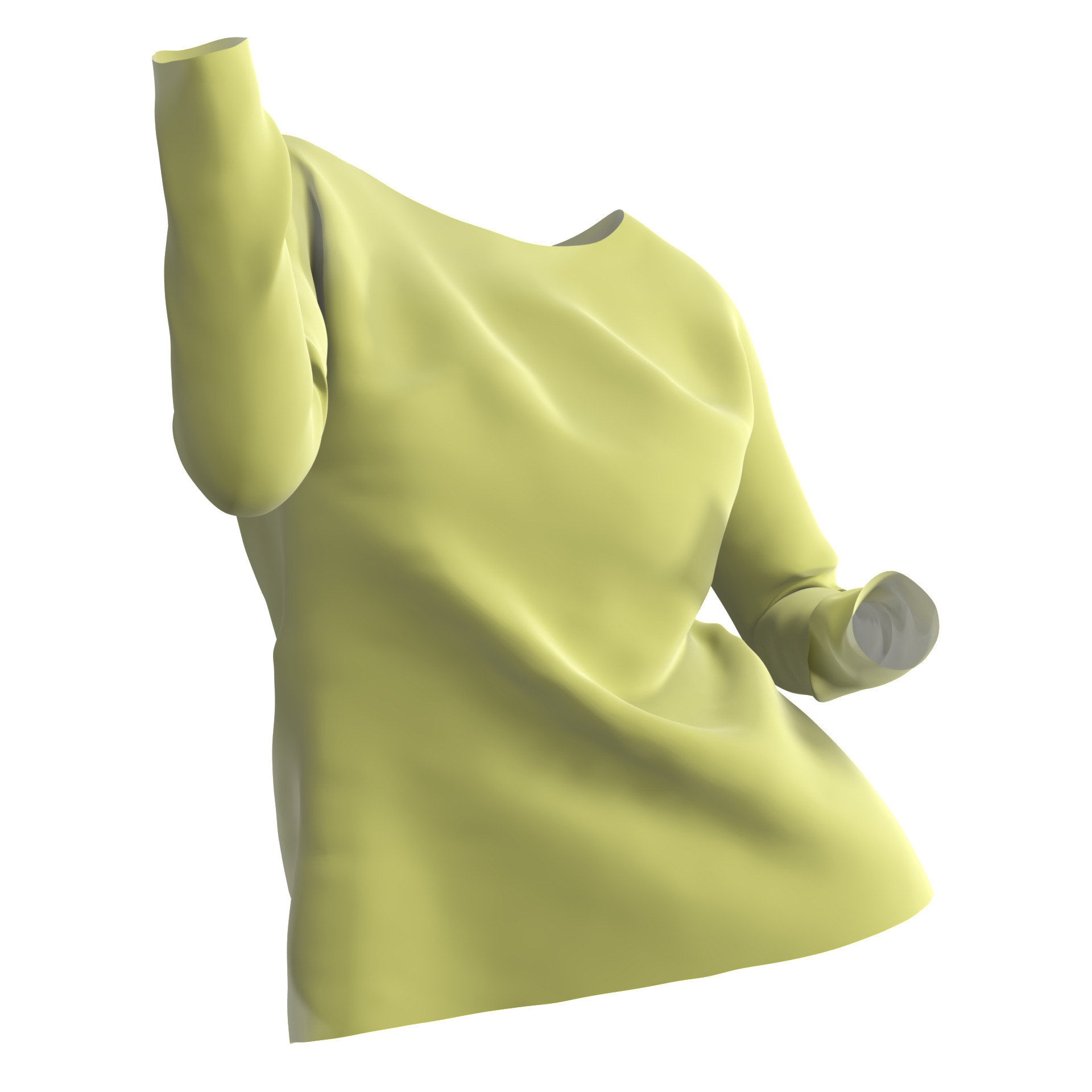} \\[-1.5ex]
			\footnotesize{(b)} & \footnotesize{(d)}
		\end{tabular}
	\end{minipage}
	\caption[PCA projection of the latent trajectory with corresponding mesh reconstructions produced by the autoencoder for four selected poses.]{PCA projection of the latent trajectory highlighting four selected poses (left) with the corresponding mesh reconstructions (right), illustrating how the latent dimensions capture meaningful and smooth variations in pose.}
	\label{fig:PCALatentTrajectoryWithPred}
\end{figure}

\subsection{Latent Garment Prior via VAE Pre-Training}
\label{sec:latent_VAE}
A recurring failure mode of direct pose-to-mesh regression is that the network must simultaneously infer global deformations and synthesize high-frequency wrinkle geometry in the correct spatial configuration. This end-to-end mapping remains high-dimensional and under-constrained: many mesh deformations are locally consistent with the pose descriptor but implausible as garments. To improve generalization, we therefore factor the problem into two steps: we first learn a \emph{pose-agnostic} prior over rest-pose garment deformations using a Variational Auto-Encoder (VAE), and then learn a pose-conditioned encoder that predicts latent codes inside the learned space (\Sec{mesh_aware_model}).

The VAE consists of an encoder $\mathcal{E}_{\psi}(\latent \mid \deformationDelta)$ and a decoder $\mathcal{D}_{\theta}(\deformationDelta \mid \latent)$, where $\latent$ is a low-dimensional latent code. The decoder $D_{\theta}(\latent)$ is also implemented as a mesh-aware Convolutional Neural Network that outputs a displacement field on the original mesh resolution. We optimize the standard evidence lower bound:

\begin{equation}
\begin{split}
\min_{\psi,\theta}\
\mathbb{E}_{\deformationDelta}\Big[
\mathbb{E}_{\latent \sim \mathcal{E}_{\psi}(\latent \mid \deformationDelta)}
\big[\,\mathcal{L}_{\mathrm{rec}}(D_{\theta}(\deformationDelta \mid \latent), \deformationDelta)\,\big]
\;+\; \\ \qquad \qquad
w_\text{KL}\ \mathrm{KL}\!\left(\mathcal{E}_{\psi}(\latent \mid \deformationDelta)\,\|\,\mathcal{N}(0,I)\right)
\Big],
\label{eq:vae_objective}
\end{split}
\end{equation}
where the supervised reconstruction loss (e.g., vertex $L^2$ and normal terms) $\mathcal{L}_{\mathrm{rec}}$ forces the decoder to represent high-frequency wrinkles while the Kullback–Leibler (KL) term shapes the latent space into a smooth, well-filled distribution aligned with an isotropic Gaussian prior.
Note that Eq.~\eqref{eq:vae_objective} is trained on rest-space deformation residuals rather than absolute vertex positions.
At inference time, obtaining these residuals requires inverse skinning, so the pose is implicitly involved through the skinning operation.

During VAE training, the decoder learns a strong deformation prior that captures the multiscale statistics of plausible rest-pose cloth shapes, while the KL term encourages a smooth, well-conditioned latent space in which nearby codes decode to coherent, nearby deformations. In practice, this leads to more stable wrinkle patterns under pose distribution shift and improved generalization across motions and characters, while preserving the deterministic behavior required for production inference. As shown in Figure~\ref{fig:PCALatentTrajectoryWithPred}, encoding ground-truth rest-pose deformations along an animated pose sequence produces a latent trajectory that varies smoothly over time. This supports our motivating assumption that the space of plausible deformations is locally low-dimensional and well-behaved: continuous changes in pose induce continuous changes in latent deformation, making it feasible to learn a pose-to-latent mapping that generalizes beyond the training distribution.

\subsection{Progressive Transition to Pose Conditioning}
\label{sec:progressive_training}
The VAE pre-training in the previous section yields a garment-valid decoder that maps a compact latent code to detailed rest-pose displacements. An immediate next step could be to freeze the decoder and replace the VAE encoder with a pose-conditioned encoder operating on mesh-aligned pose features. In practice, we found this direct replacement to be too abrupt: the decoder has only been exposed to latent codes produced from geometric inputs, whereas a pose-conditioned encoder would initially produce latents from a markedly different input distribution. This mismatch leads to unstable optimization and weaker generalization, often manifesting as overfitting to the pose distribution seen during training or as high-frequency artifacts under pose variation.

\begin{figure}
	\centering
    \setlength{\tabcolsep}{0.0010\linewidth}
    \begin{tabular}{
          >{\centering\arraybackslash}m{0.1\linewidth}
          >{\centering\arraybackslash}m{0.27\linewidth}
          >{\centering\arraybackslash}m{0.27\linewidth}
          >{\centering\arraybackslash}m{0.27\linewidth}
        }
        \footnotesize{Input} &
        \includegraphics[width=\linewidth]{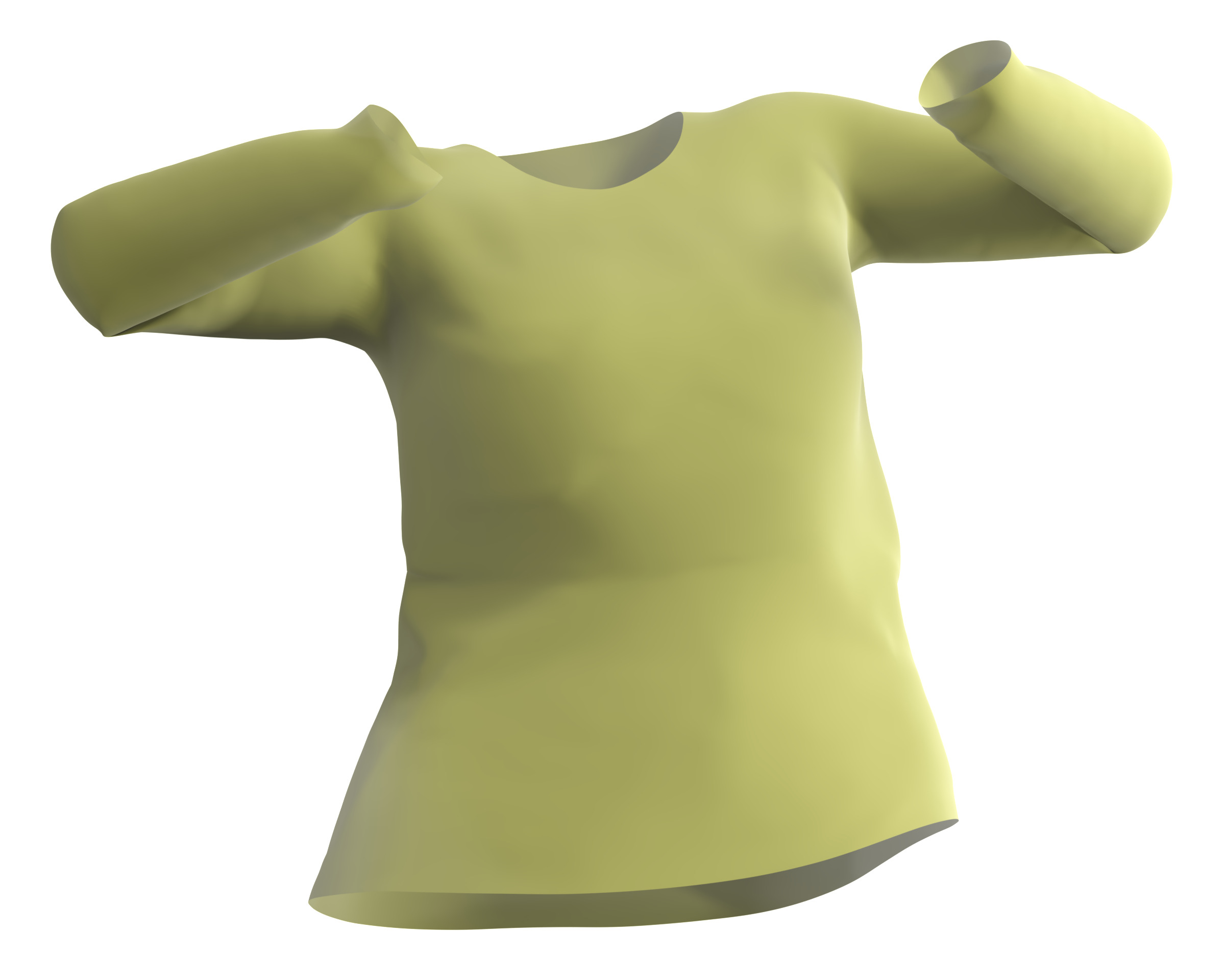} &
        \includegraphics[width=\linewidth]{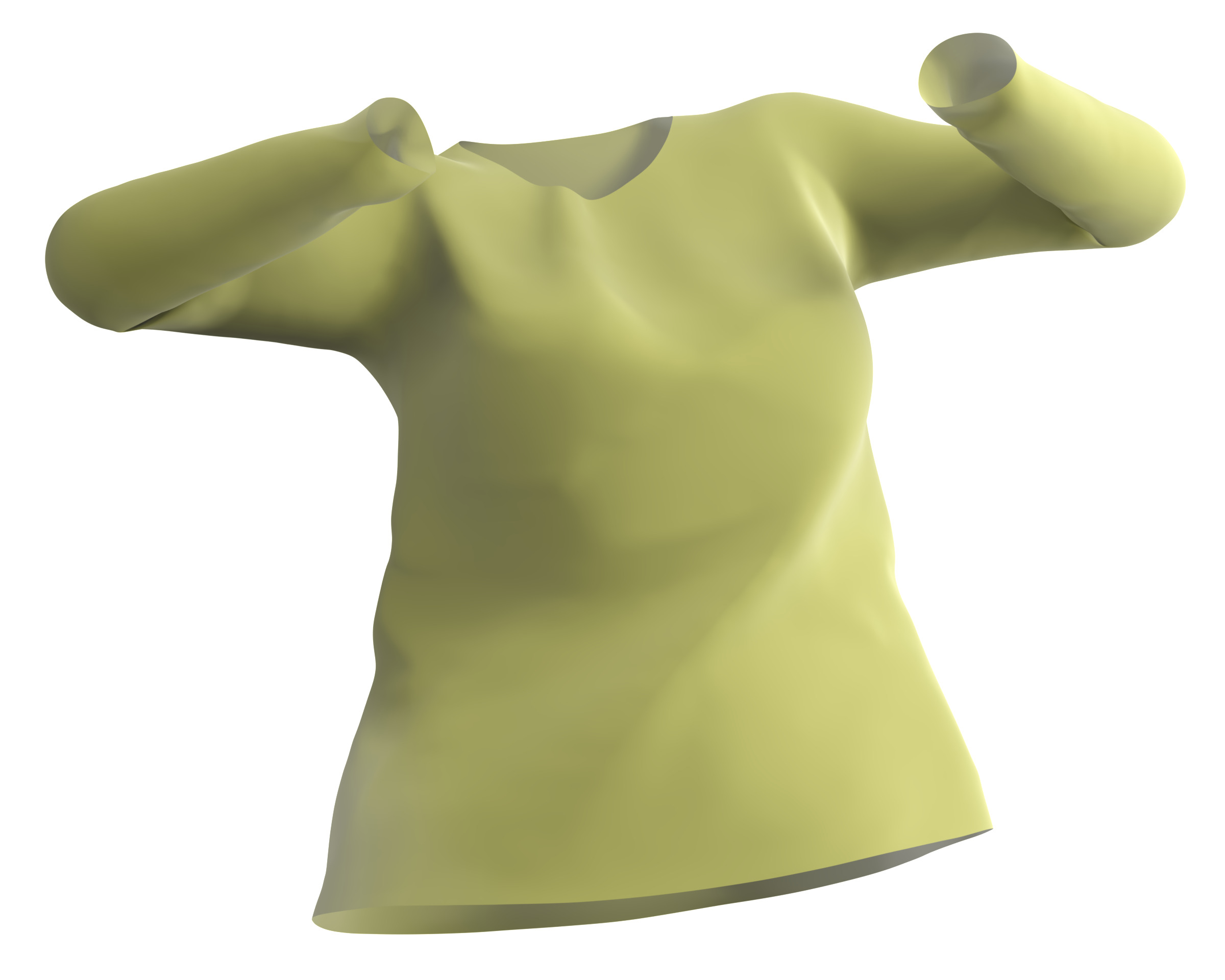} &
        \includegraphics[width=\linewidth]{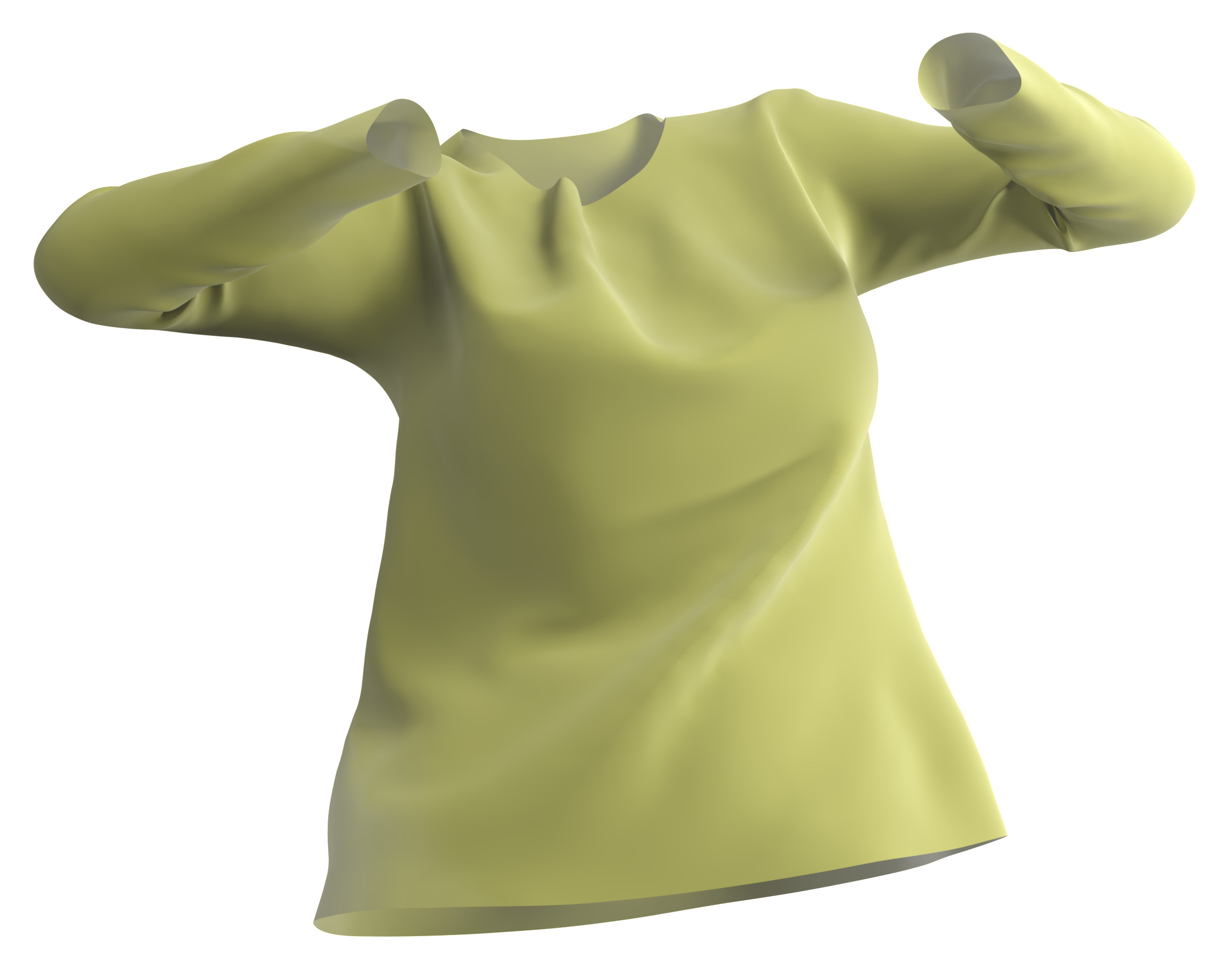} \\

        \footnotesize{Recon.} &
        \includegraphics[width=\linewidth]{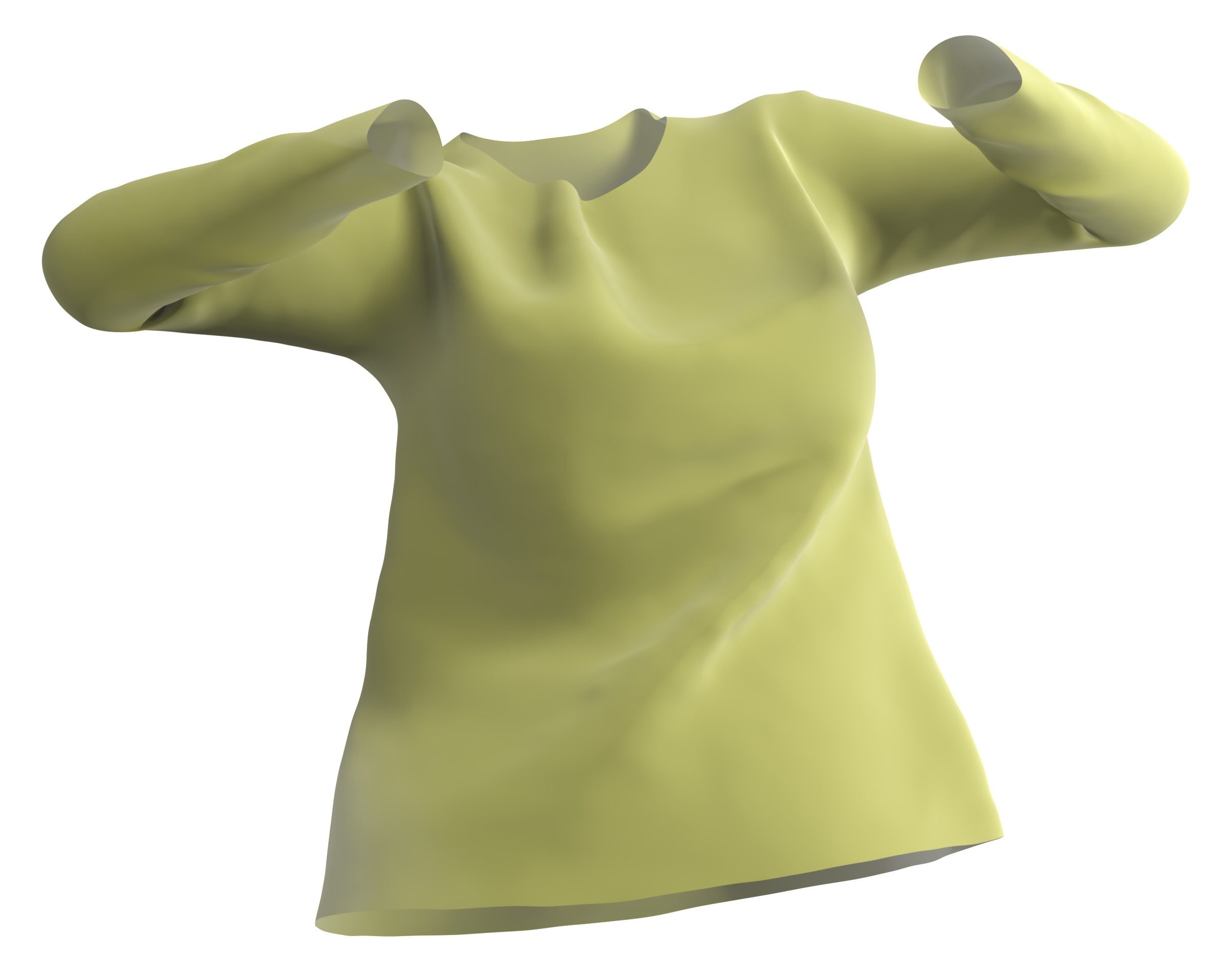} &
        \includegraphics[width=\linewidth]{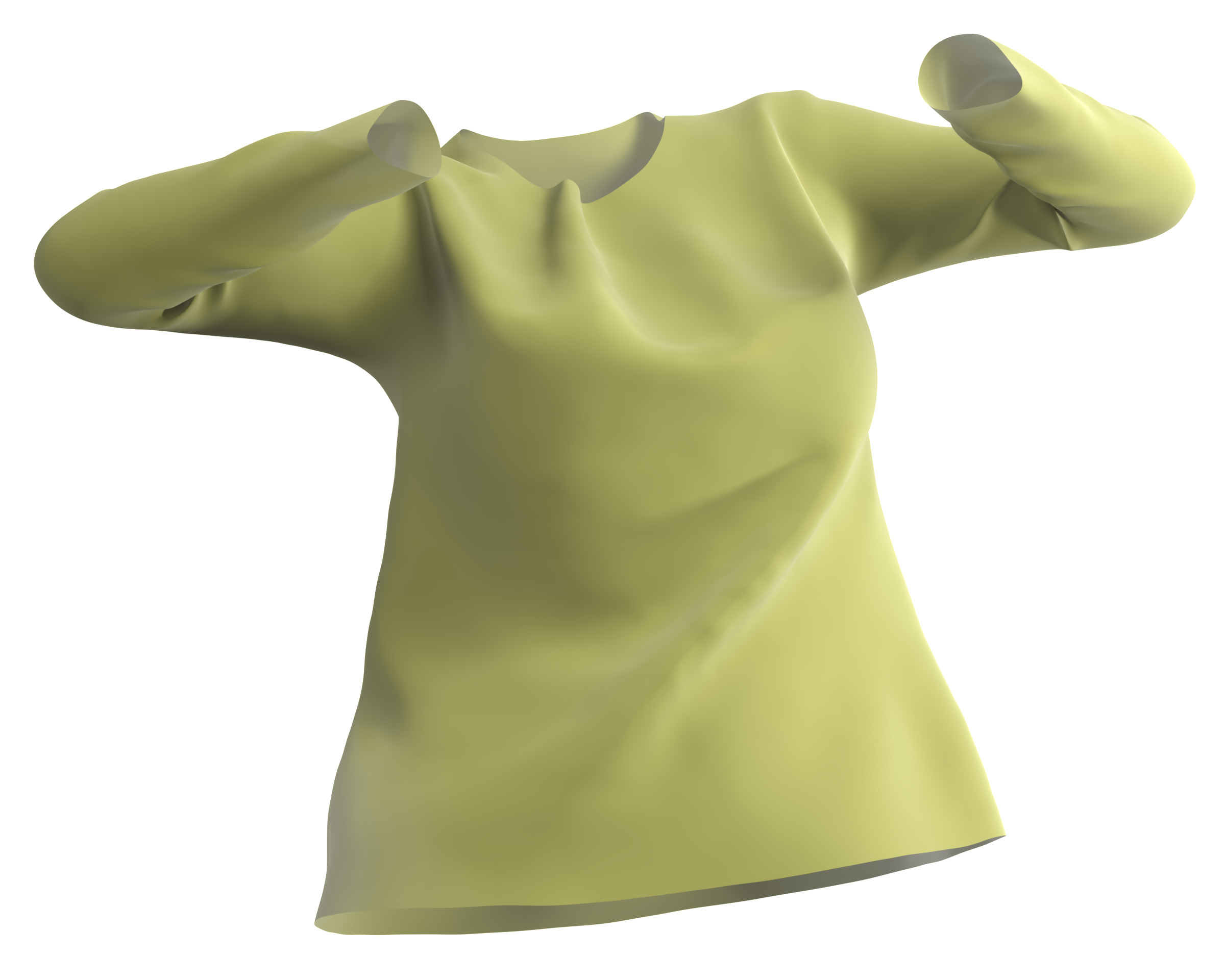} &
        \includegraphics[width=\linewidth]{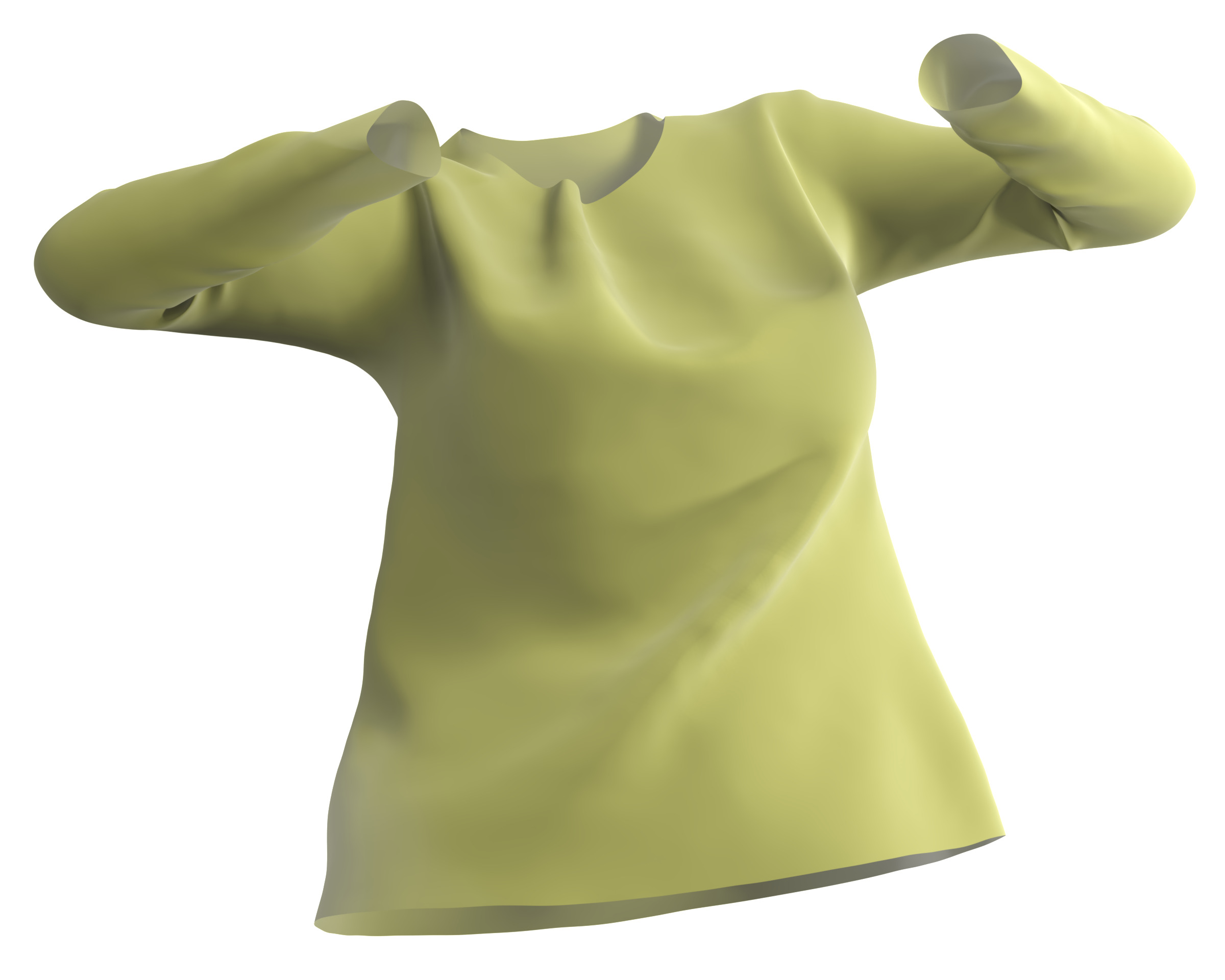} \\
         & \footnotesize{(a) $L=10$} & \footnotesize{(b) $L=1$} & \footnotesize{(c) $L=0$} \\
	\end{tabular}
	\caption{Progressive training stages: (a) reconstructing from heavily smoothed deformations ($10$ Laplacian iterations); (b) reconstructing from lightly smoothed deformations ($1$ Laplacian iteration); and (c) reconstructing the original deformations.}
    \label{fig:UpresSmoothedInputs}
\end{figure}

To smooth out this transition, we introduce a progressive curriculum that gradually morphs the auto-encoding task into an ``upsampling-from-coarse'' task before training with pose features. Concretely, after training the VAE in the standard fine-to-fine setting, we continue training while progressively coarsening the encoder input through a Laplacian smoothing operator. The progressively smoothed inputs are defined by
\begin{equation}
\deformationDelta^{(0)} = \deformationDelta,\qquad
\deformationDelta^{(l)} = \left(\mat{I}-\lambda\mat{L}\right)^{-1}\deformationDelta^{(l-1)},\ \ l=1,\dots,L,
\label{eq:progressive_smoothing}
\end{equation}
where $\mat{L}$ is the cotangent Laplacian and $\lambda$ controls the smoothing strength. Increasing $l$ progressively removes high-frequency content (wrinkles and fine folds), producing a sequence of increasingly low-detail inputs that remain defined on the original mesh connectivity. A visualization of our smoothing operator, showing progressive removal of geometric detail, is shown in \Fig{UpresSmoothedInputs} (top row).

During this refinement phase, we optimize the reconstruction term of Eq.~\eqref{eq:vae_objective} with the encoder conditioned on $\deformationDelta^{(l)}$ and the decoder supervised against the full-detail displacement field $\deformationDelta$. We disable the KL term by setting $w_\text{KL}=0$, ensuring a deterministic refinement that avoids imposing additional constraints on the latent distribution. In the meanwhile, the smoothing level $l$ is increased according to a fixed curriculum (\Sec{results}).

As illustrated in \Fig{UpresSmoothedInputs}, recovering wrinkle details from Laplacian-smoothed inputs is comparatively well-conditioned: the VAE consistently reconstructs high-frequency structures that closely matches the ground truth. This observation motivates our subsequent pose-conditioning strategy: if the decoder can reliably synthesize fine details from weak geometric cues, it can also be finetuned to predict deformations from mesh-aligned pose features.

\subsection{Mesh-Aware Deformation Model}
\label{sec:mesh_aware_model}

Once the autoencoder has been trained to decode detailed deformations from increasingly weaker geometric inputs, we replace its pose-agnostic encoder with a pose-conditioned encoder while keeping the decoder fixed. This curriculum substantially stabilizes the handoff: the pose-conditioned encoder can rely on the well-behaved decoder space already used for reconstruction, rather than learning the entire deformation synthesis process end-to-end. We implement pose conditioning through three complementary mechanisms: joint-to-mesh splatted pose features, fine-scale pose injection via FiLM modulation, and a coarse-level cross-attention block between pose embeddings and mesh latent features.

\paragraph*{Joint-to-mesh splatted transforms.}
Prior mesh-based deformer models often compress joint transforms into a global latent vector and inject it into the mesh decoder via fully connected layers ~\cite{chentanez_cloth_2020}. This coupling can create a bottleneck for pose information to get through.
Instead, we construct a \emph{mesh-aligned pose field} by splatting joint information directly onto garment vertices using skinning weights. Each joint transform is embedded into a feature vector, and vertex-level pose features are obtained by smoothly blending the joint features according to their LBS influences.
Additionally, the rest body deformations from the ground-truth data are also splatted to the cloth mesh, and concatenated to the splatted joint transforms as the input of the network.
This produces a continuous per-vertex conditioning signal over the garment surface, enabling subsequent mesh convolutions to process pose information locally while still propagating effects globally through the multi-resolution hierarchy.

\paragraph*{Improving pose parameters conditioning.} To further condition the network on pose parameters, we adopt two complementary mechanisms. For the finer-resolution encoder blocks, we apply FiLM modulation \cite{perez_film_2017}, where pose information is pooled across joints using an attention mechanism and then used to generate per-channel scaling and shifting parameters that modulate the outputs of each SpiralConv layer.
For the coarsest mesh level, we additionally employ cross-attention \cite{vaswani_attention_2023}, where the encoded pose parameters serve as keys and values and the current feature map
serves as the query.
This enables global feature interactions with the pose embeddings.
\begin{figure*}[t!]
	\centering
	\setlength{\tabcolsep}{0.00125\linewidth}
    {\footnotesize
	\begin{tabular}{ccccccc}
        \includegraphics[width=0.164125\linewidth]{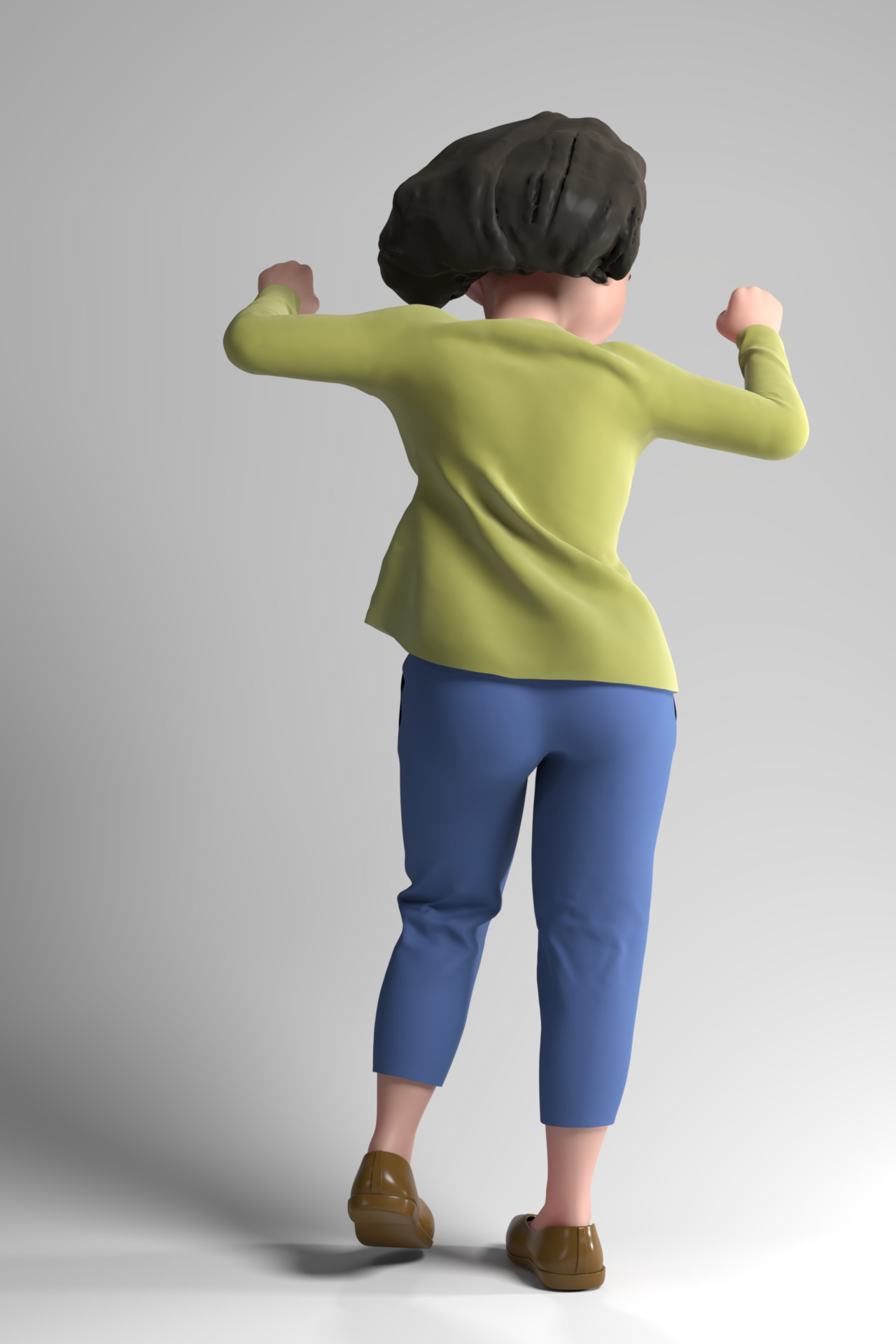} &
        \includegraphics[width=0.164125\linewidth]{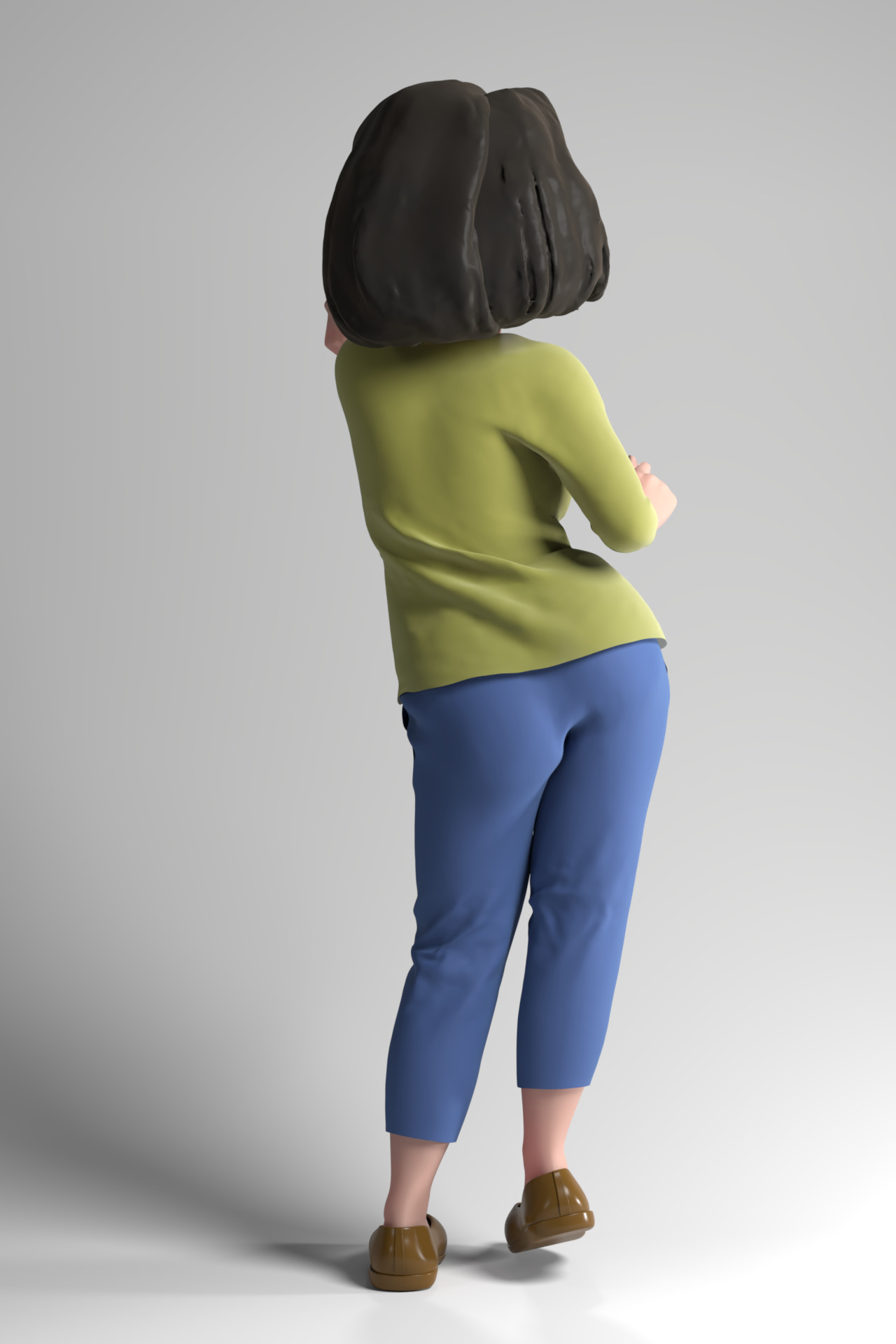} &
        \includegraphics[width=0.164125\linewidth]{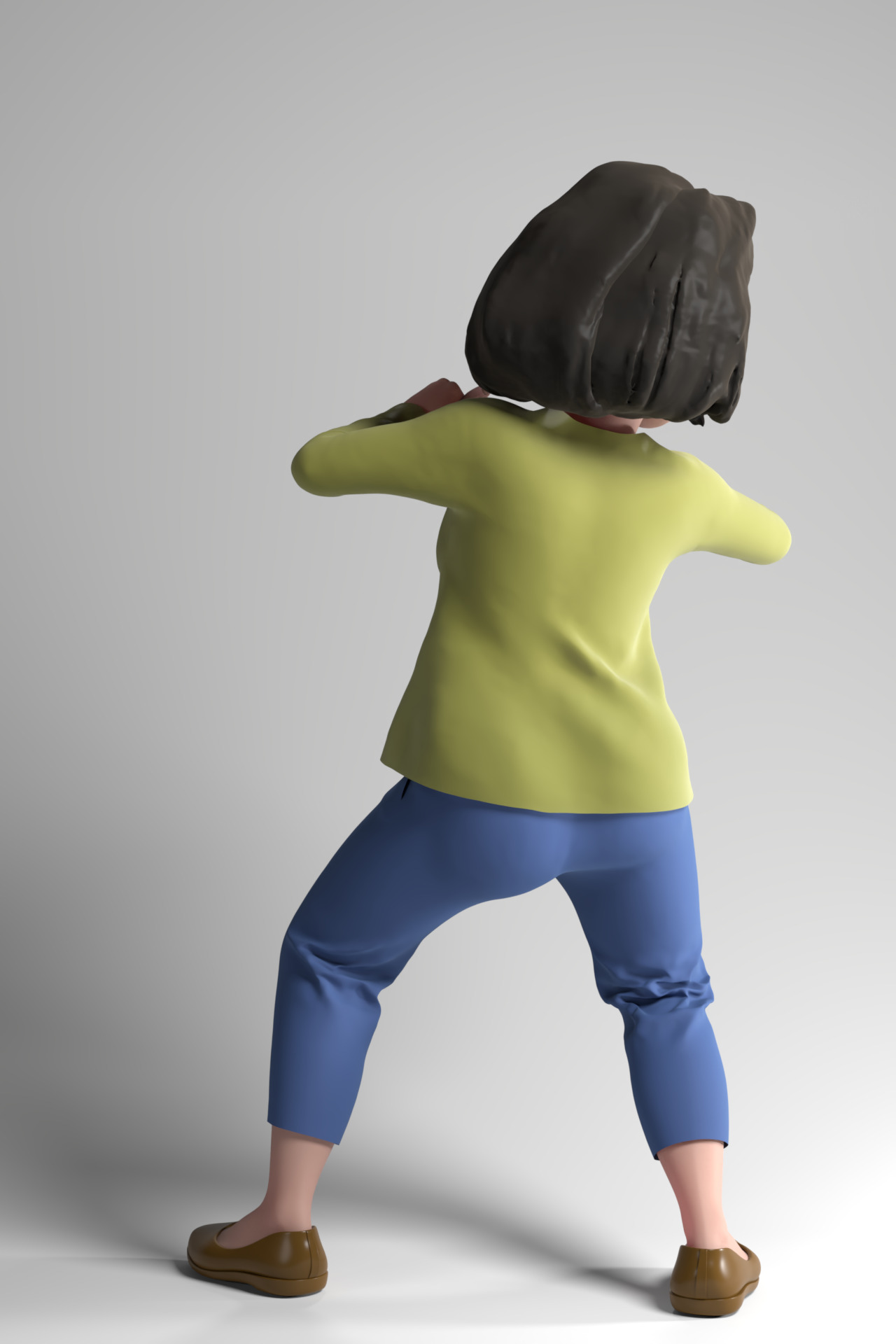} &
        \includegraphics[width=0.164125\linewidth]{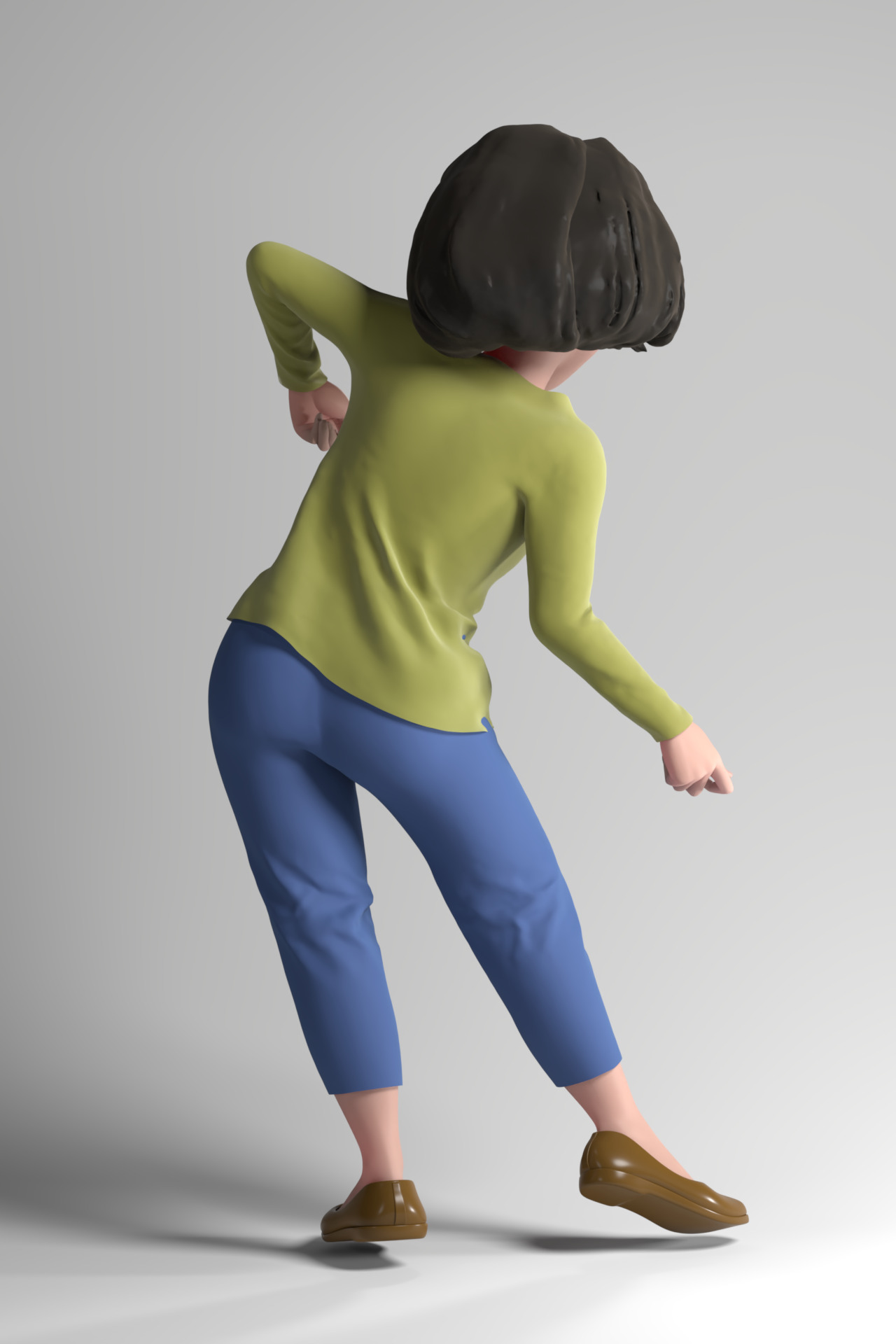} &
        \includegraphics[width=0.164125\linewidth]{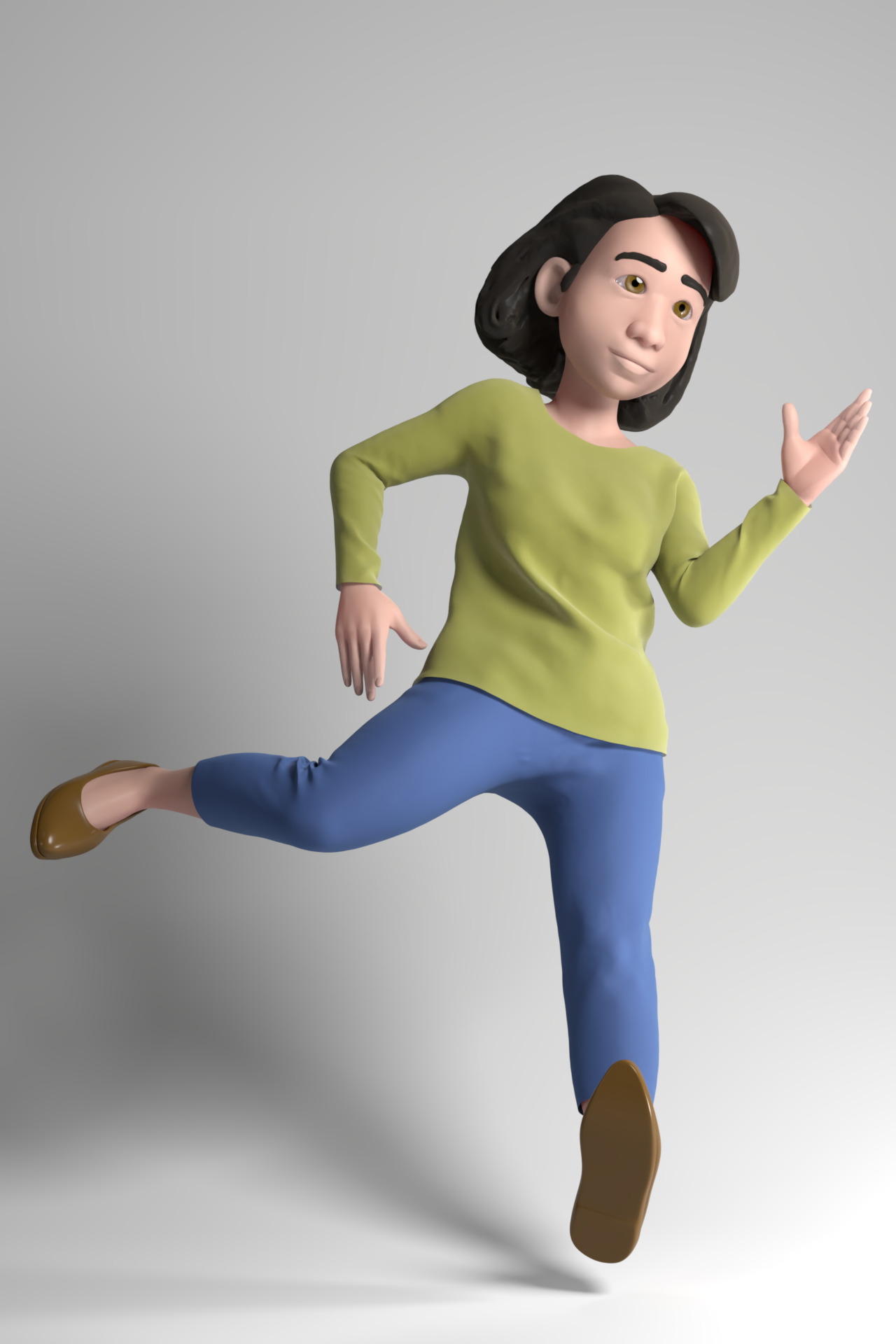} &
        \includegraphics[width=0.164125\linewidth]{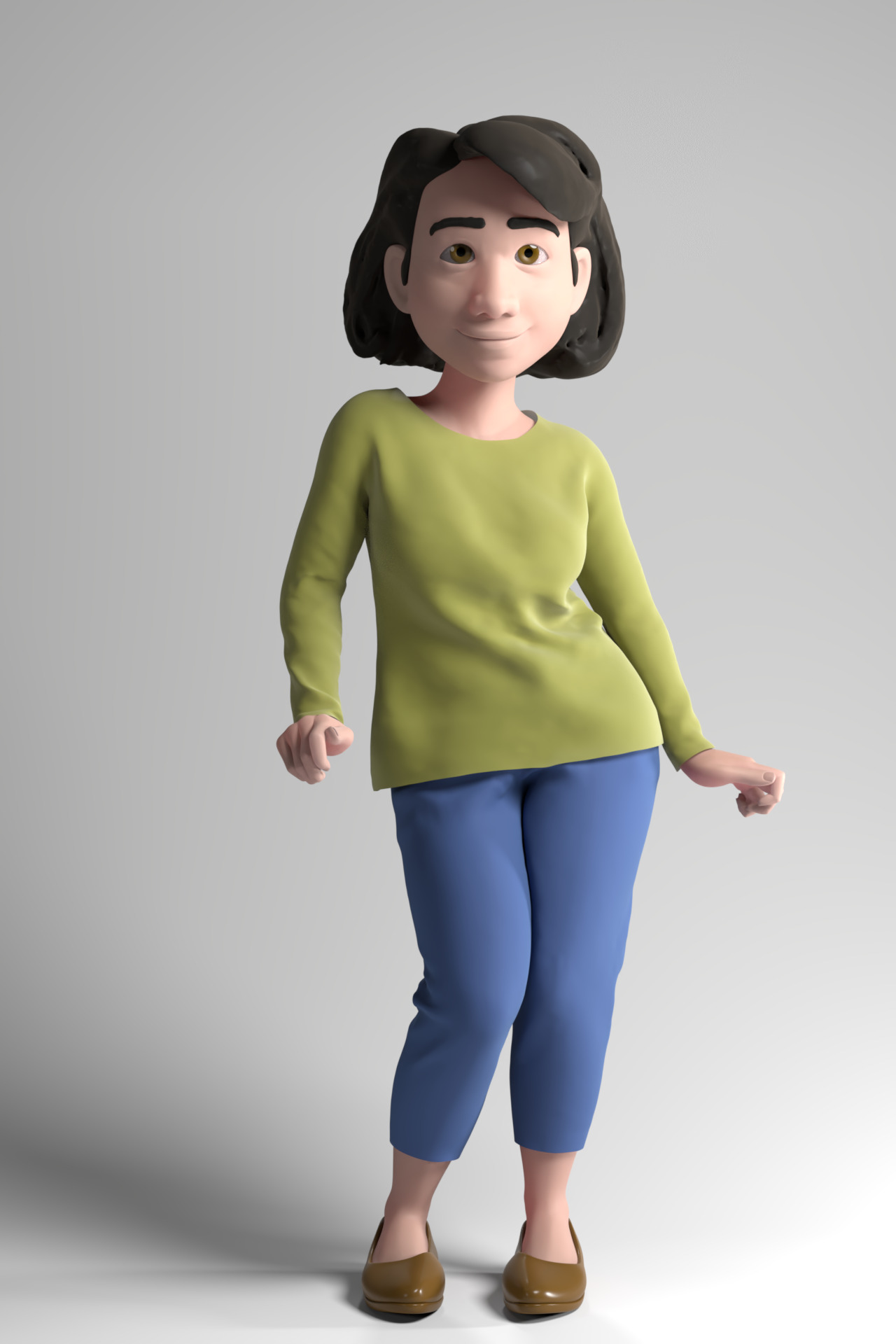} \\

        \includegraphics[width=0.164125\linewidth]{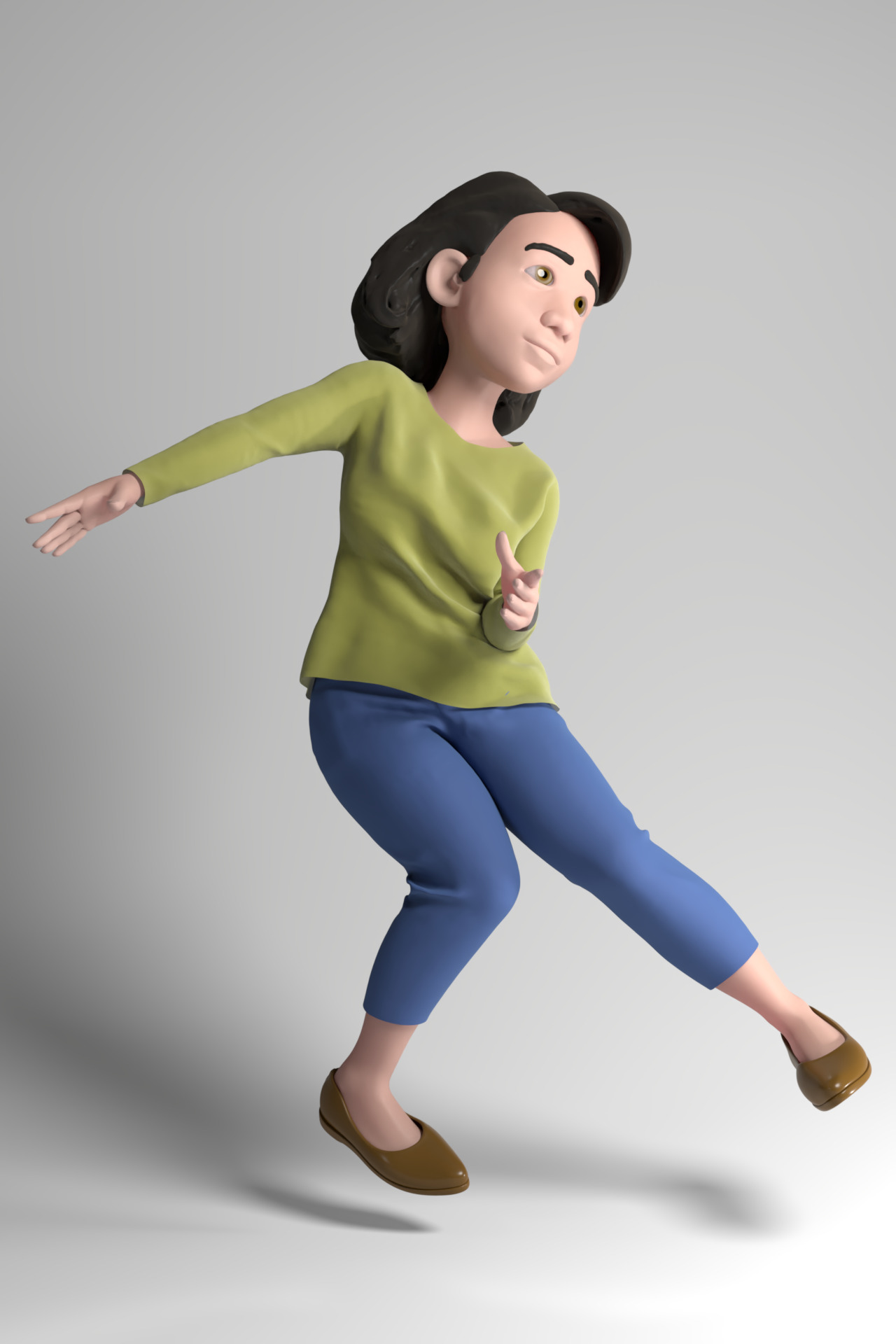} &
        \includegraphics[width=0.164125\linewidth]{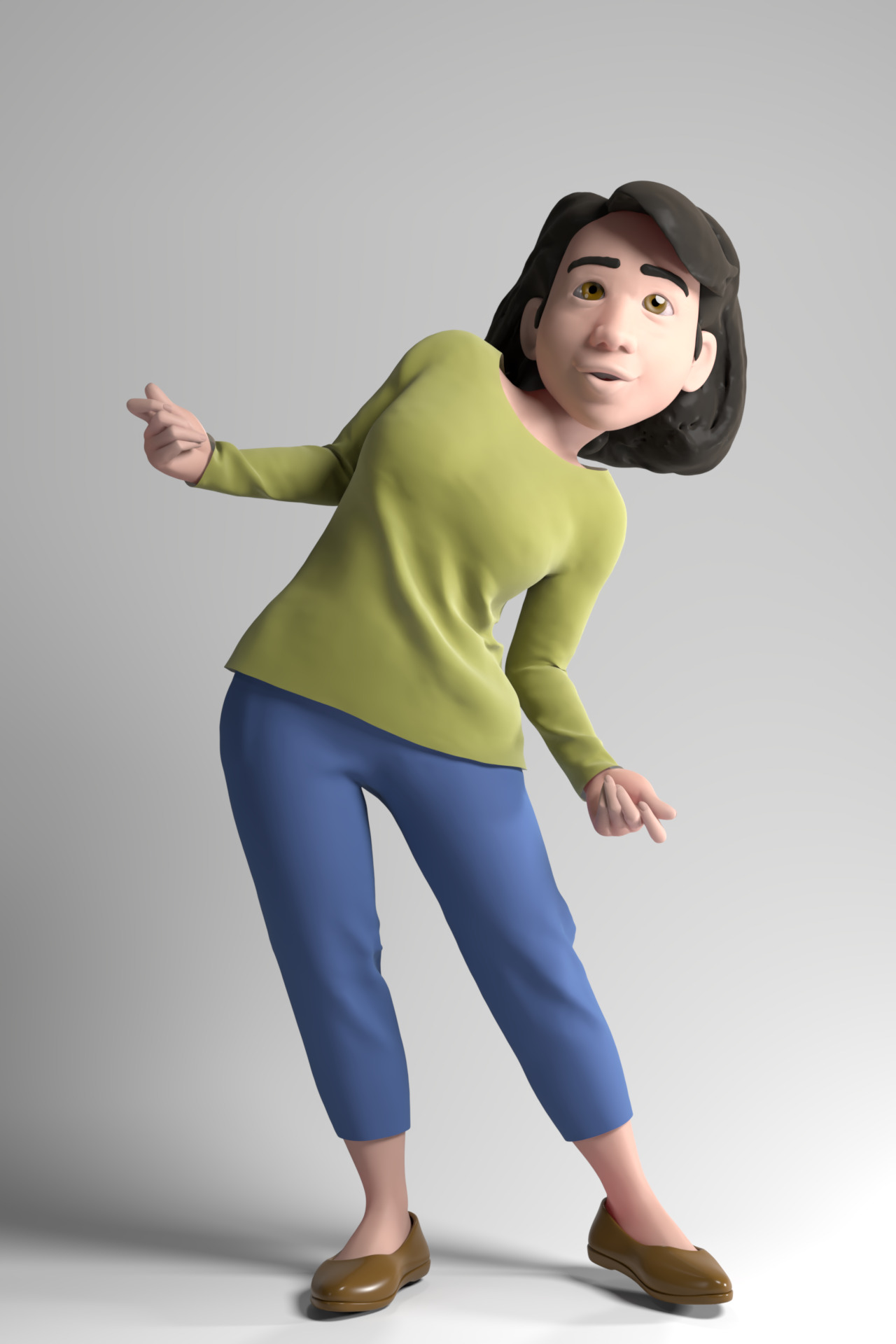} &
        \includegraphics[width=0.164125\linewidth]{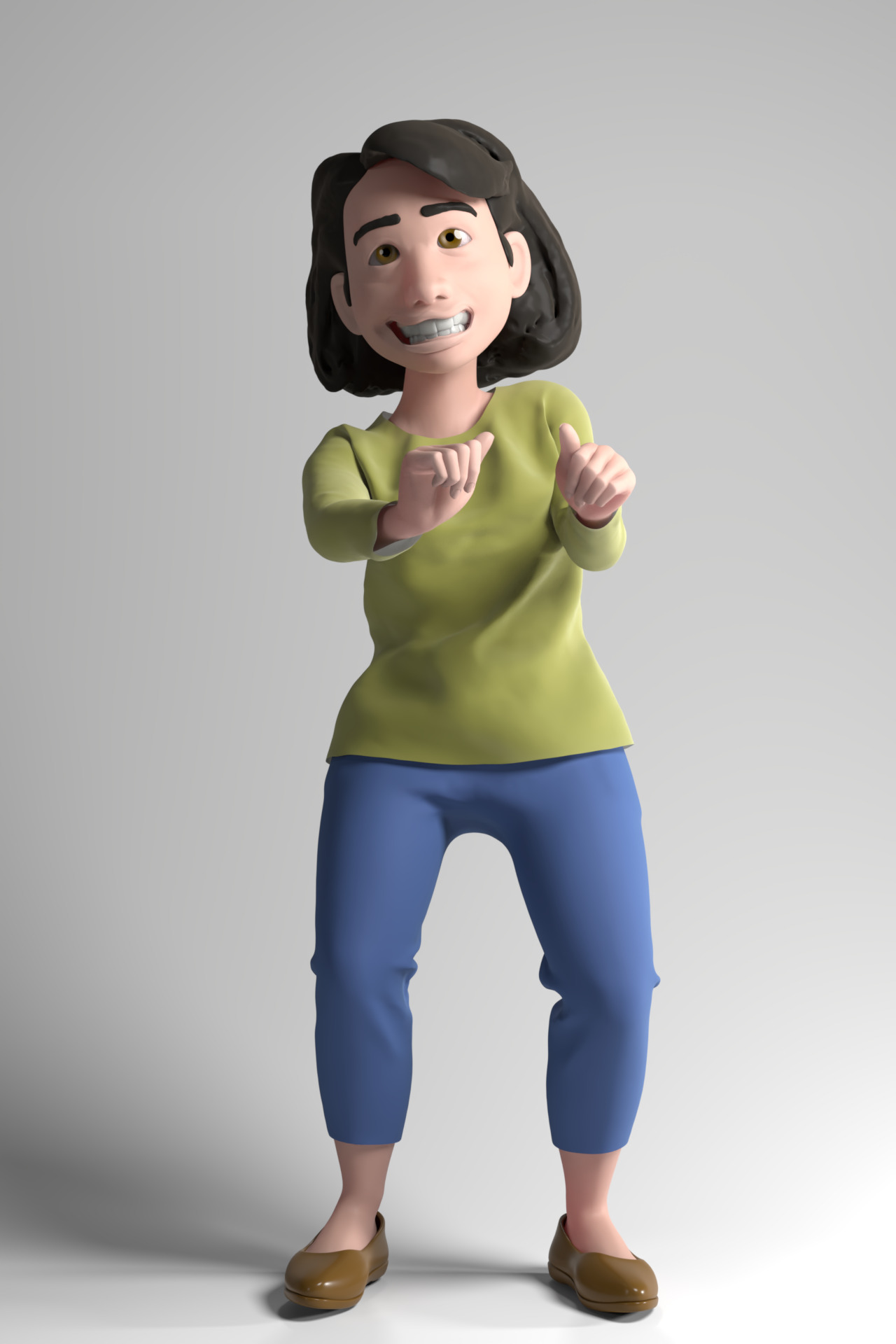} &
        \includegraphics[width=0.164125\linewidth]{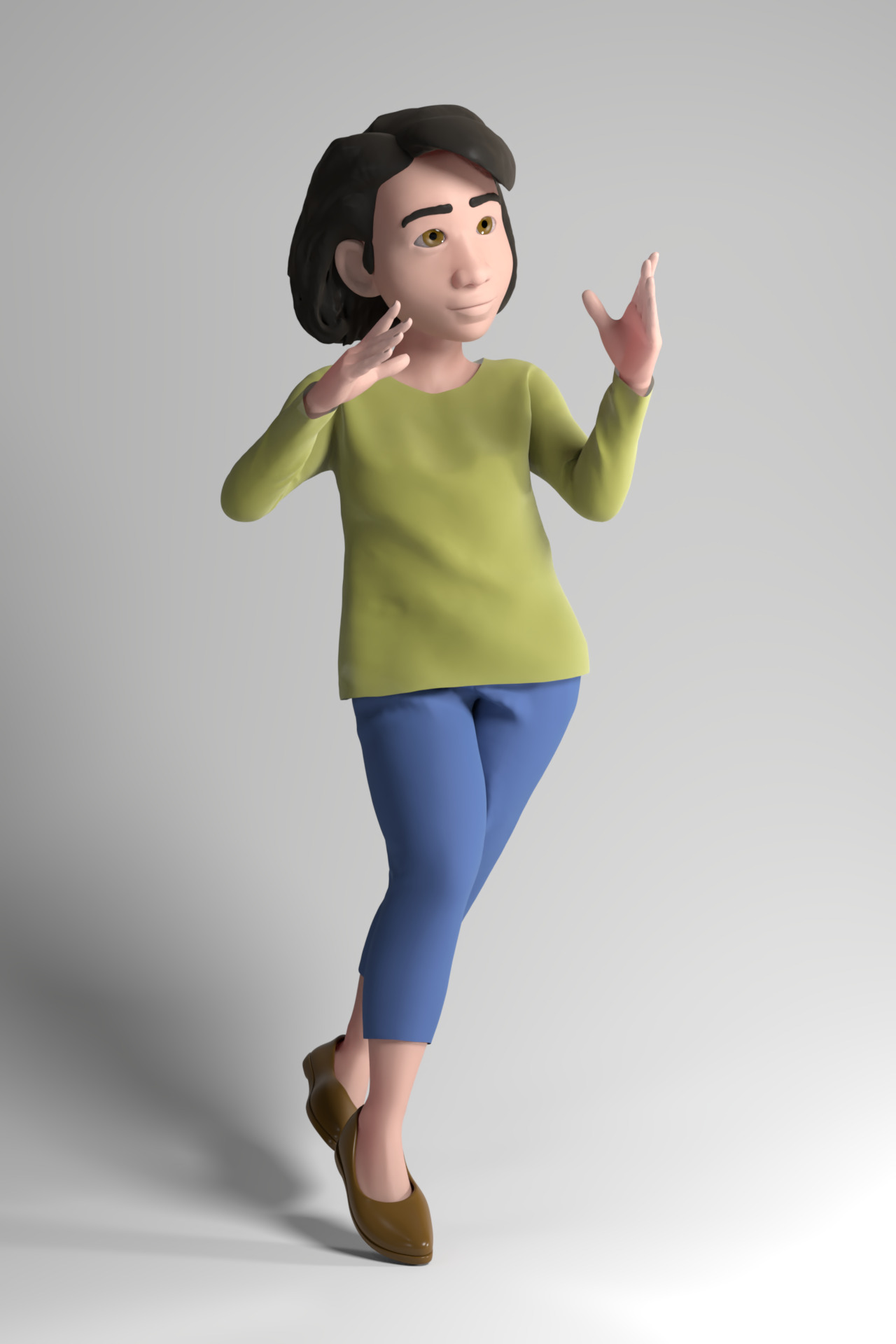} &
        \includegraphics[width=0.164125\linewidth]{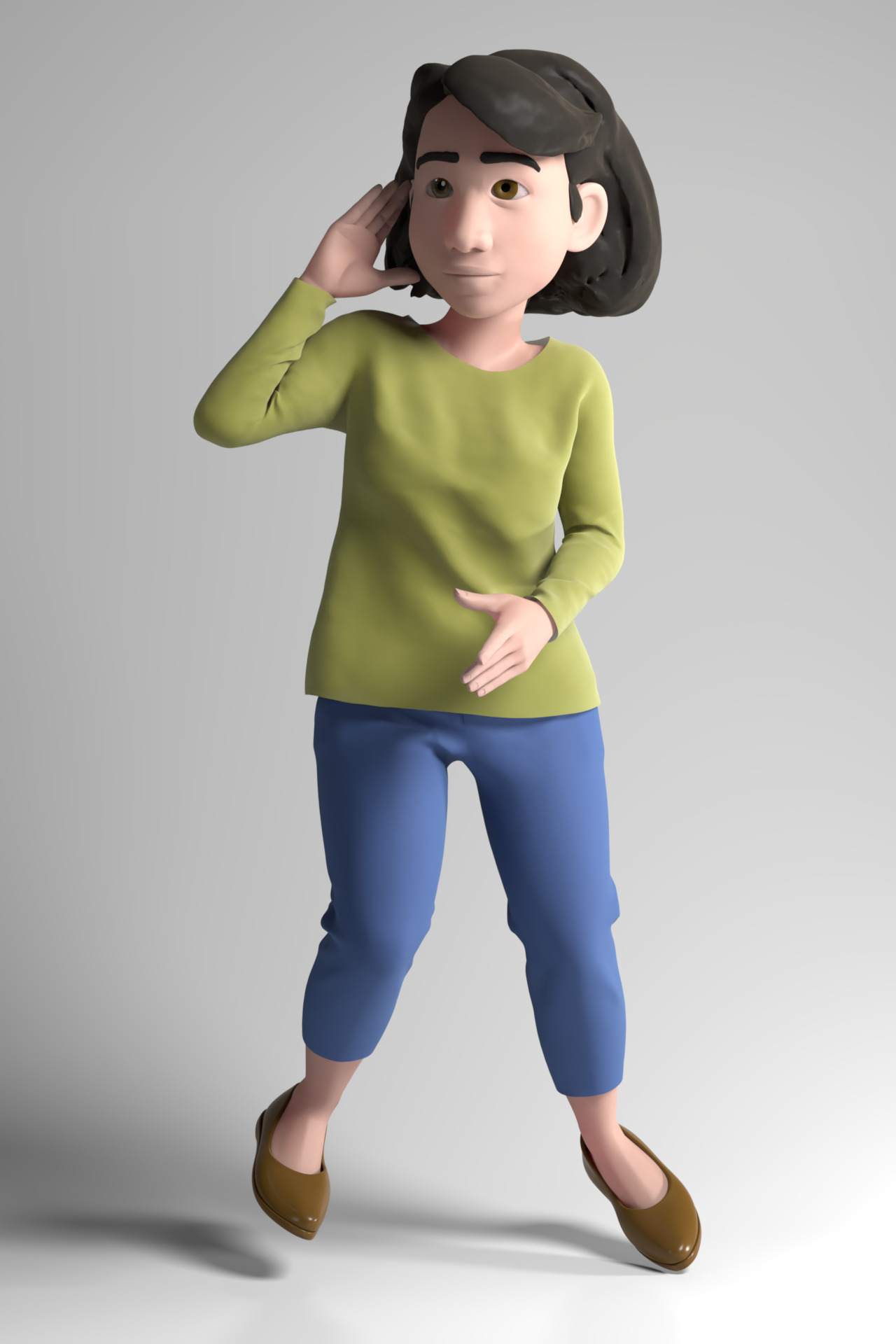} &
        \includegraphics[width=0.164125\linewidth]{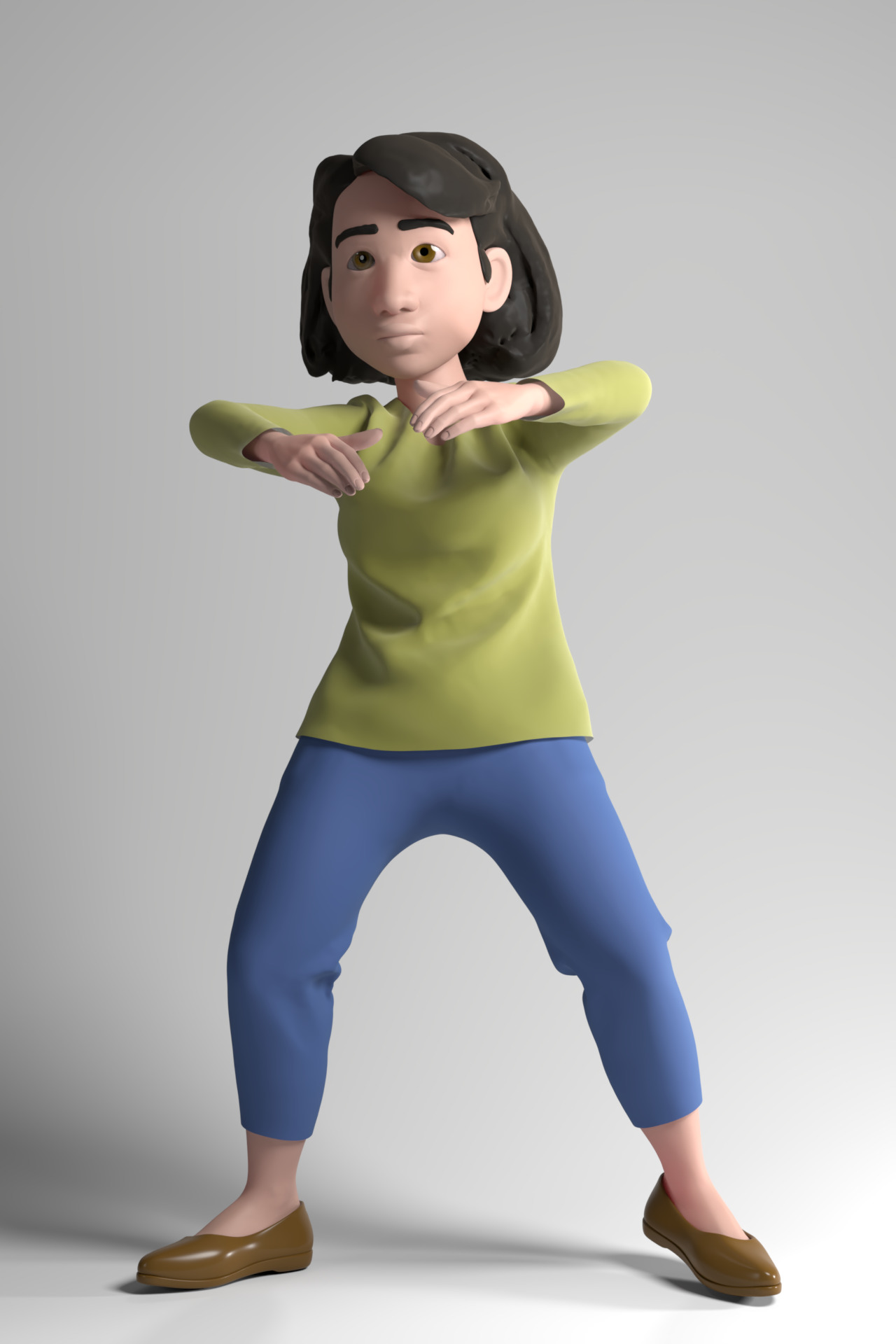}

	\end{tabular}}
  \caption{Validation frames from the \jenShirt and \jenPants datasets. The examples show a range of poses and corresponding garment deformations. Our method captures both overall garment shape and fine-scale wrinkle structure across these examples.}
  \label{fig:ValidationFrames}
\end{figure*}

\begin{figure*}[t!]
	\centering
	\setlength{\tabcolsep}{0.00125\linewidth}
    {\footnotesize
	\begin{tabular}{ccccccc}
        \includegraphics[width=0.1405\linewidth]{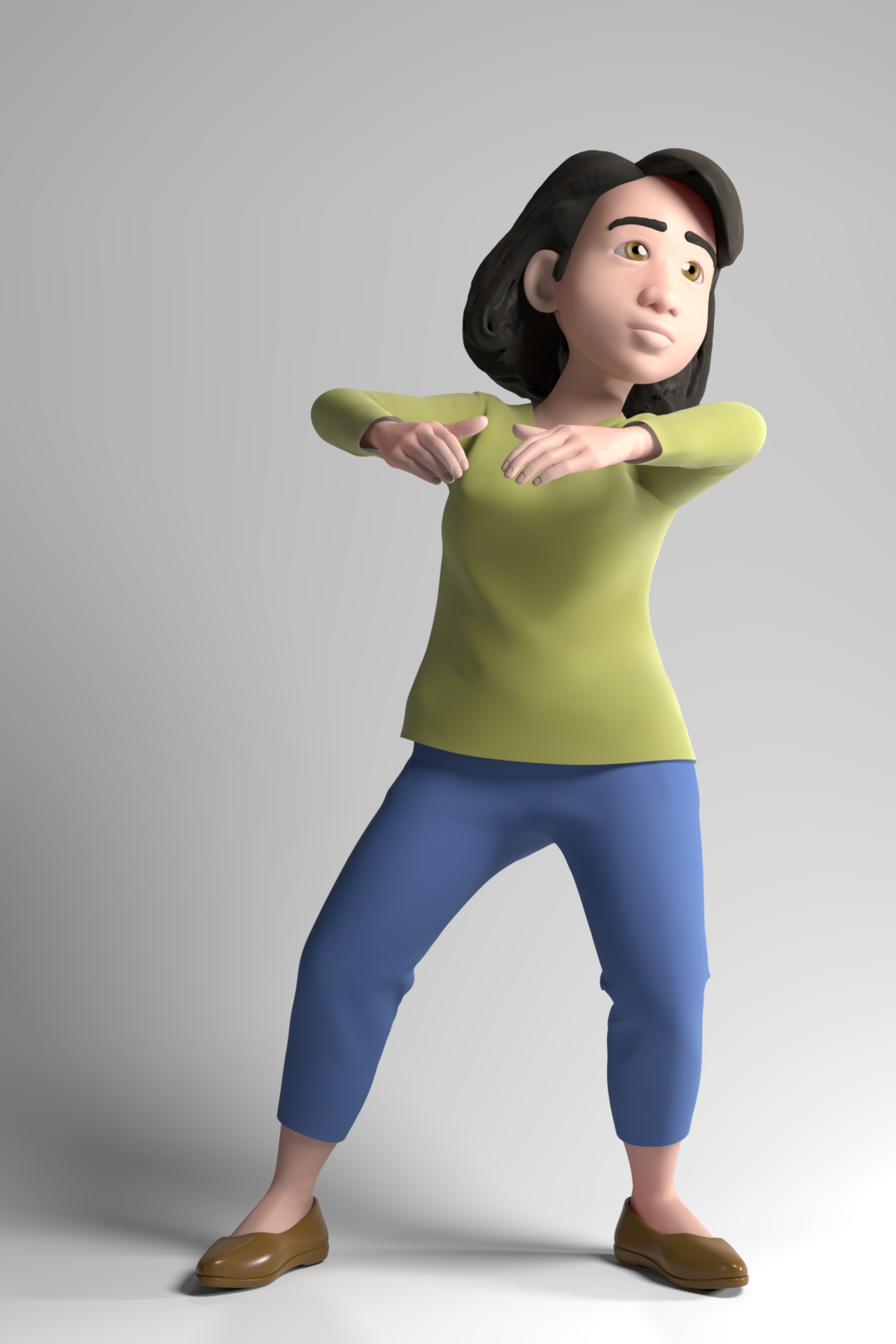} &
        \includegraphics[width=0.1405\linewidth]{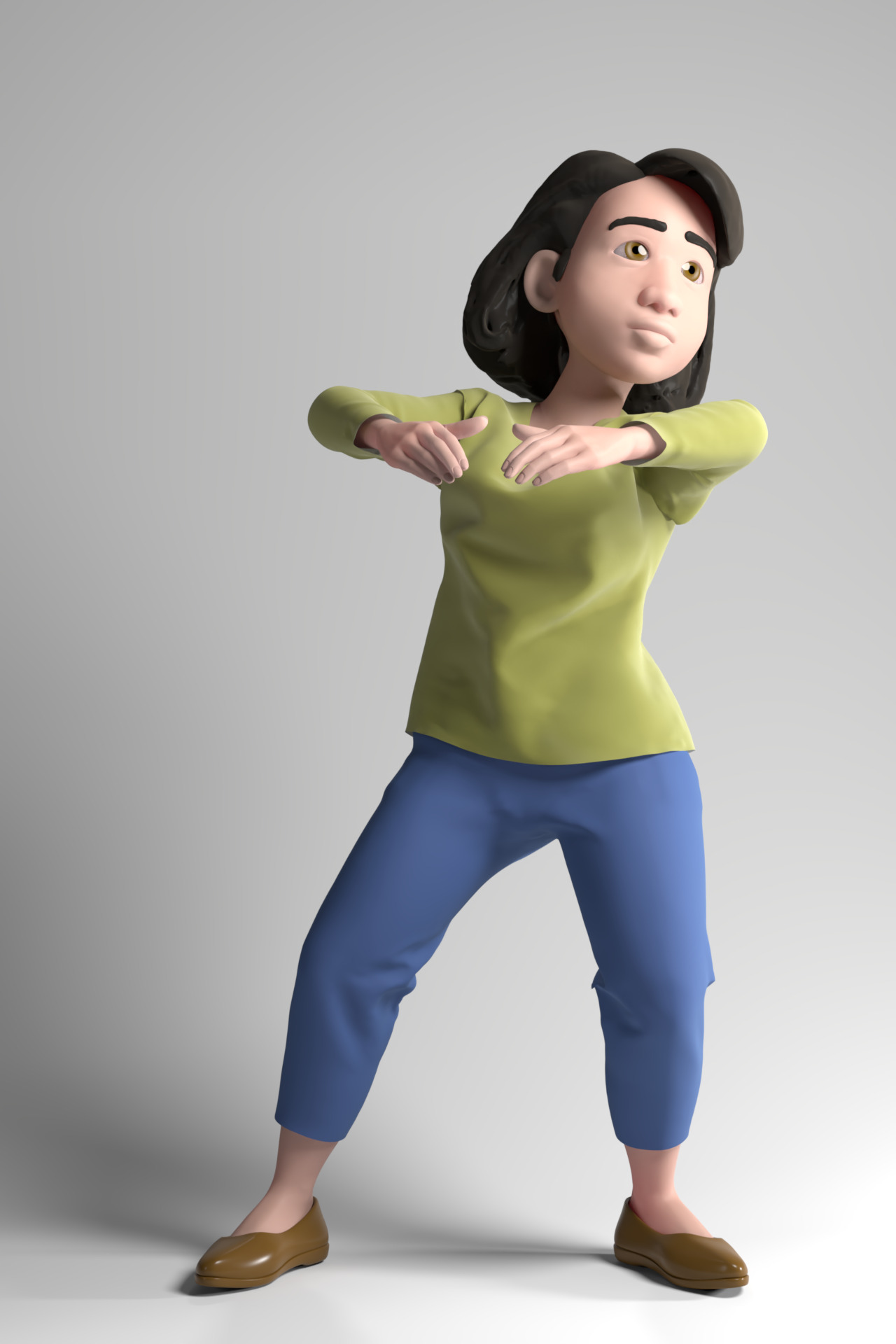} &
        \includegraphics[width=0.1405\linewidth]{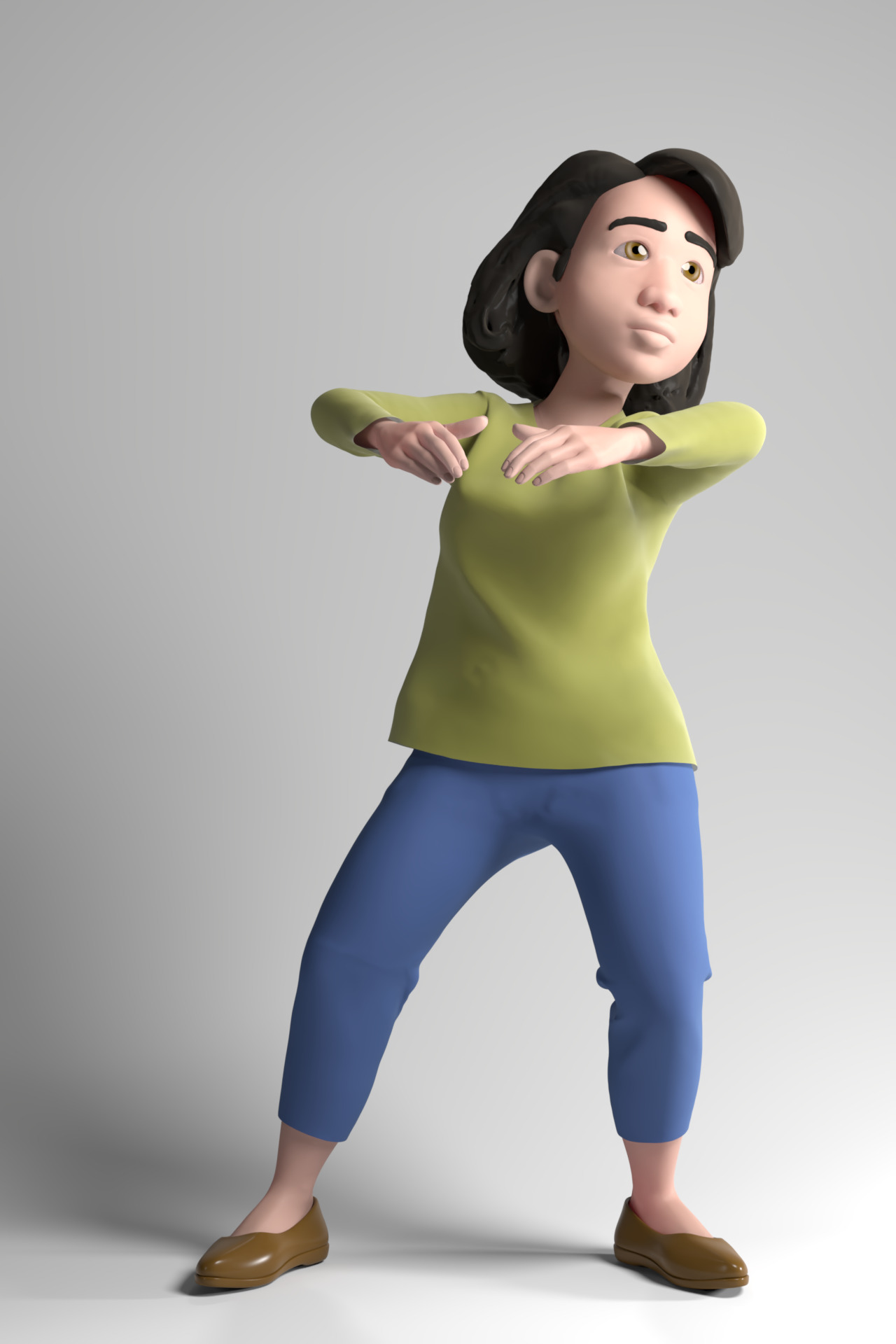} &
        \includegraphics[width=0.1405\linewidth]{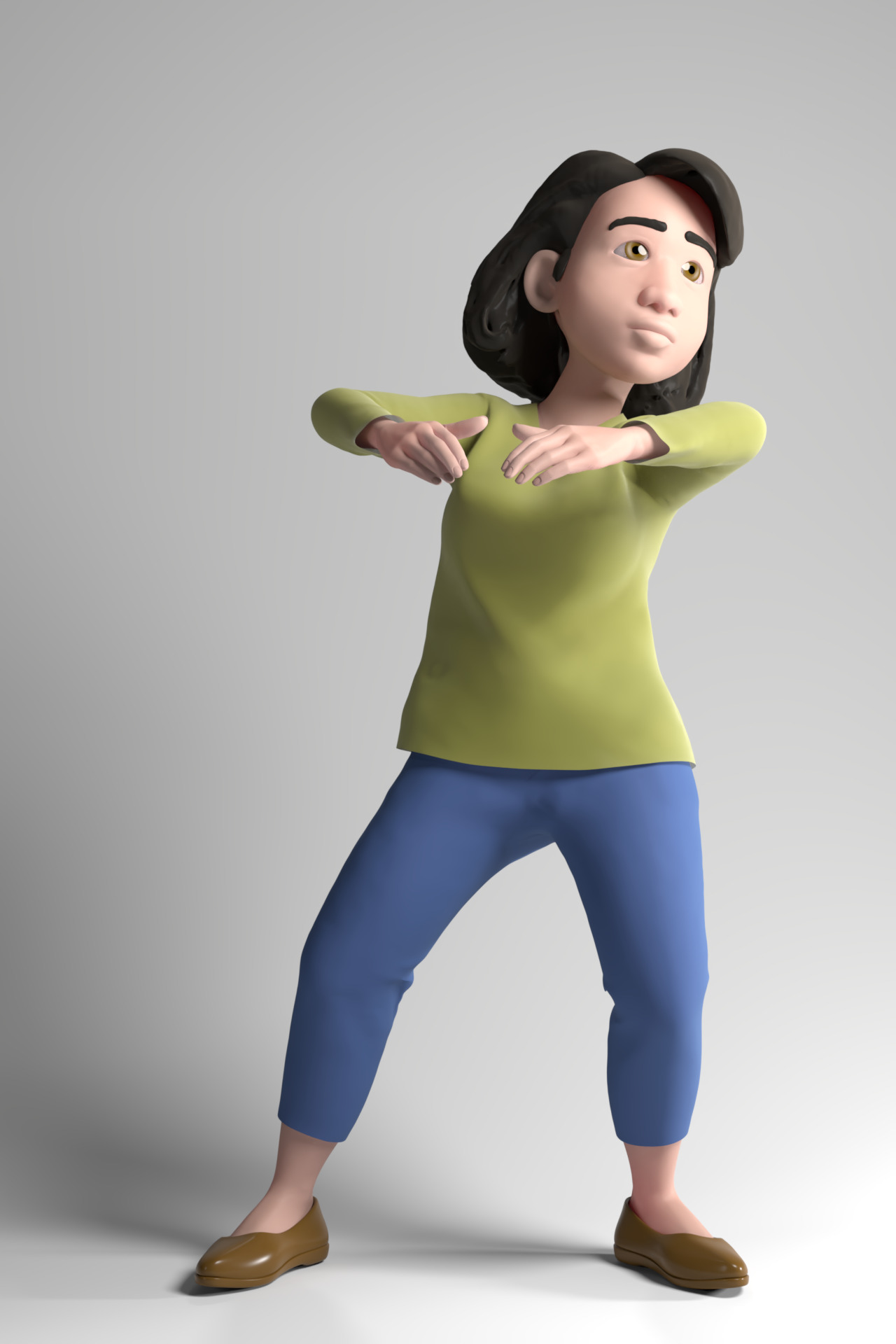} &
        \includegraphics[width=0.1405\linewidth]{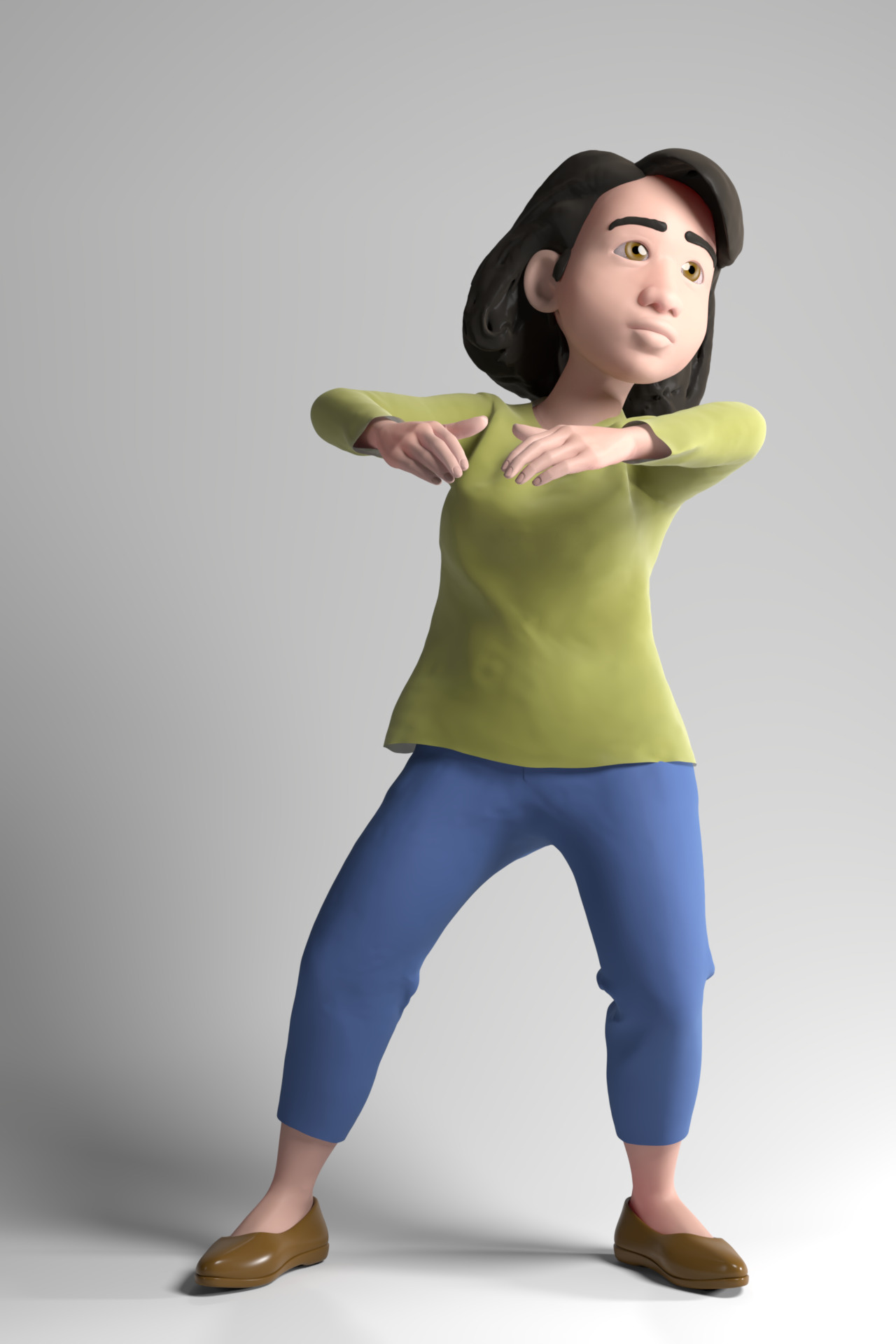} &
        \includegraphics[width=0.1405\linewidth]{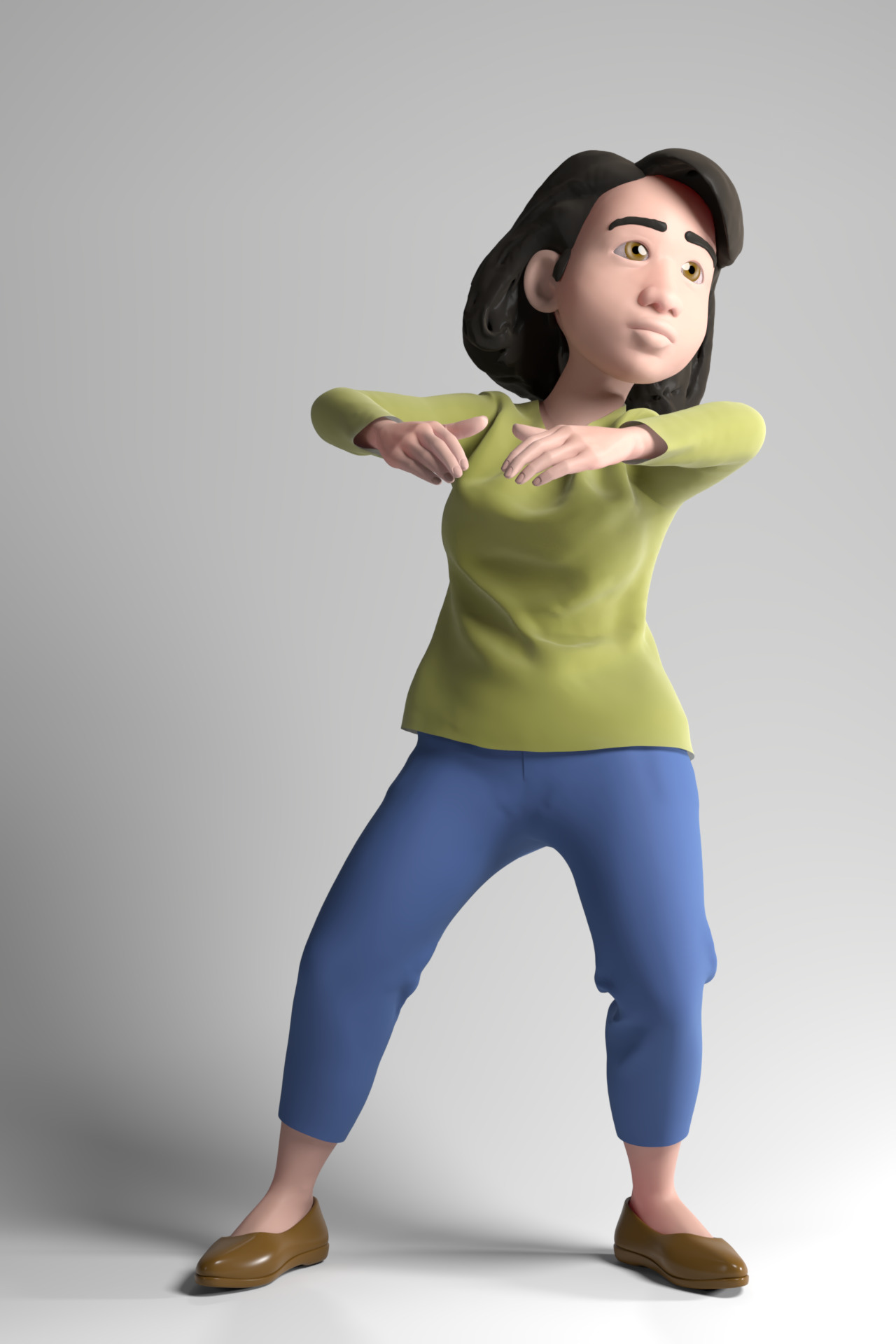} &
        \includegraphics[width=0.1405\linewidth]{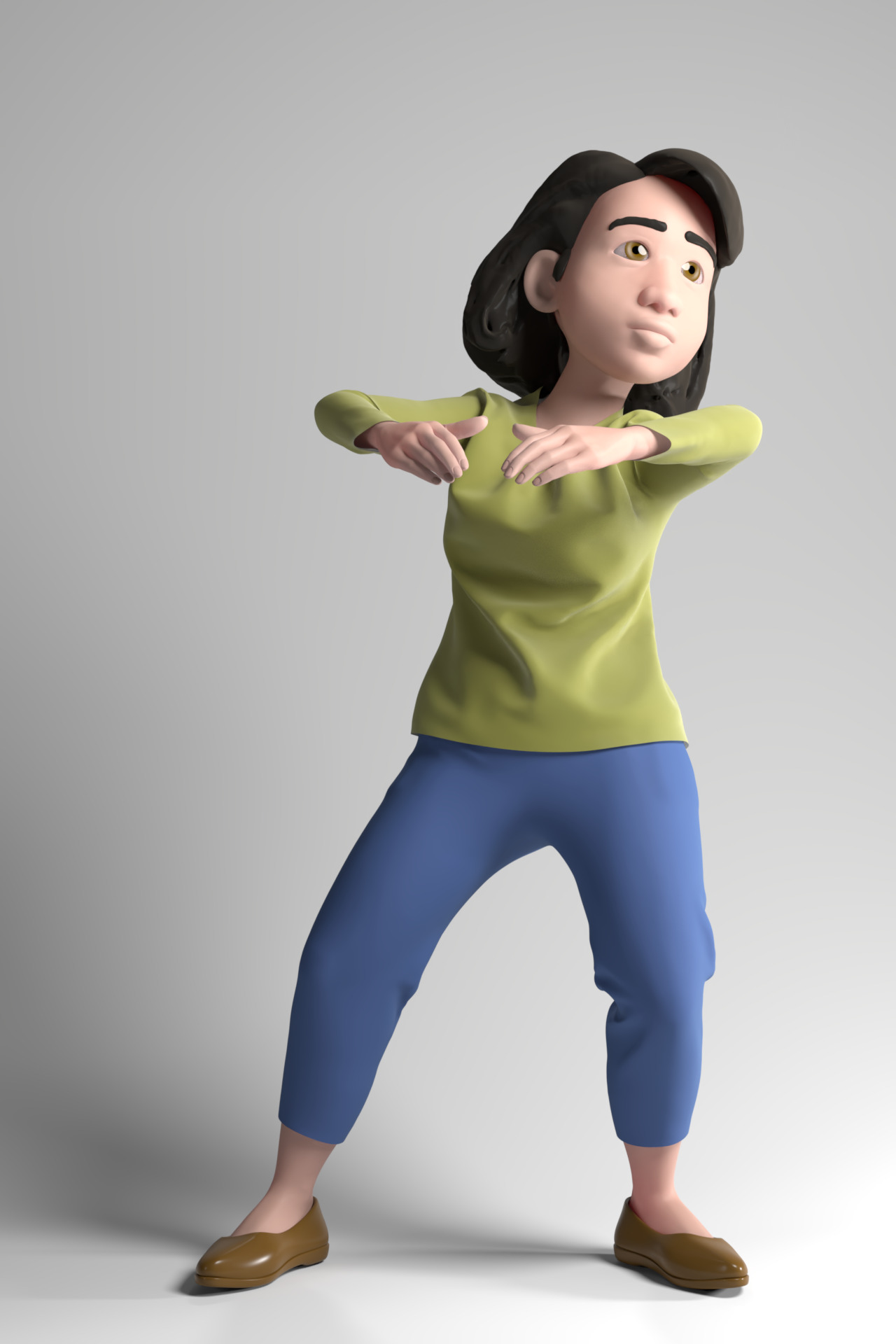}\\

        \includegraphics[width=0.1405\linewidth]{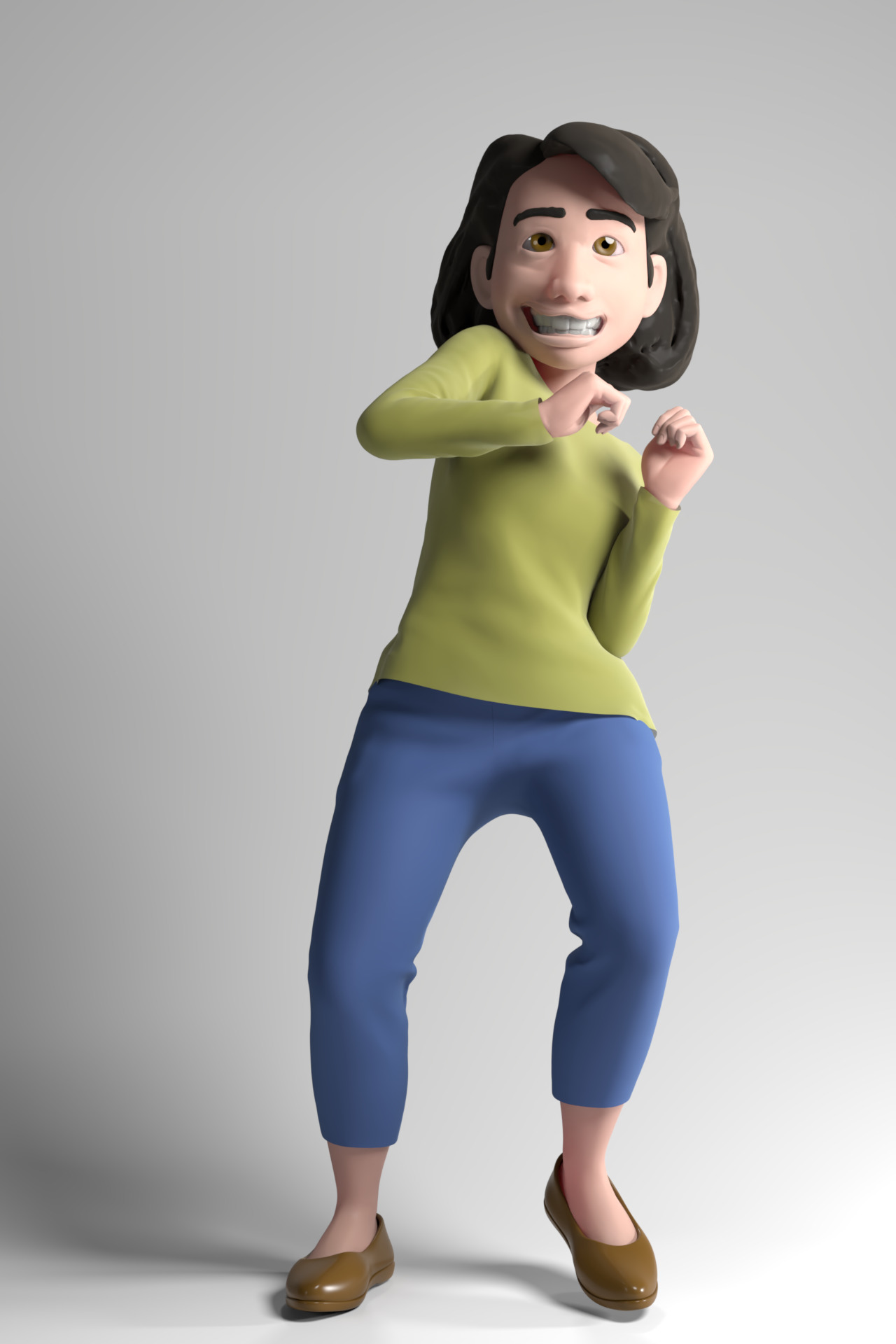} &
        \includegraphics[width=0.1405\linewidth]{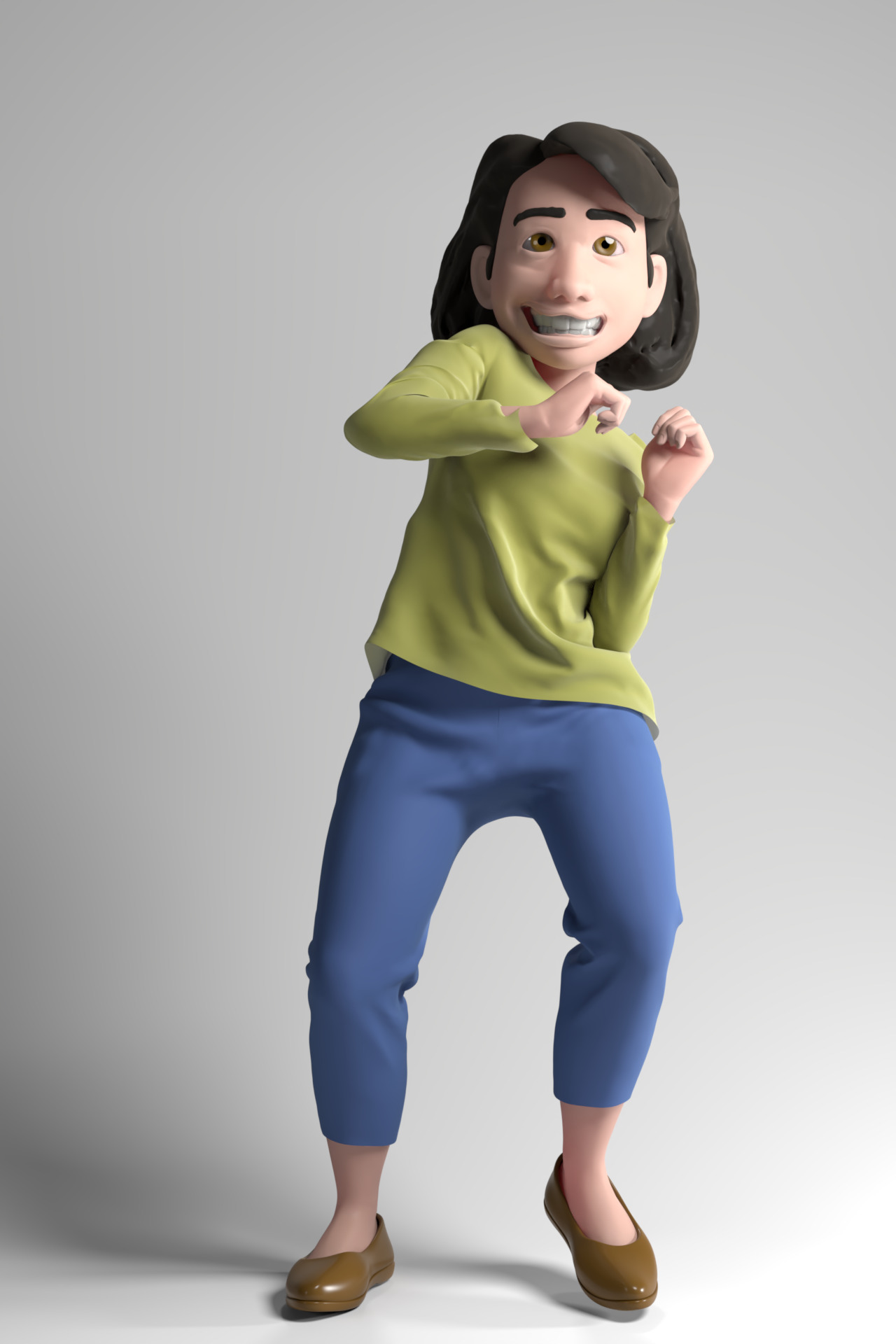} &
        \includegraphics[width=0.1405\linewidth]{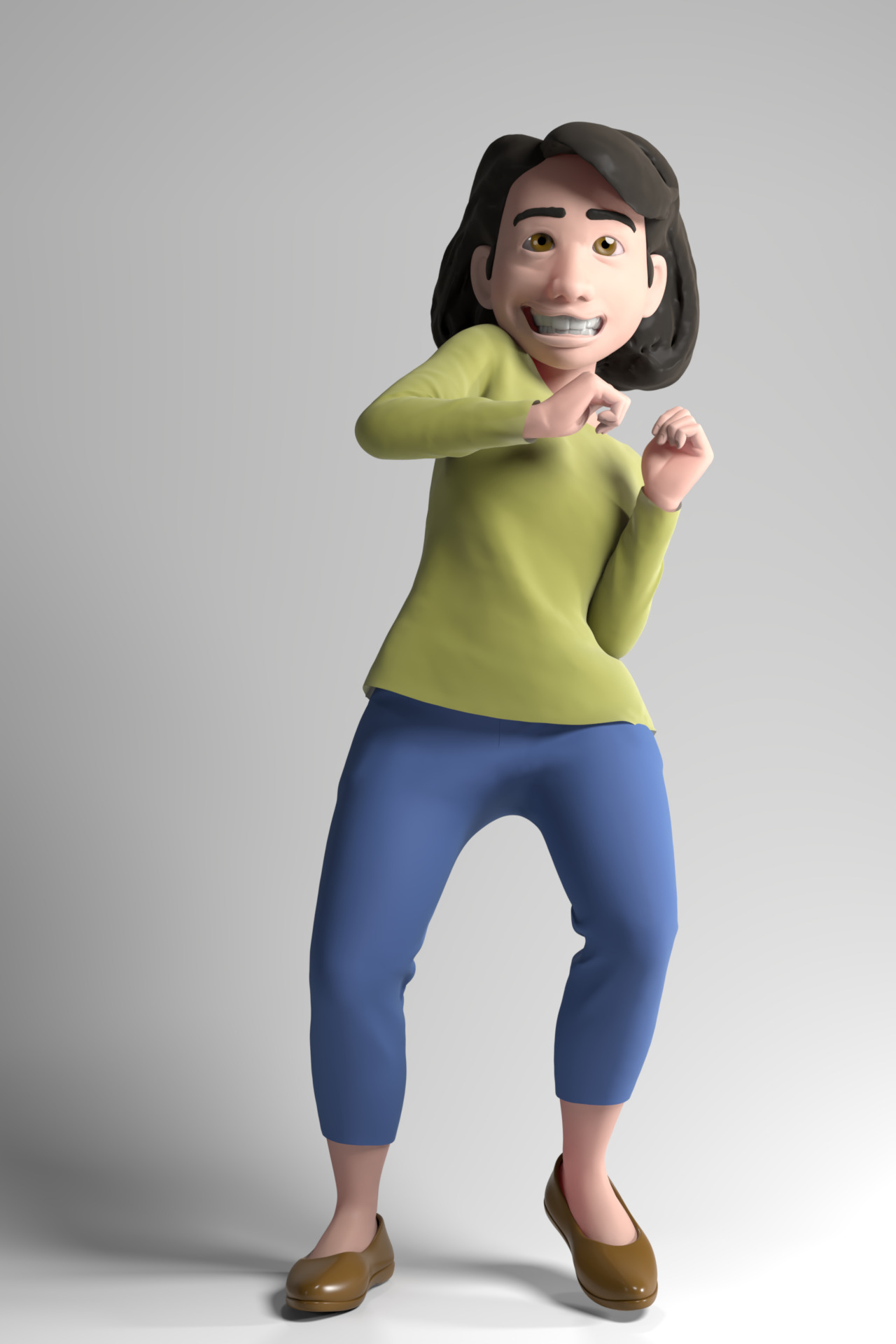} &
        \includegraphics[width=0.1405\linewidth]{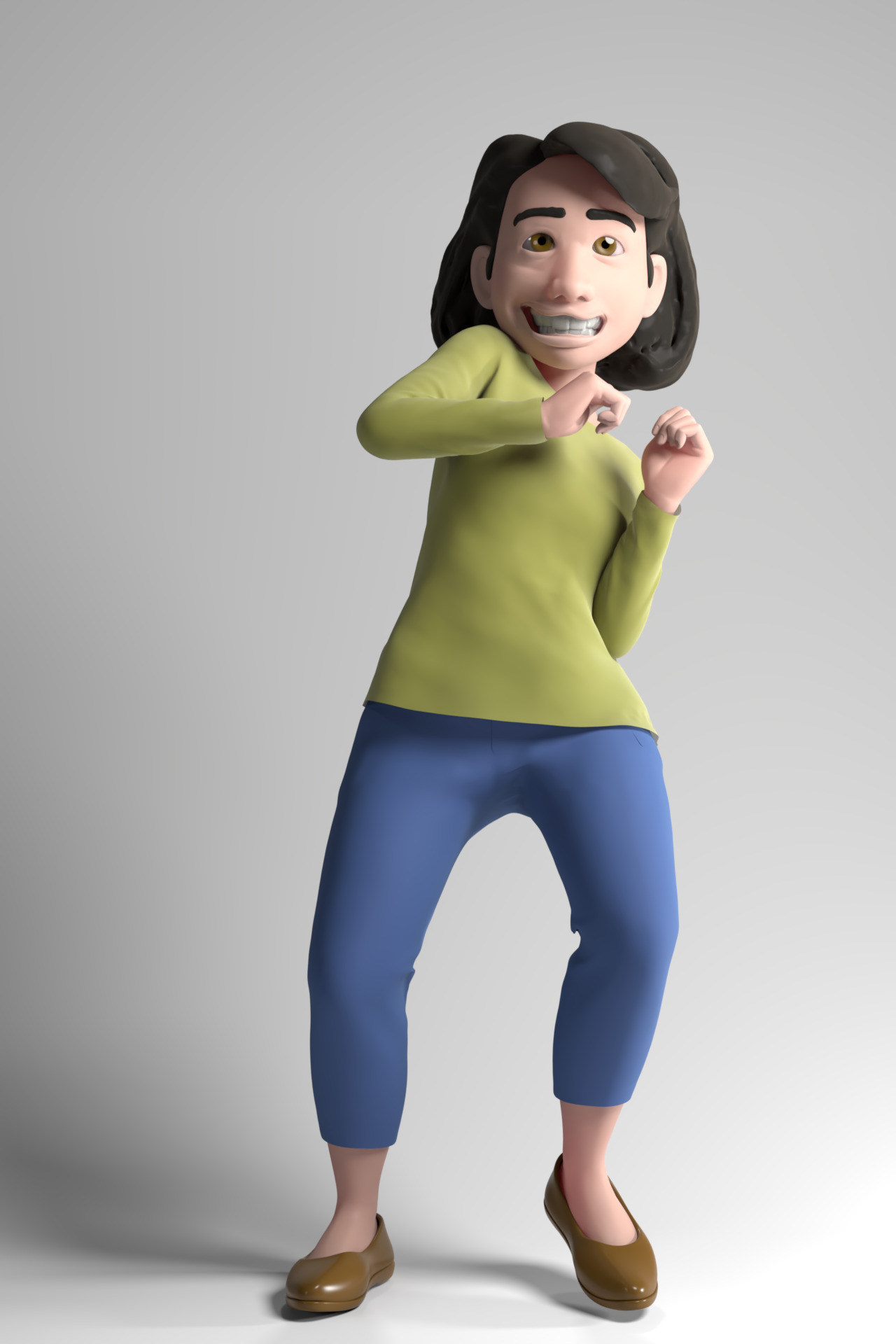} &
        \includegraphics[width=0.1405\linewidth]{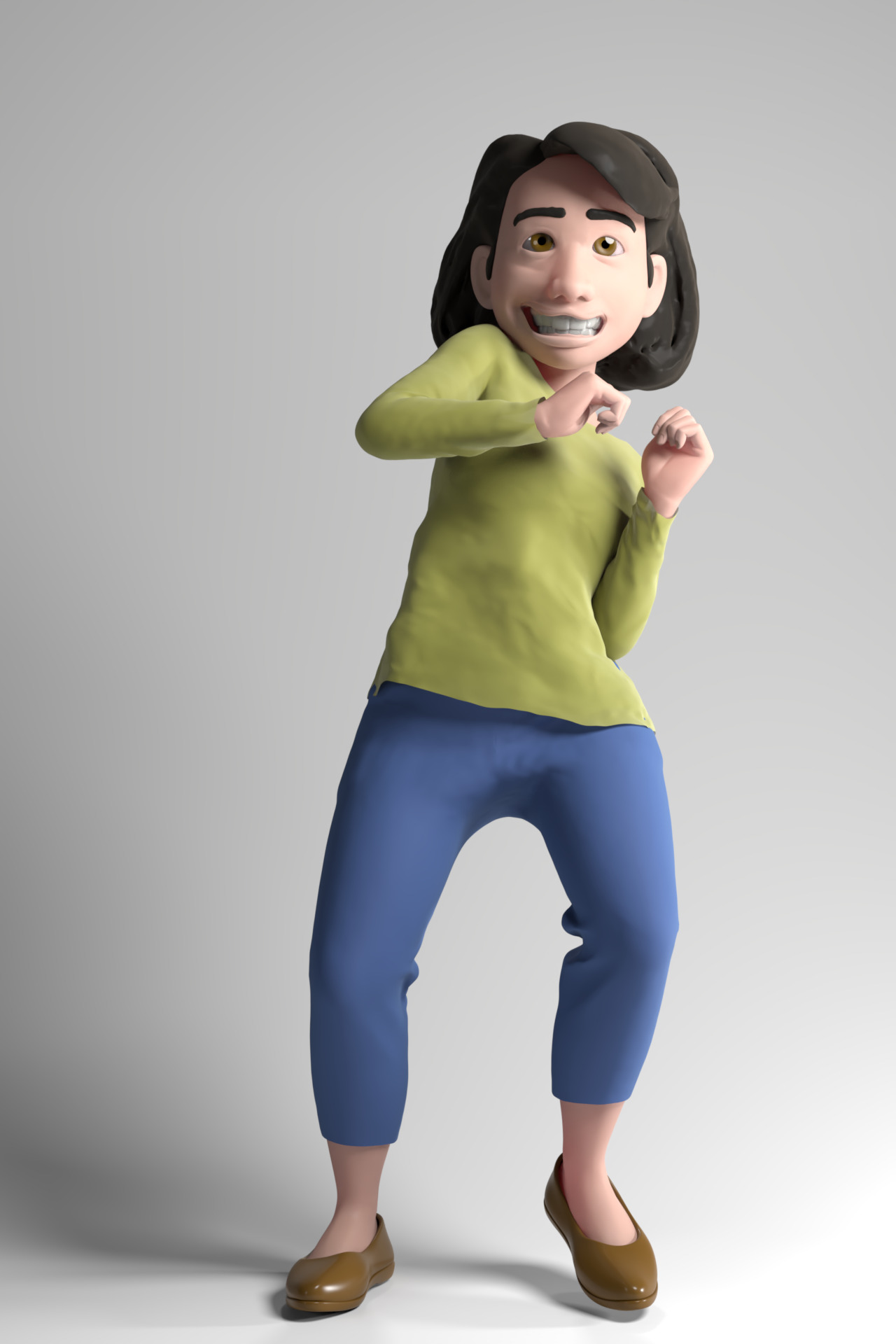} &
        \includegraphics[width=0.1405\linewidth]{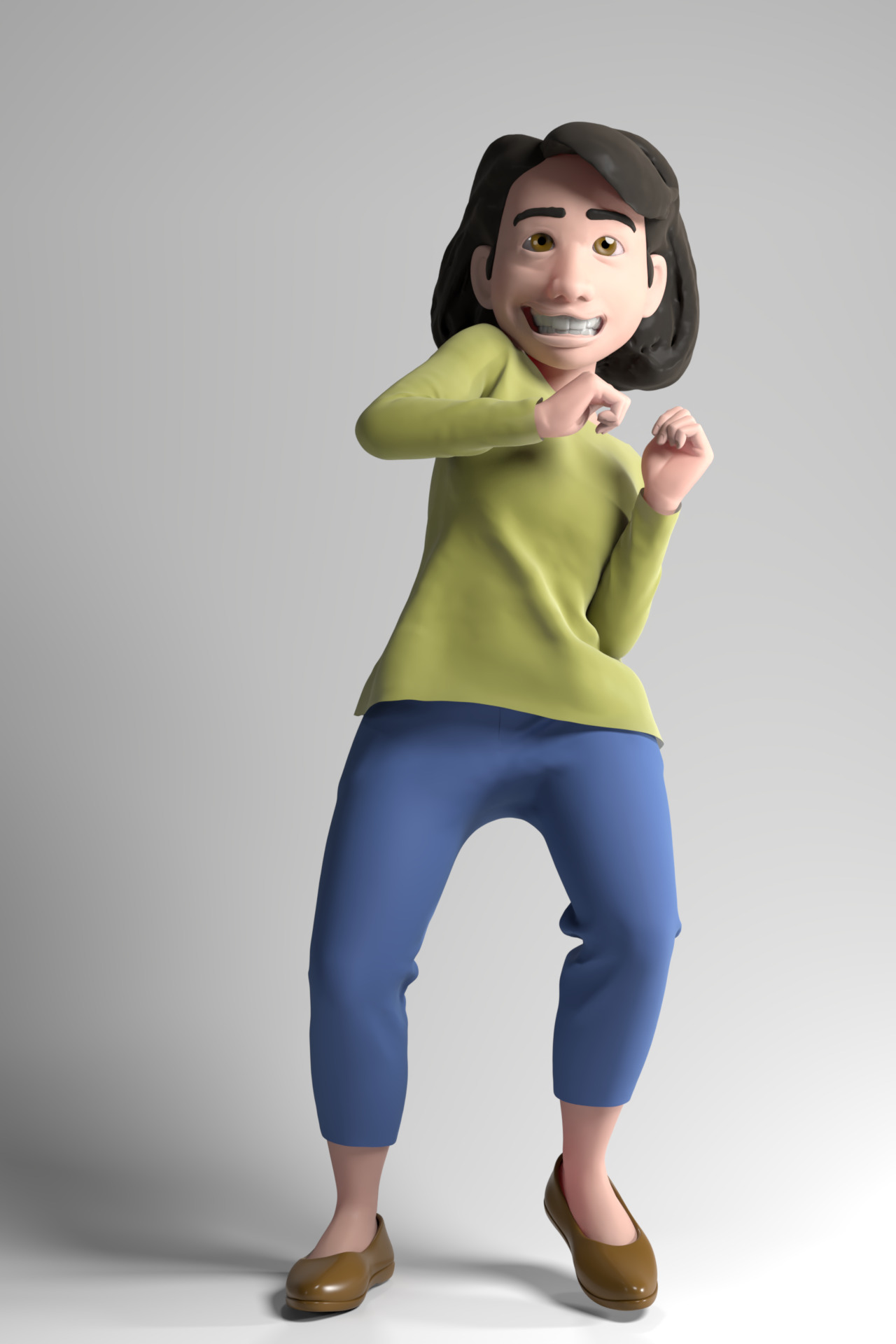} &
        \includegraphics[width=0.1405\linewidth]{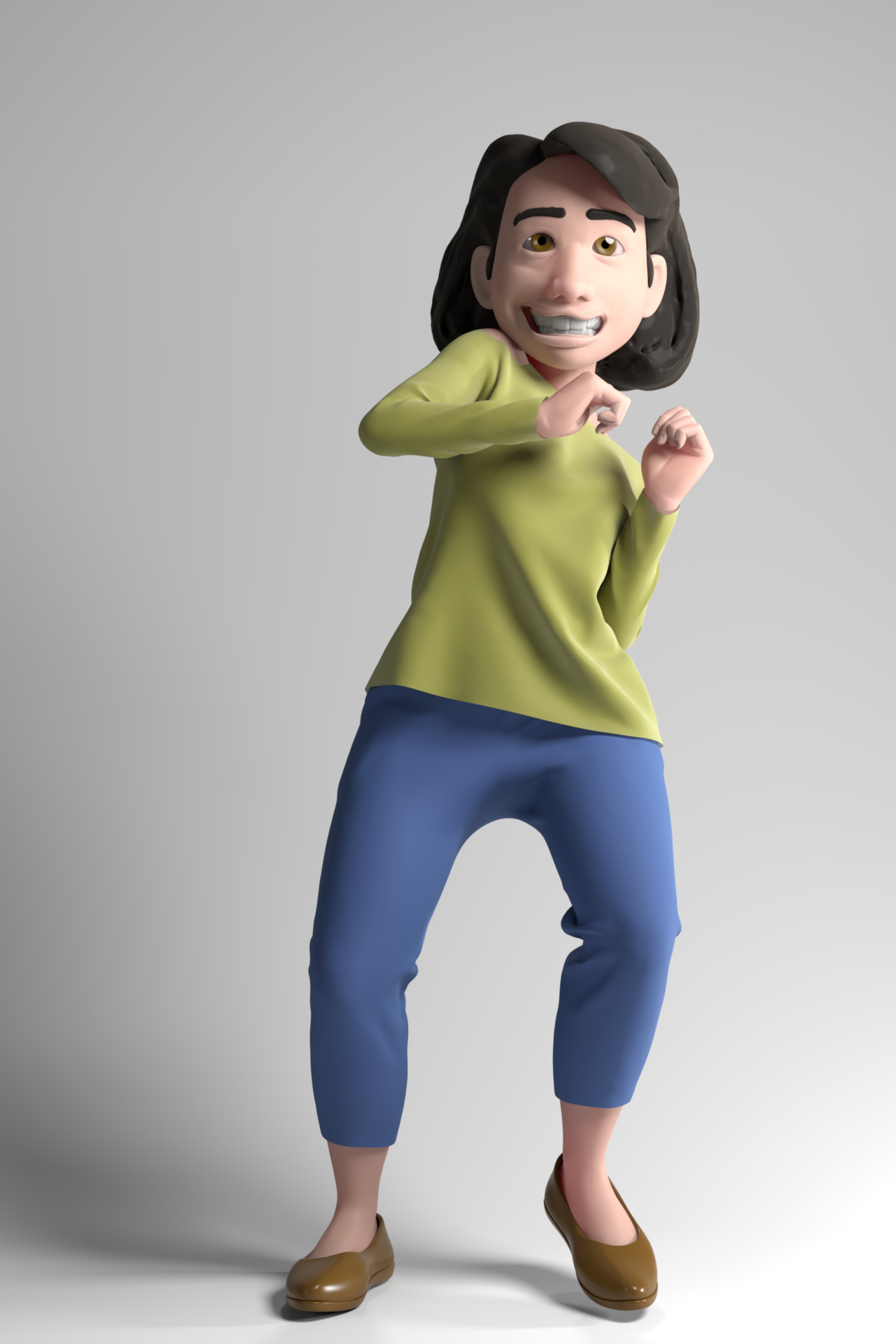}\\

        \includegraphics[width=0.1405\linewidth]{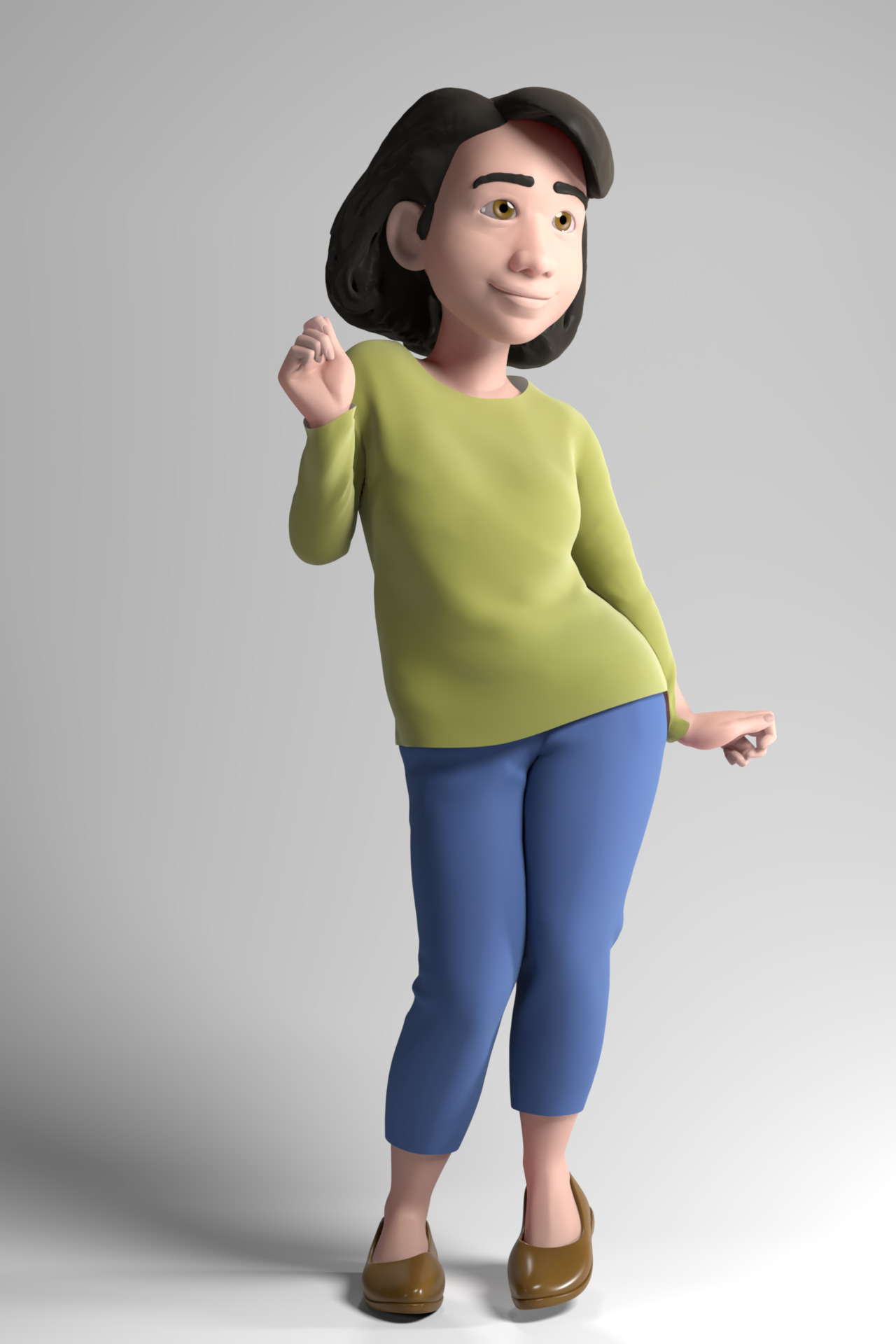} &
        \includegraphics[width=0.1405\linewidth]{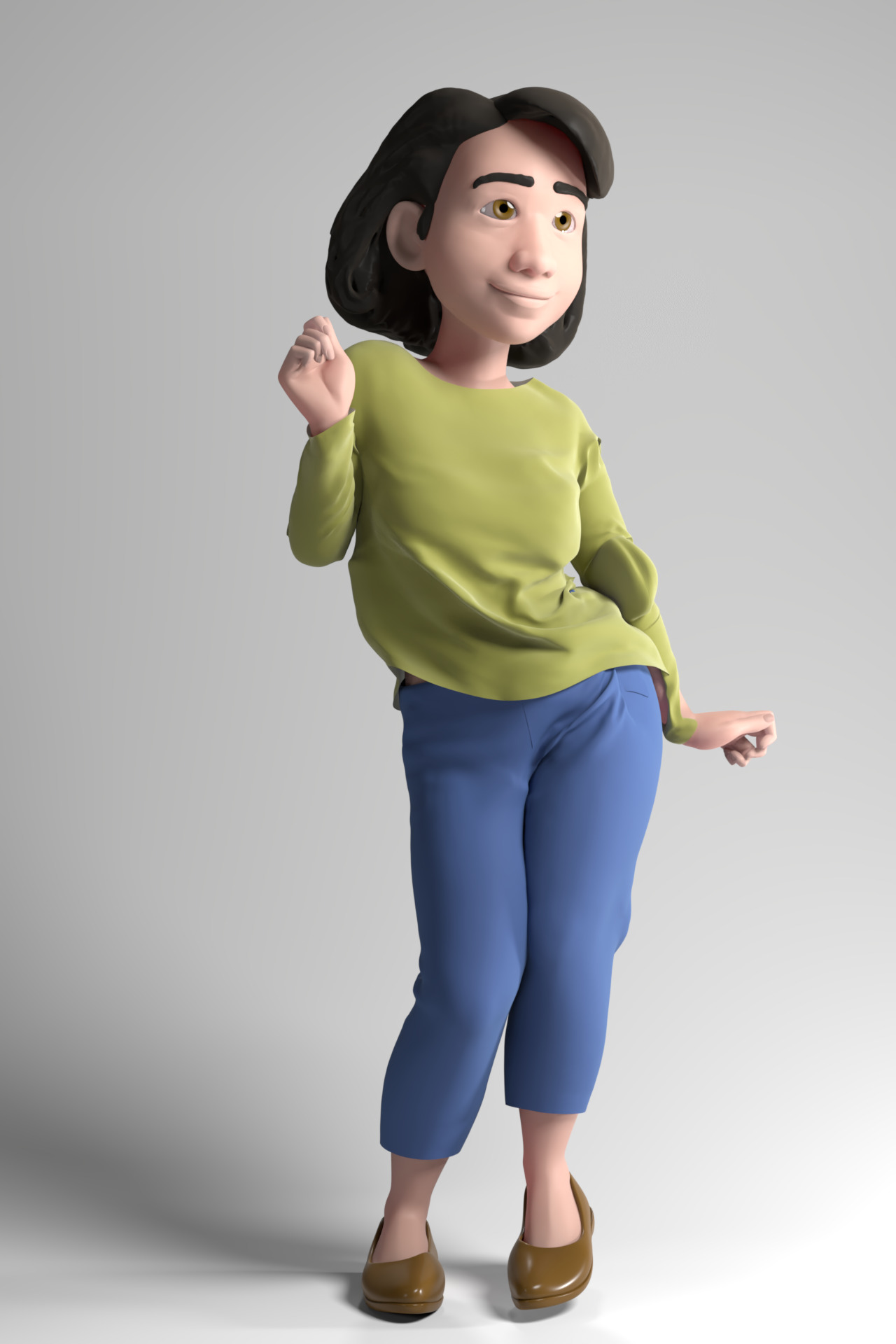} &
        \includegraphics[width=0.1405\linewidth]{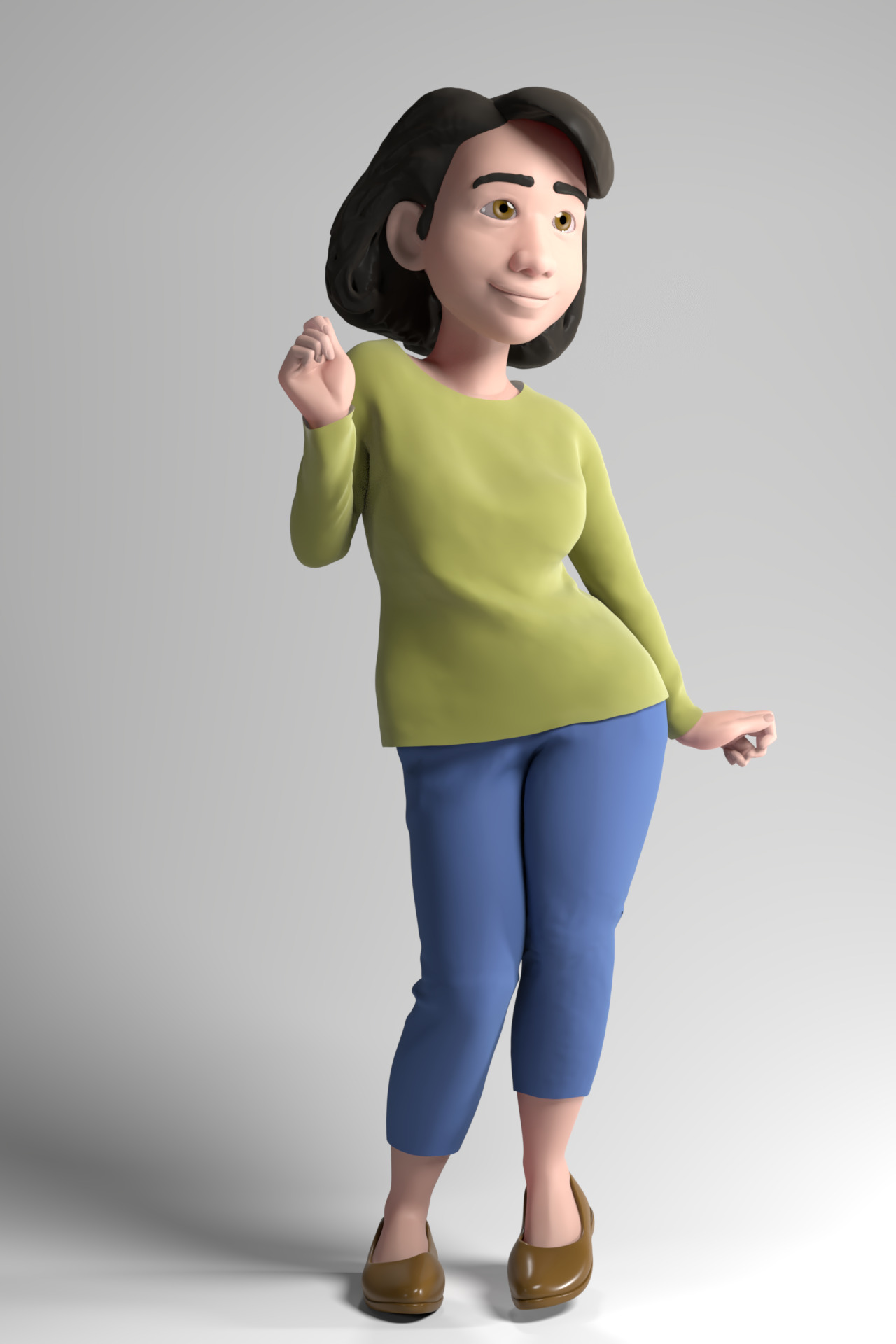} &
        \includegraphics[width=0.1405\linewidth]{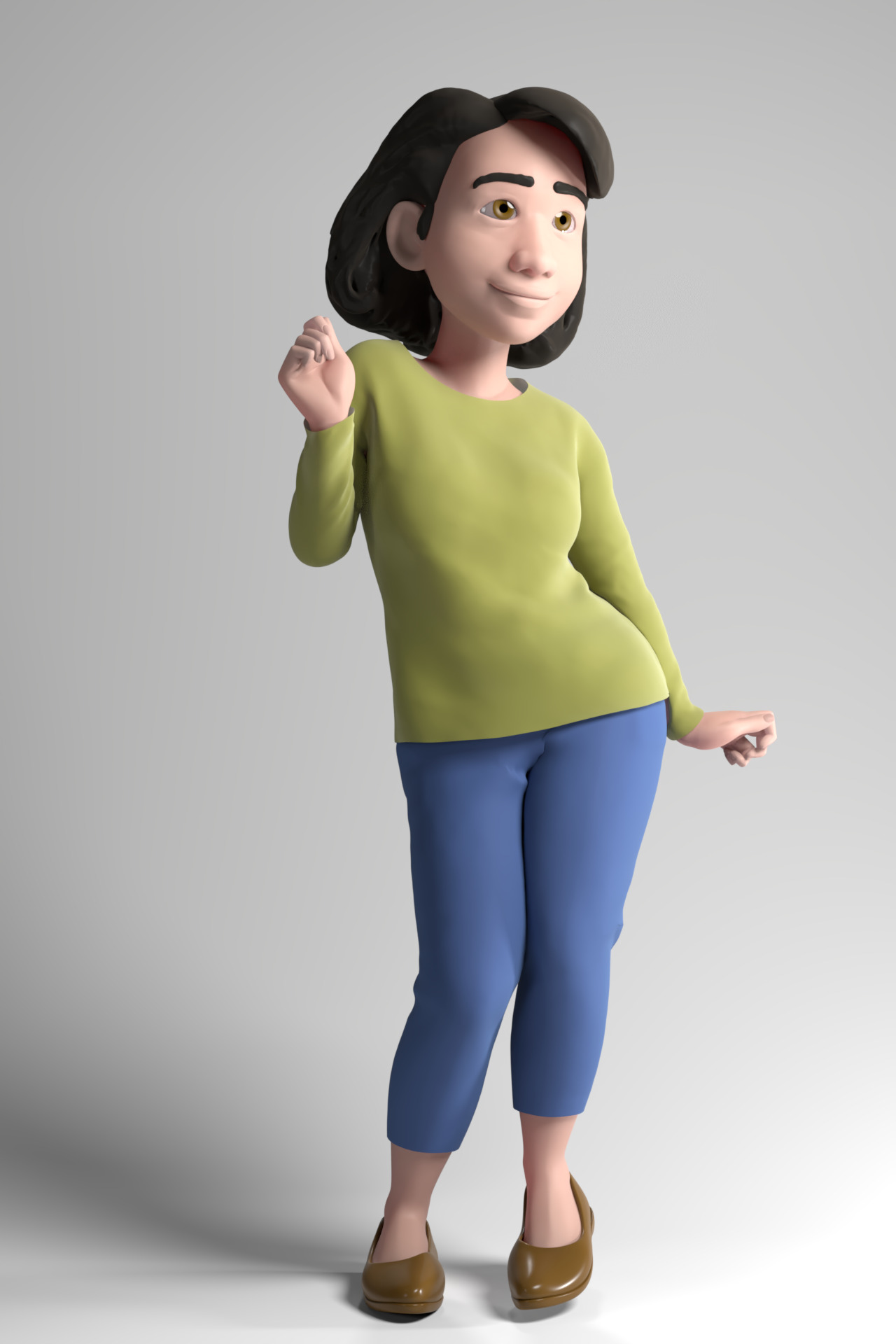} &
        \includegraphics[width=0.1405\linewidth]{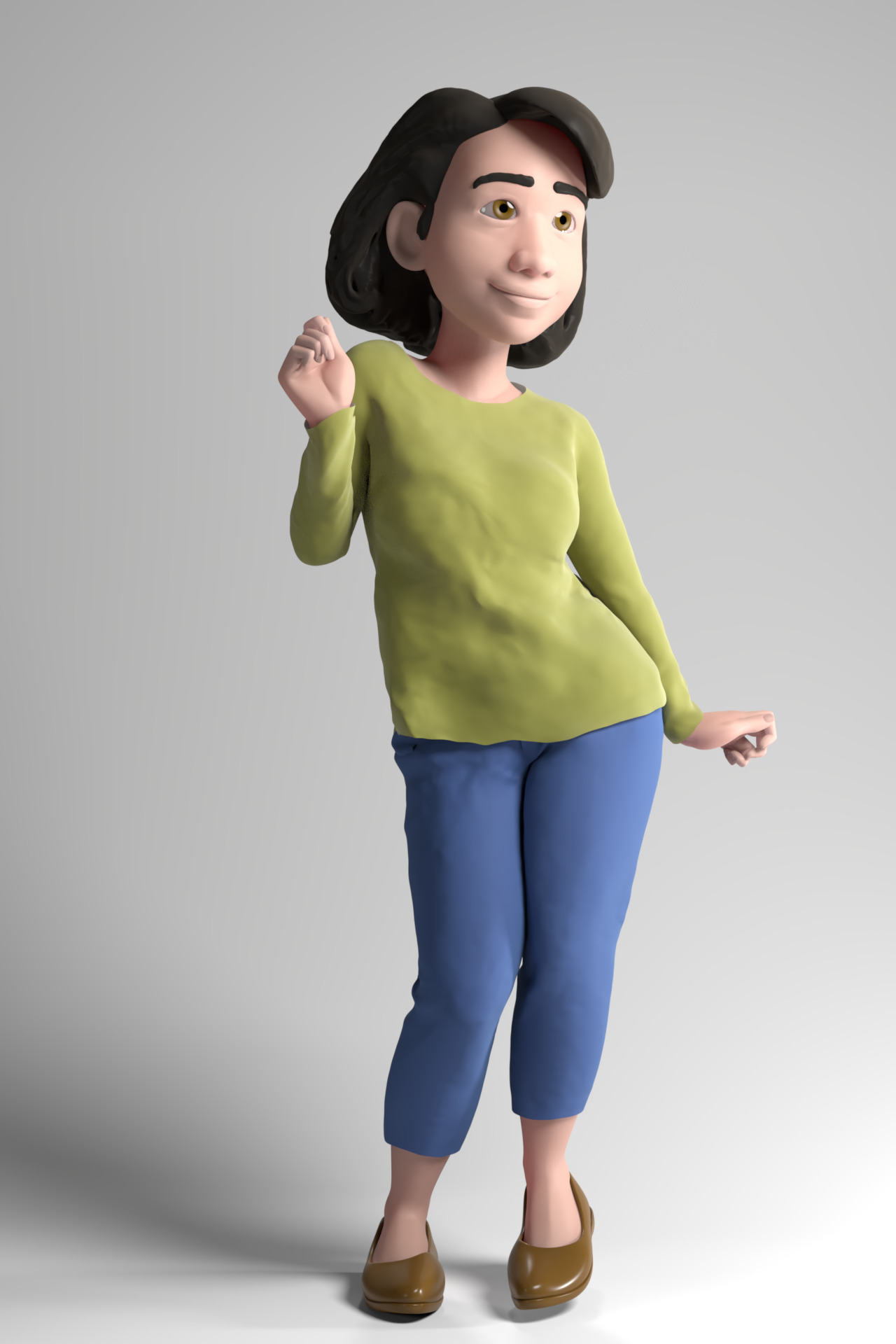} &
        \includegraphics[width=0.1405\linewidth]{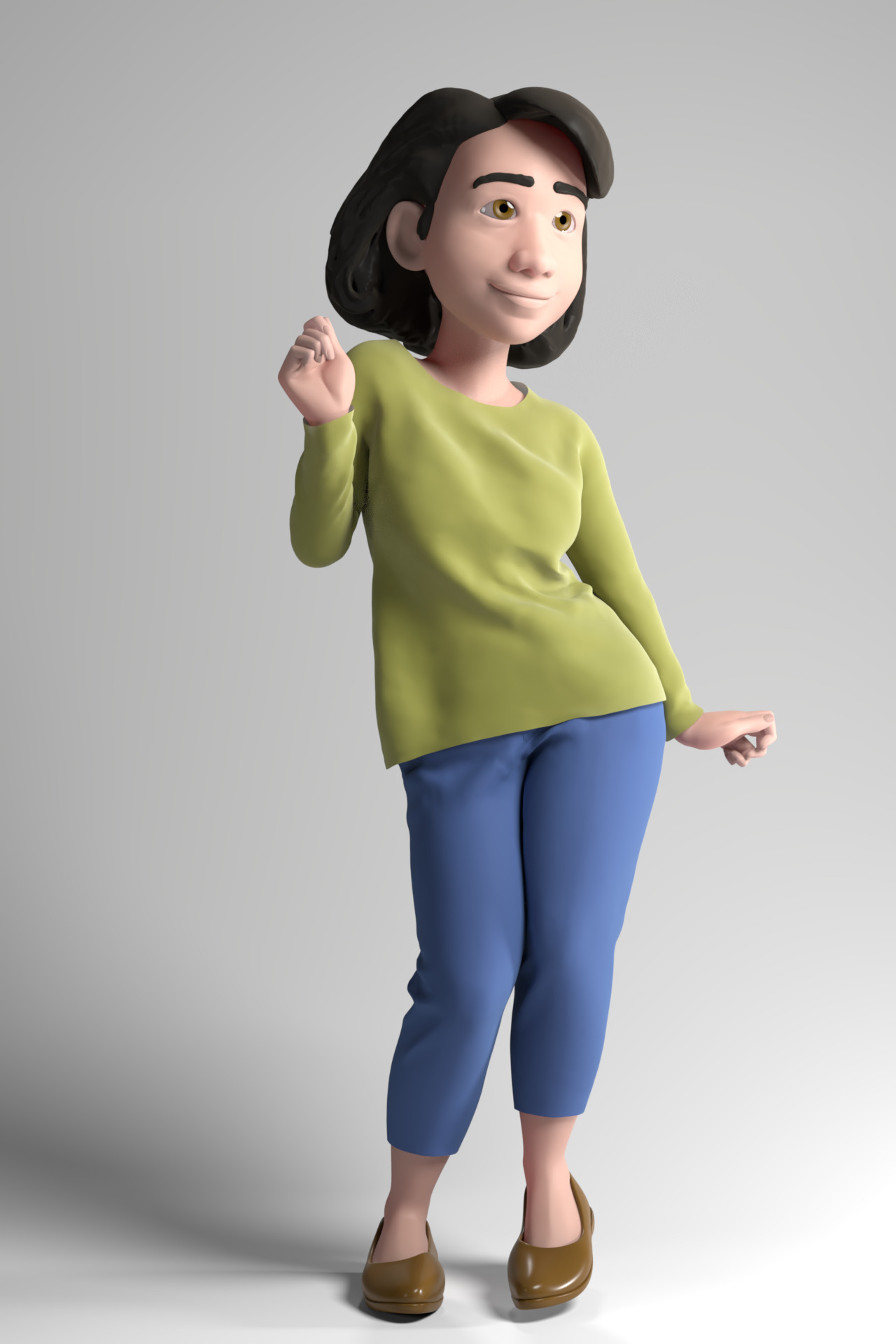} &
        \includegraphics[width=0.1405\linewidth]{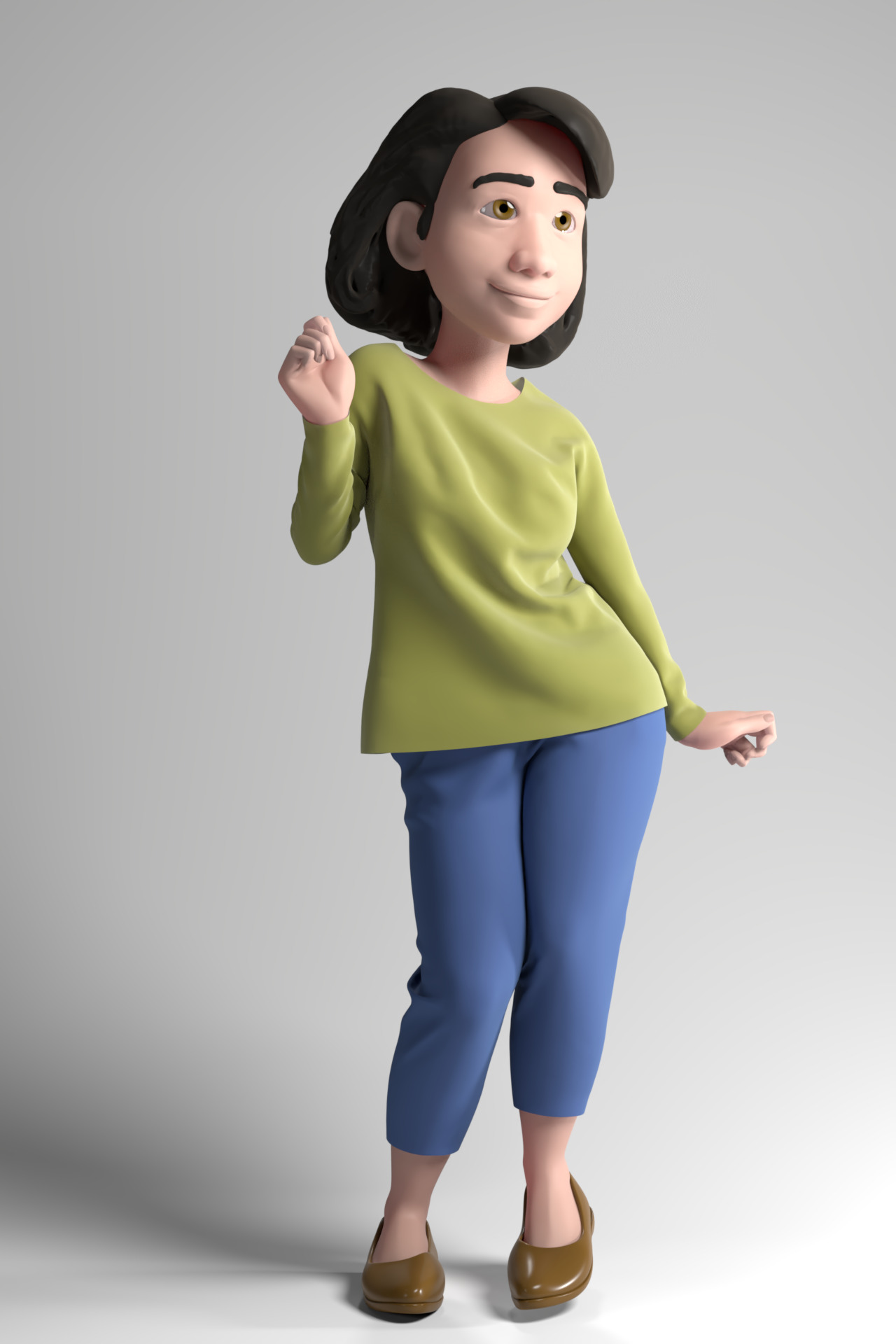}\\

		Maya 2025.2 & Unreal Engine 5.6 & \cite{bailey_fast_2018} & \cite{patel_tailornet_2020} & \cite{chentanez_cloth_2020} & Ours & Ground Truth
	\end{tabular}}
  \caption{Qualitative comparison of our method against two industrial tools and three real-time academic baselines on an unseen animation from \jenShirt and \jenPants datasets. Maya \protect{\cite{autodesk_autodesk_2025}}, \protect{\cite{bailey_fast_2018}}, and \protect{\cite{patel_tailornet_2020}} produce overly smoothed wrinkles; Unreal Engine \protect{\cite{epic_games_unreal_2025}} shows temporal discontinuities and deformations unrelated to the pose; \protect{\cite{chentanez_cloth_2020}} exhibits high-frequency artifacts. In contrast, our method produces sharper, more coherent wrinkle patterns while maintaining temporal stability.}
  \label{fig:ComparisonBaselineModelsOurs}
\end{figure*}

\subsection{Local Normal Distribution Loss}
\label{sec:lnd}
Pointwise reconstruction losses (e.g., vertex $L^2$) penalize small spatial misalignments of wrinkles and folds, even they exhibit visually plausible high-frequency details. To reduce sensitivity to such phase shifts while still supervising fine-scale appearance, we introduce a distributional loss that matches \emph{local surface orientation statistics} between prediction and ground truth. Given predicted and ground-truth meshes, we compute per-vertex unit normals and represent each normal by spherical angles $(\varphi,\omega)$, with $\varphi \in [0,\pi]$ and $\omega \in [0,2\pi)$. For each vertex $i$, we define a local neighborhood $\mathcal{\Omega}_k(i)$ using $k$-nearest neighbors in the bind-pose geometry; we optionally evaluate multiple neighborhood sizes $k \in \mathcal{K}$ to capture statistics at different spatial scales. Within each neighborhood, we differentiably construct a two-dimensional histogram over $(\varphi,\omega)$ using Gaussian kernel accumulation around fixed bin centers in angular space. The resulting histogram is normalized to sum to one, yielding a local probability distribution over surface orientations. We use $8$ bins for $\varphi$ and $16$ bins for $\omega$, with a fixed kernel bandwidth.

Let $\hat{H}_{i,k}$ and $H_{i,k}$ denote the predicted and ground-truth normalized orientation histograms for vertex $i$ and neighborhood size $k$, respectively. We define the local normal distribution loss as the mean squared error between corresponding histograms, averaged over vertices and neighborhood scales:
\begin{equation}
\mathcal{L}_{\text{hist}} =
\frac{1}{|\mathcal{K}|} \sum_{k \in \mathcal{K}}
\frac{1}{|\mathcal{V}|} \sum_{i \in \mathcal{V}}
\left\lVert \hat{H}_{i,k} - H_{i,k} \right\rVert_2^2,
\end{equation}
where $\mathcal{V}$ is the set of vertices. This loss provides an alternative to pointwise supervision by encouraging realistic local wrinkle energy and directional variability.

\section{Experiments and Results}
\label{sec:results}

\begin{figure*}[t]
	\centering
	\setlength{\tabcolsep}{0.00125\linewidth}
    {\footnotesize
    \begin{tabular}{cccccccc}
        \includegraphics[width=0.123\linewidth]{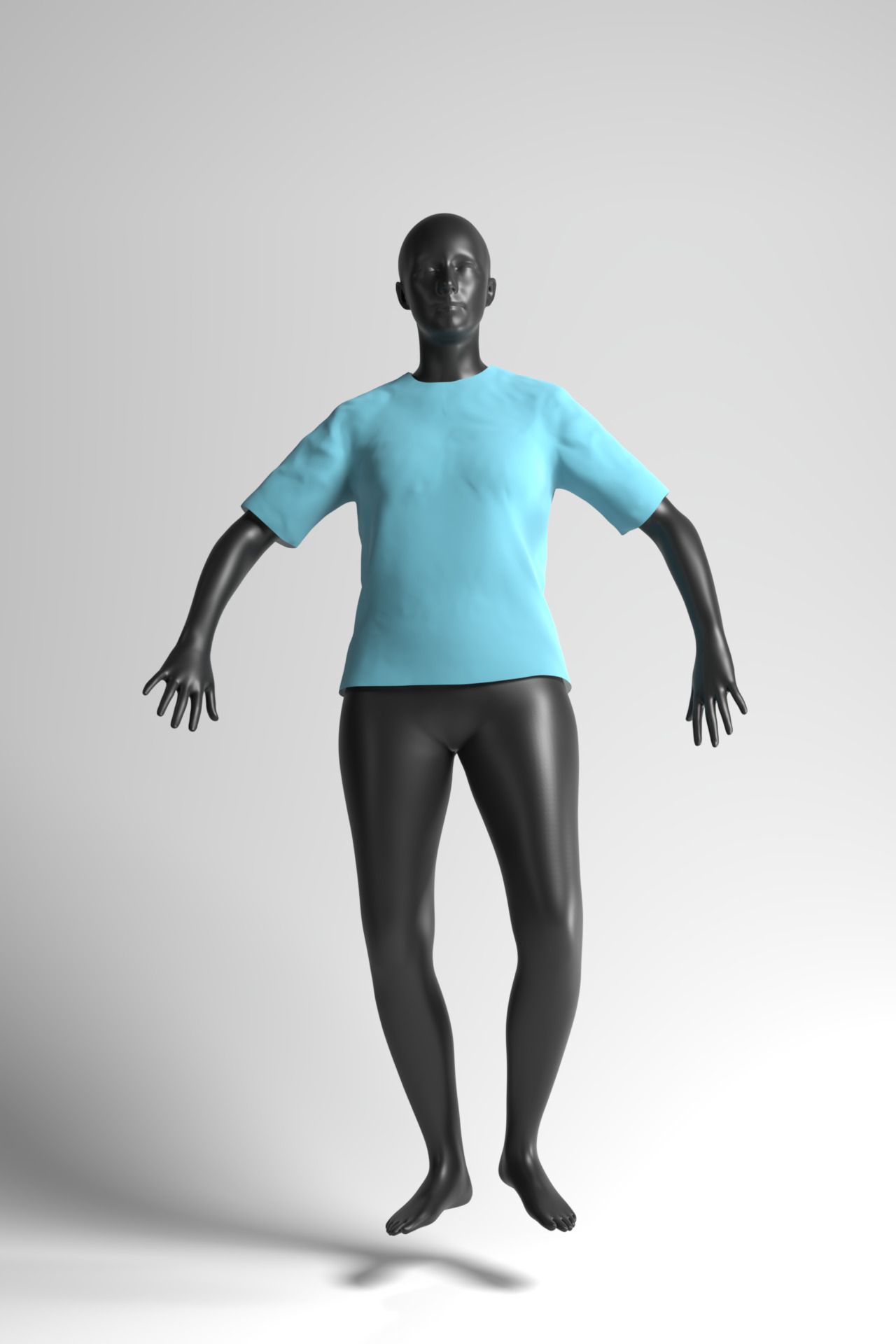} &
        \includegraphics[width=0.123\linewidth]{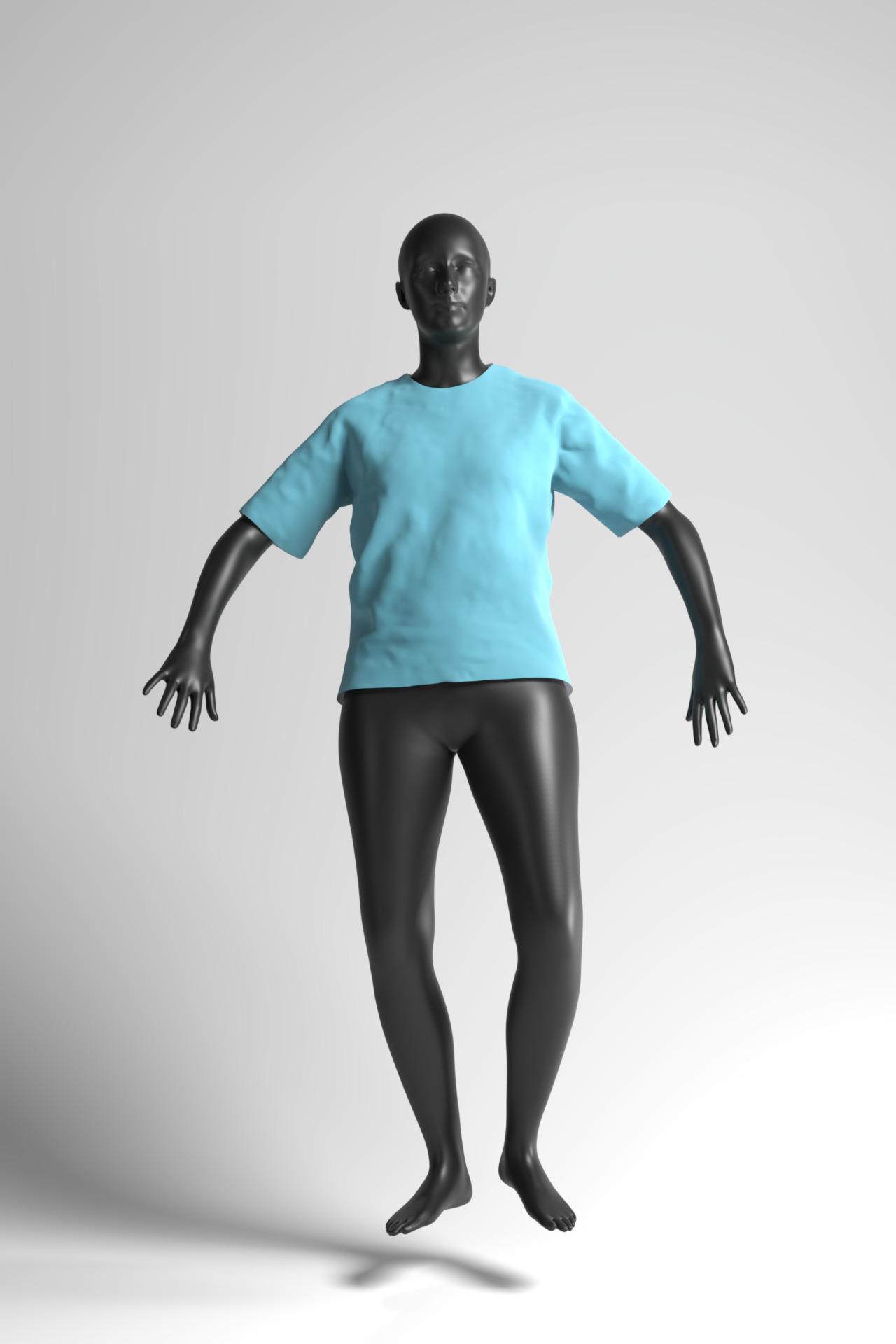} &
        \includegraphics[width=0.123\linewidth]{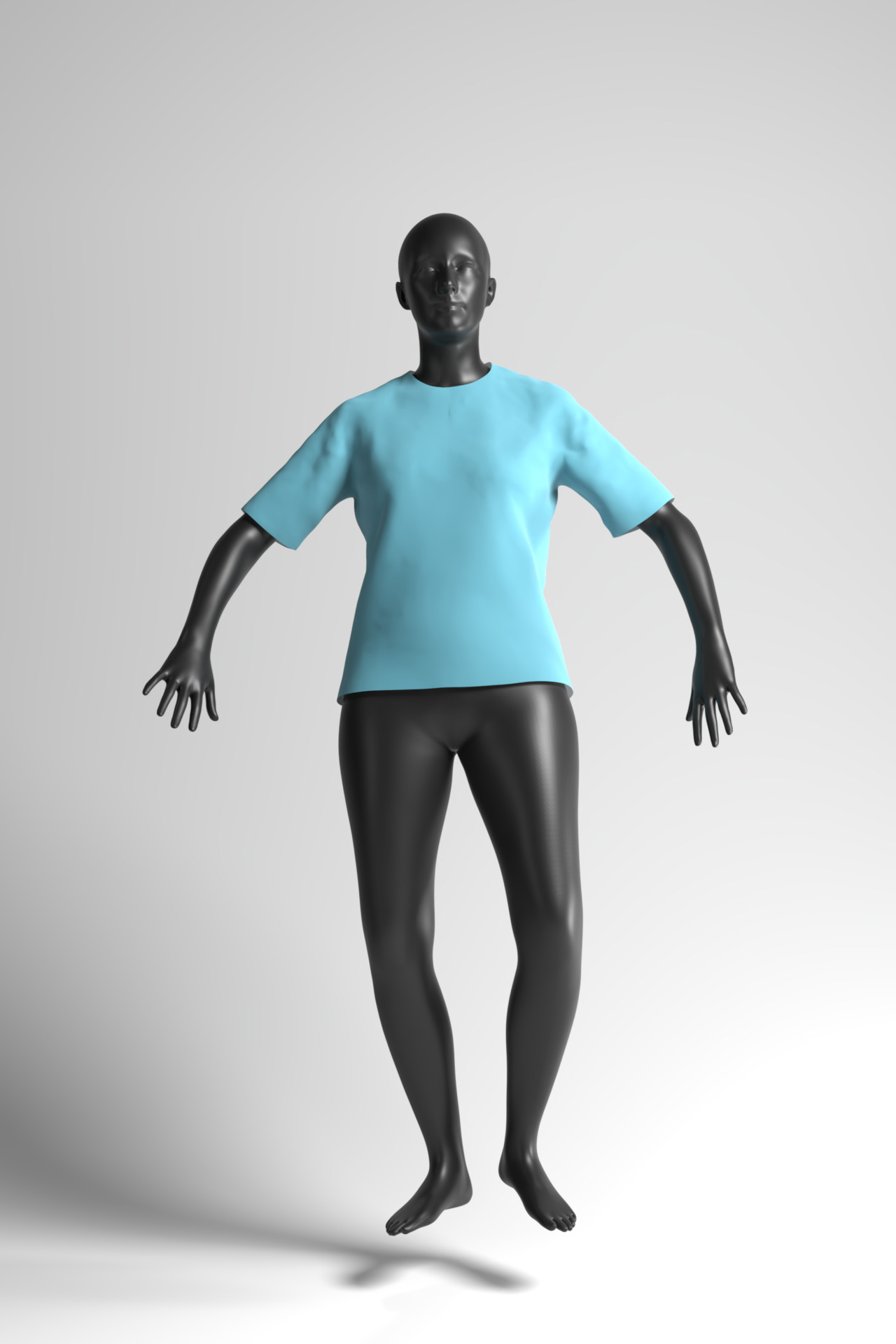} &
        \includegraphics[width=0.123\linewidth]{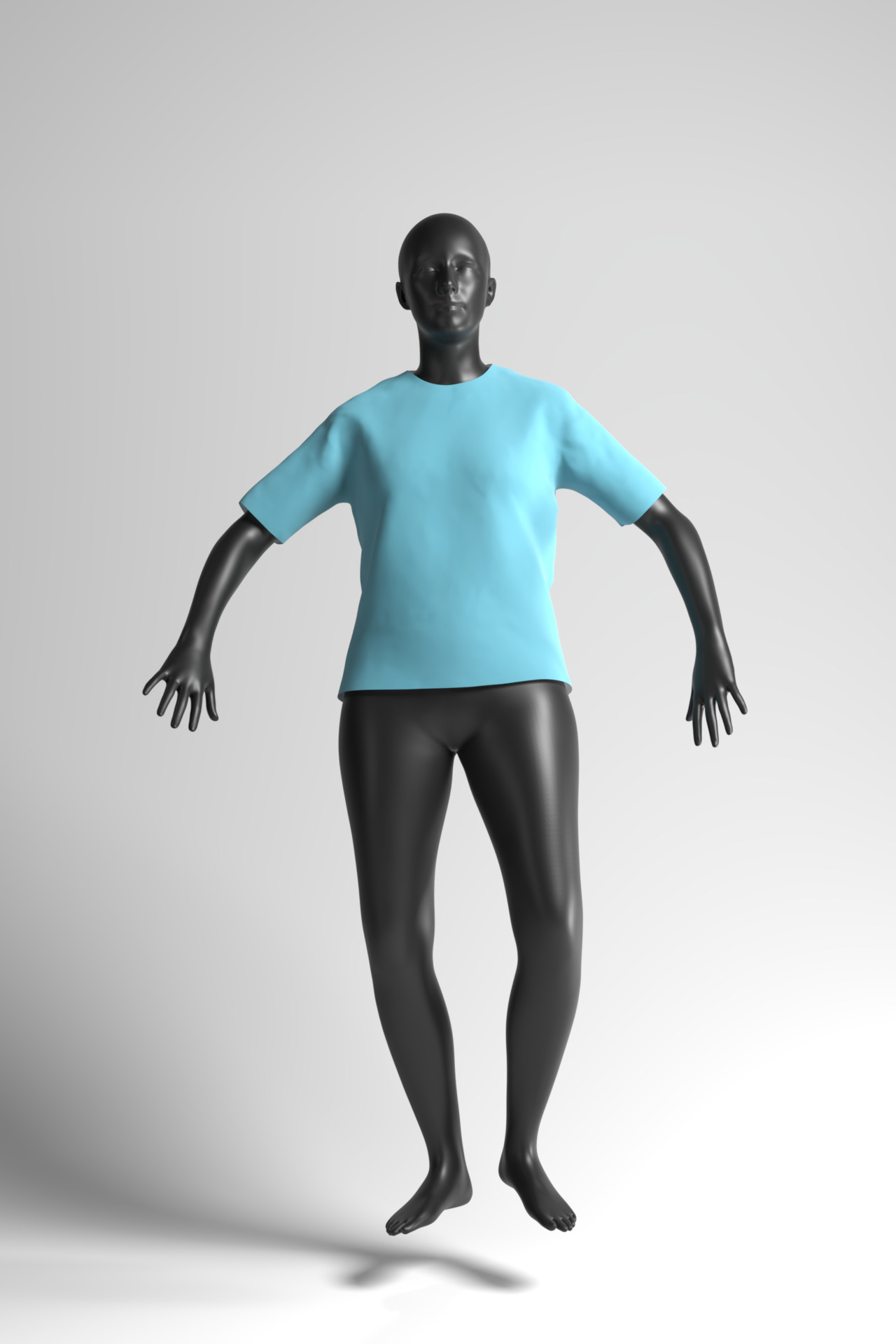} &
        \includegraphics[width=0.123\linewidth]{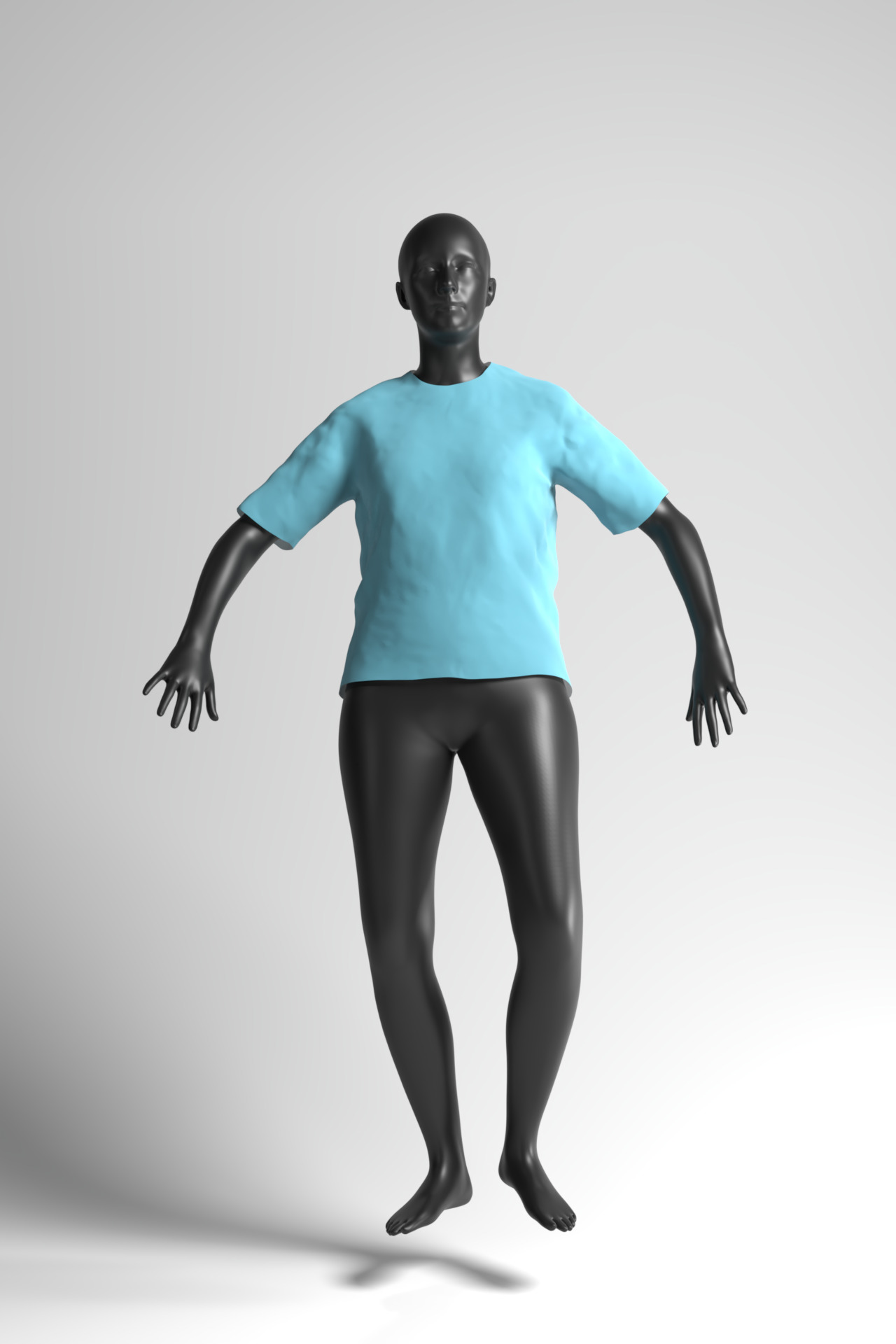} &
        \includegraphics[width=0.123\linewidth]{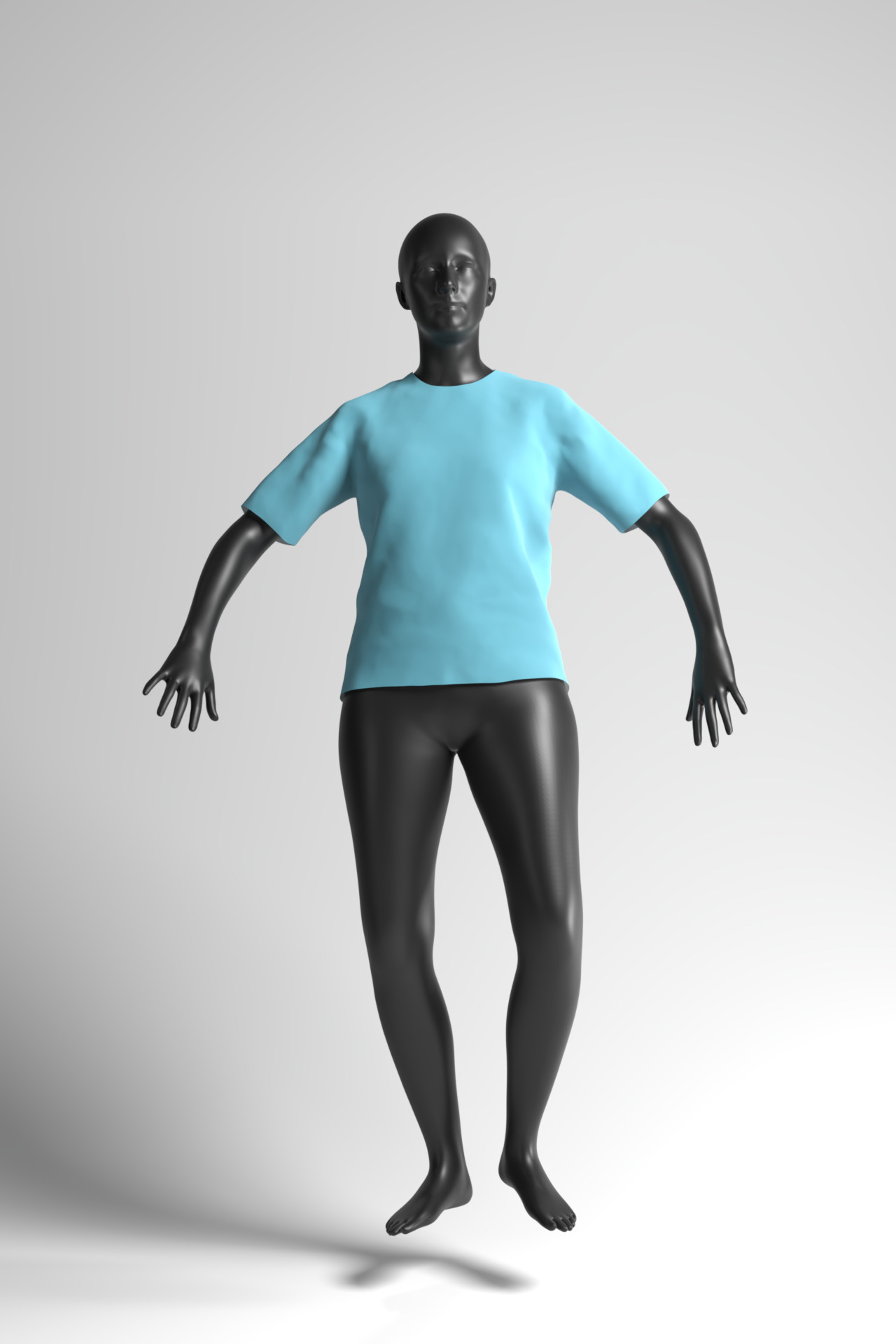} &
        \includegraphics[width=0.123\linewidth]{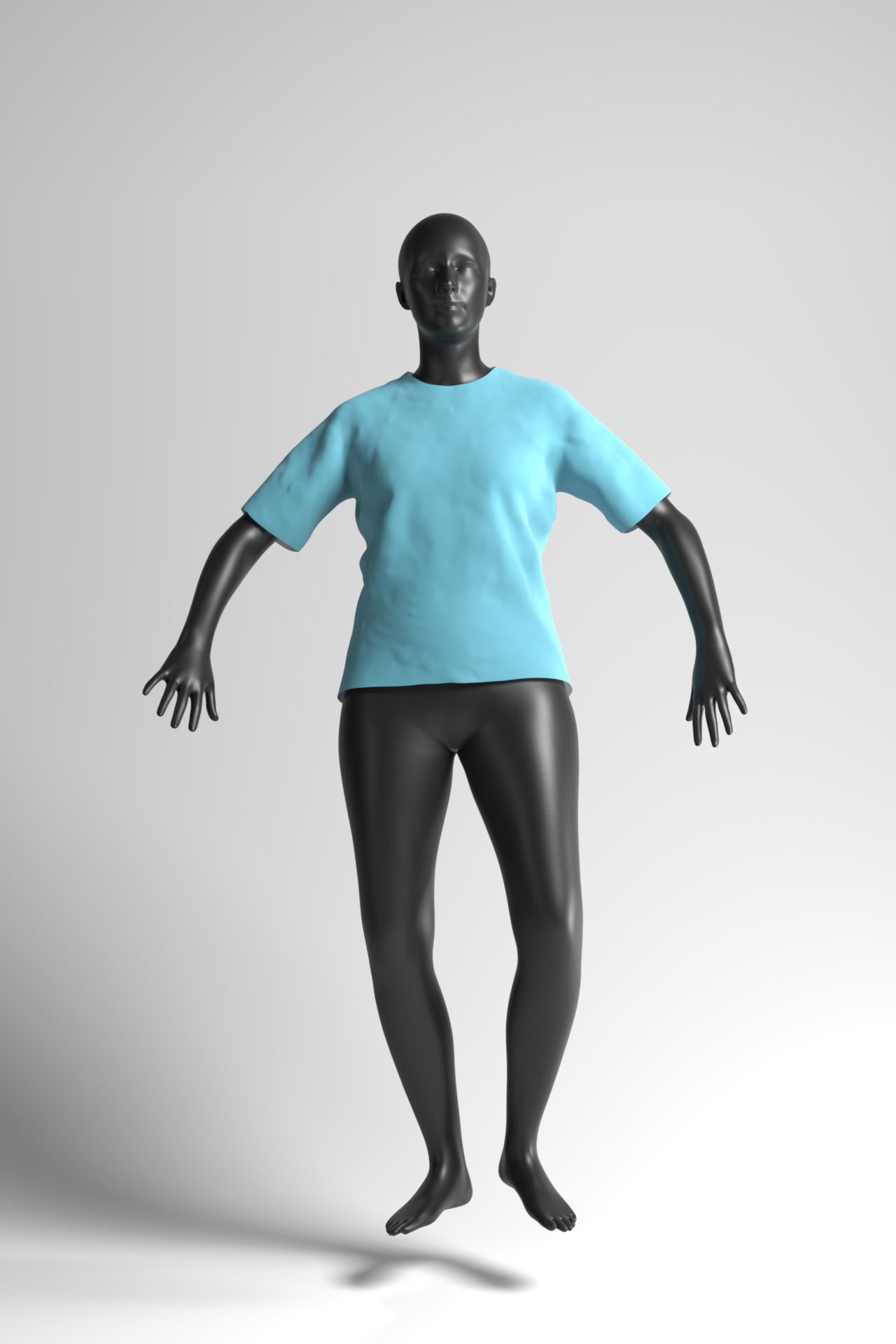} &
        \includegraphics[width=0.123\linewidth]{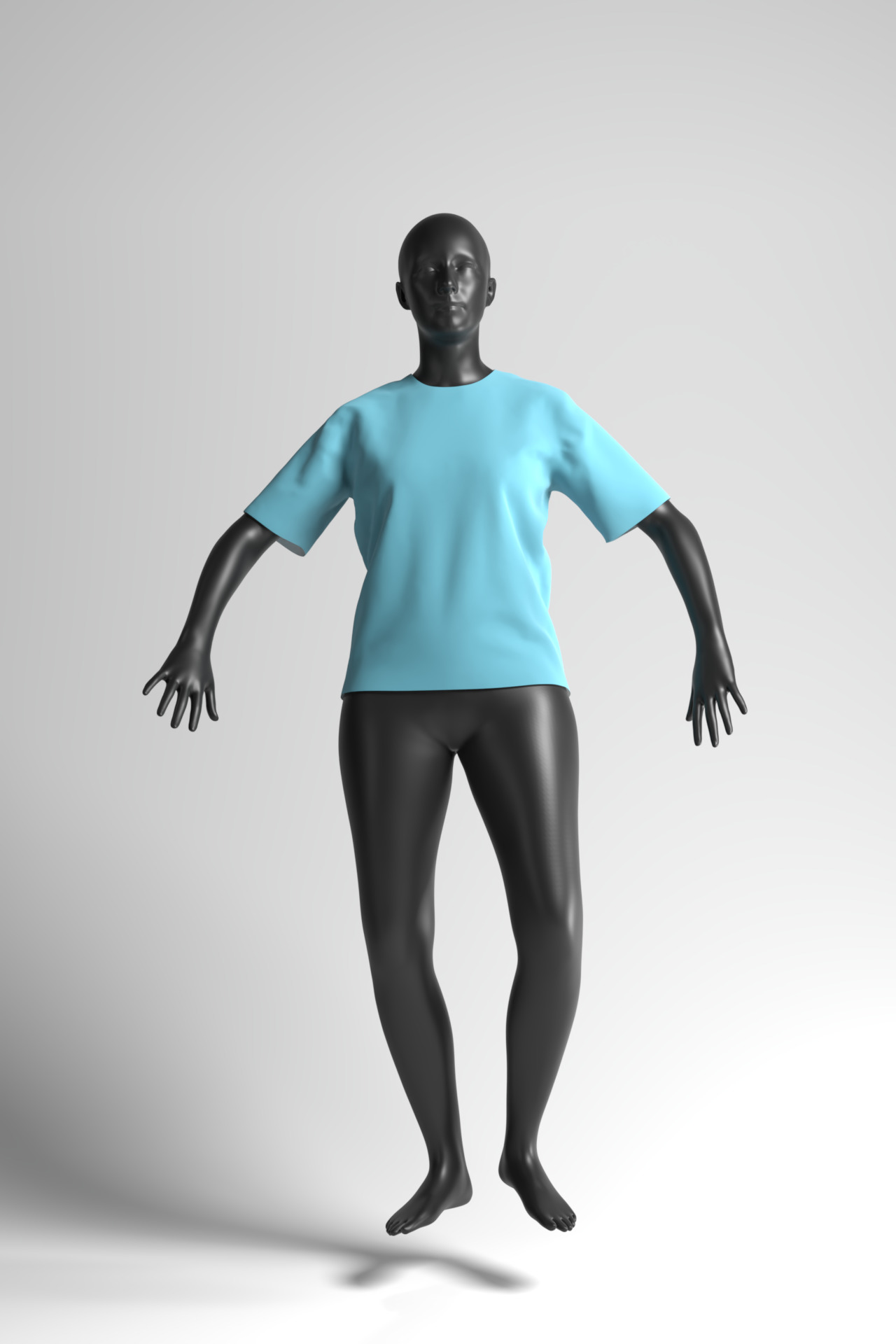}\\

        \includegraphics[width=0.123\linewidth]{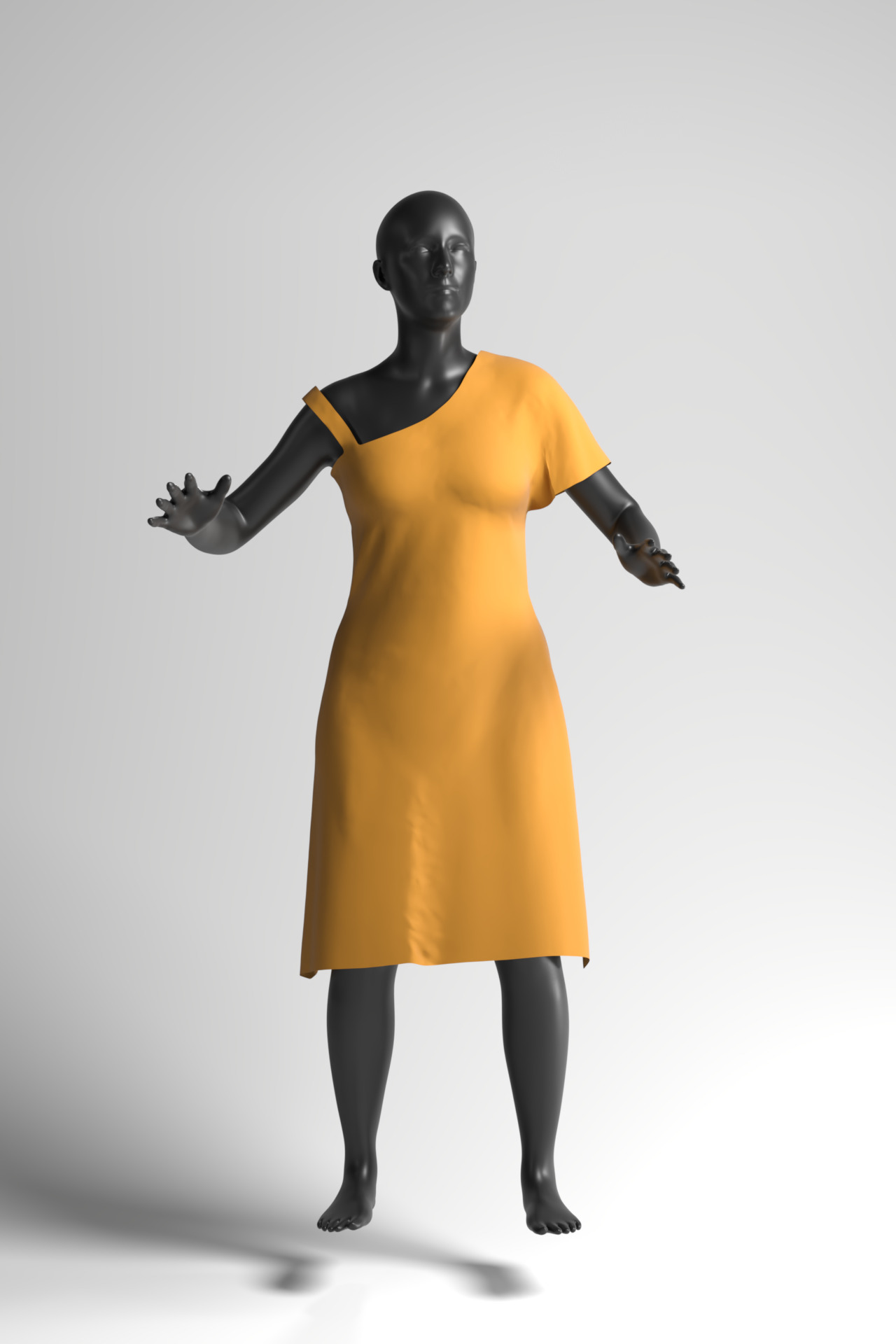} &
        \includegraphics[width=0.123\linewidth]{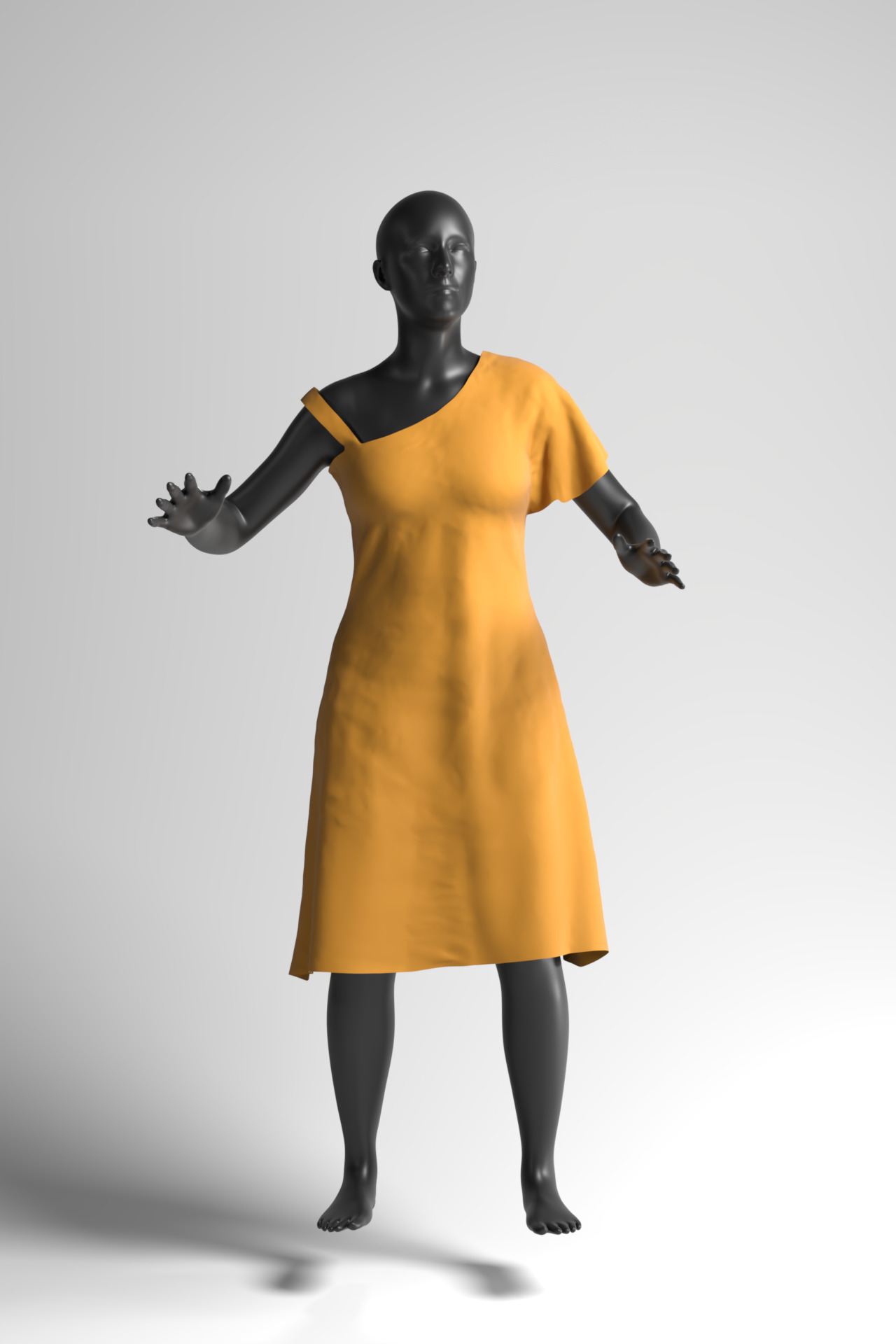} &
        \includegraphics[width=0.123\linewidth]{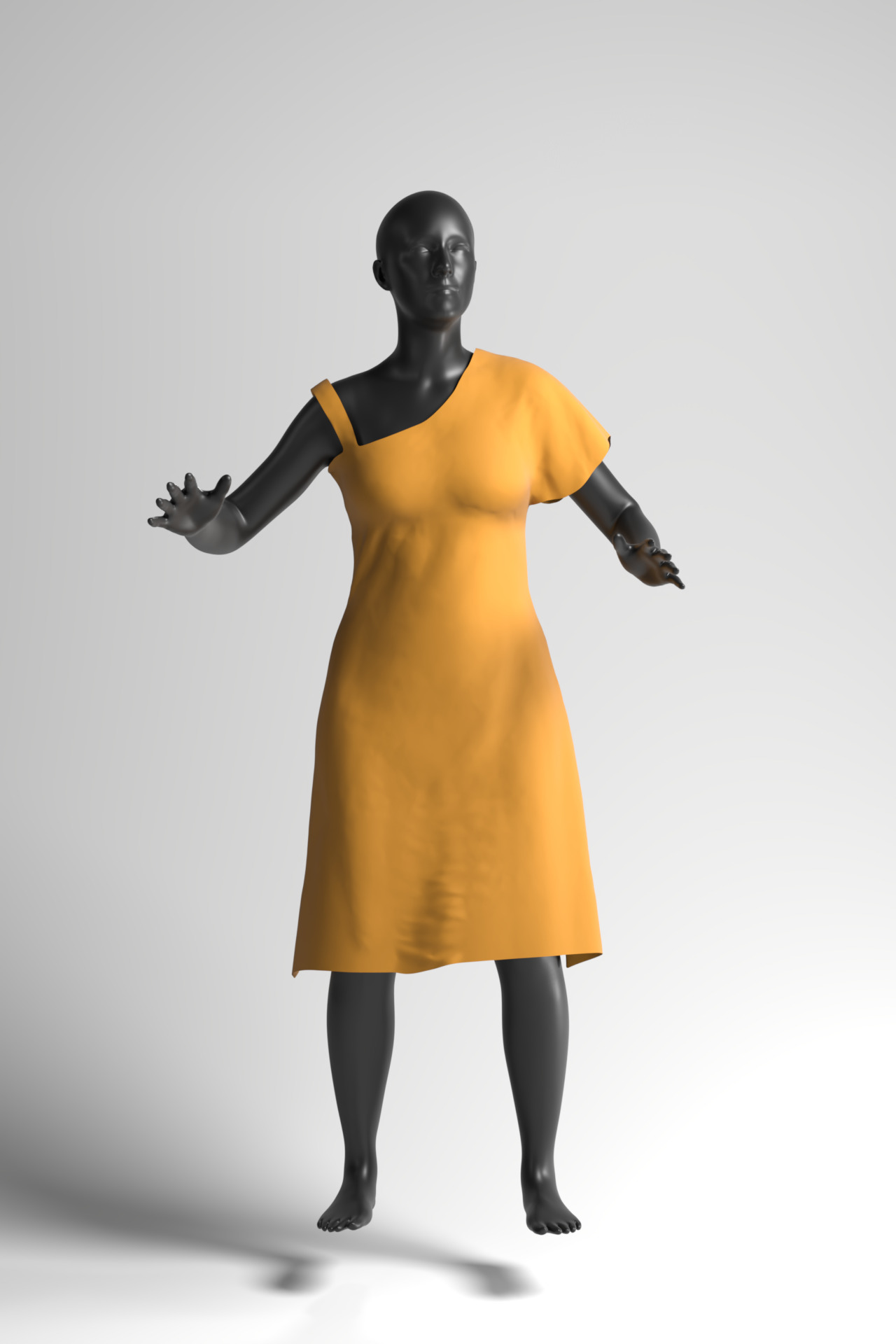} &
        \includegraphics[width=0.123\linewidth]{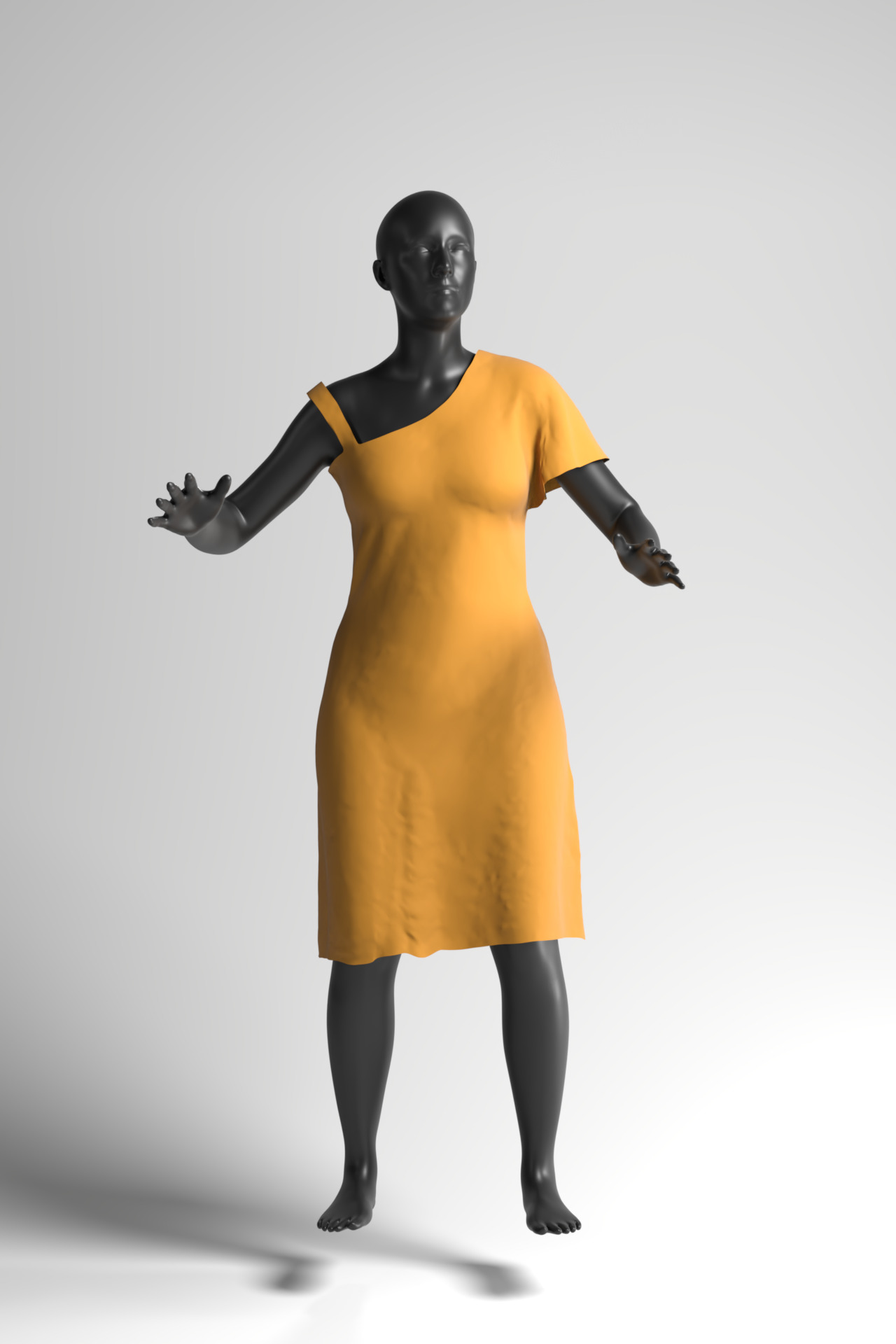} &
        \includegraphics[width=0.123\linewidth]{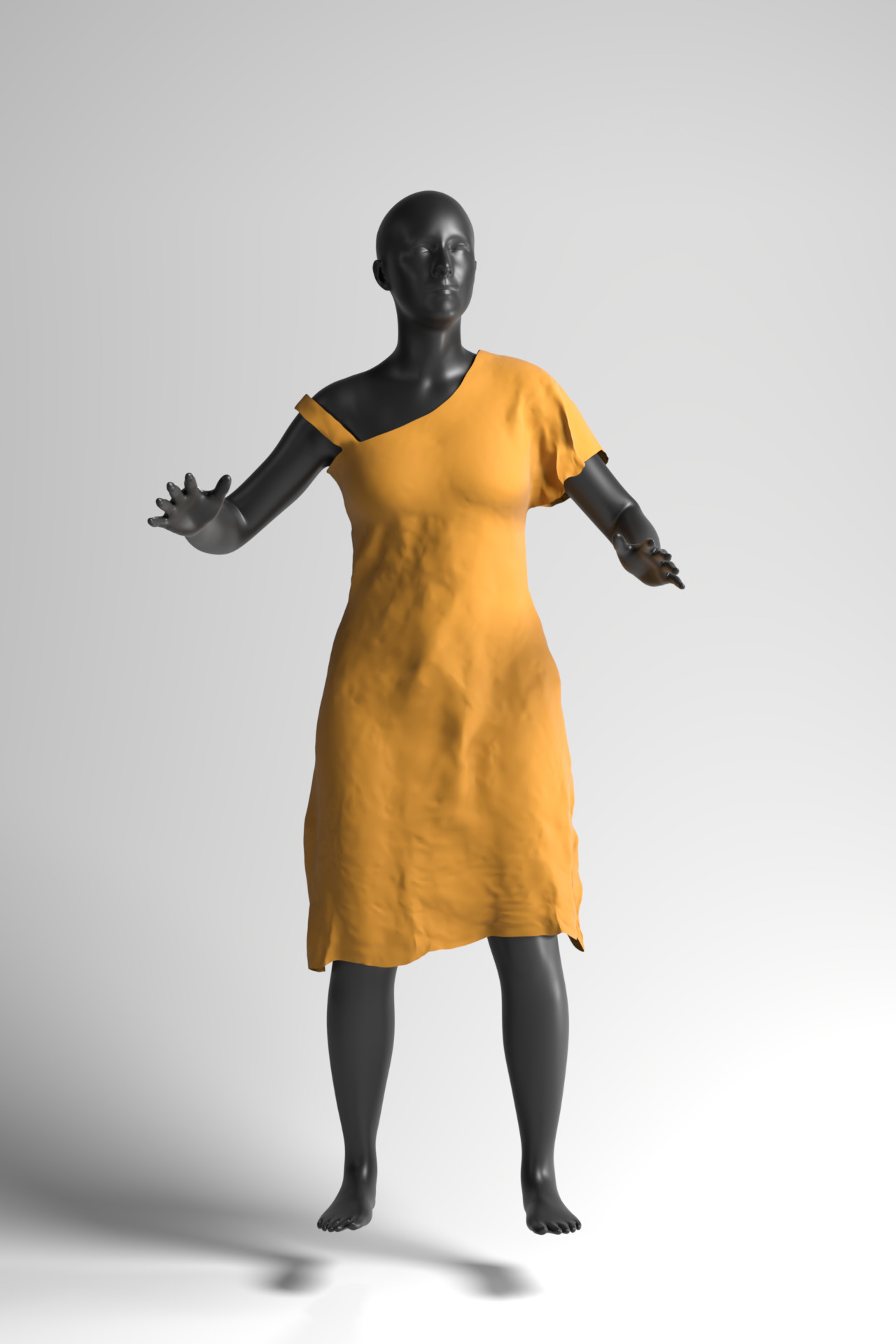} &
        \includegraphics[width=0.123\linewidth]{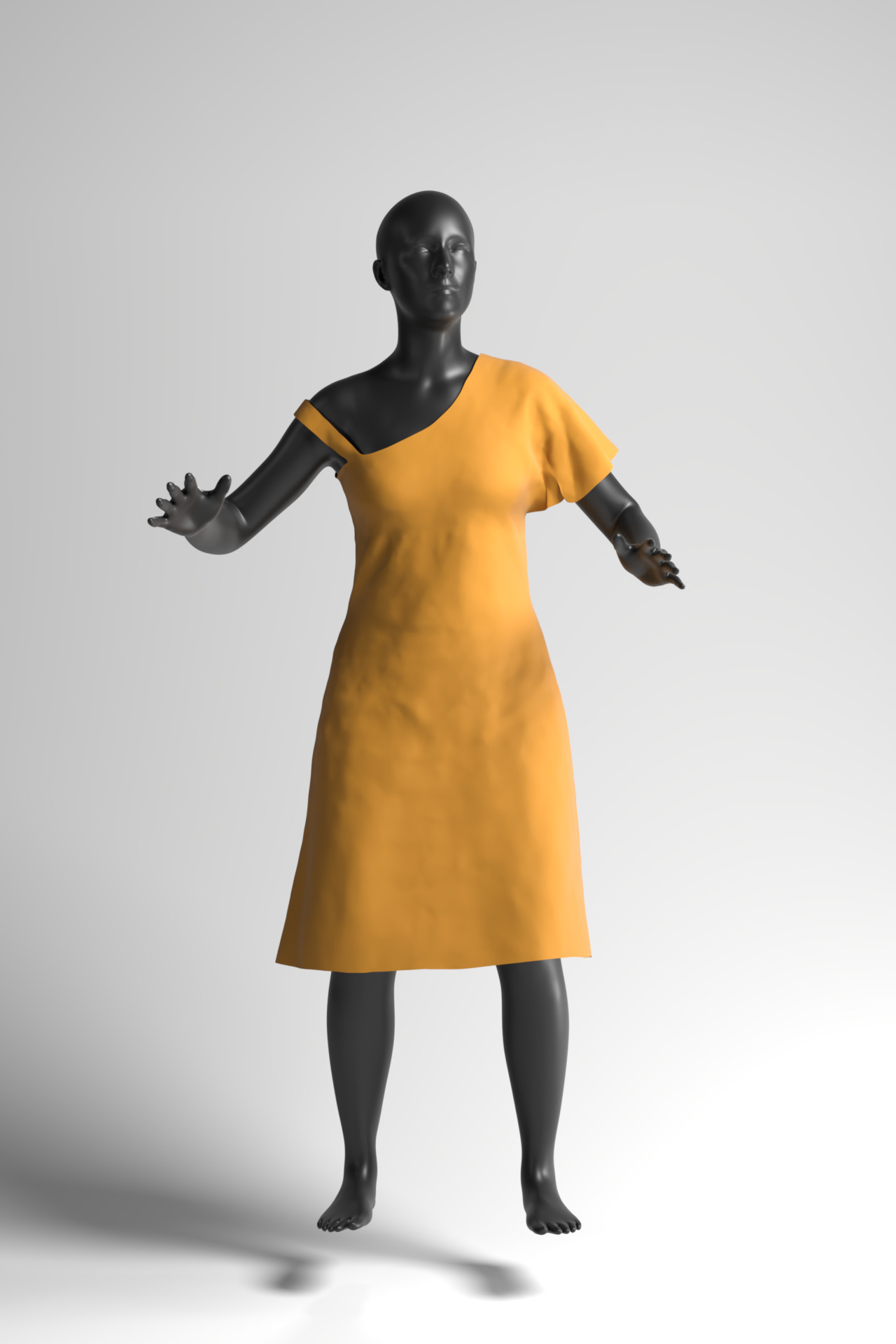} &
        \includegraphics[width=0.123\linewidth]{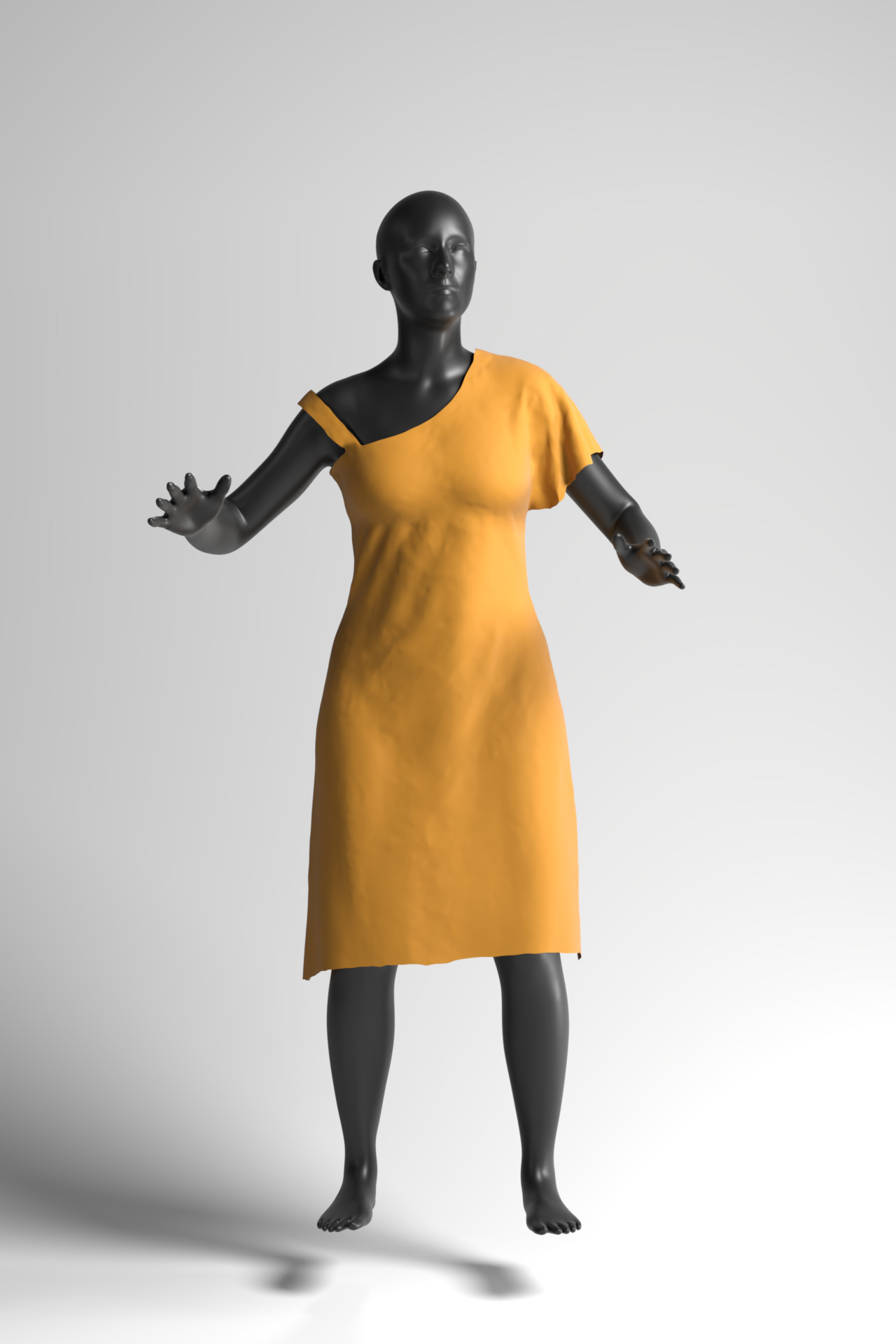} &
        \includegraphics[width=0.123\linewidth]{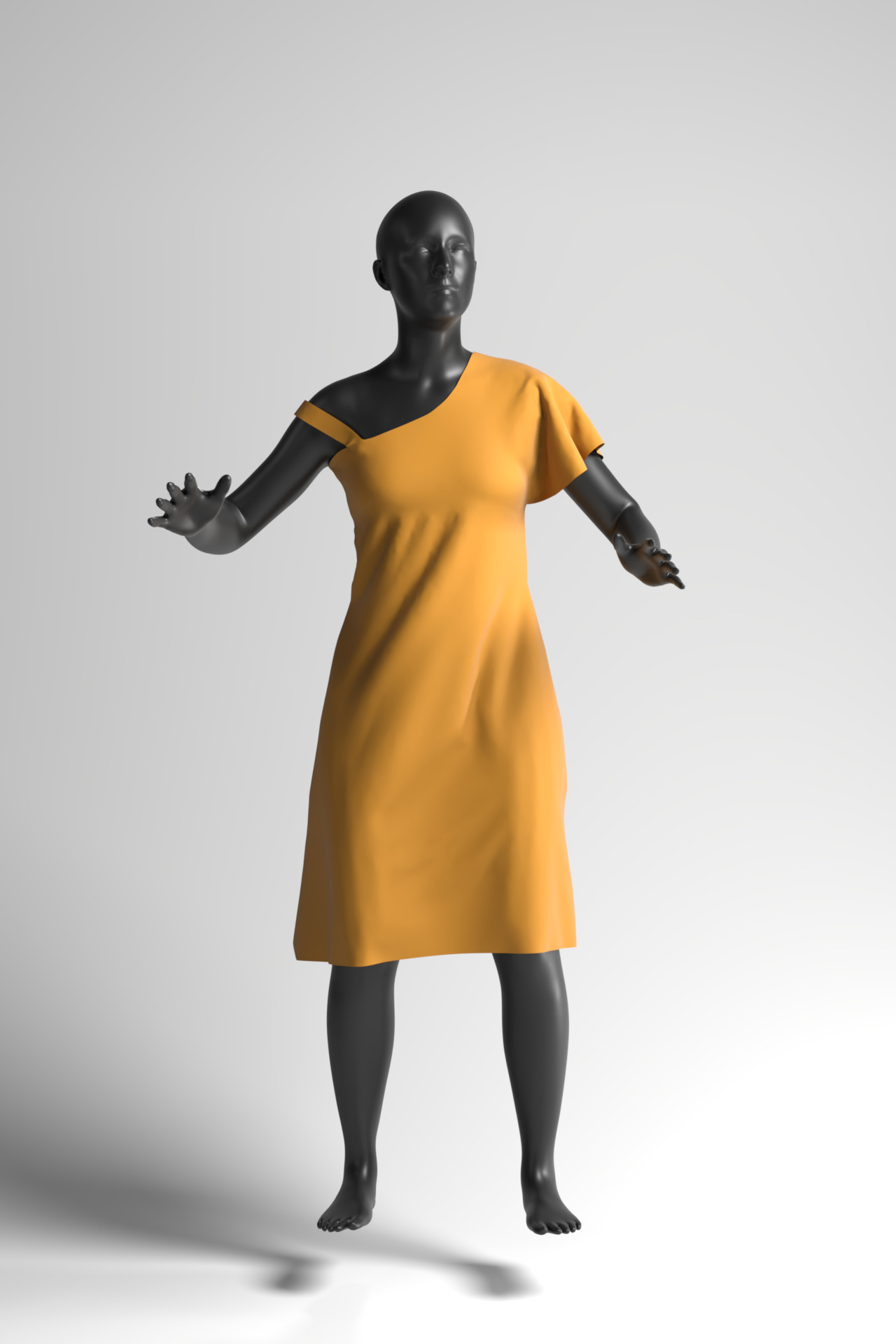}\\

		Maya 2025.2 & Unreal Engine 5.6 & \cite{bailey_fast_2018}& \cite{patel_tailornet_2020}& \cite{chentanez_cloth_2020}& \cite{santesteban_self-supervised_2021}& Ours & Ground Truth
	\end{tabular}}
    \caption{Qualitative comparison of our method against two industrial tools and four real-time prior works on the \vto validation dataset, following the same setup as Figure~\ref{fig:ComparisonBaselineModelsOurs}. Baseline models exhibits overly smoothed results or high-frequency artifacts. Our model presents the best visual qualities.}
  \label{fig:ComparisonBaselineModelsVTO}
\end{figure*}

\paragraph*{Architecture details.}
Our VAE architecture operates on mesh vertex features using a mesh-aware residual SpiralConv \cite{gong_spiralnet_2019} backbone. Each resolution level consists of an encoder block followed by a symmetric decoder block. The encoder applies a residual SpiralConv stack (GELU + SpiralConv + LayerNorm) to provide a strong locality bias directly on the surface while remaining more structure-aware than MLPs and transformers \cite{chandran_shape_2022}. We then reduce resolution using Quadric Error Metric (QEM) pooling; the decoder mirrors this structure with the corresponding unpooling operator. To mitigate resampling artifacts, we apply a Laplacian smoothing operator before each QEM pooling and unpooling. In all experiments we use a hierarchy with 3 levels of mesh resolutions and QEM pooling matrices. The channel widths increase from 64 at the finest level to 512 at the coarsest. Within each block, we use 5 residual sub-layers per level and pose conditioning is injected through the mechanisms described in \Sec{mesh_aware_model}, while the rest-space displacement field is decoded at full resolution.

\paragraph*{Training.}
We train all models with AdamW~\cite{loshchilov_decoupled_2019} with $\beta_1=0.9$, $\beta_2=0.999$ and batch size 16. The VAE pre-training is run for 400{,}000 optimization steps, and each  progressive curriculum  stage is trained for 100{,}000 steps, followed by the pose deformer training for another 100{,}000 steps.
We use a cosine annealing schedule that decays the learning rate from $1{\times}10^{-4}$ to $1{\times}10^{-5}$ over each training horizon.
Poses $\theta$ are represented as $4\times4$ homogeneous transformation matrices of each joint, including rotation, scaling and translation which are often all animated in the production setting.
In addition to the local normal distribution loss, we use a weak latent supervision term  ($w_{\mathrm{lat}}=0.1$), computing MAE between predicted latent codes and the ground-truth ones from the pre-trained VAE. Additional architectural and training details are provided in \App{implementation}.

\subsection{Datasets}

Our dataset consists of a rigged human female character \dataset{Jen}, with three types of motions: (i) long artist-authored dancing animations; (ii) short motion clips sampled from a larger animation library; (iii) randomly sampled poses \cite{bailey_fast_2018}. Partition (i) consists of 1{,}264 frames over 6 sequences, and is used for validation. Partition (ii) has 9{,}090 frames and captures common actions such as walking, turning, or falling. To improve coverage of poses encountered at inference time, we extensively sample random joint transforms in Partition (iii), comprising 12{,}230 frames. This sampling is designed to cover a wide range of poses with large joint transforms.
Partitions (ii) and (iii) are used for training.

Partition (iii) and the majority of Partition (ii) consist of discontinuous poses, generated with quasi-static simulation. While susceptible to temporal inconsistencies and reduced secondary dynamics, this often remains the only practical option, as dynamic simulation between poses frequently triggers self-intersections that cause the simulator to fail. We also evaluate our method on the \vto dataset \cite{santesteban_learning-based_2019}, a simulated virtual try-on benchmark with a T-shirt and a long dress modeled on SMPL \cite{loper_smpl_2015} human body models. A single body shape from the dataset was used. Various motion captured sequences of talking, walking, jumping sum up to 6{,}665 frames of training and 452 frames for validation.

\subsection{Quantitative and Qualitative Evaluations}
\label{sec:quantitative_qualitative_evaluations}
\Fig{TeaserImage} and \Fig{ValidationFrames} show our model on representative validation frames from the \dataset{Jen} dataset. Despite complex motions and challenging wrinkling patterns, our model generates physically plausible garment deformations in real-time.
We further compare against state-of-the-art baselines from both academia and available commercial software. \Fig{ComparisonBaselineModelsOurs} summarizes qualitative comparisons using our internal dataset with the ML Deformers in Autodesk Maya 2025 \cite{autodesk_autodesk_2025} and Unreal Engine 5.6 \cite{epic_games_unreal_2025}, a per-joint MLP baseline adapted to garments \cite{bailey_fast_2018}, a two-stage MLP model that predicts coarse and fine details~\cite{patel_tailornet_2020}, and a mesh-convolutional model~\cite{chentanez_cloth_2020}. Across all above methods, our approach produces the most realistic fine-scale wrinkles and the best temporal coherence. \Fig{ComparisonBaselineModelsVTO} shows the same comparisons on the \vto datasets.
In some examples in \Fig{ComparisonBaselineModelsVTO}, both our model and several baselines exhibit minor high-frequency artifacts or reduced smoothness relative to the simulation ground truth, suggesting that these artifacts are not specific to a single loss formulation or architecture.
We report Mean Absolute Error (MAE) of all methods in \Tab{ComparisonBaselineModels}. We also report the measurements from training our method with the traditional reconstruction losses (Ours w/o NormDist).
Our approach outperforms the other methods in terms of MAE across different datasets. Inference runtimes on \jenShirt dataset are also reported in \Tab{ExtendedEvaluation}. Our method can run inference with 212.2 FPS. We demonstrate its real-time capability through a posing application in \Fig{TeaserImage} (left) and in the supplemental video.

\begin{table}
    \centering
    {\footnotesize
    \rowcolors{2}{grayband}{white}
    \resizebox{\linewidth}{!}{
	\begin{tabular}{ccccc}
		\toprule
		Model    &  \jenShirt   & \jenPants & VTO T-Shirt & VTO Dress \\
		\midrule
		Maya 2025.2 & \mn{0.4740} & \mn{0.2076} & \mn{0.0063} & \mb{0.0117} \\
		Unreal Engine 5.6 & \mn{0.7617} & \mn{0.2487} & \mn{0.0063} & \mn{0.0148} \\
        \cite{bailey_fast_2018} & \mn{0.4202} & \mb{0.1676} & \mb{0.0061} & \mn{0.0144} \\
        \cite{patel_tailornet_2020} & \mb{0.4133} & \mn{0.2177} & \mn{0.0068} & \mg{0.0108} \\
        \cite{chentanez_cloth_2020} & \mn{0.4479} & \mn{0.2105} & \mb{0.0061} & \ms{0.0111} \\
        \cite{santesteban_self-supervised_2021} & \mn{-} & \mn{-} & \mg{0.0057} & \ms{0.0111} \\
		Ours w/o NormDist & \mg{0.3196} & \mg{0.0845} & \ms{0.0060} & \mg{0.0108} \\
        Ours & \ms{0.3605} & \ms{0.0849} & \mn{0.0063} & \mg{0.0108} \\
		\bottomrule
	\end{tabular}
    }
    }
	\caption{Mean Absolute Error (MAE) comparisons with previous work across four scenes. Our method achieves highest quality measurements on the validation dataset.
    Colored dots denote the \protect\mg{best} and \protect\ms{second best} results.}
	\label{tab:ComparisonBaselineModels}
\end{table}

\begin{table}
    \centering
    {\footnotesize
    \setlength{\tabcolsep}{2.pt}
    \rowcolors{2}{grayband}{white}
    \resizebox{\linewidth}{!}{
	\begin{tabular}{ccccccc}
		\toprule
		Model & MAE & Curv  & Contour & Collision & LPIPS & FPS \\
		\midrule
		Maya 2025.2 & \mn{0.4740} & \mn{0.00406} & \mn{4.4137} & \mn{0.0660} & \mn{0.1745} & \mg{1352.6}\\
		Unreal Engine 5.6 & \mn{0.7617} & \mn{0.00242} & \mn{11.8196} & \mn{0.1180} & \mn{0.2436} & \mb{\phantom{1}660.8}\\
		\cite{bailey_fast_2018}& \mn{0.4202} & \mn{0.00318} & \mb{4.1687} & \mb{0.0467} & \mb{0.1620} & \mn{\phantom{1}595.8}\\
		\cite{patel_tailornet_2020}& \mb{0.4133} & \mn{0.00360} & \ms{3.6534} & \mn{0.0524} & \mn{0.1661} & \ms{1295.5}\\
		\cite{chentanez_cloth_2020}& \mn{0.4479} & \ms{0.00186} & \mn{4.6055} & \mn{0.0541} & \mn{0.1783} & \mn{\phantom{1}336.7}\\
		Ours w/o NormDist & \mg{0.3196} & \mb{0.00211} & \mg{2.8876} & \mg{0.0147} & \mg{0.1458} & \mn{\phantom{1}212.2}\\
		Ours & \ms{0.3605} & \mg{0.00173} & \mn{4.5713} & \ms{0.0156} & \ms{0.1478} & \mn{\phantom{1}212.2}\\
		\bottomrule
	\end{tabular}}}
	\caption{Extended evaluation for \jenShirt validation dataset, measuring Mean Absolute Error (MAE), mean curvature (Curv), distances of mesh boundries (Contour), collision rate (Collision), perceptual difference (LPIPS), and inference speed in frame per second (FPS). All experiments are performed on an NVIDIA GeForce RTX 4070Ti GPU. Colored dots denote the \protect\mg{best} and \protect\ms{second best} results.}
	\label{tab:ExtendedEvaluation}
\end{table}

\begin{figure}[t]
	\centering
	\setlength{\tabcolsep}{0.0020\linewidth}
	\begin{tabular}{ccc}
        \includegraphics[width=0.30\linewidth]{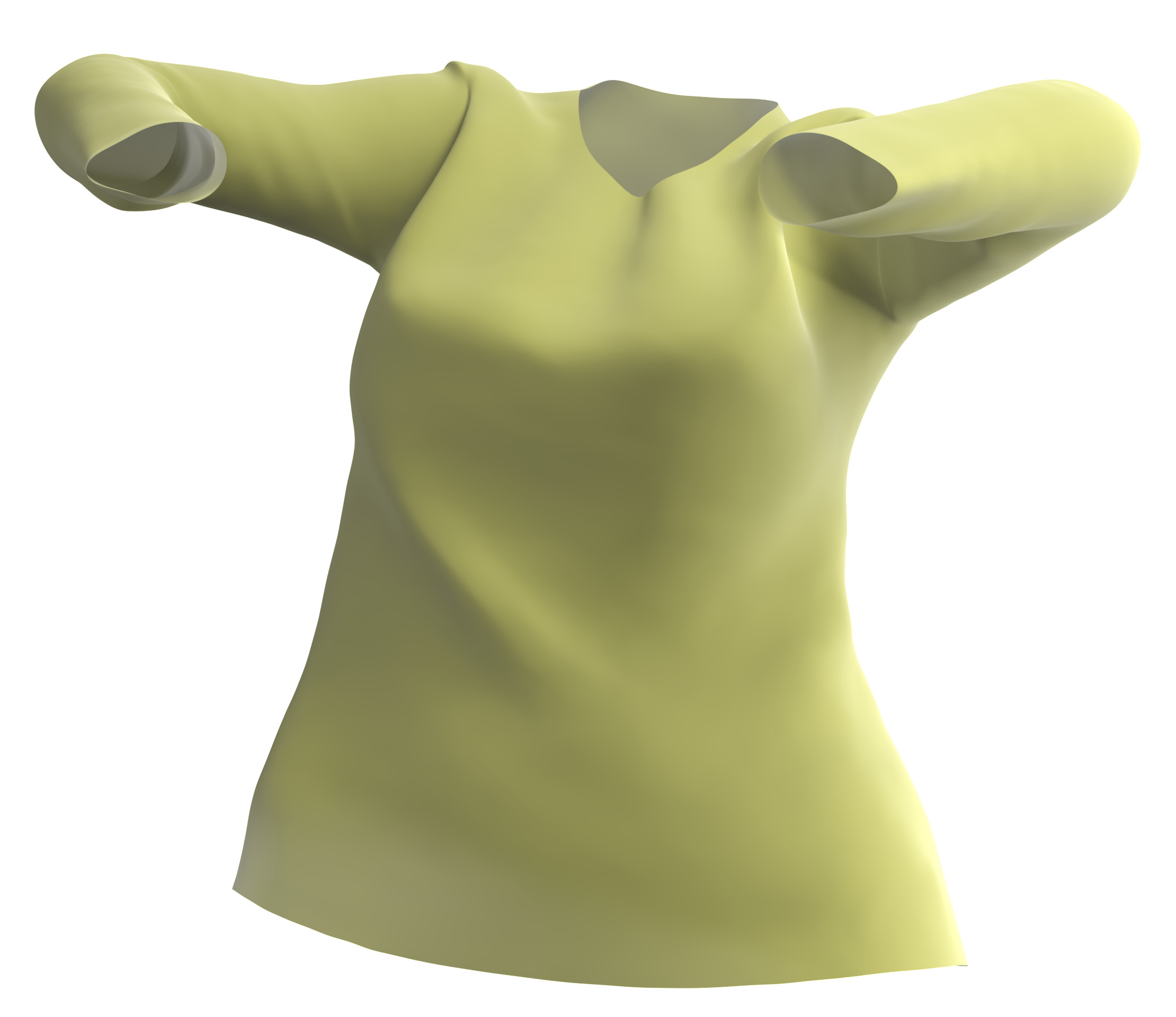} &
        \includegraphics[width=0.30\linewidth]{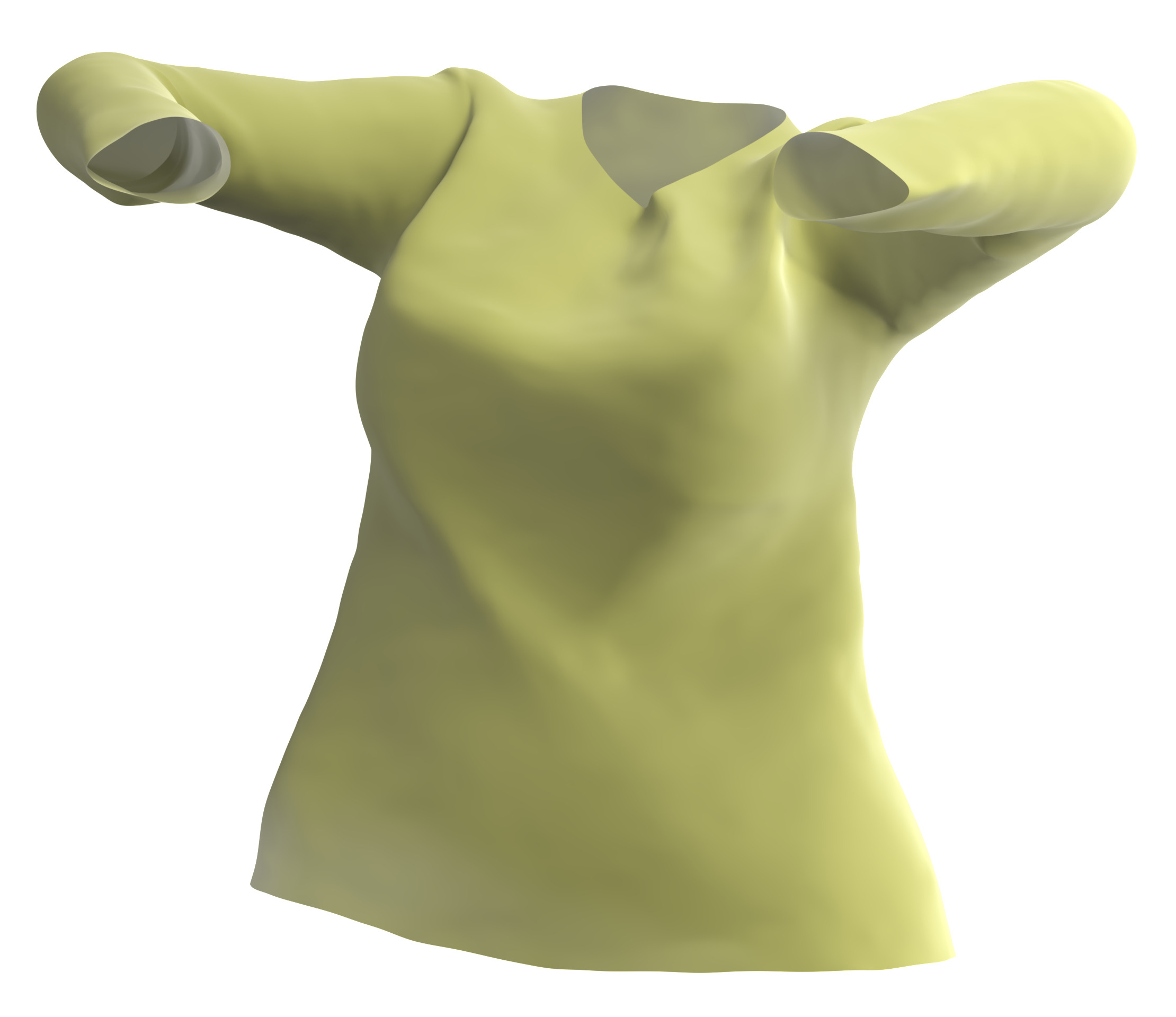} &
        \includegraphics[width=0.30\linewidth]{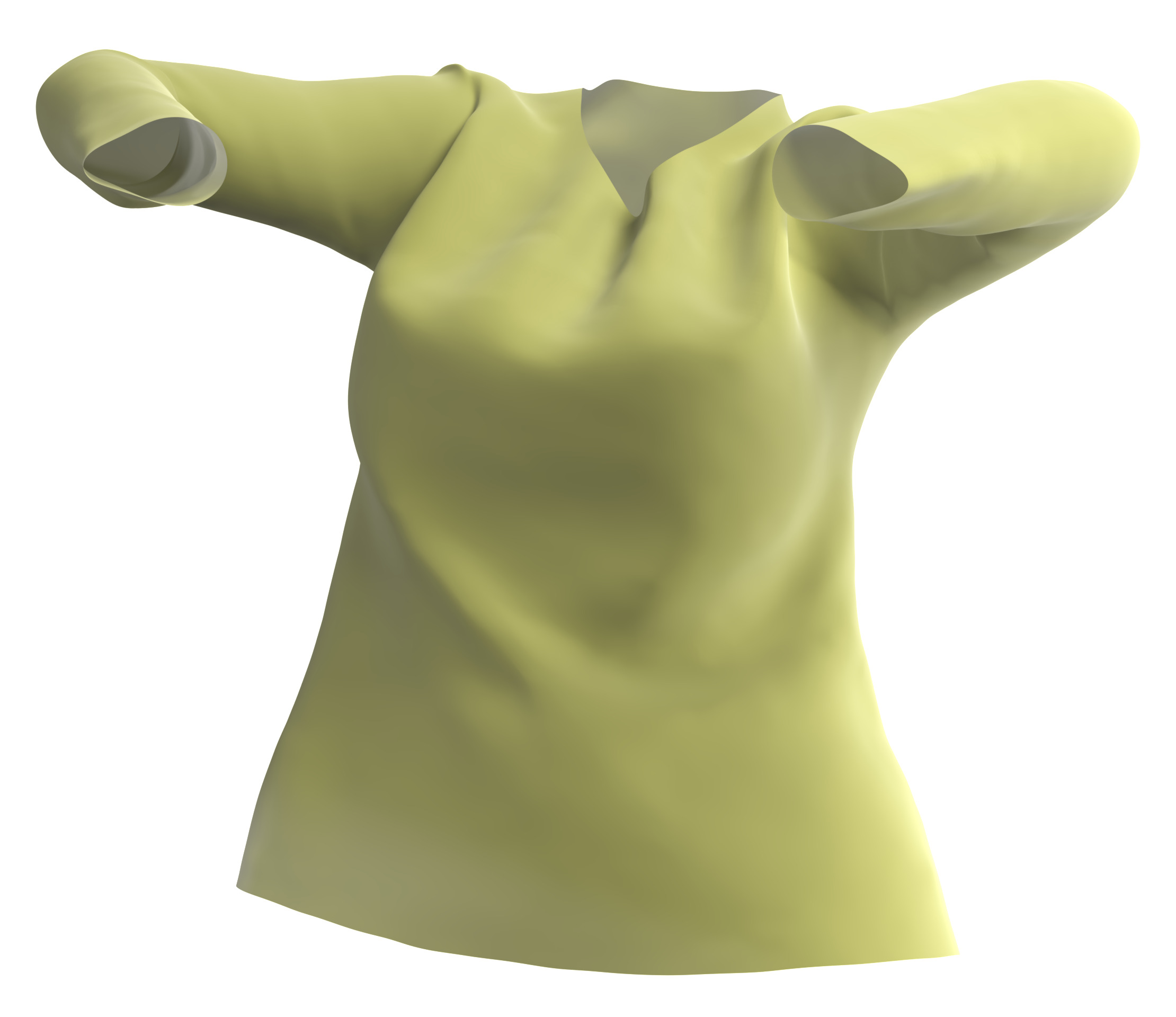} \vspace{-0.2cm}\\

        \footnotesize{(a) Per-joint MLP} &\footnotesize{(b) Splatted Mesh Features} & \footnotesize{(c) VAE Pre-training}\\

        \includegraphics[width=0.30\linewidth]{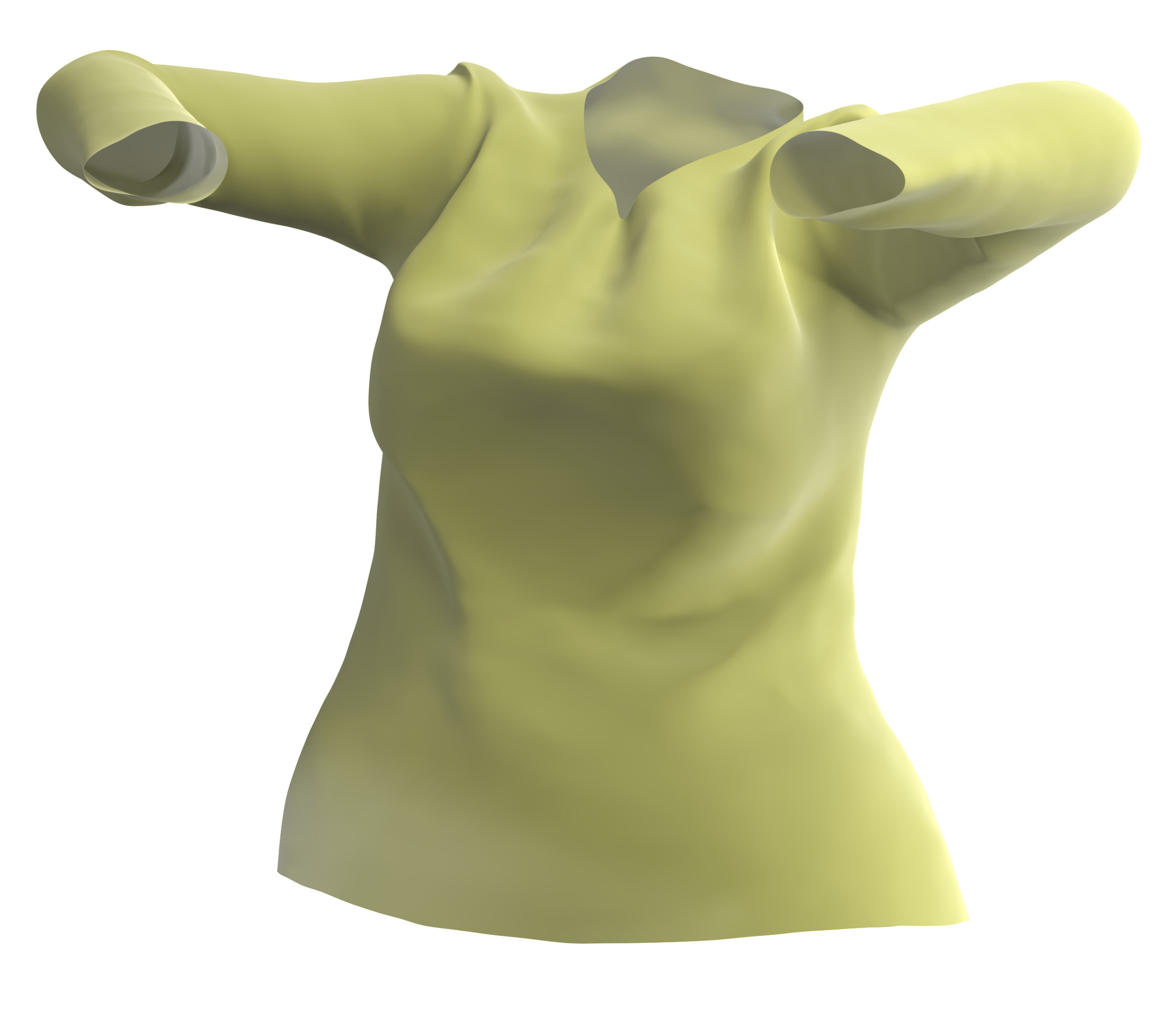} &
        \includegraphics[width=0.30\linewidth]{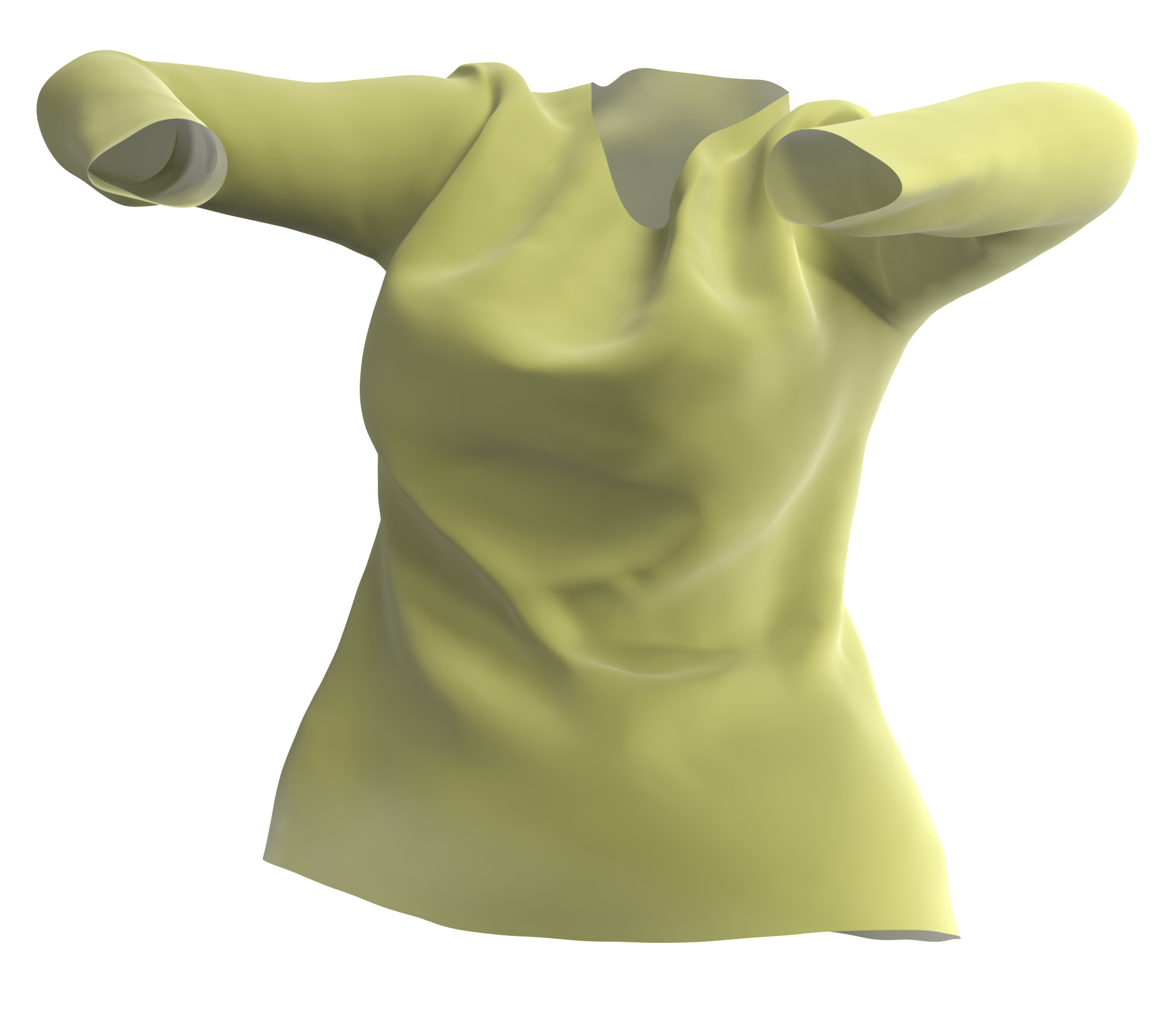} &
        \includegraphics[width=0.30\linewidth]{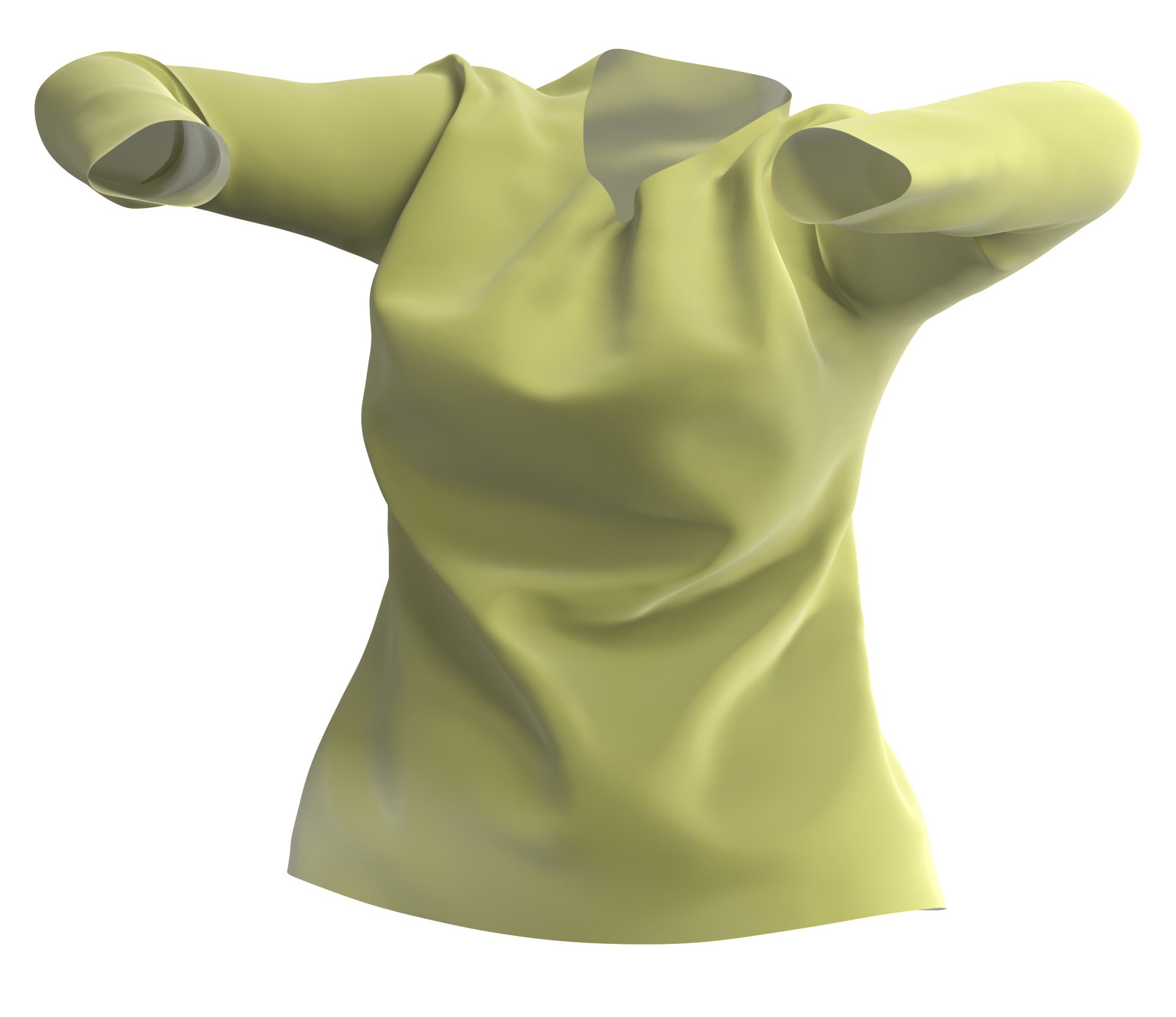} \vspace{-0.2cm}\\
        \footnotesize{(d) Progressive Training} & \footnotesize{(e) NormDist Loss} & \footnotesize{(f) Ground Truth}
	\end{tabular}
	\caption{Visual ablations of the main components of our pipeline on \jenShirt dataset. From (a) to (f), the component denoted in the caption is added to the previous model. Each stage consistently improves wrinkle sharpness and perceptual details.}
    \label{fig:ModelImprovements}
\end{figure}

\begin{figure}[t]
    \begin{overpic}[width=0.92\linewidth]{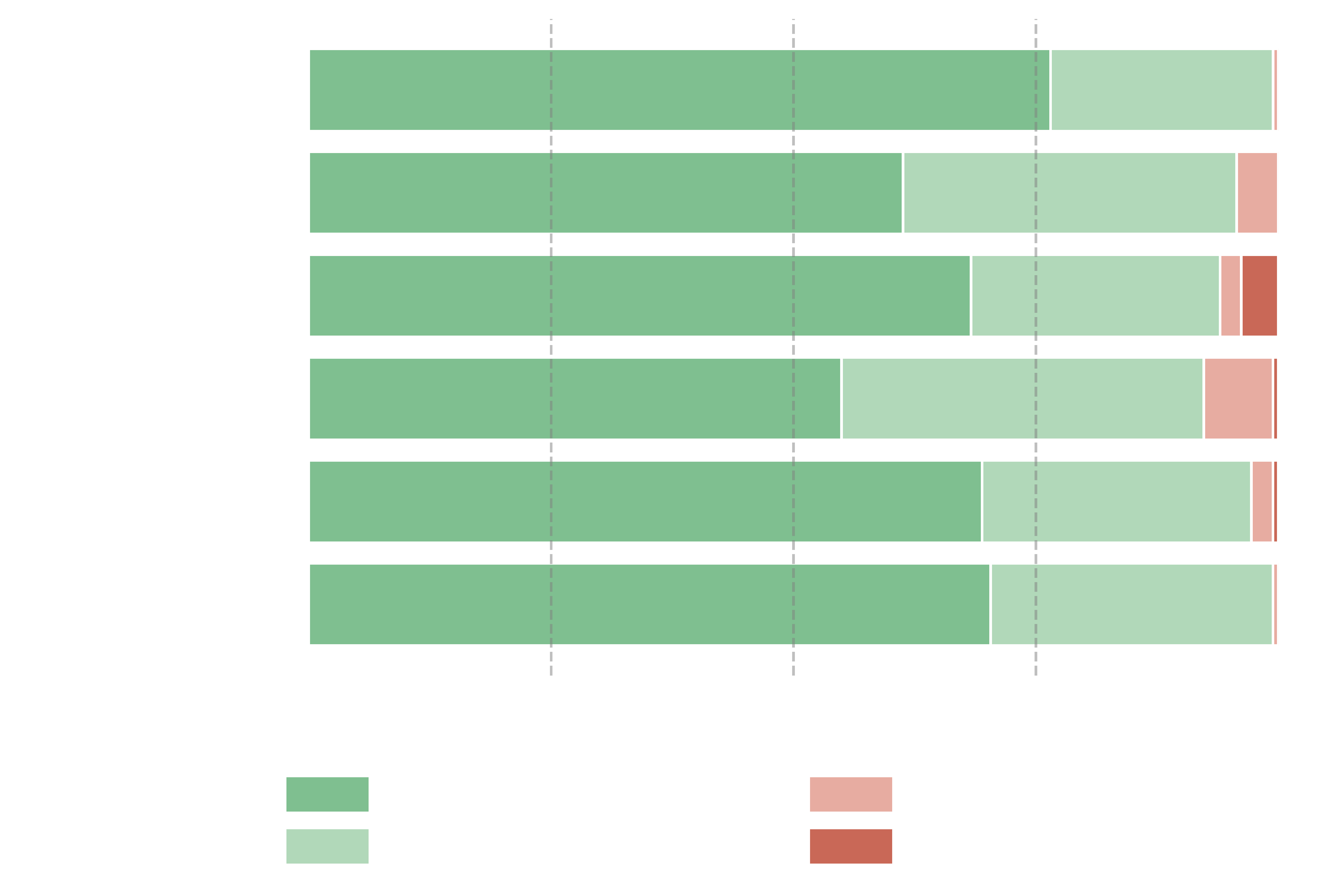}
        \put(21,60.5){\makebox(0,0)[r]{{\overlaylabel LBS}}}
        \put(21,52.78){\makebox(0,0)[r]{{\overlaylabel Maya 2025.2}}}
        \put(21,45.06){\makebox(0,0)[r]{{\overlaylabel Unreal Engine 5.6}}}
        \put(21,37.34){\makebox(0,0)[r]{{\overlaylabel \cite{bailey_fast_2018}}}}
        \put(21,29.62){\makebox(0,0)[r]{{\overlaylabel \cite{patel_tailornet_2020}}}}
        \put(21,21.9){\makebox(0,0)[r]{{\overlaylabel \cite{chentanez_cloth_2020}}}}

        \put(41.3,14.7){\makebox(0,0){{\overlaylabel \textcolor{mygray}{25\%}}}}
        \put(59.3,14.7){\makebox(0,0){{\overlaylabel \textcolor{mygray}{50\%}}}}
        \put(77.7,14.7){\makebox(0,0){{\overlaylabel \textcolor{mygray}{75\%}}}}

        \put(29.5,7.5){\makebox(0,0)[l]{{\overlaylabel strong preference ours}}}
        \put(29.5,3.6){\makebox(0,0)[l]{{\overlaylabel weak preference ours}}}
        \put(68.5,7.5){\makebox(0,0)[l]{{\overlaylabel weak preference others}}}
        \put(68.5,3.6){\makebox(0,0)[l]{{\overlaylabel strong preference others}}}
    \end{overpic} \hfill
    \caption{User study on cloth prediction quality. Each horizontal stacked bar shows the distribution of responses when comparing our method against a baseline or commercial system. Green indicates preference for our method, with dark and light shades denoting strong and weak preference respectively; red indicates preference for the alternative. Across all comparisons, the majority of participants preferred our results.}
    \label{fig:UserStudy}
\end{figure}

\paragraph*{Additional metrics.}
Pointwise distances often fail to reflect perceptual deformation quality: small spatial shifts of wrinkles can inflate MAE despite visually plausible results.
Additionally, low MAE does not guarantee a correct silhouette and physical~plausibility.
We therefore report complementary metrics that target specific failure modes. The \emph{Contour} metric evaluates global shape by computing a symmetric Chamfer distance between boundary vertex sets of the predicted and ground-truth garments.
The \emph{Collision} quantifies garment-body interpenetration: we measure the minimum displacement required to move garment vertices outside the body surface and report the mean.
\emph{LPIPS}~\cite{zhang_unreasonable_2018} is also reported by rendering garment images because perceptual metrics correlate more closely with human judgments than pixel-level errors.
Finally, to assess wrinkle qualities invariant to small phase shifts, we compare distributions of surface detail using mean curvature (\emph{Curv}): we form soft histograms over mean curvature magnitude of the vertices using 64 logarithmically spaced bins, and report the MAE between predicted and ground-truth histograms.

\Tab{ExtendedEvaluation} summarizes evaluations of additional metrics on \jenShirt dataset. Our method not only improves MAE over previous approaches, but also exhibits fewer collisions, better contour agreement, and higher perceptual similarity to the ground truth.
The local normal distribution loss exhibits a slight increase in MAE. This reflects a deliberate trade-off: by supervising local deformation statistics rather than exact positions, we encourage richer wrinkle structures and high-frequency details, even when deformations are not perfectly phase-aligned with the reference. However, this can occasionally over-emphasize sharp normal variations in ambiguous fold regions, which may appear as localized crease-like wrinkles that are less smooth than the physics-based ground truth. Additional results are provided in \App{additional_results}.

\paragraph*{User study.}
We conducted user study with 28 participants and collected 1096 votes. Participants were shown side-by-side comparisons of rendered frames of our method and baselines. They were asked to indicate their preference in terms of visual quality without knowing the source of images. The results are shown in \Fig{UserStudy}. In total, 96.3\% of the votes are in favor of our method.

\begin{table}[t]
	\centering
    {\footnotesize
    \resizebox{\linewidth}{!}{
	\begin{tabular}{ccccccc}
        \toprule
        \multicolumn{2}{c}{Input} &
        \multicolumn{2}{c}{Model} &
        \multicolumn{2}{c}{Training} &
        Metric \\
        \cmidrule(lr){1-2}
        \cmidrule(lr){3-4}
        \cmidrule(lr){5-6}
        \cmidrule(lr){7-7}
		Splatted\vphantom{g} & Body\vphantom{g} & Latent\vphantom{g} & VAE Pre-\vphantom{g} & Progressive & Loss\vphantom{g} & MAE\vphantom{g}\\
        Transforms & Deltas & Size & Trained & Training & \\
		\midrule

        $\checkmark$ & $\times$ &  86$\times$1 & $\checkmark$ & $\times$ & Pointwise & 0.4335\\
        \rowcolor{lightgreen}
        $\checkmark$ & $\times$ & 229$\times$1 & $\checkmark$ & $\times$ & Pointwise & 0.4170\\
        $\checkmark$ & $\times$ & 624$\times$3 & $\checkmark$ & $\times$ & Pointwise & 0.4185\\
        \midrule

        $\checkmark$ & $\times$     & 229$\times$1 & $\times$     & $\times$ & Pointwise & 0.4216\\
        $\times$     & $\checkmark$ & 229$\times$1 & $\times$     & $\times$ & Pointwise & 0.4220\\
        $\checkmark$ & $\checkmark$ & 229$\times$1 & $\times$     & $\times$ & Pointwise & 0.3713\\
        $\checkmark$ & $\times$     & 229$\times$1 & $\checkmark$ & $\times$ & Pointwise & 0.4170\\
        $\times$     & $\checkmark$ & 229$\times$1 & $\checkmark$ & $\times$ & Pointwise & 0.4361\\
        \rowcolor{lightgreen}
        $\checkmark$ & $\checkmark$ & 229$\times$1 & $\checkmark$ & $\times$ & Pointwise & 0.3566\\
        \midrule

        $\checkmark$ & $\checkmark$ & 229$\times$1 & $\checkmark$ & $\times$ & Pointwise & 0.3566\\
		$\checkmark$ & $\checkmark$ & 229$\times$1 & $\checkmark$ & $\checkmark$ & Pointwise & 0.3196\\
        $\checkmark$ & $\checkmark$ & 229$\times$1 & $\checkmark$ & $\times$     & NormDist & 0.3553\\
        \rowcolor{lightgreen}
        $\checkmark$ & $\checkmark$ & 229$\times$1 & $\checkmark$ & $\checkmark$ & NormDist & 0.3605\\
		\bottomrule
	\end{tabular}}}
	\caption{Ablation studies on the \jenShirt dataset. The top rows ablate on the latent size. The middle rows demonstrate that VAE pretraining and both types of condition splatting improve performance. In the bottom rows we show the advantage of progressive training, and the trade-off between wrinkle sharpness and MAE accuracy of our losses. }
    \label{tab:Ablation}
\end{table}

\subsection{Ablation Studies}

\Tab{Ablation} analyzes the contributions of the main design choices in our pipeline on \jenShirt dataset. We first study the effect of the VAE latent dimensionality by varying bottleneck sizes, measured in MAE (right column). Starting from \jenShirt mesh with $4820$ vertices ($4820\times 3$ degrees of freedom), we evaluate three bottlenecks: $624\times 3$ ($\sim 7.8\times$ compression), $229\times 1$ ($\sim 64\times$ compression), and $86\times 1$ ($\sim 228\times$ compression). Size $229\times 1$ provides the best trade-off between compression and reconstruction quality. This setting is used in all subsequent experiments. Then, we ablate the pose-conditioning components and training strategies (middle rows).
Incorporating joint-to-mesh splatted pose features, body displacement cues, and VAE pre-training each improves performance. The best result is obtained with all three together (in green).
Finally, progressive training consistently reduces MAE, indicating more stable optimization and improved generalization.
As noted in \Sec{quantitative_qualitative_evaluations}, the histogram loss slightly increases MAE but yields sharper, more perceptually convincing high-frequency details.
\Fig{ModelImprovements} visually provides evidence on how each stage progressively improves wrinkle sharpness and perceptual details.

\section{Conclusions}

We presented a learning-based pipeline for quasi-static garment deformation tailored to production settings, where future usage requires reasonable predictions in large pose ranges and the target appearance demands high-quality, plausible wrinkles rather than merely low MAE metrics.
To our knowledge, this is the first approach to explicitly address core production constraints in a unified system:
generating a production-driven dataset that combines artist-authored motion snippets with carefully filtered random poses, and emphasizing perceptual wrinkle realism alongside physical plausibility.
Our method combines mesh-aware pose conditioning with VAE pre-training and a staged training strategy that progressively transitions from the pre-training to pose-driven prediction, enabling robust generalization. Across multiple datasets, we show that each stage contributes measurably to performance, and that these improvements persist under a stricter validation protocol using held-out motion trajectories rather than near-duplicate training frames. In contrast to prior work where generalization is often assessed on samples closely related to the training distribution, our evaluation highlights the importance of validation on genuinely unseen sequences when targeting production deployment.

\begin{figure}
	\centering
	\setlength{\tabcolsep}{0.0050\linewidth}
    {\footnotesize
    \begin{tabular}{
          >{\centering\arraybackslash}m{0.12\linewidth}
          >{\centering\arraybackslash}m{0.25\linewidth}
          >{\centering\arraybackslash}m{0.25\linewidth}
          @{\hspace{0.05\linewidth}}
          >{\centering\arraybackslash}m{0.18\linewidth}
        }
        Ours &
		\includegraphics[width=\linewidth]{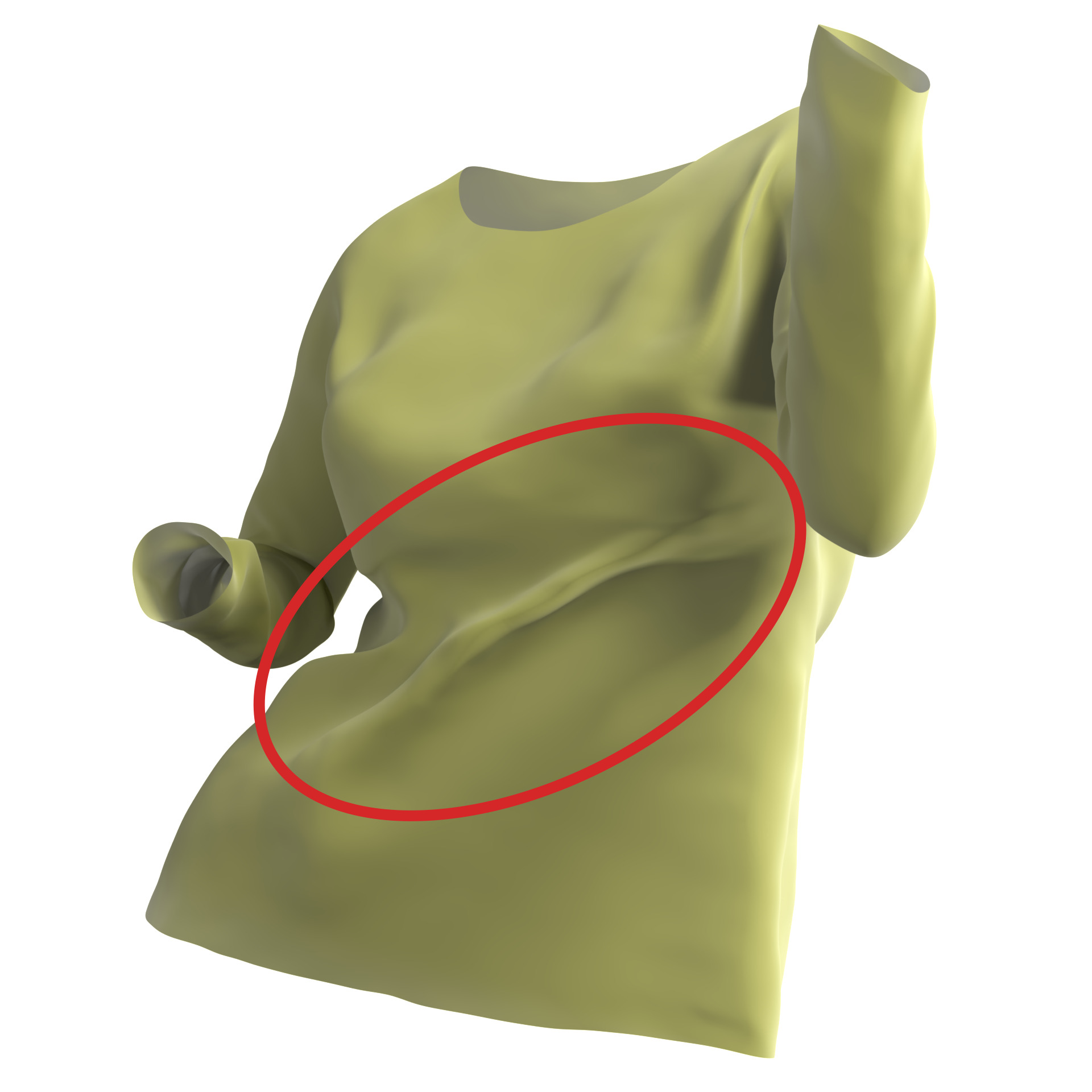} &
		\includegraphics[width=\linewidth]{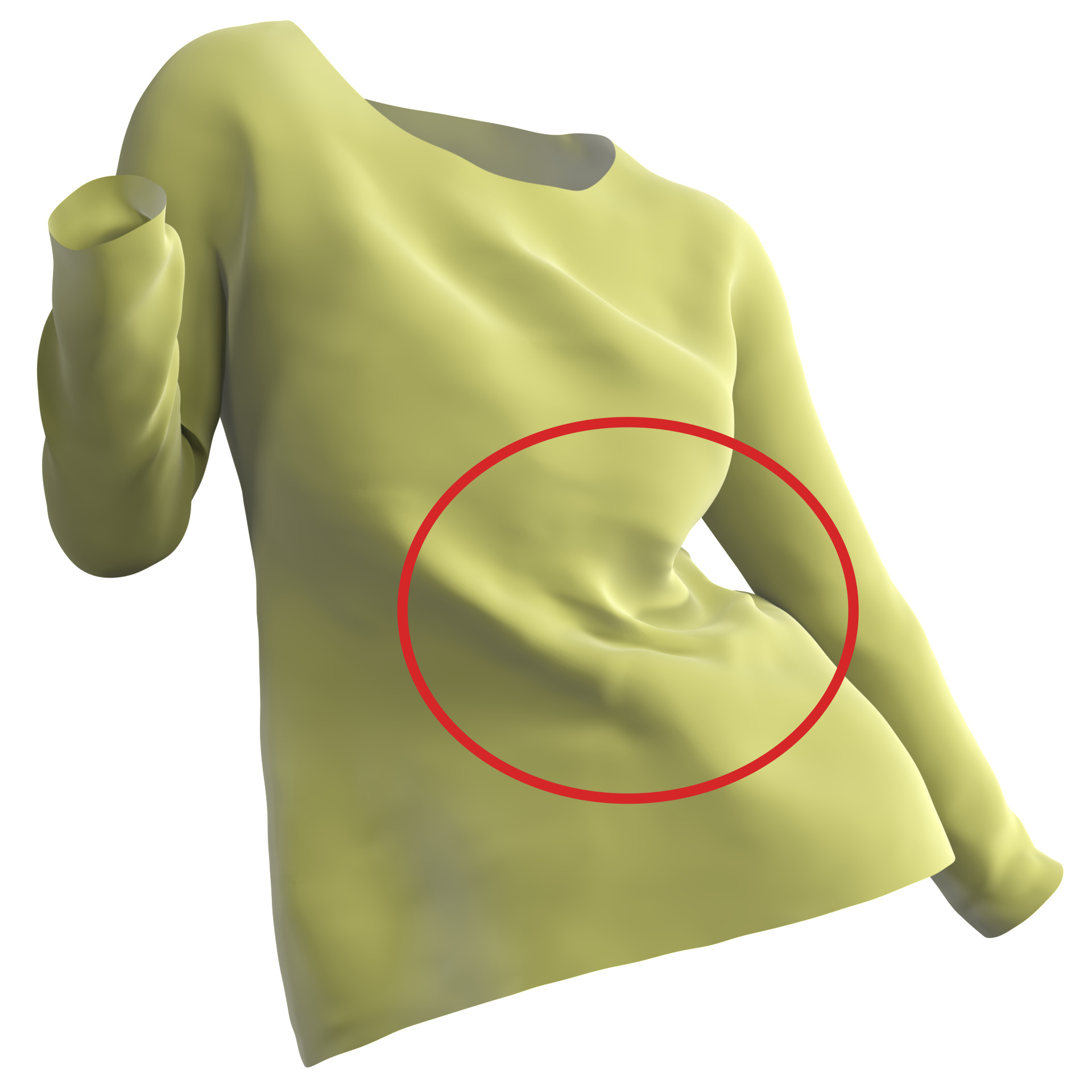} &
		\includegraphics[width=\linewidth]{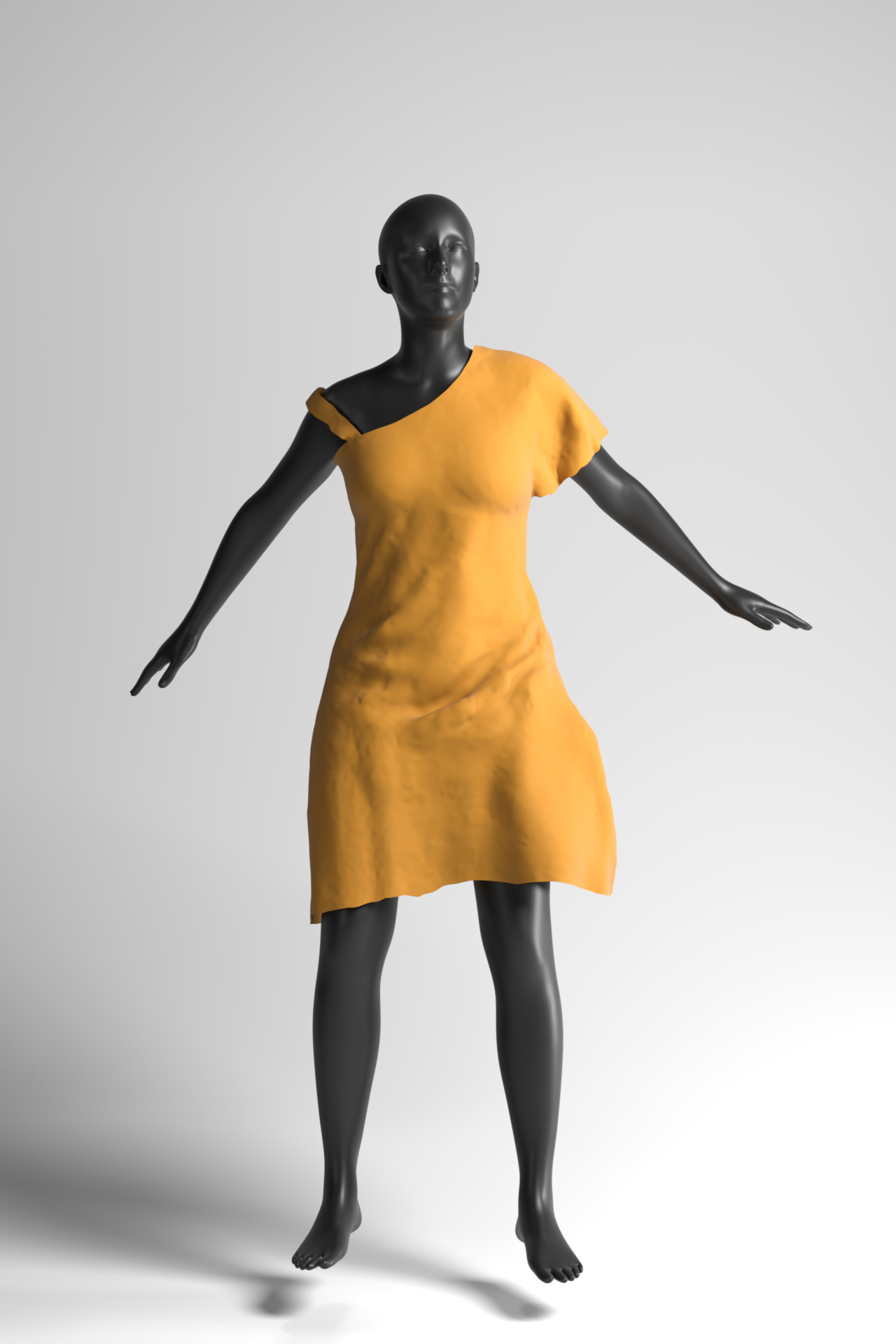} \\

        \shortstack{Ground\\Truth} &
        \includegraphics[width=\linewidth]{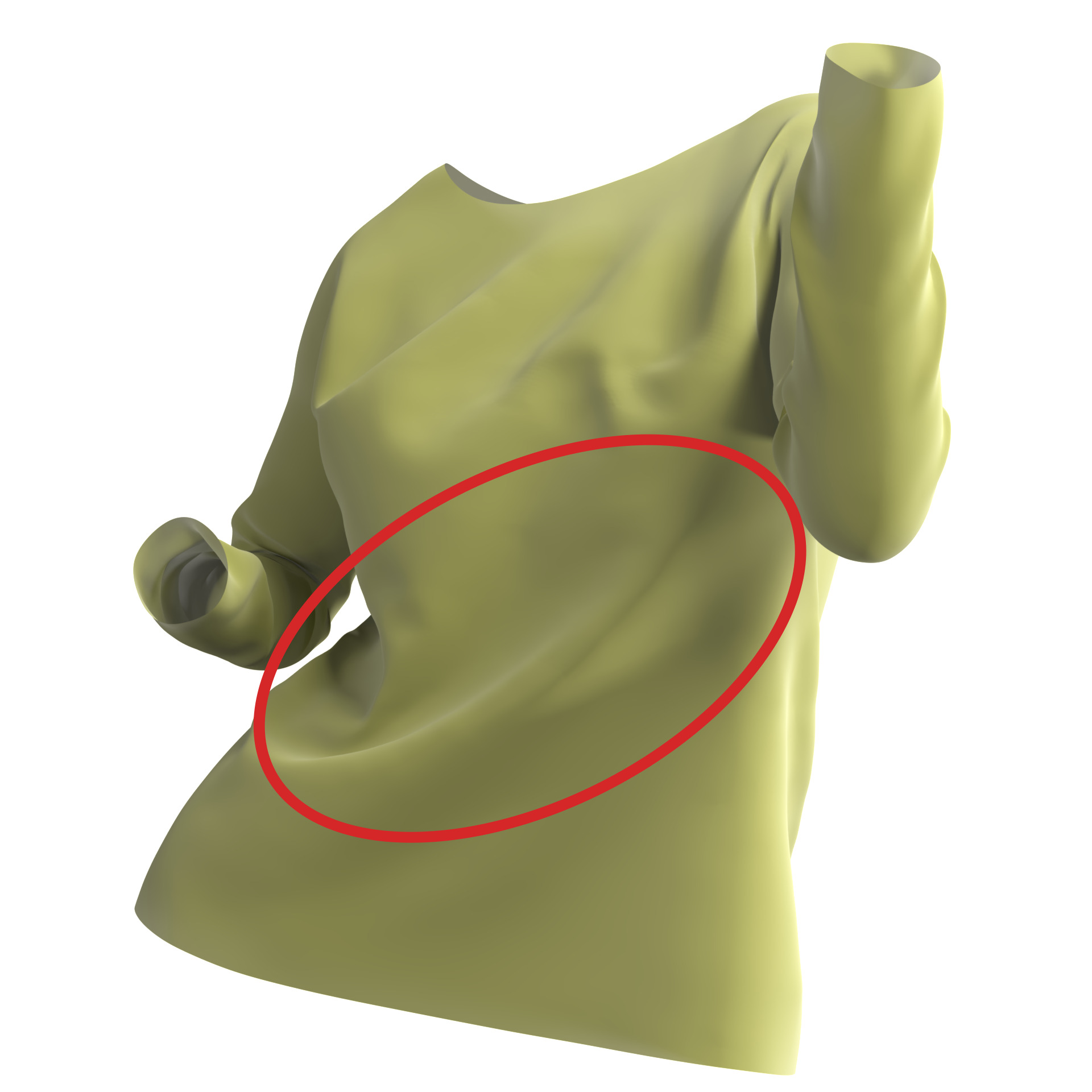} &
		\includegraphics[width=\linewidth]{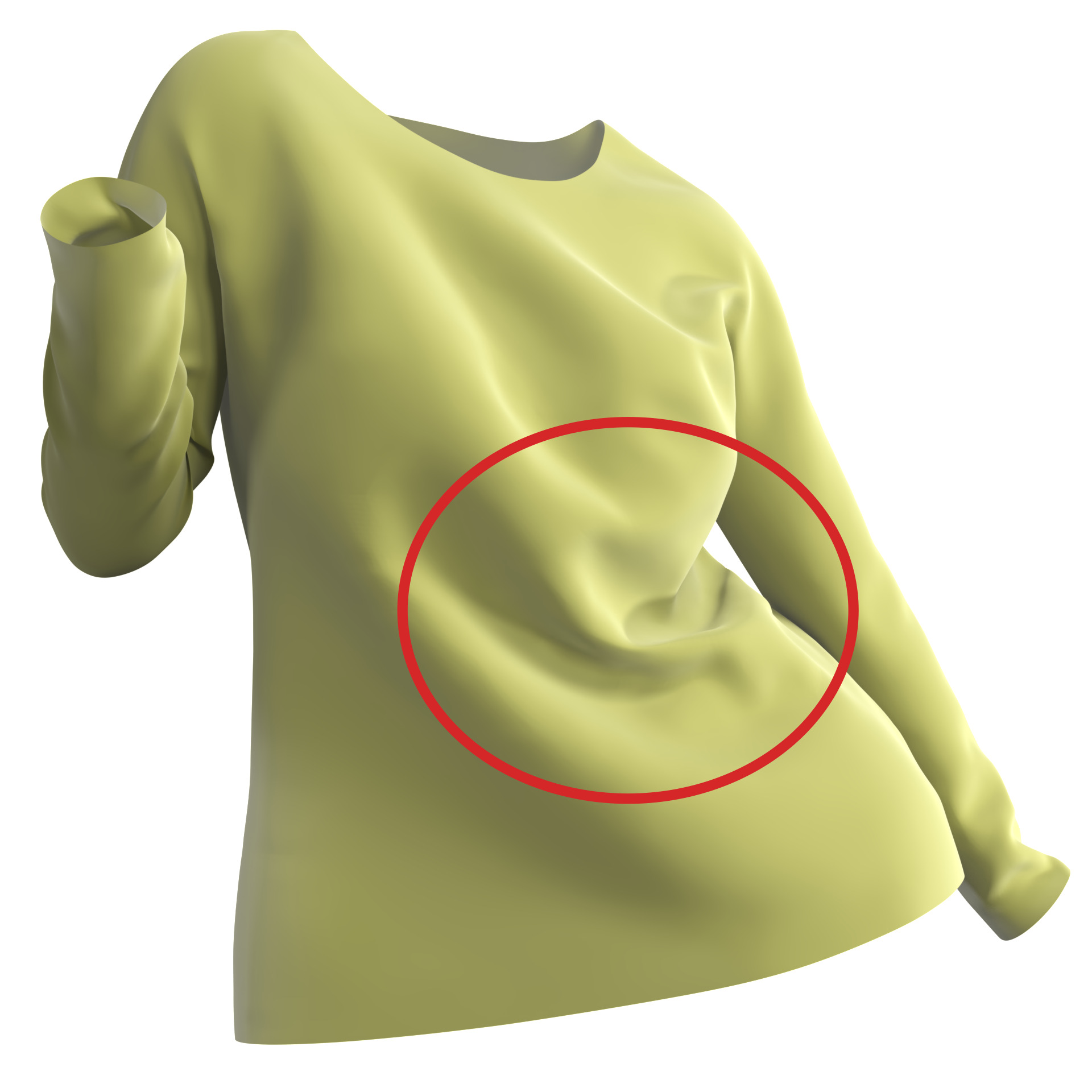} &
		\includegraphics[width=\linewidth]{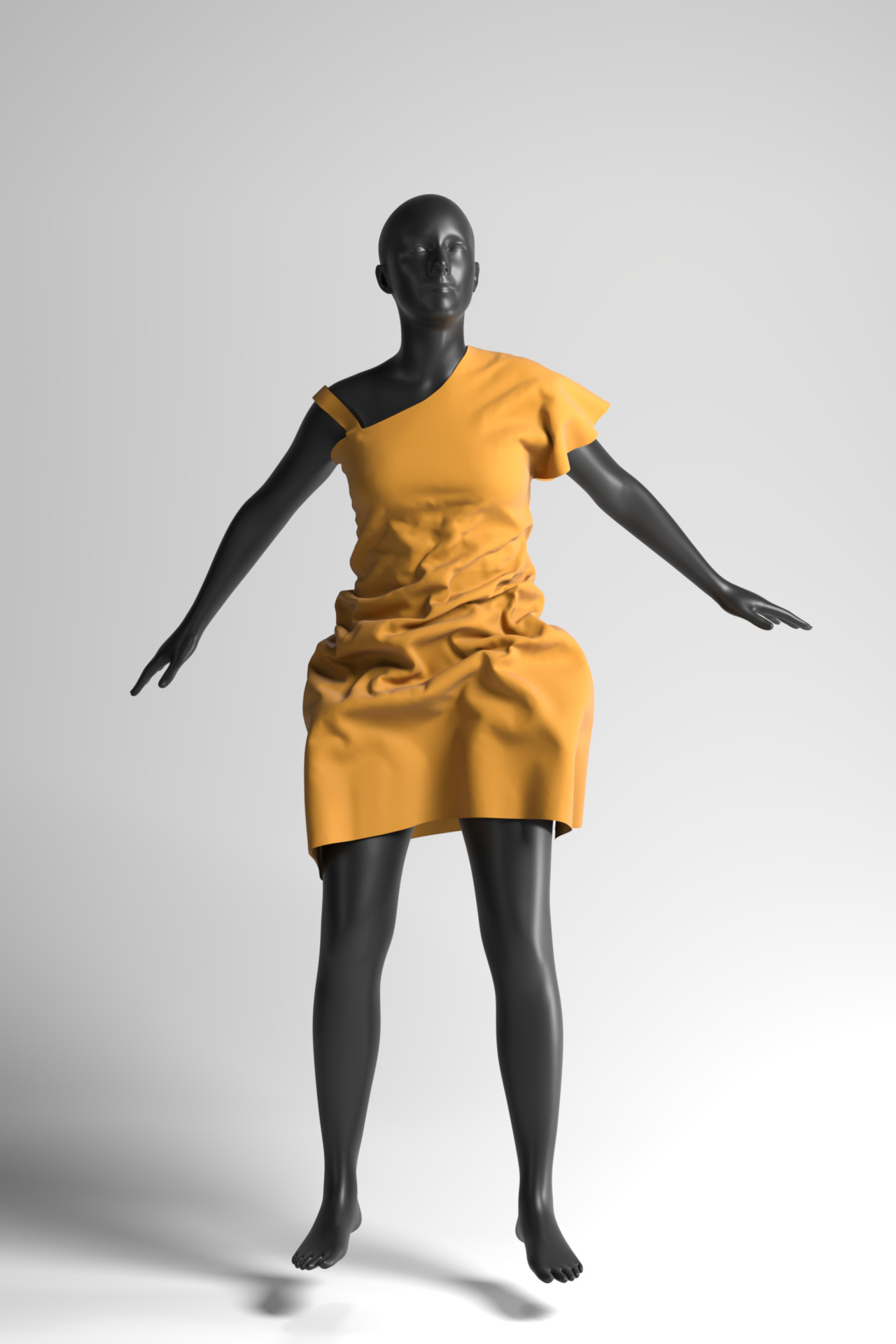} \\
		&(a) & (b)&(c)\\
	\end{tabular}
    }
	\caption{Limitations of our framework. (a) the non-localized normal distribution loss can create wrinkles with an arc bending upwards instead of downwards. (b) several wrinkles the should compose into a single larger one remain separated. (c) extensive sliding during jumping motion is not modeled with our constant skinning weights paradigm.}
    \label{fig:Limitations}
\end{figure}

\paragraph*{Limitations and future work.}
Despite the encouraging results, several limitations remain. First, cloth deformation behavior is inherently history-dependent: similar poses with different inertia and history can correspond to different wrinkle configurations. This temporal ambiguity renders the regression problem ill-posed and encourages deterministic predictors to average over multiple plausible outcomes, especially under highly dynamic motions (\Fig{Limitations}). Incorporating temporal information
and/or modeling latent trajectories is a natural next step.
Closely related, a deterministic formulation cannot represent the multi-modality of cloth at a fixed pose; probabilistic predictors with stochastic latent variables or diffusion models may better capture this variability and reduce over-smoothing. Second, our pose conditioning relies on skinning-derived correspondences to project joint transformations or body cues onto the garment. While effective for fitted clothing, garment--body sliding (common in loose garments and large motions) diverges from this correspondence and lead to
systematic artifacts. Dynamic correspondences or motion fields that account for sliding are interesting future avenues.
Third, our local normal distribution supervision illustrates a broader trade-off in deformation learning: matching local orientation statistics encourages expressive high-frequency detail and tolerates small spatial shifts, but can occasionally yield overly pronounced wrinkles or sharp creases in ambiguous regions, as well as globally inconsistent fold organization (\Fig{Limitations}). Hybrid objectives that preserve distributional robustness while adding lightweight global or continuity constraints may alleviate these cases.
Finally, evaluation remains challenging: no single metric reliably captures perceptual quality and physical plausibility across modeling choices. Though user studies help alleviate this situation, developing better quantitative criteria for wrinkle coherence, plausible material behavior, and perceptual realism remains an open problem.

\bibliographystyle{eg-alpha-doi}
\bibliography{references}

\appendix
\begin{figure*}[htbp]
	\centering
    \includegraphics[width=0.98\linewidth]{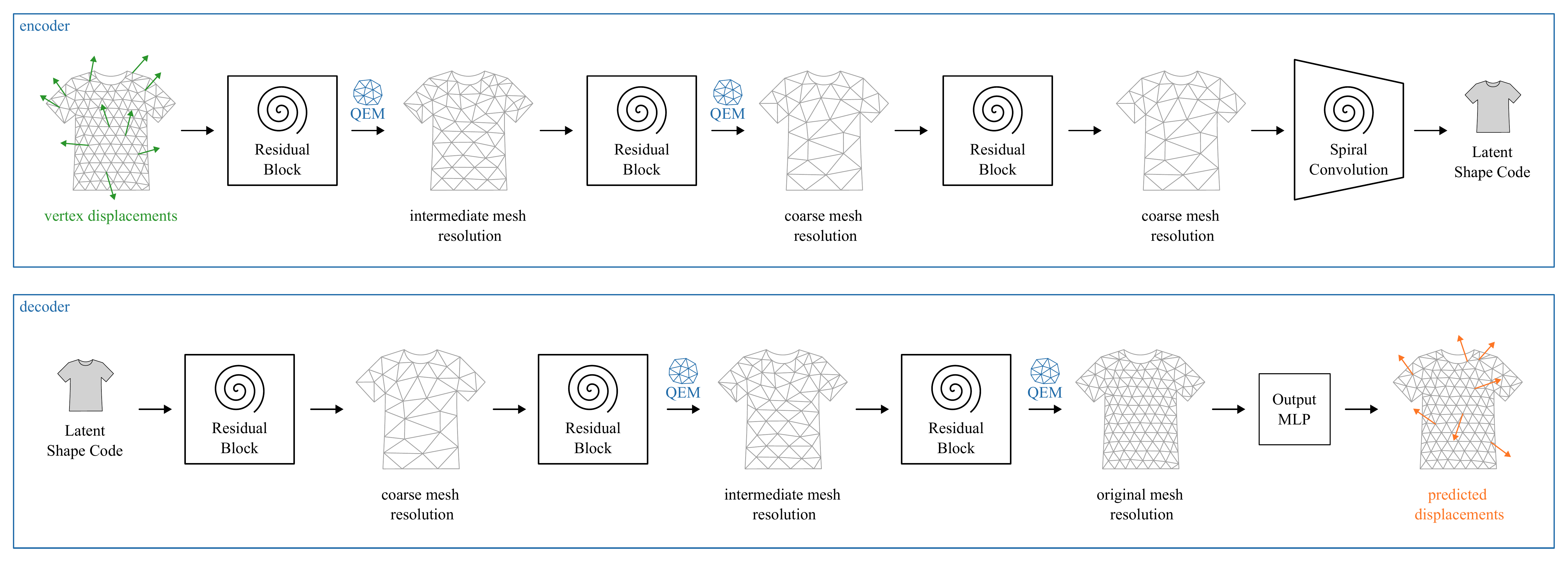}
	\caption{VAE architecture. The encoder extracts a compact latent representation from the input vertex displacements using spiral convolutions and hierarchical QEM-based mesh downsampling. The decoder mirrors this structure and reconstructs the full-resolution mesh deformation from the latent shape code through upsampling and spiral convolution blocks.}
	\label{fig:SpiralConvVAE}
\end{figure*}

\begin{figure*}[t]
	\centering
	\setlength{\tabcolsep}{0.0130\linewidth}
	\includegraphics[width=0.98\linewidth]{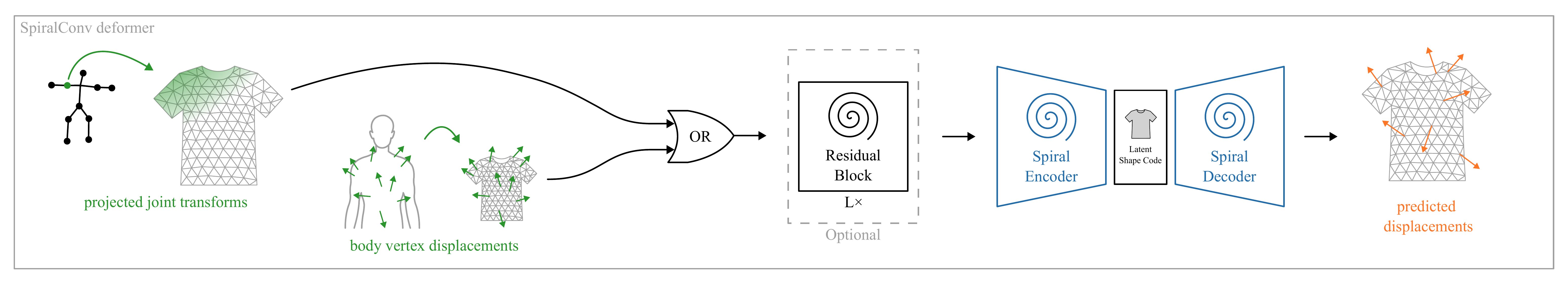}
    \caption{Overview of the mesh-based deformer architecture. Pose parameters are encoded per joint and projected onto the garment mesh using skinning weights to form mesh-aligned pose features. These features, optionally augmented with transferred body displacements, are processed by a spiral-convolution deformer that mirrors the hierarchical structure of the pretrained autoencoder.}
	\label{fig:MeshBasedDeformer}
\end{figure*}

\section{Model Implementation Details}
\label{app:implementation}

\subsection{Mesh Auto-Encoder Architecture}
Our VAE is implemented as a mesh-aware convolutional network based on spiral convolutions and a fixed multi-resolution mesh hierarchy. Spiral convolutions operate on ordered one-ring neighborhoods, enabling convolutions directly on irregular mesh surfaces while preserving geometric locality. This formulation provides a strong inductive bias for surface-based deformation modeling.

We construct a hierarchy of three mesh resolutions using Quadric Error Metrics (QEM) simplification, yielding a set of fixed pooling and unpooling operators. Coarser levels capture global shape variation, while finer levels model localized wrinkle detail. Both the encoder and decoder operate on this hierarchy, with feature dimensionality increasing towards coarser levels and decreasing symmetrically during decoding.

\begin{figure}[htbp]
	\centering
    \begin{tabular}{c}
        \includegraphics[height=0.45\linewidth]{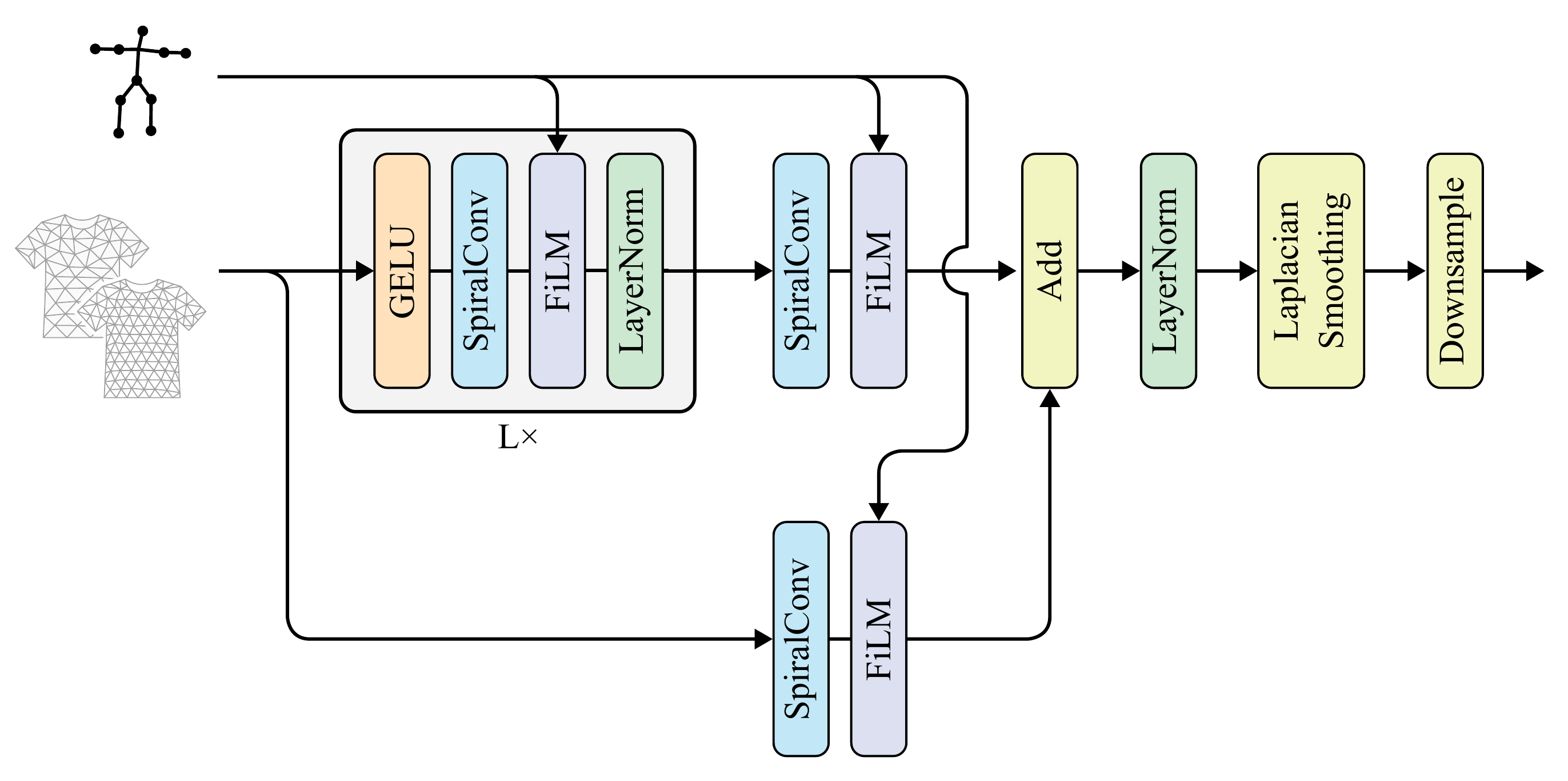}\\
        (a) \\
        \includegraphics[height=0.45\linewidth]{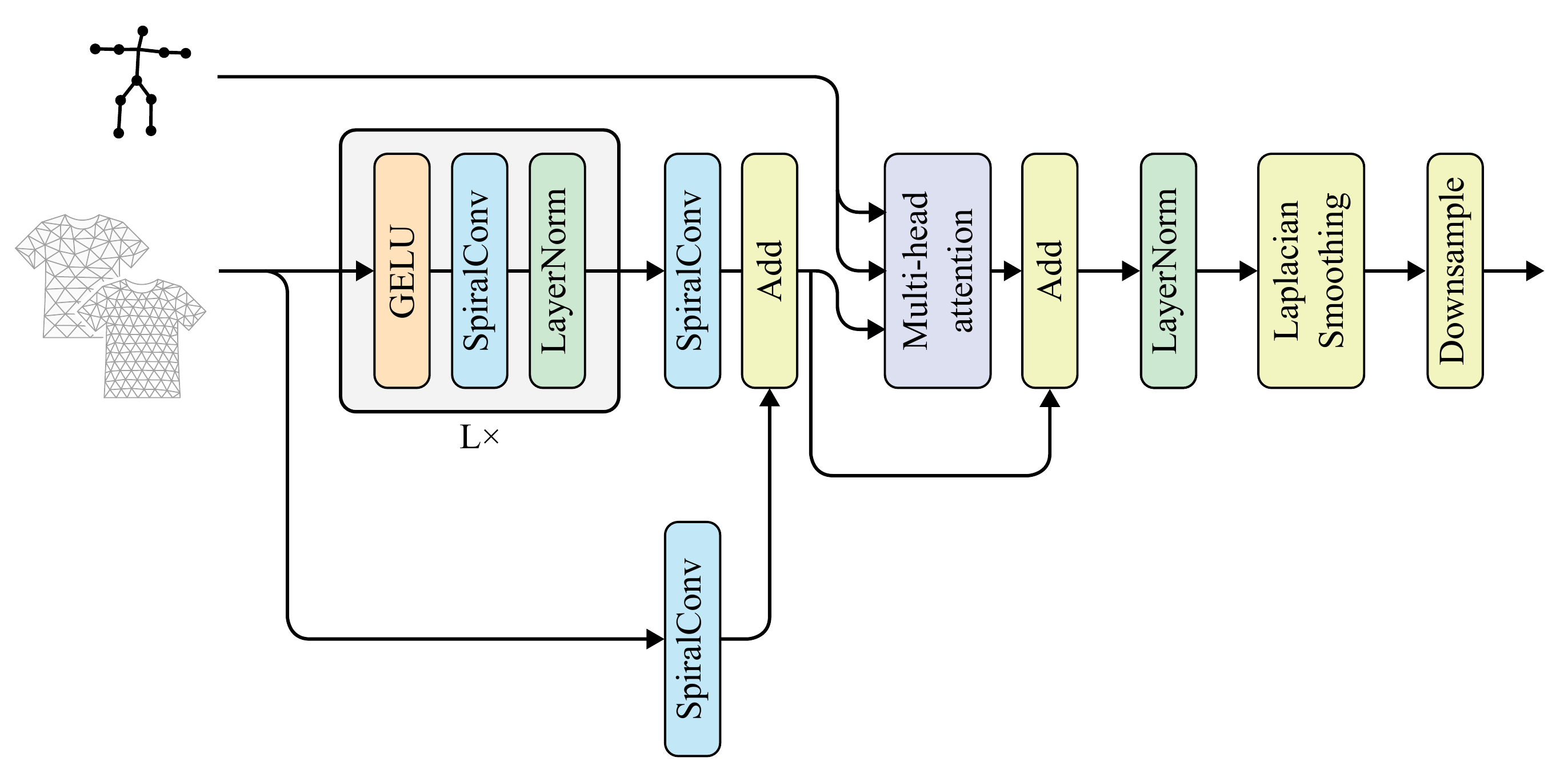}\\
        (b)\\
	\end{tabular}
	\caption{Detailed structure of a spiral convolution encoder block. Each block consists of a residual spiral convolution path and a skip connection, followed by normalization, optional Laplacian smoothing, and mesh downsampling. The encoder supports two additional pose-conditioning mechanisms: (a) FiLM modulation and (b) cross-attention.}
	\label{fig:EncoderBlock}
\end{figure}

\paragraph*{Residual spiral convolution blocks.} Each encoder level consists of a residual spiral convolution block, followed by optional Laplacian smoothing and QEM-based downsampling. The decoder mirrors this structure, applying unpooling, smoothing, and residual spiral convolution blocks to reconstruct higher-resolution features. The final decoder output is a per-vertex displacement field defined on the original mesh resolution. Each residual block processes vertex features through two parallel paths: a residual path and a skip path. The residual path consists of $L$ stacked sublayers, each composed of a GELU activation, a spiral convolution, and Layer Normalization. Formally, a single sublayer is defined as (see \Fig{EncoderBlock})
\begin{equation}
	f(x) = \text{LayerNorm}\left(\text{SpiralConv}\left(\text{GELU}\left(x\right)\right)\right)\ .
\end{equation}

The residual path applies this function $L$ times in sequence. In parallel, the skip path applies a single spiral convolution to the input features. The two paths are combined and normalized to produce the block output. To mitigate aliasing artifacts introduced by mesh resolution changes, we apply a lightweight Laplacian smoothing operator to the feature maps after pooling or unpooling. This promotes smooth feature variation across the surface while preserving high-frequency detail through the residual connections.

\subsection{Mesh-Based Deformer Architecture}
The mesh-based deformer reuses the encoder architecture of the pre-trained autoencoder to predict latent codes rather than vertex displacements directly. Pose parameters are first encoded per joint and then transferred to the garment mesh using skinning weights, producing a mesh-aligned pose field. This splatting operation ensures that vertices influenced by similar joints receive similar pose descriptors, providing spatially structured conditioning aligned with garment geometry.

In addition to joint transforms, we optionally concatenate transferred body displacement features to the mesh input, enabling the model to exploit correlations between garment and body deformation. The resulting per-vertex feature maps are processed through the hierarchical spiral convolution encoder, and the final feature embedding is projected into the latent space of the pre-trained autoencoder. The frozen decoder then maps this latent code to the final rest-space displacement field. By operating within the learned latent space, the deformer remains constrained to a garment-valid shape manifold, improving robustness and geometric consistency.

\paragraph*{Additional pose conditioning mechanisms.}
To incorporate pose information at multiple spatial scales, we employ two complementary conditioning mechanisms within the mesh-based deformer. At finer mesh resolutions, we apply FiLM modulation. Joint pose features are pooled across joints using an attention mechanism to produce a global pose embedding, which is mapped to per-channel scaling and shifting parameters ($\gamma$, $\beta$). These parameters modulate the outputs of spiral convolution layers, enabling pose-dependent feature adaptation while maintaining computational efficiency at high resolutions. At the coarsest mesh resolution, we additionally apply a cross-attention block. Here, the current mesh feature map serves as the query, while the encoded pose features serve as keys and values. This allows global interactions between pose and mesh features, enabling the model to capture long-range pose effects. Cross-attention is applied only at the coarsest level to avoid the quadratic memory and compute cost at higher resolutions. Body displacements are transferred to the garment mesh via distance-weighted interpolation. For each garment vertex, we identify the 6 nearest body vertices in the rest-pose and compute normalized inverse-distance weights. The garment-level body displacement is then obtained as a weighted sum of the corresponding body vertex displacements. This interpolation yields a smooth, mesh-aligned body displacement field that can be concatenated with the splatted pose features as additional input to the deformer.

\paragraph*{Implementation details.}
All spiral neighborhood orderings, QEM pooling and unpooling operators, and mesh hierarchies are precomputed and fixed across all experiments. The same backbone architecture is used for VAE pre-training, progressive refinement, and pose-conditioned training; only the encoder inputs and supervision terms differ between stages.

We consider two training variants within the same progressive curriculum, both optimized using the objective in Eq.~(5). In the pointwise variant, reconstruction is supervised using a weighted combination of vertex $L^1$, vertex $L^2$, and normal losses. In the distributional variant, these pointwise losses are replaced by the proposed local normal distribution loss (Section~3.4). In both cases, we apply an additional weak latent supervision term to encourage consistency with the pre-trained VAE latent space. Architectural hyperparameters, optimization settings, and loss weights are summarized in \Tab{Hyperparams}.

\begin{table}[htbp]
    \centering
    {\footnotesize
    \resizebox{\linewidth}{!}{
    \begin{tabular}{ll}
        \toprule
        \textbf{Parameter} & \textbf{Value} \\
        \midrule
        \multicolumn{2}{l}{\textbf{Mesh hierarchy}} \\
        Number of resolution levels & 3 \\
        Vertices per level &
        \begin{tabular}[t]{@{}l@{}}
            \jenShirt: 4820 / 1041 / 229 \\
            \jenPants: 7564 / 1633 / 352 \\
            \vto T-Shirt: 4424 / 954 / 205 \\
            \vto Dress: 12146 / 2614 / 564
        \end{tabular} \\
        Pooling method & QEM \\
        Laplacian smoothing & before downsampling / after unpooling \\
        \midrule
        \multicolumn{2}{l}{\textbf{Spiral convolution encoder/decoder}} \\
        Channels per level &
        \begin{tabular}[t]{@{}l@{}}
            \jenShirt: 64 / 128 / 512 \\
            \jenPants: 64 / 128 / 512 \\
            \vto T-Shirt: 32 / 64 / 128 \\
            \vto Dress: 32 / 64 / 128
        \end{tabular} \\
        Residual layers per block ($L$) & 5 \\
        Spiral lengths (fine$\rightarrow$coarse) & 13 / 13 / 13 \\
        Activation & GELU \\
        Normalization & LayerNorm \\
        \midrule
        \multicolumn{2}{l}{\textbf{Latent space}} \\
        Latent dimensionality &
        \begin{tabular}[t]{@{}l@{}}
            \jenShirt: $229 \times 1$ \\
            \jenPants: $352 \times 1$ \\
            \vto T-Shirt: $205 \times 1$ \\
            \vto Dress: $564 \times 1$
        \end{tabular} \\
        KL weight ($w_\text{KL}$) & 0.01 \\
        Latent supervision weight & 0.1 \\
        \midrule
        \multicolumn{2}{l}{\textbf{Pose conditioning}} \\
        Joint embedding dimension & 64 \\
        FiLM levels & fine, mid \\
        Cross-attention level & coarsest \\
        Attention heads & 4 \\
        \midrule
        \multicolumn{2}{l}{\textbf{Optimization}} \\
        Optimizer & AdamW \\
        Batch size & 16 \\
        Learning rate & $1\!\times\!10^{-4} \rightarrow 1\!\times\!10^{-5}$ \\
        \midrule
        \multicolumn{2}{l}{\textbf{Loss weights}} \\
        Vertex $L^1$ loss ($w_{L^1}$) & 0.1 \\
        Vertex $L^2$ loss ($w_{L^2}$) & 1.0 \\
        Normal loss ($w_{n}$) & 2.0 \\
        KL divergence ($w_{\text{KL}}$) & 0.01 \\
        Local normal distribution loss ($w_{\text{hist}}$) & 1.0 \\
        Latent supervision loss ($w_{\text{lat}}$) & 0.1 \\
        \bottomrule
    \end{tabular}}}
    \caption{Architectural and training hyper-parameters used in all experiments. Dataset-specific values are reported explicitly where they differ; all other parameters are shared across datasets.}
    \label{tab:Hyperparams}
\end{table}

\section{Collision Post-processing}
\begin{figure}
	\centering
    {\footnotesize
    \setlength{\tabcolsep}{0.0050\linewidth}
	\begin{tabular}{c c @{\hspace{0.03\linewidth}} c c}

    	\includegraphics[width=0.245\linewidth]{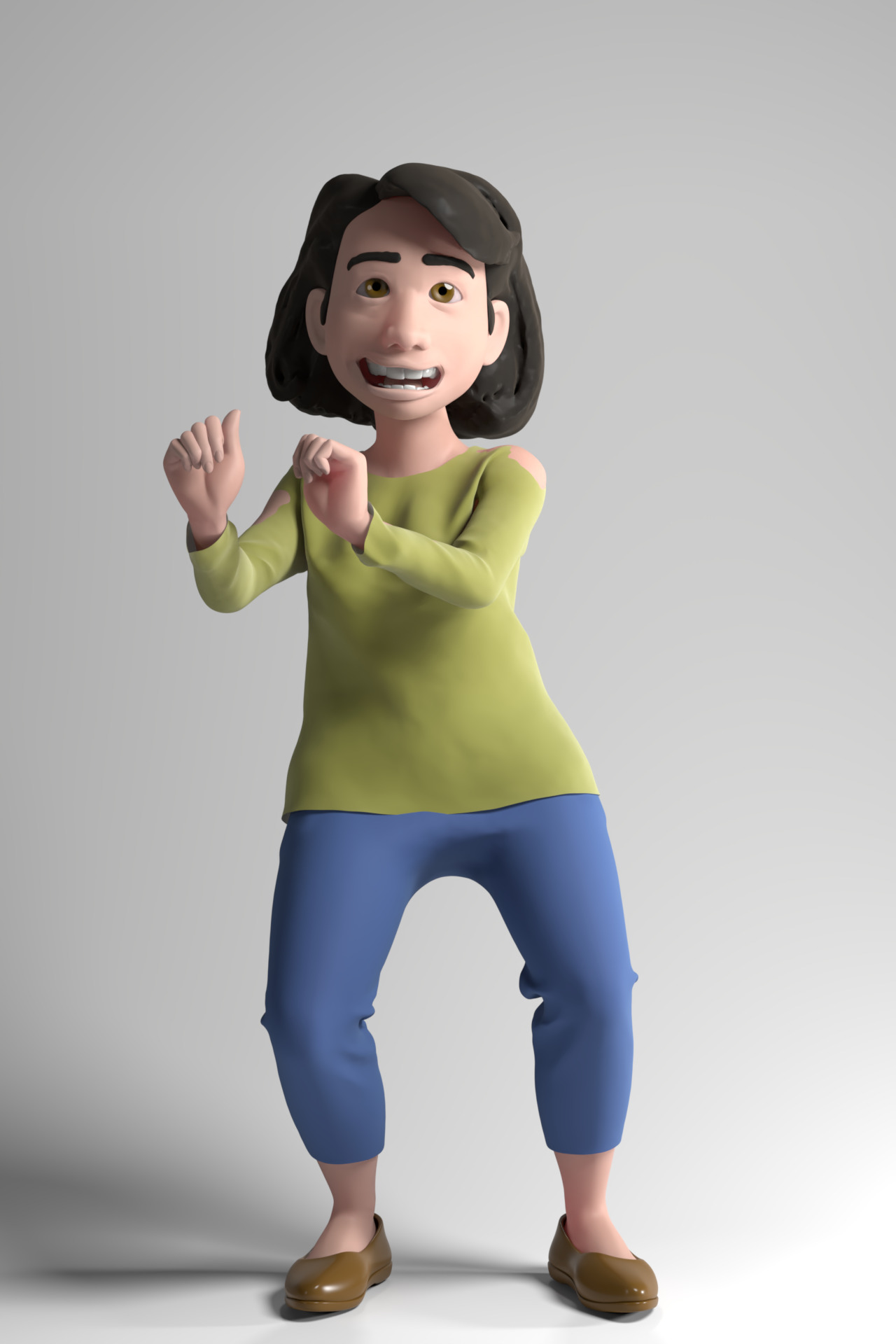} &
        \includegraphics[width=0.245\linewidth]{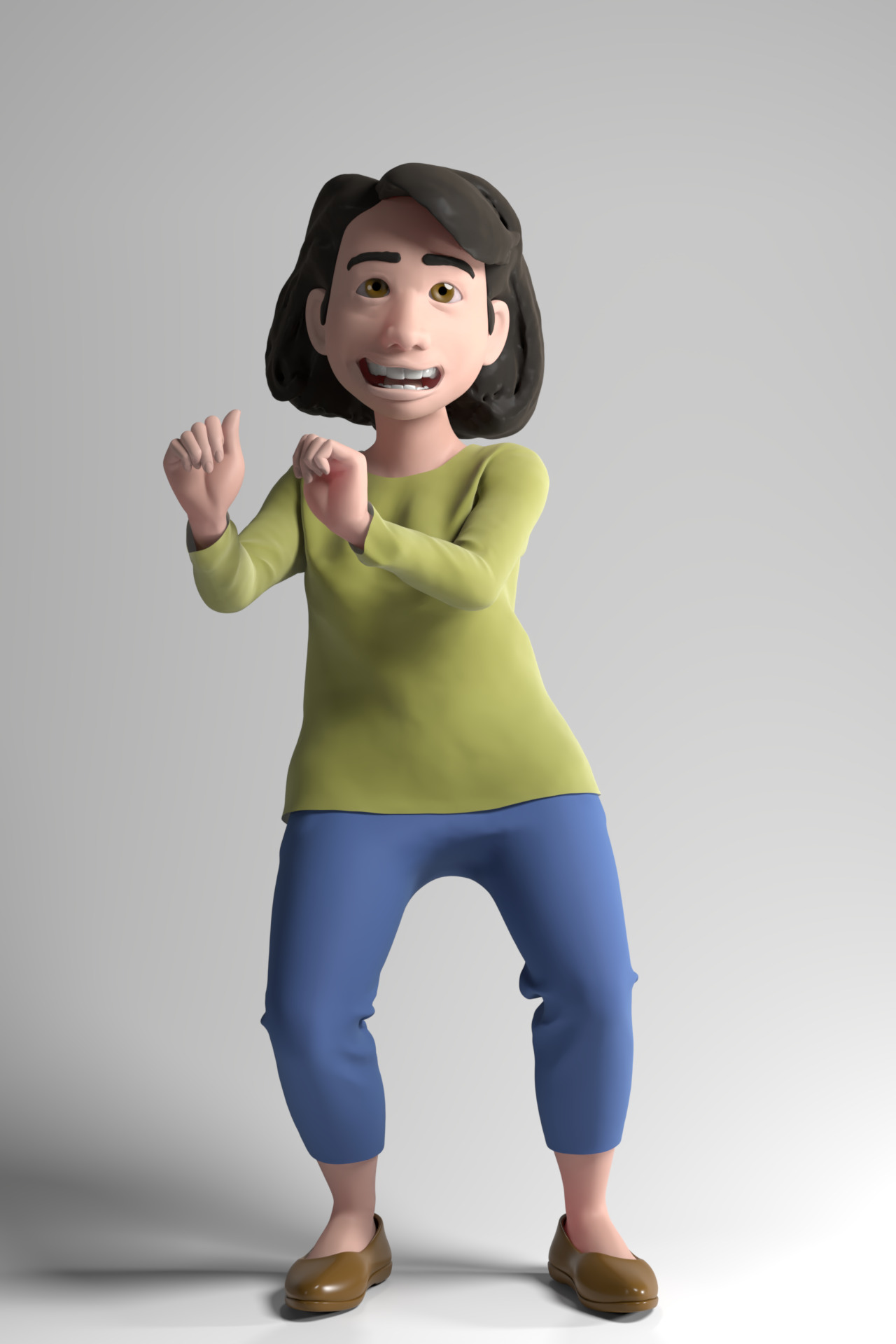} &

        \includegraphics[width=0.245\linewidth]{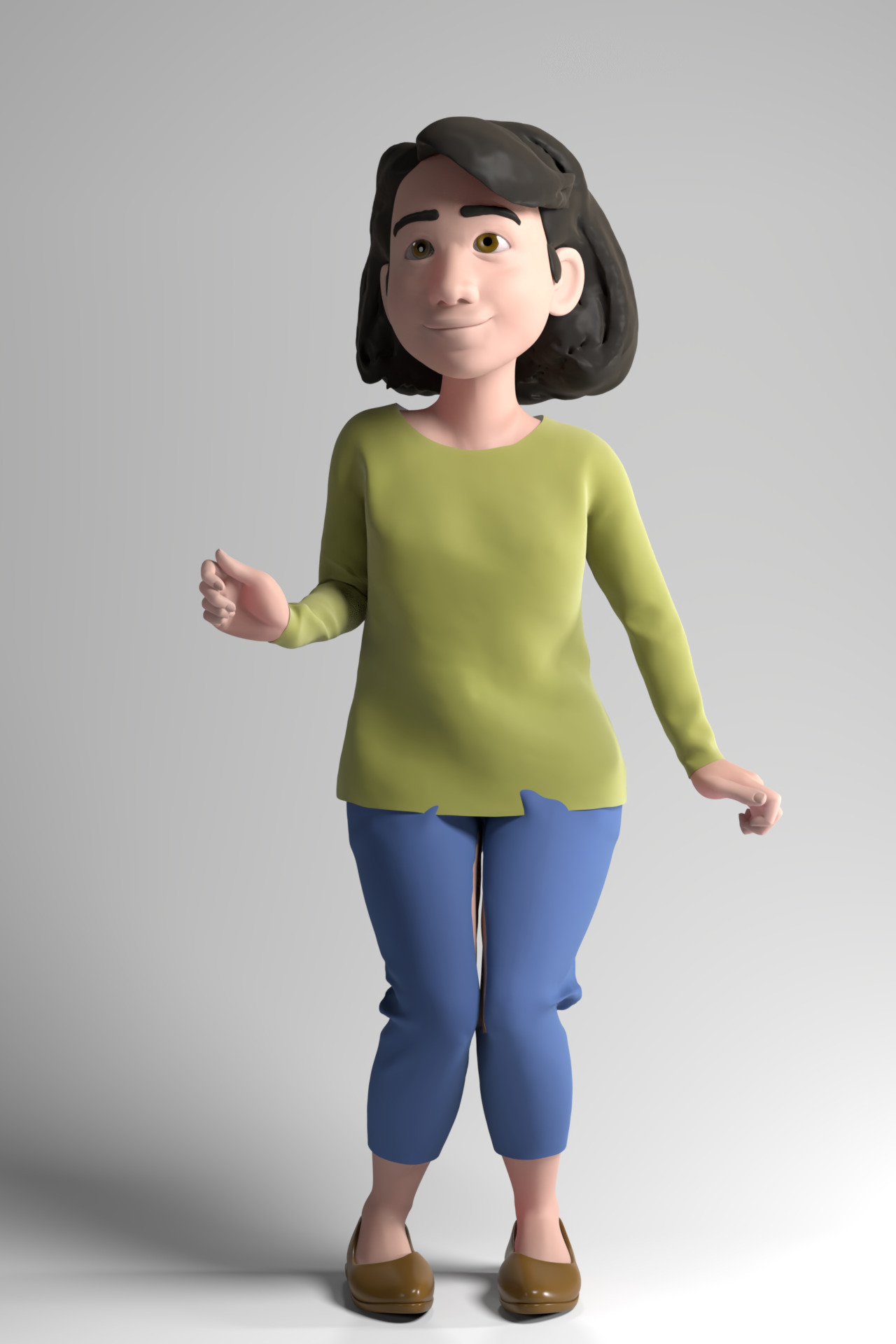} &
        \includegraphics[width=0.245\linewidth]{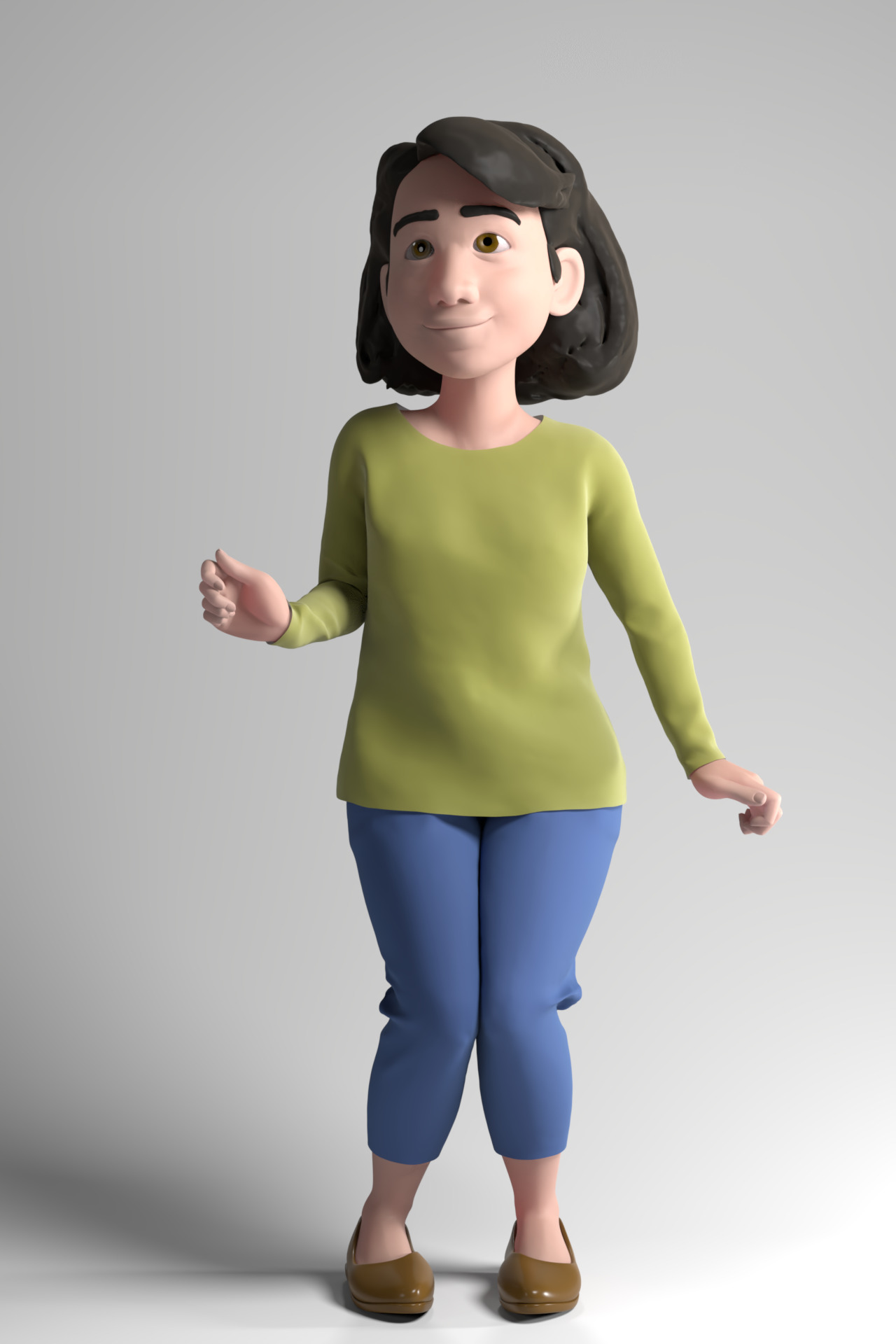} \\

		Before & After & Before & After
	\end{tabular}
    }
	\caption{Collision handling is applied at inference time after the deformation. As can be seen from before and after collision handling, this step performs minor corrections and does not effect overall appearance.}
    \label{fig:CollisionPost}
\end{figure}

Because our predictor is trained purely with supervised reconstruction losses and does not include explicit contact or collision constraints, occasional interpenetrations with the body (and between garments in layered setups) can occur. These artifacts are visually disproportionate: even small penetrations can create obvious shading and silhouette breaks in production rendering. We therefore apply a lightweight CUDA-based post-process using NVIDIA Warp \cite{macklin_warp_2022} to resolve collisions. For each garment vertex $\mathbf{x}$, we query the closest point $\mathbf{p}$ on the collider surface and its signed distance $d$ (positive outside, negative inside). If $d$ falls below a user-defined tolerance $d_{\mathrm{th}}$, we apply a corrective displacement in the outward direction $(\mathbf{x}-\mathbf{p})$, with magnitude proportional to $(d_{\mathrm{th}}-d)$ and scaled by a strength parameter; otherwise, no correction is applied. \Fig{CollisionPost} illustrates the effect of this procedure: even tiny interpenetrations (e.g., around the shoulders or the shirt hem) are clearly visible before correction. Since the post-process runs entirely on the GPU, the runtime overhead is negligible in practice.

\section{Additional Results}
\label{app:additional_results}

\begin{figure*}[t]
	\centering
	\setlength{\tabcolsep}{0.00125\linewidth}
	\begin{tabular}{ccccccc}
        \includegraphics[width=0.1405\linewidth]{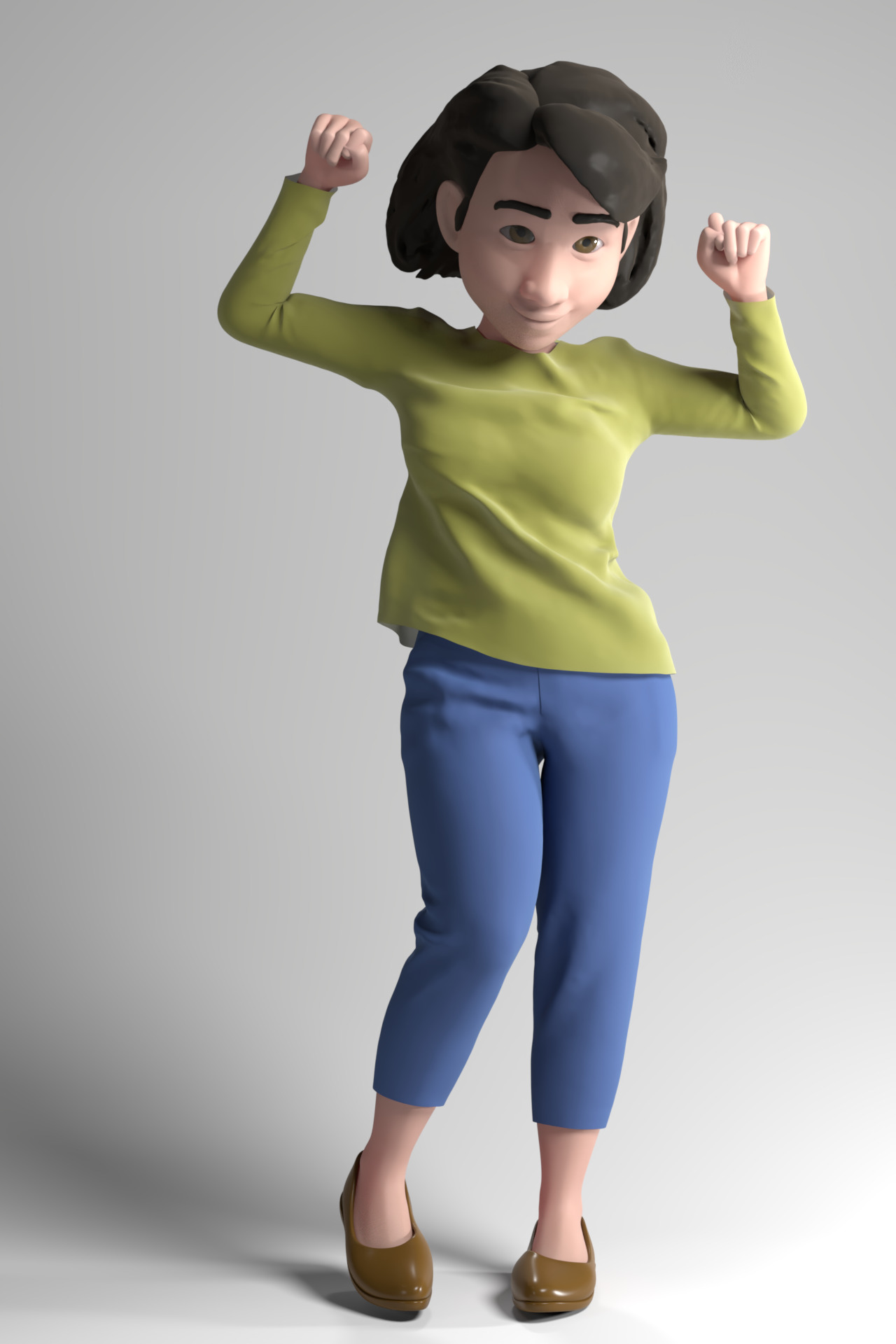} &
        \includegraphics[width=0.1405\linewidth]{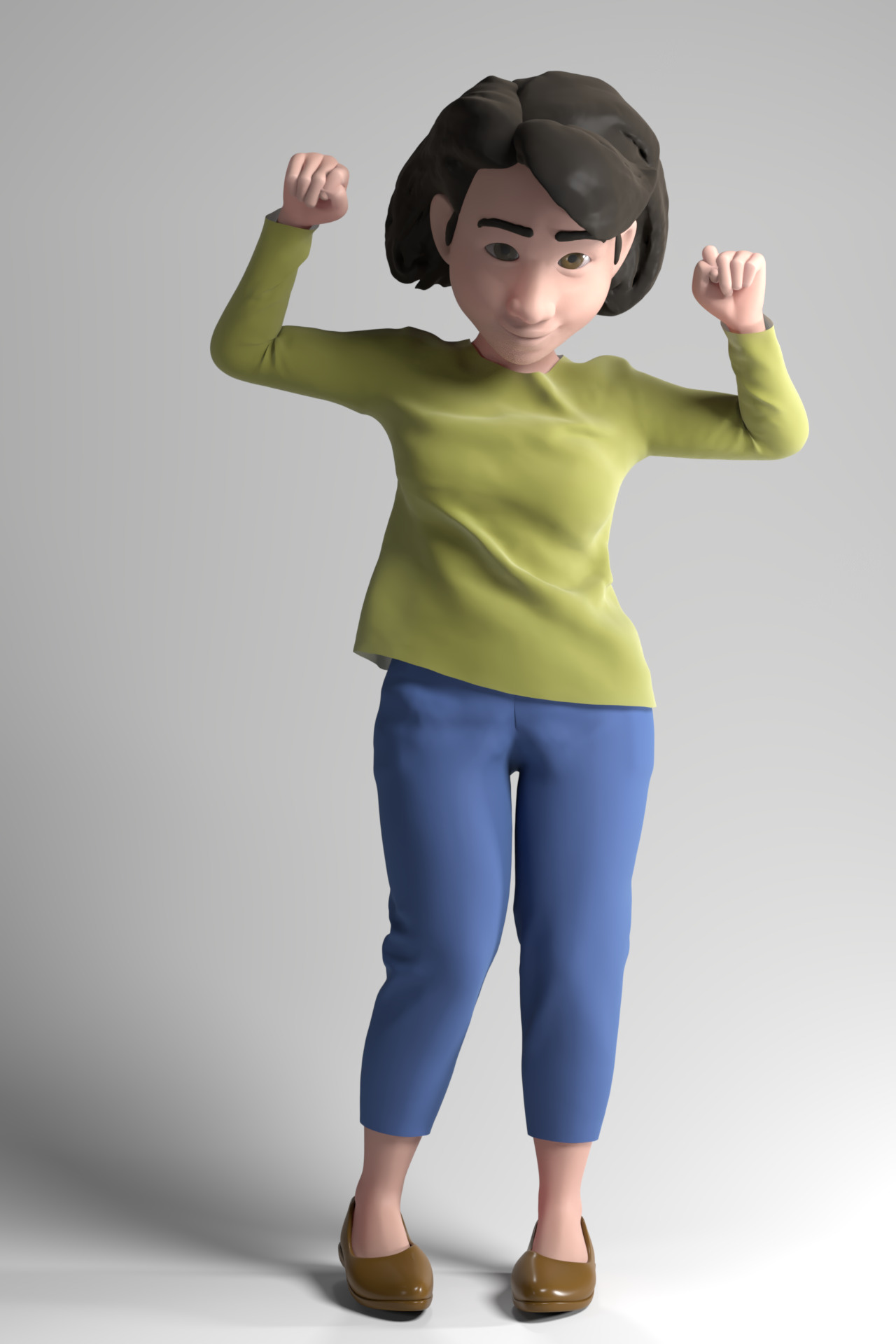} &
        \includegraphics[width=0.1405\linewidth]{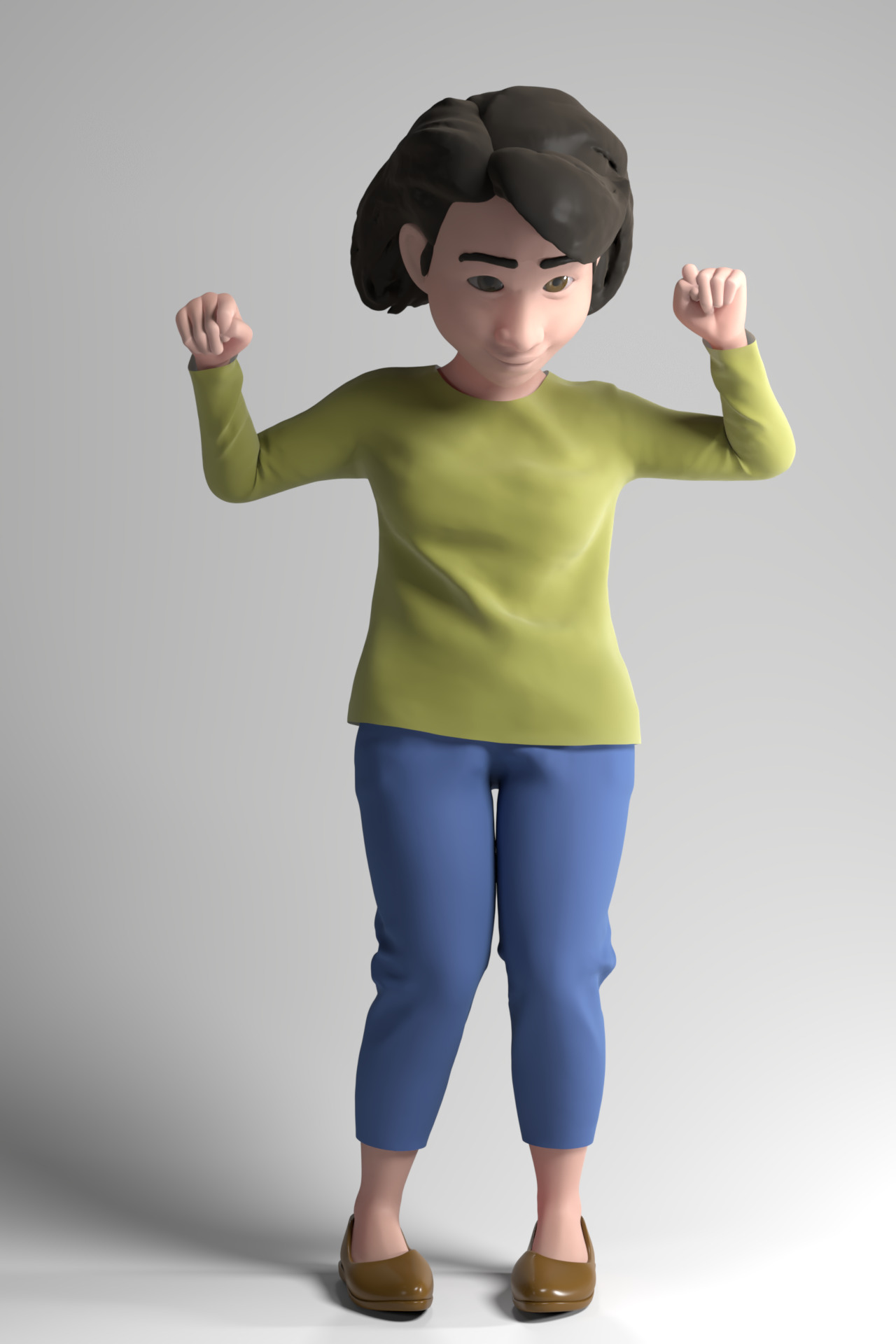} &
        \includegraphics[width=0.1405\linewidth]{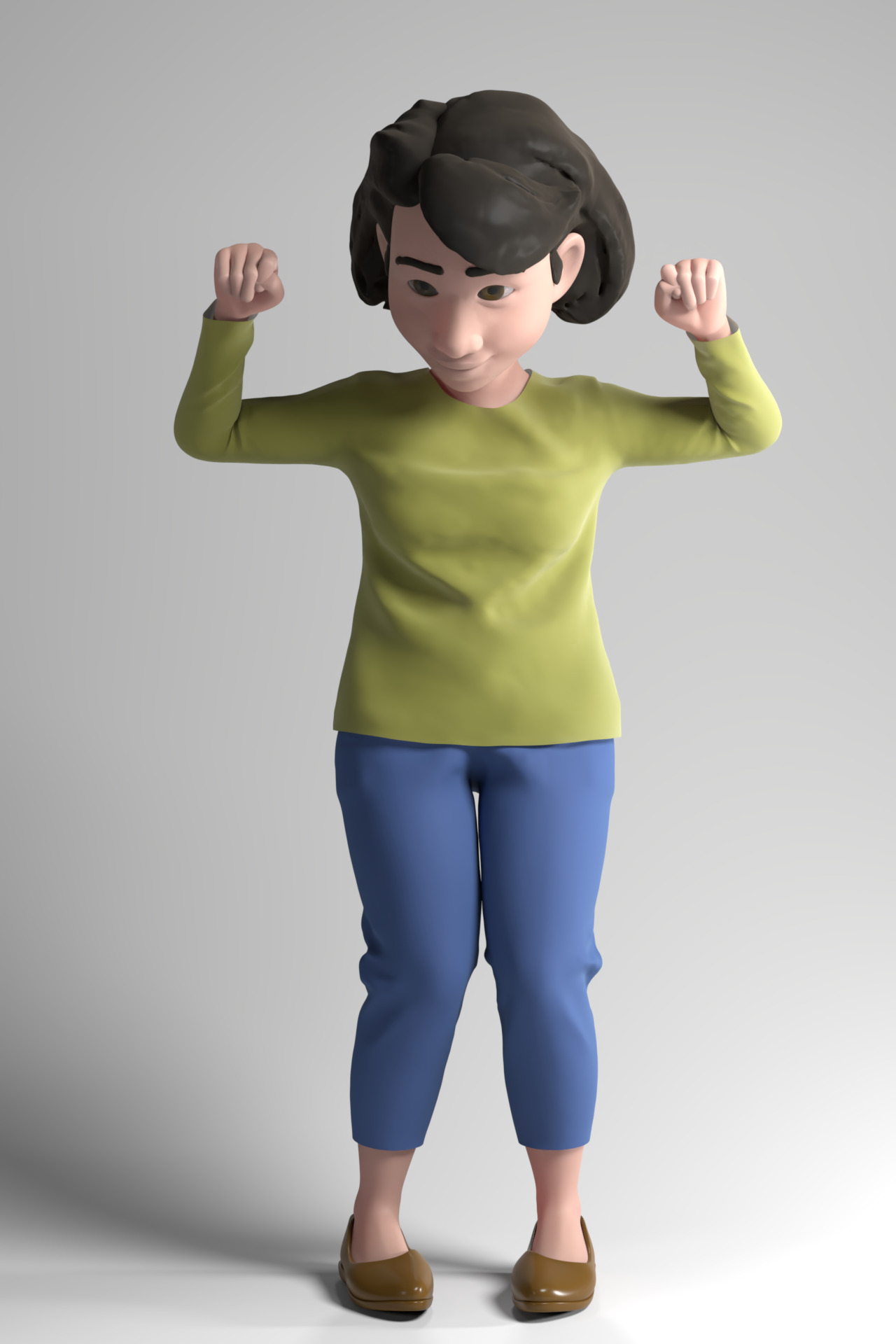} &
        \includegraphics[width=0.1405\linewidth]{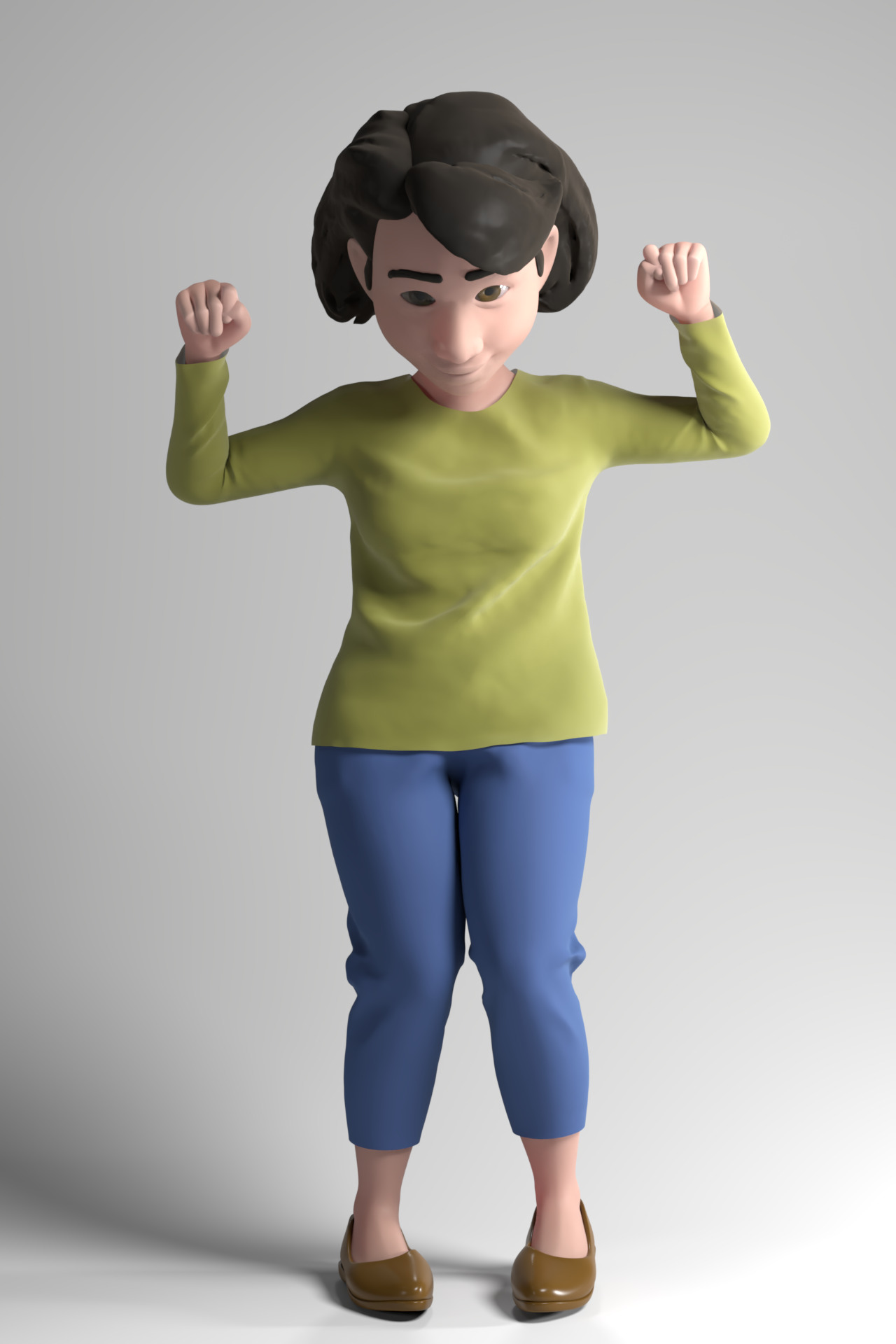} &
        \includegraphics[width=0.1405\linewidth]{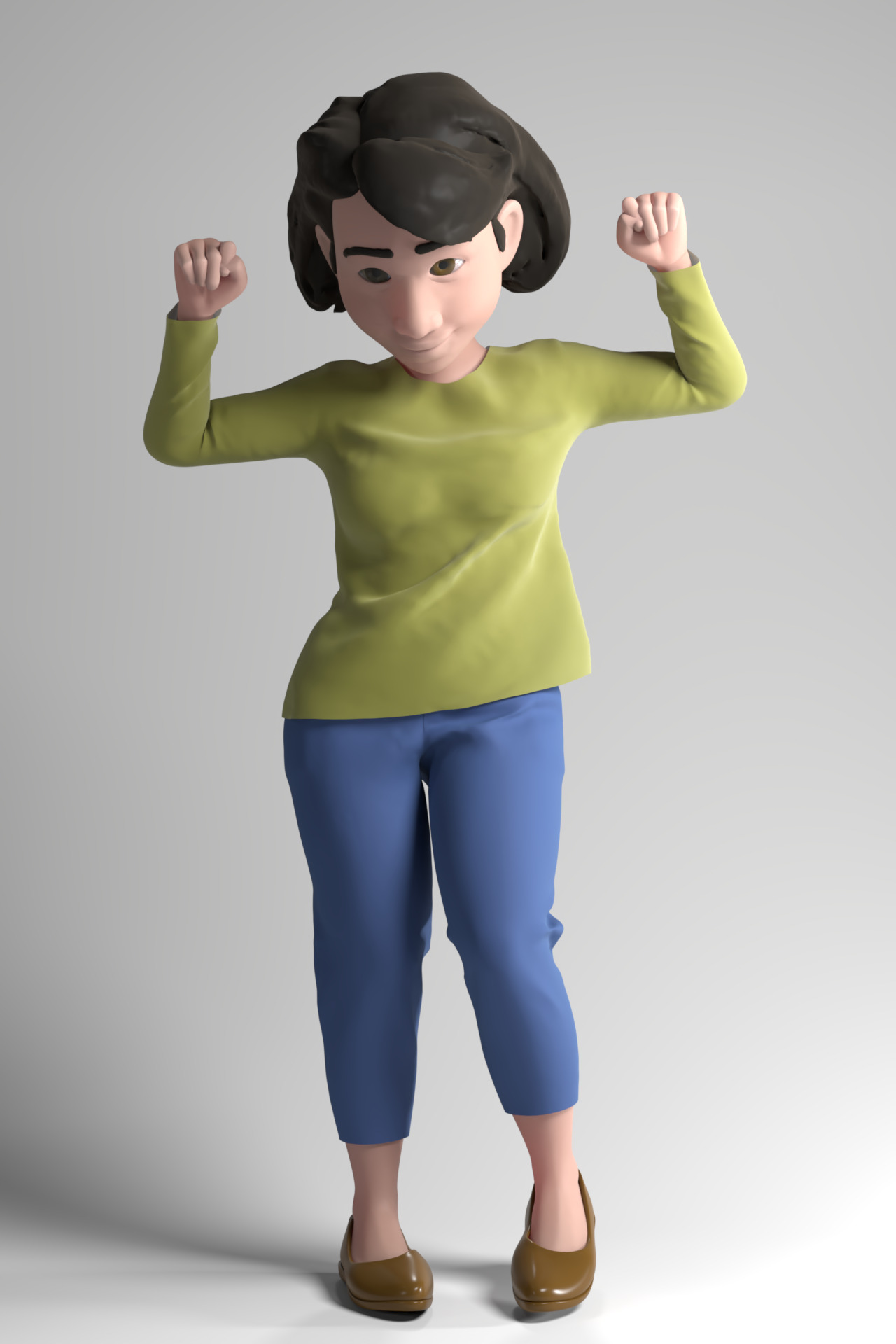} &
        \includegraphics[width=0.1405\linewidth]{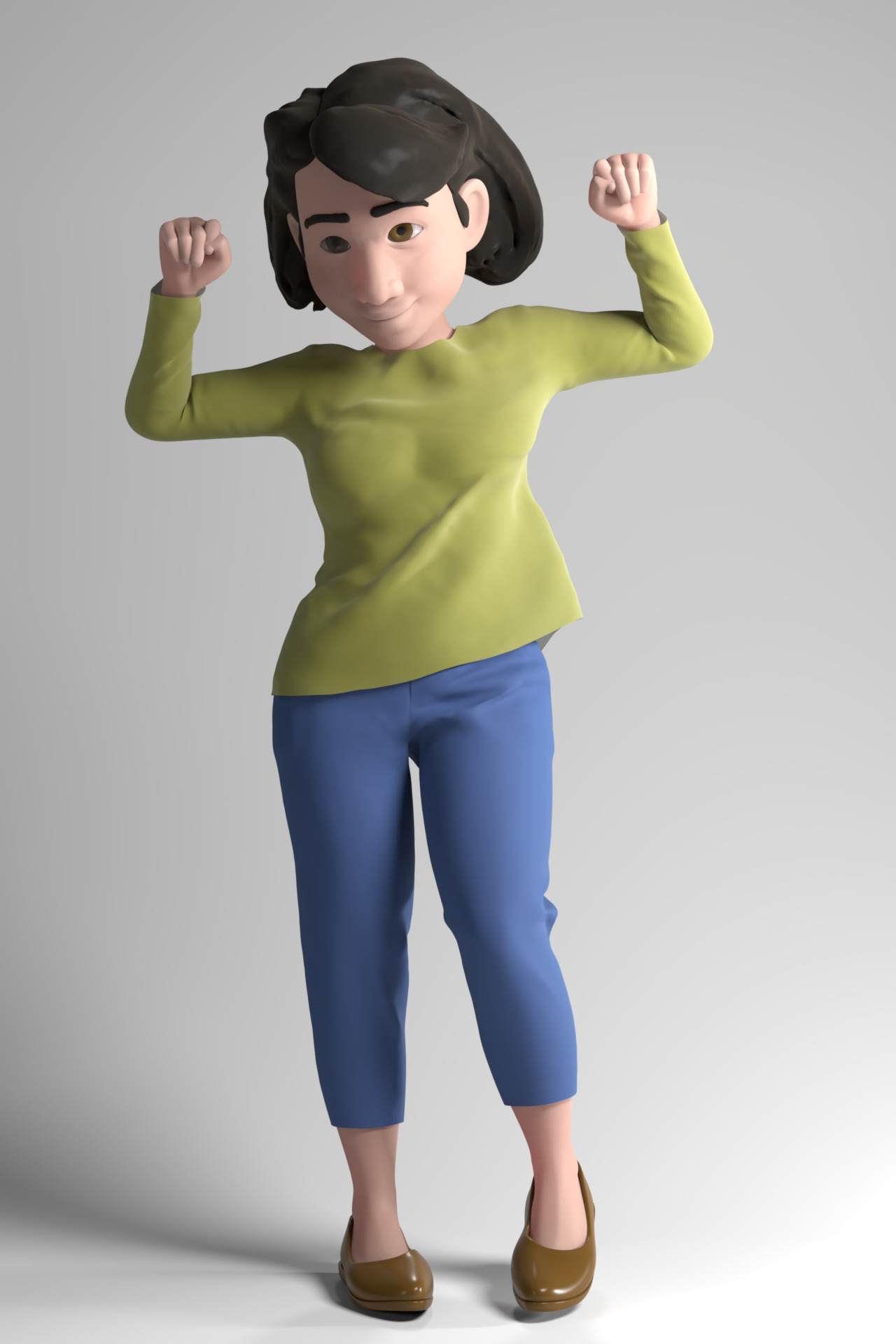} \\
        \includegraphics[width=0.1405\linewidth]{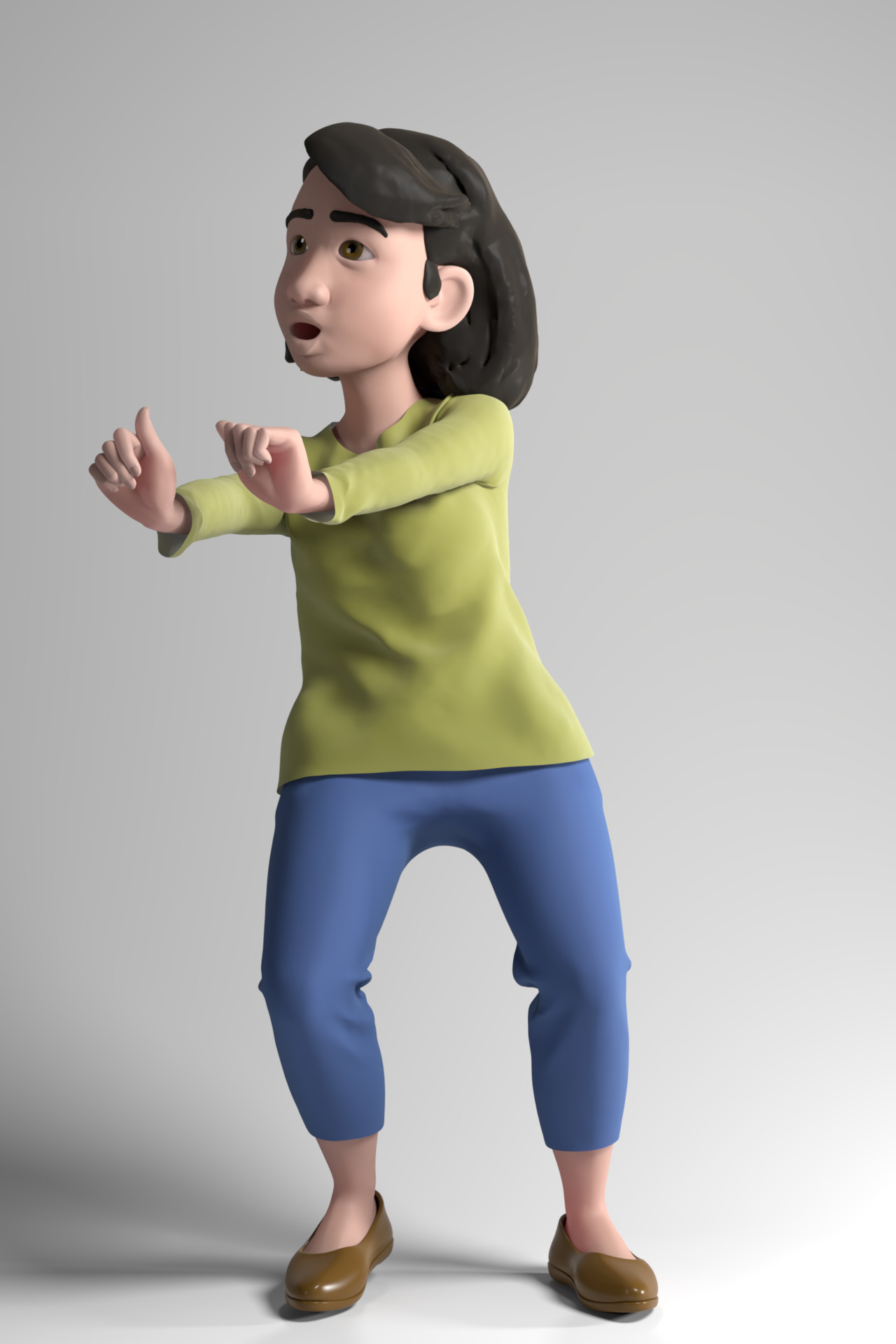} &
        \includegraphics[width=0.1405\linewidth]{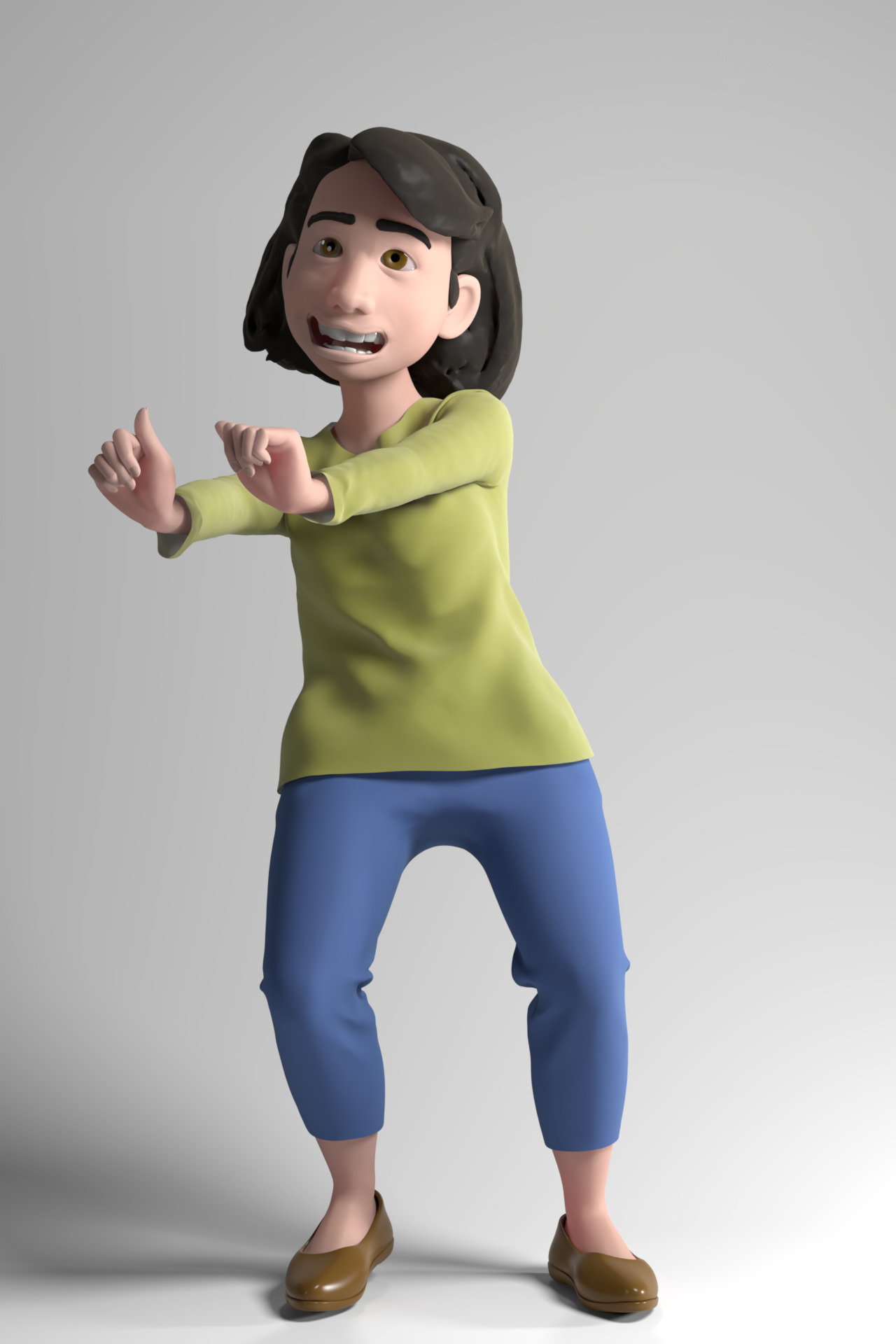} &
        \includegraphics[width=0.1405\linewidth]{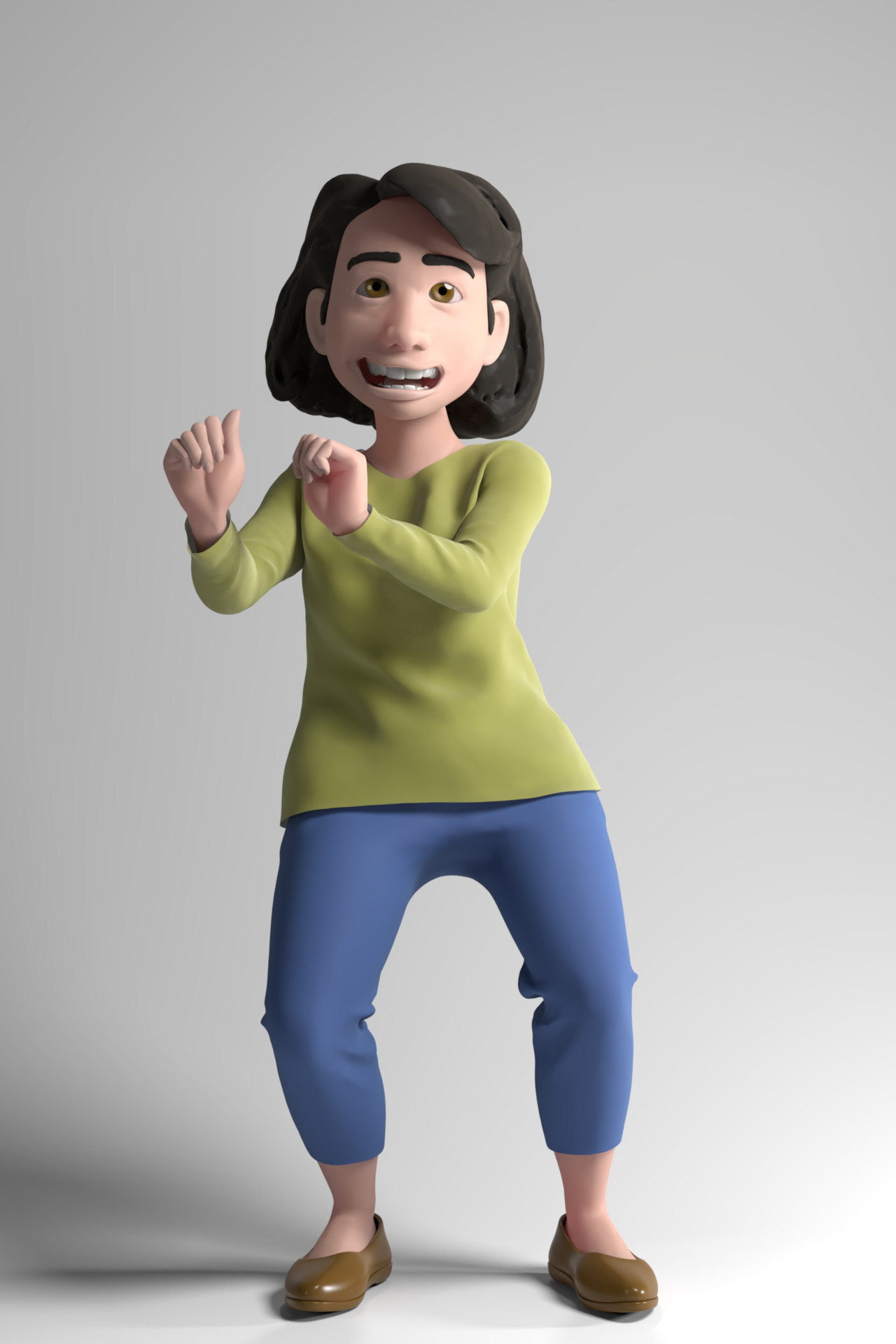} &
        \includegraphics[width=0.1405\linewidth]{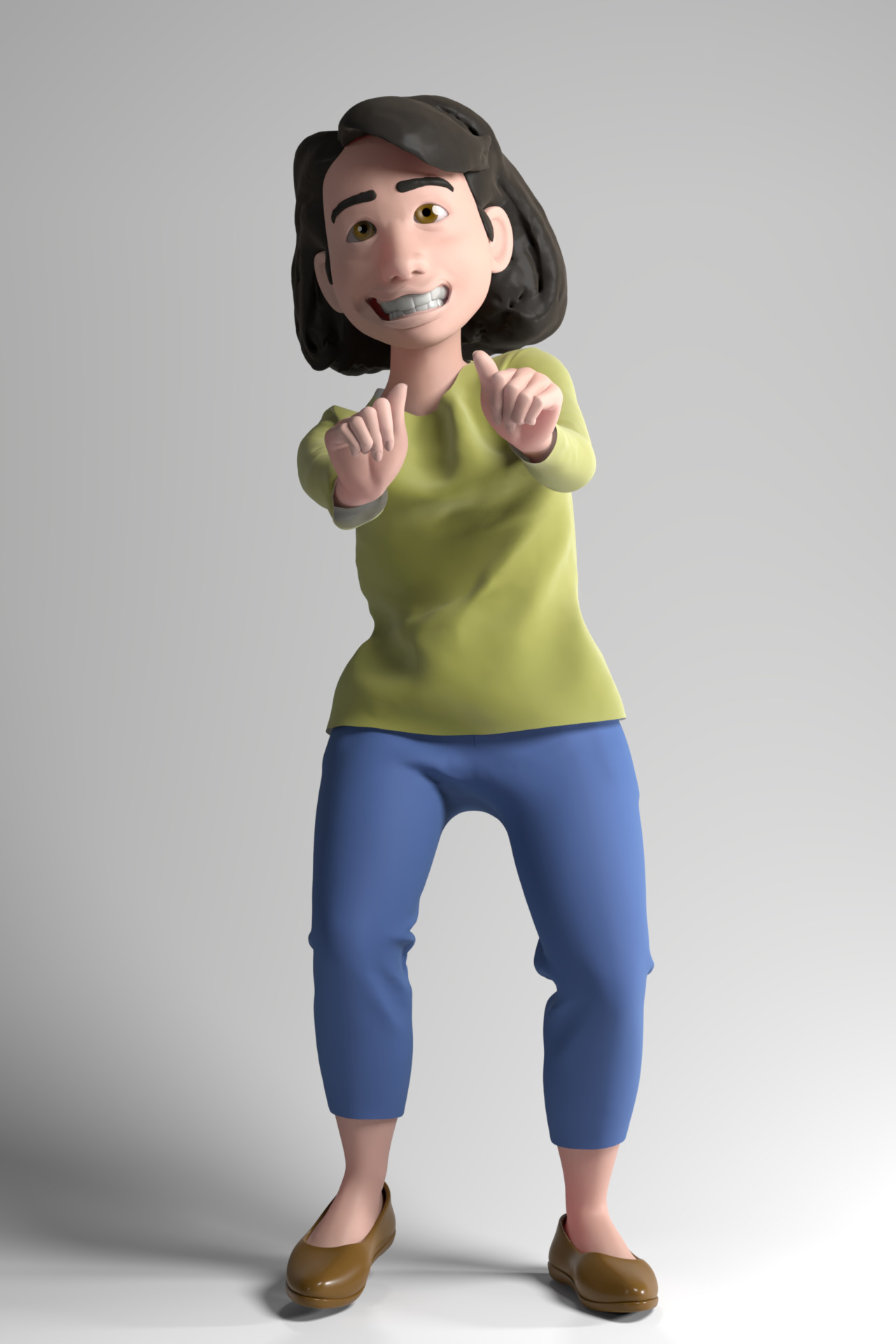} &
        \includegraphics[width=0.1405\linewidth]{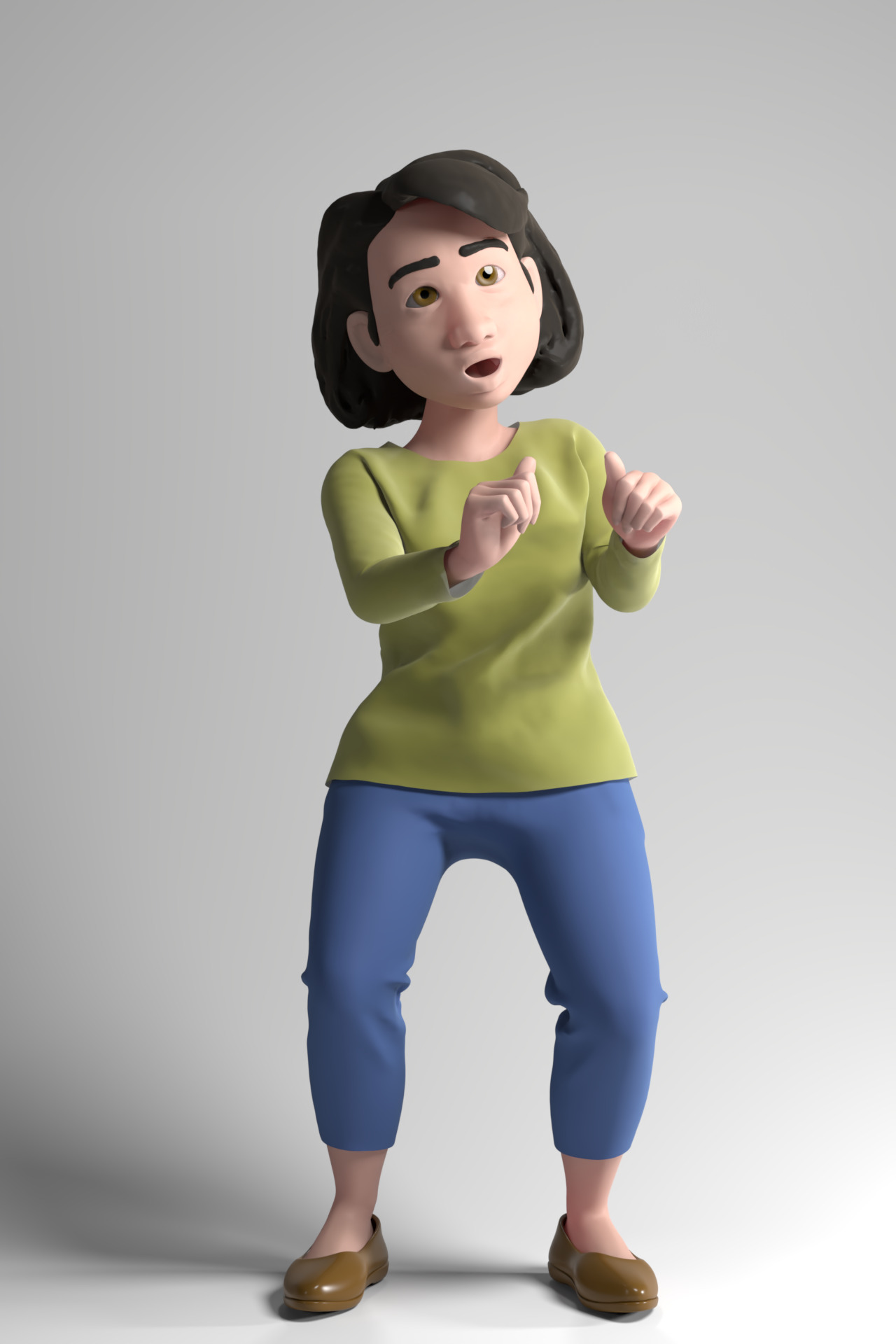} &
        \includegraphics[width=0.1405\linewidth]{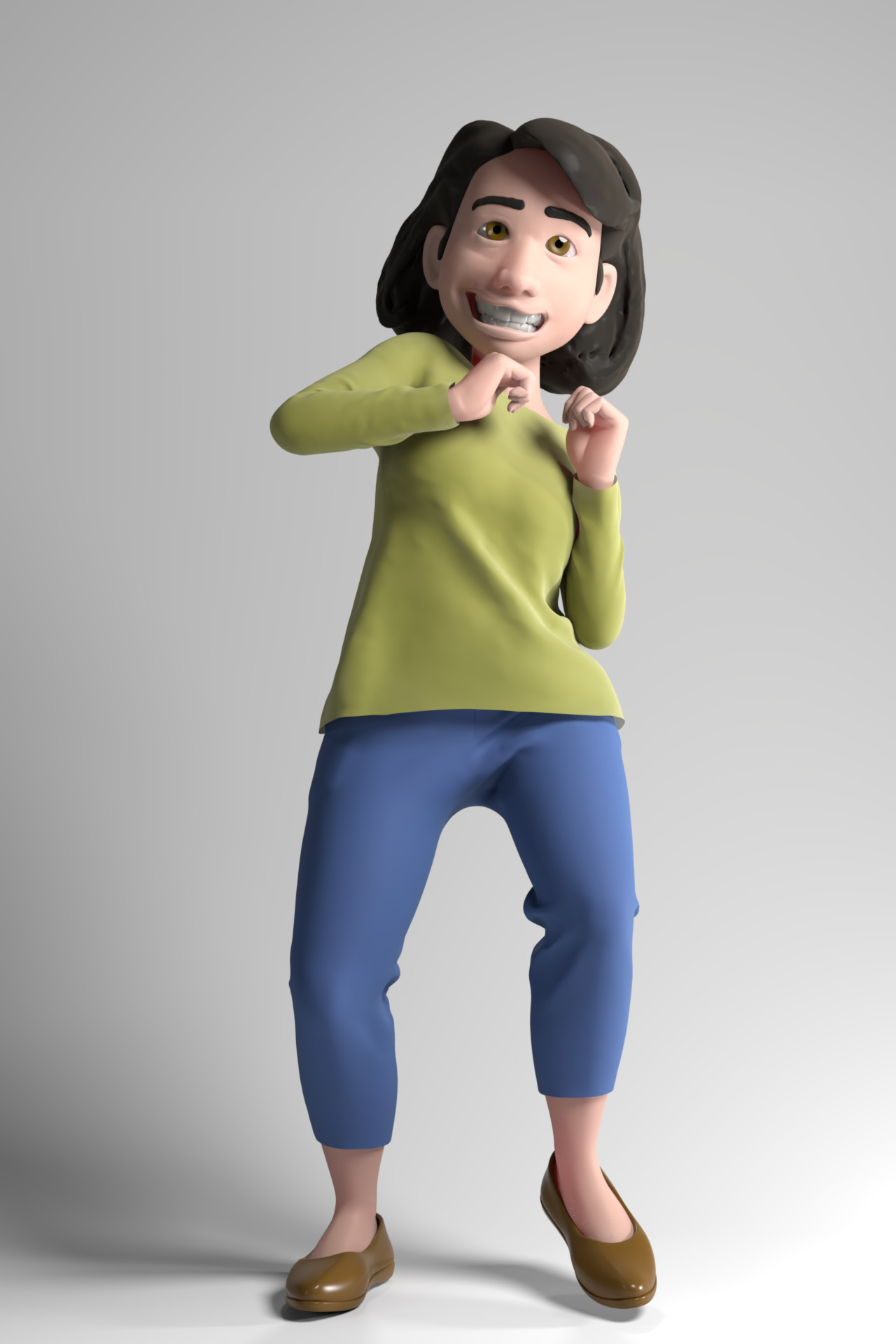} &
        \includegraphics[width=0.1405\linewidth]{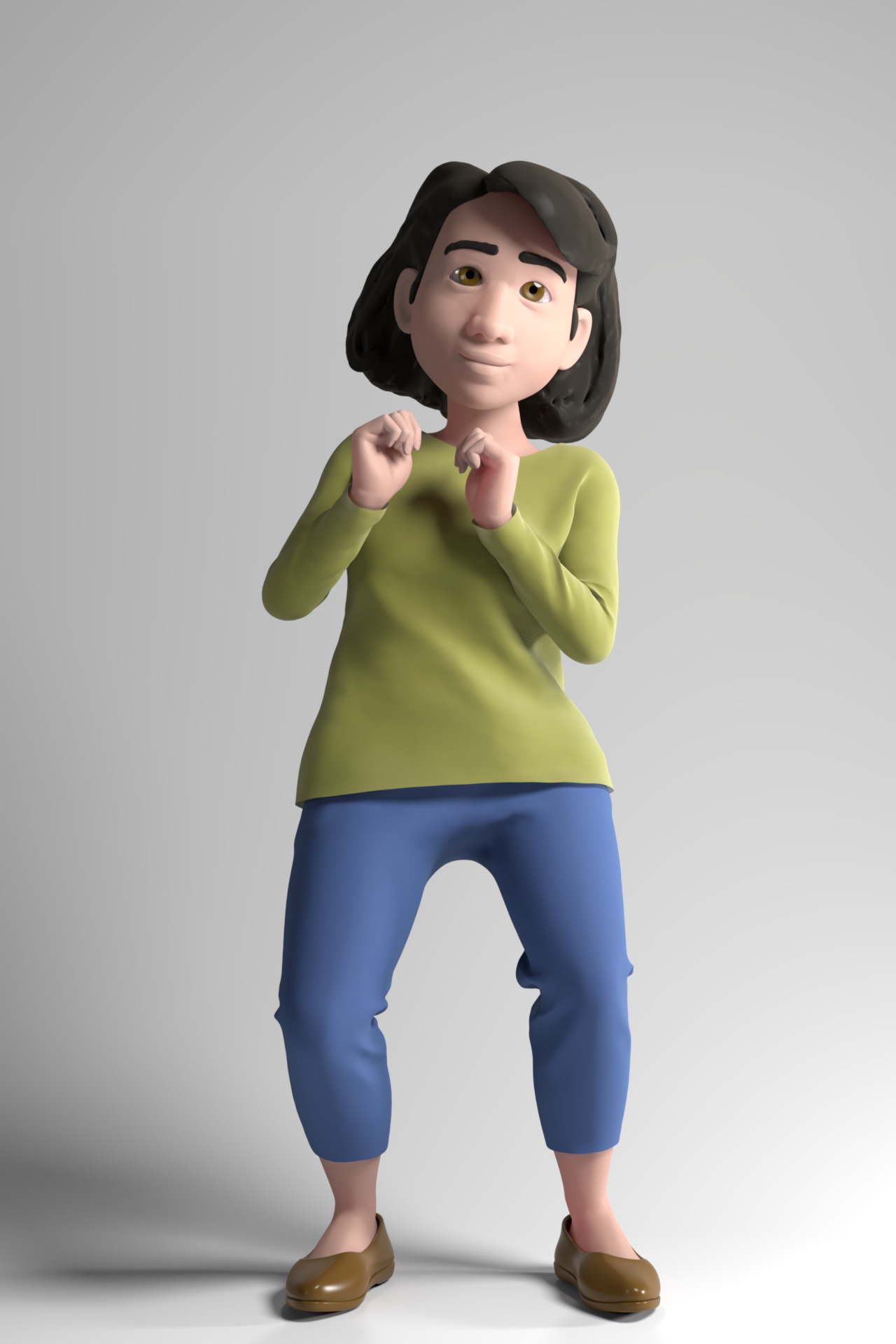} \\
        \includegraphics[width=0.1405\linewidth]{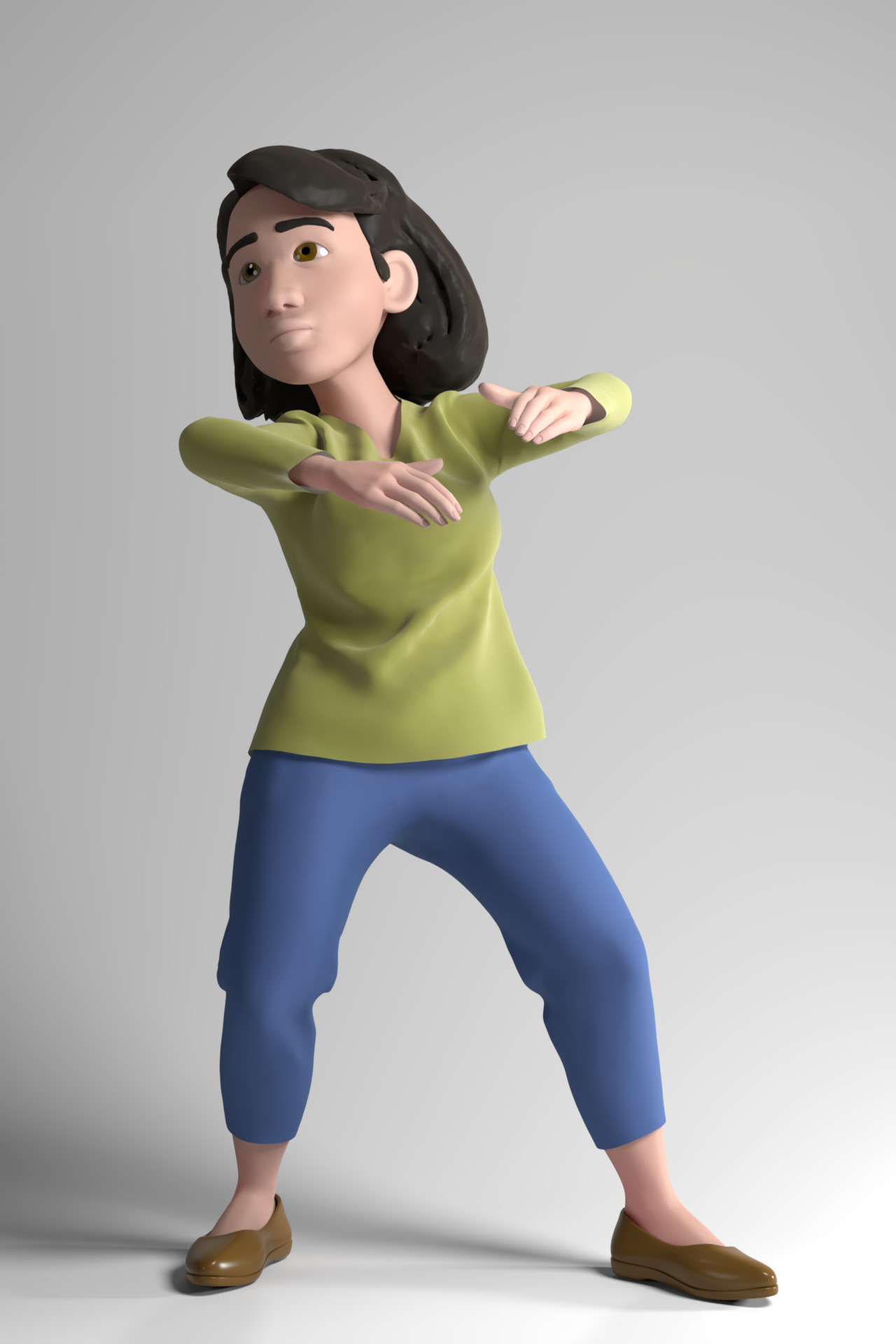} &
        \includegraphics[width=0.1405\linewidth]{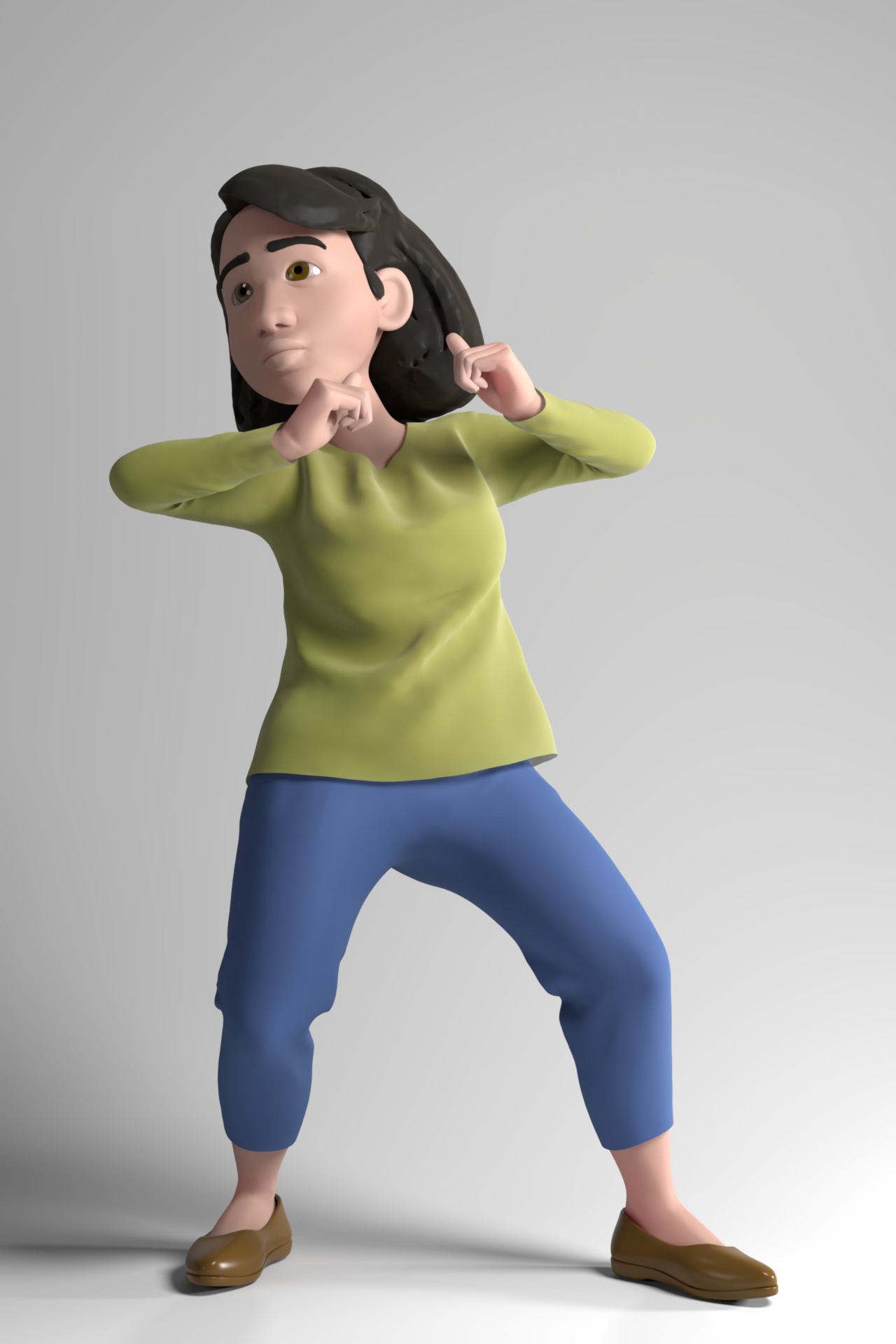} &
        \includegraphics[width=0.1405\linewidth]{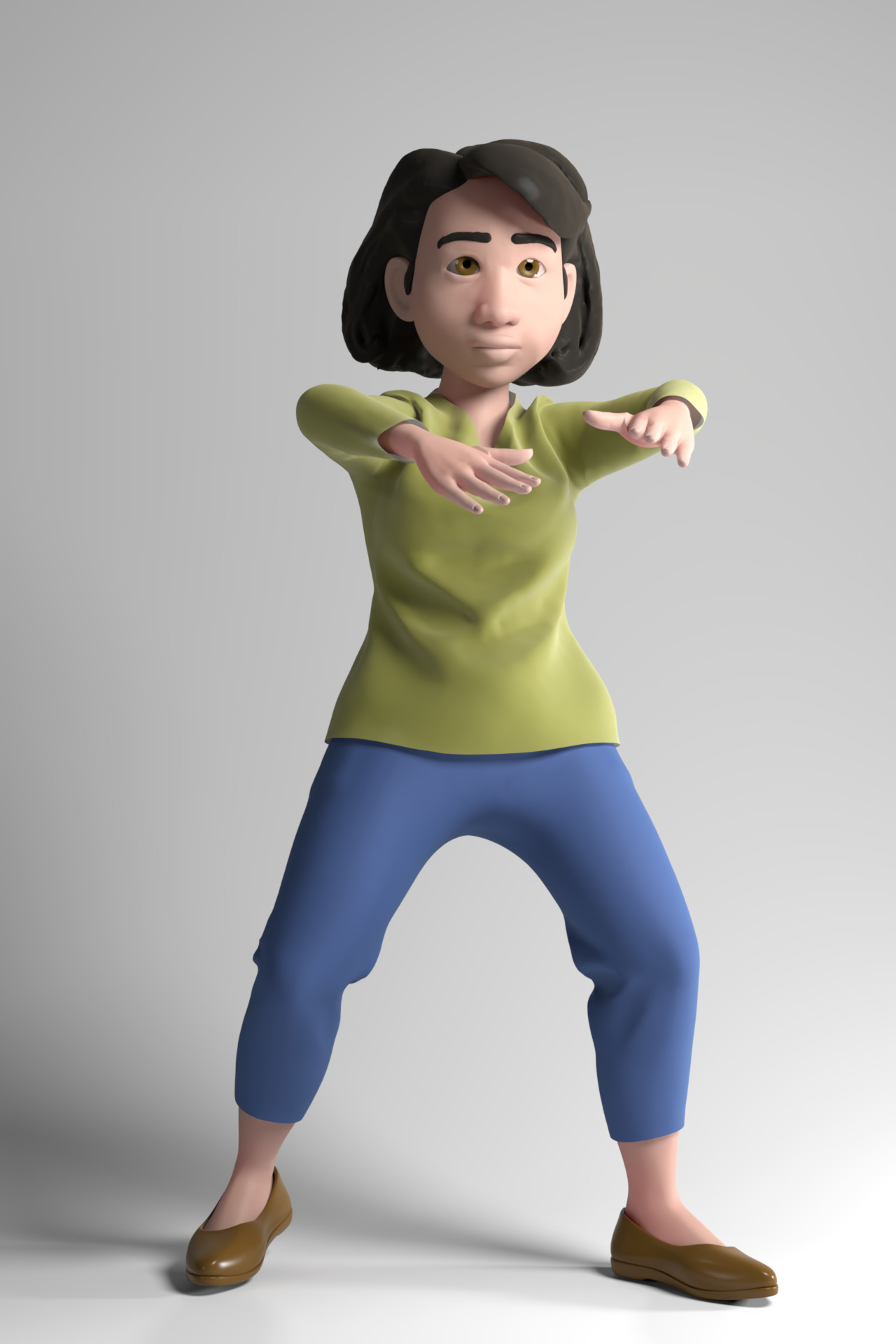} &
        \includegraphics[width=0.1405\linewidth]{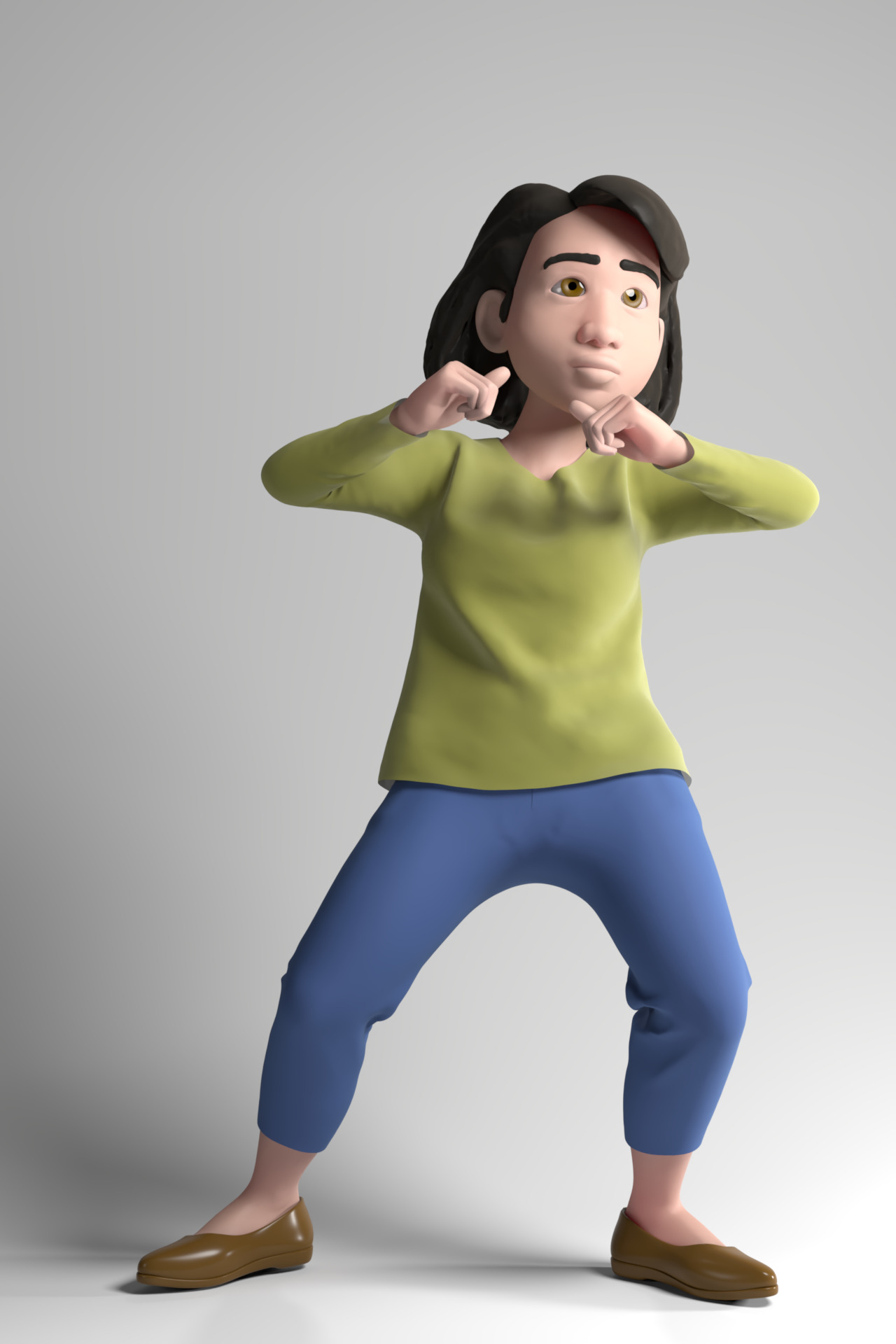} &
        \includegraphics[width=0.1405\linewidth]{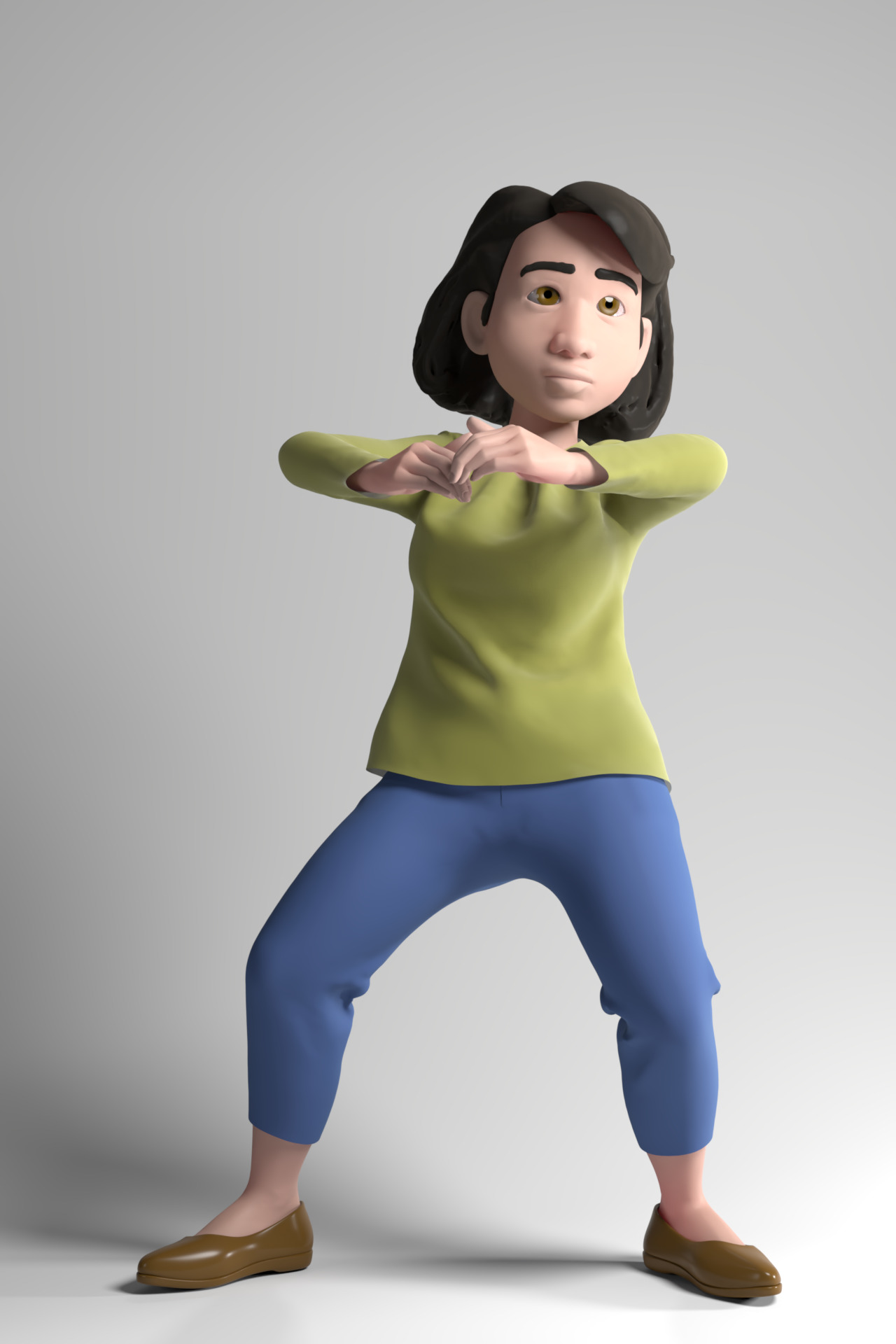} &
        \includegraphics[width=0.1405\linewidth]{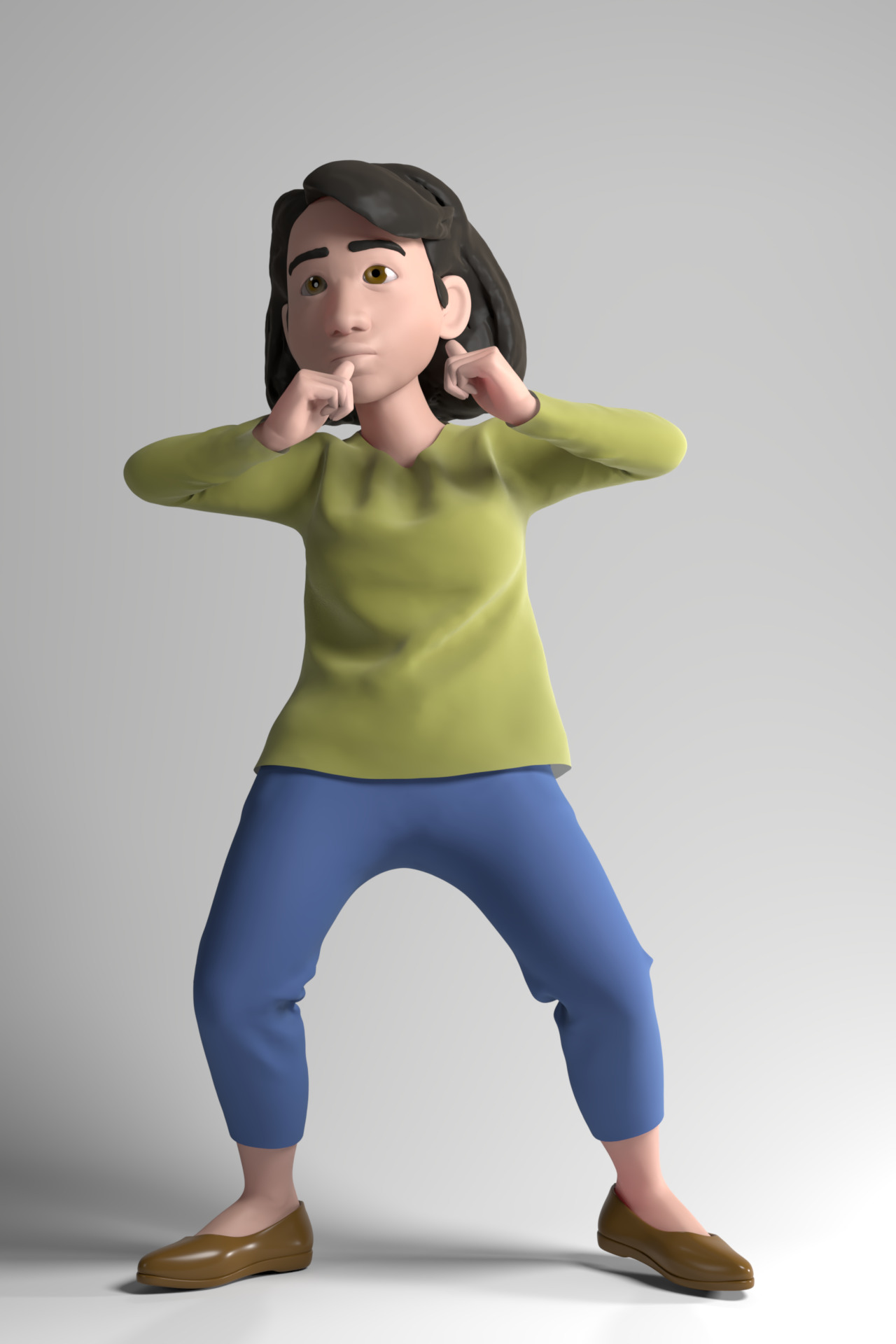} &
        \includegraphics[width=0.1405\linewidth]{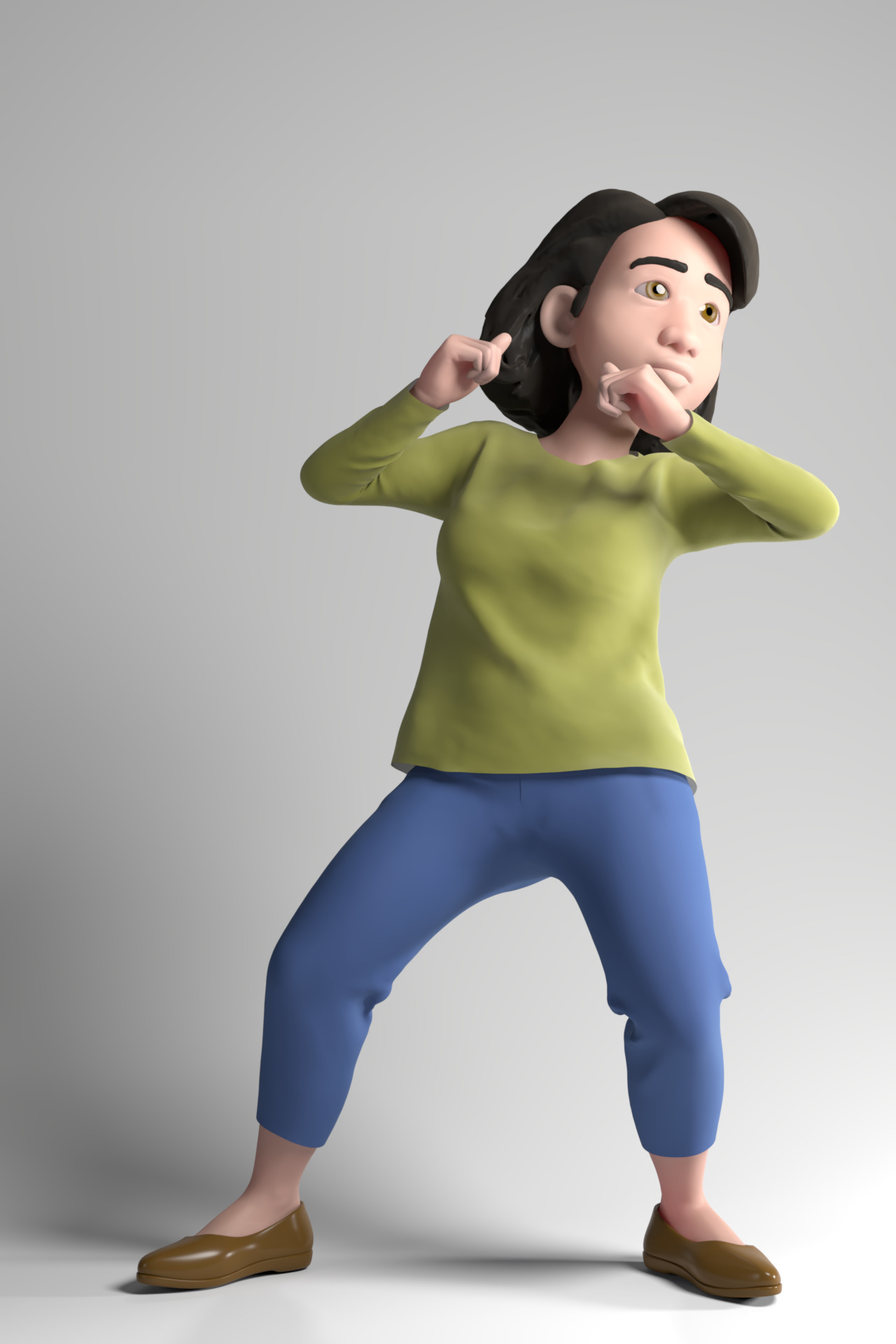} \\
        \includegraphics[width=0.1405\linewidth]{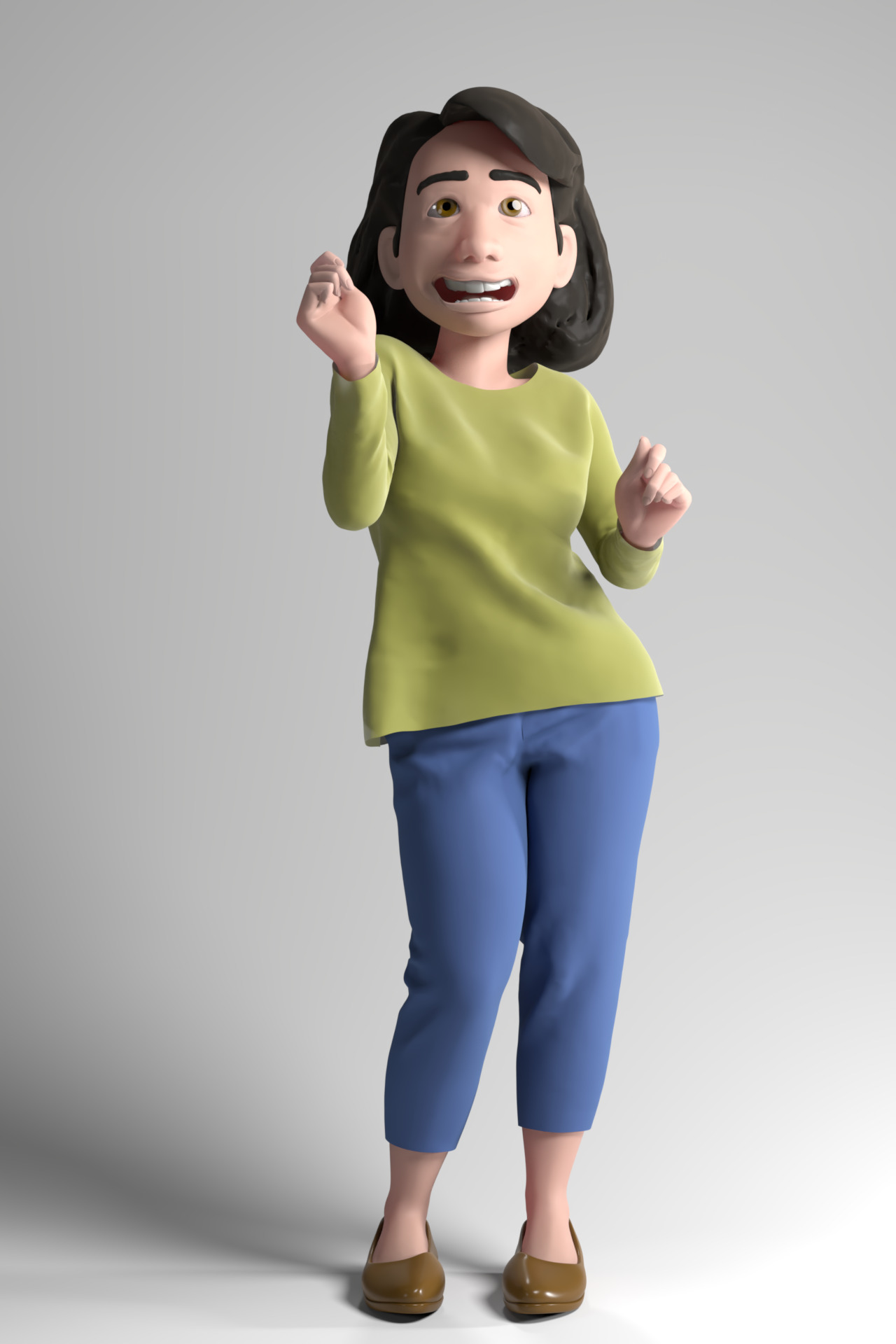} &
        \includegraphics[width=0.1405\linewidth]{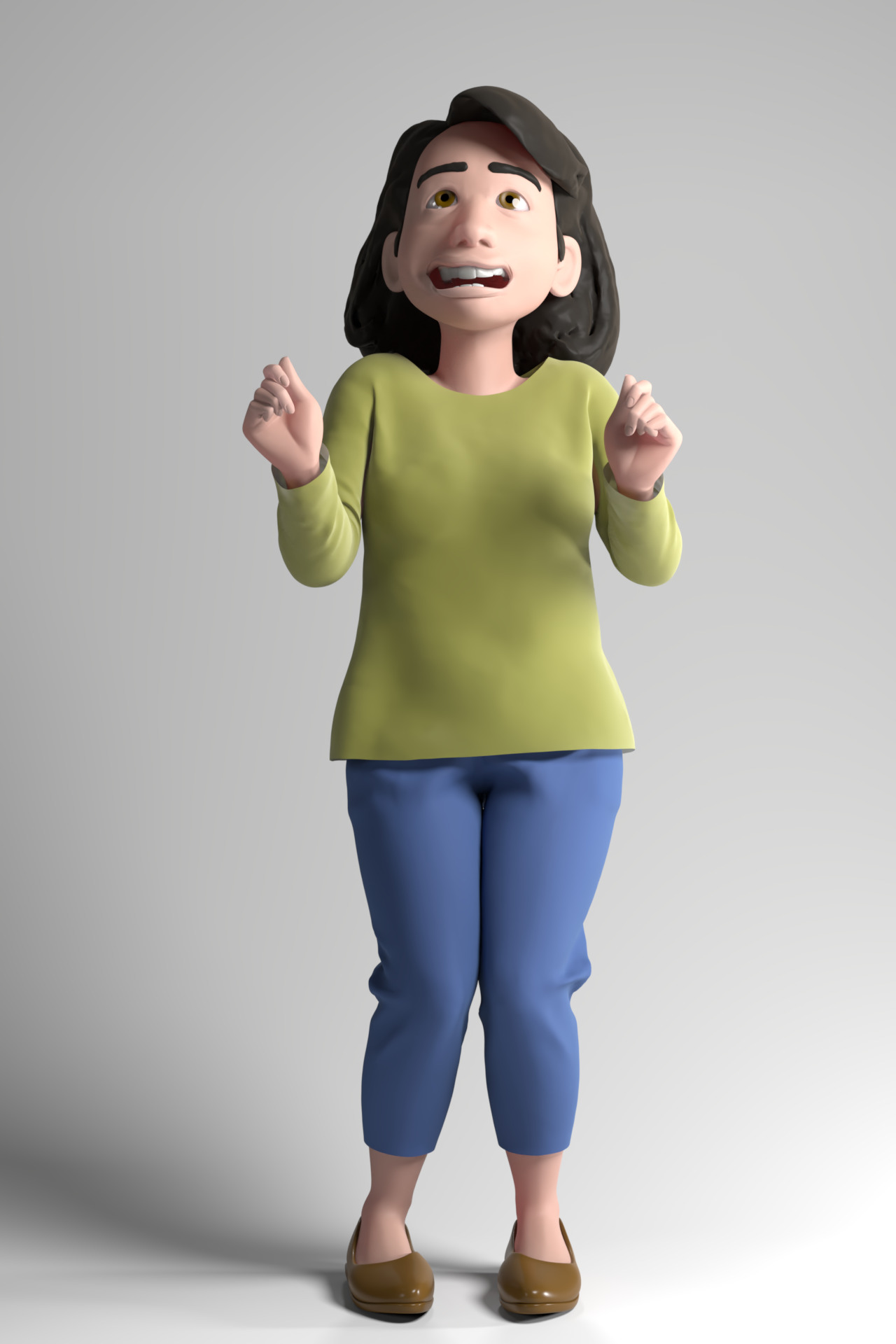} &
        \includegraphics[width=0.1405\linewidth]{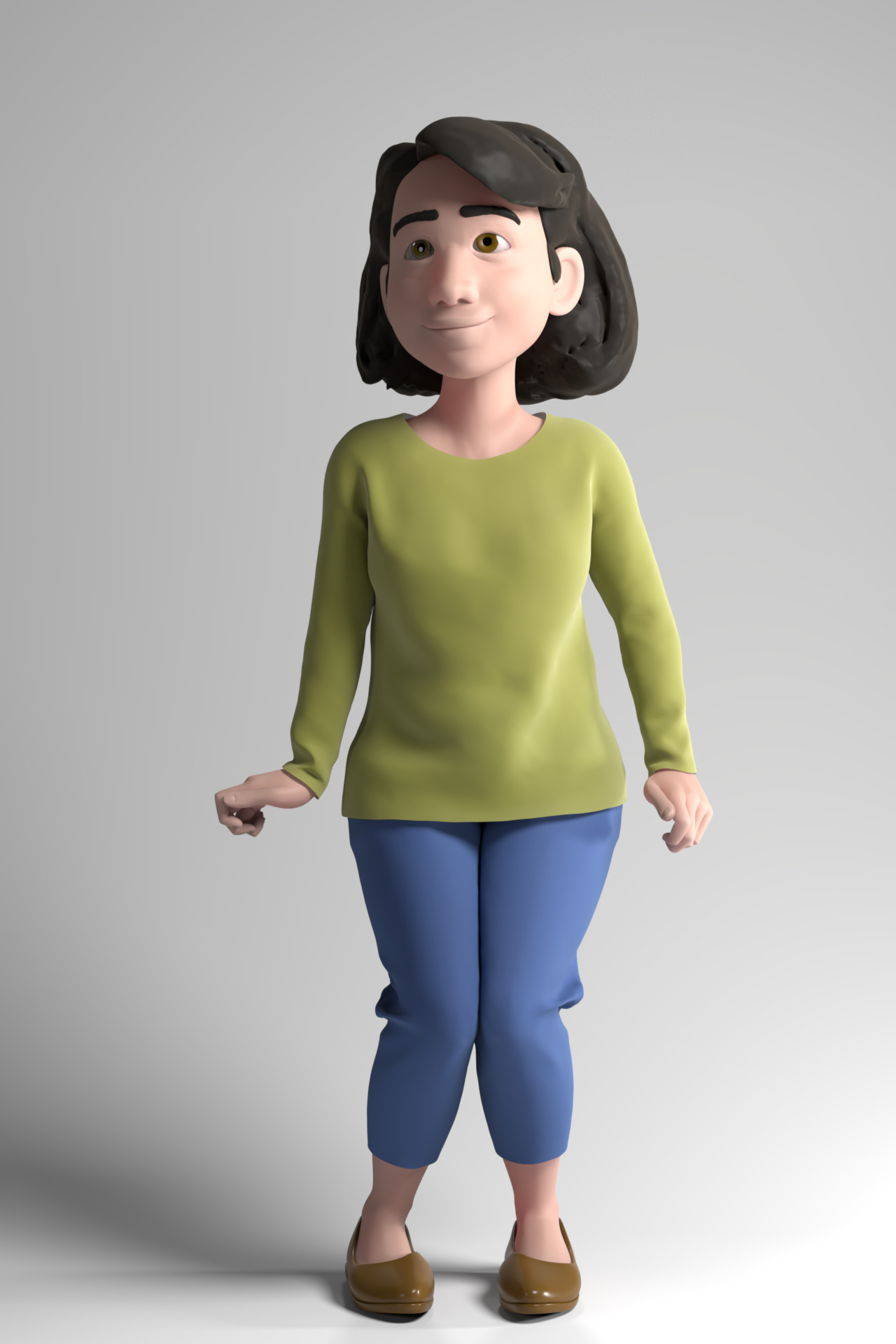} &
        \includegraphics[width=0.1405\linewidth]{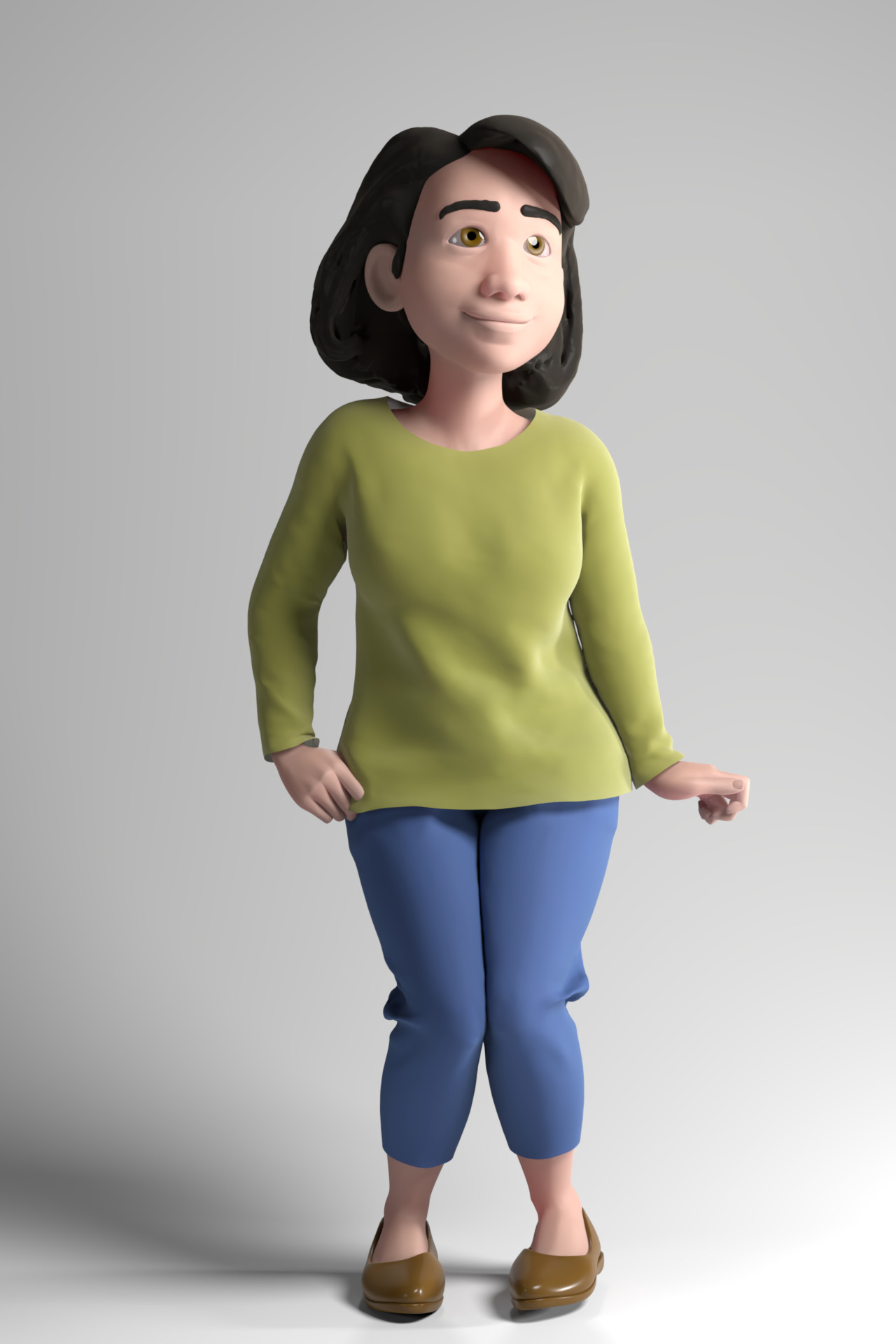} &
        \includegraphics[width=0.1405\linewidth]{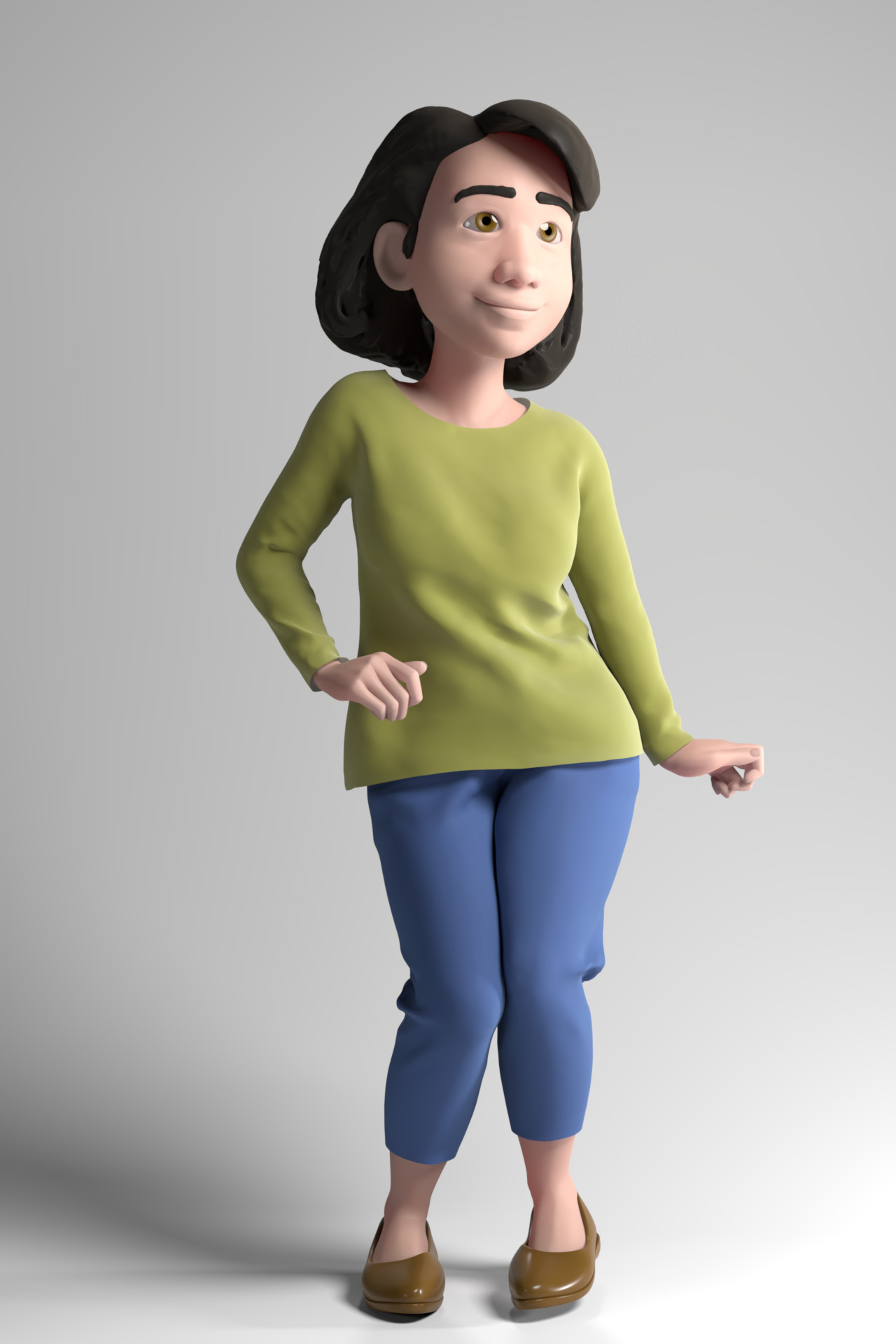} &
        \includegraphics[width=0.1405\linewidth]{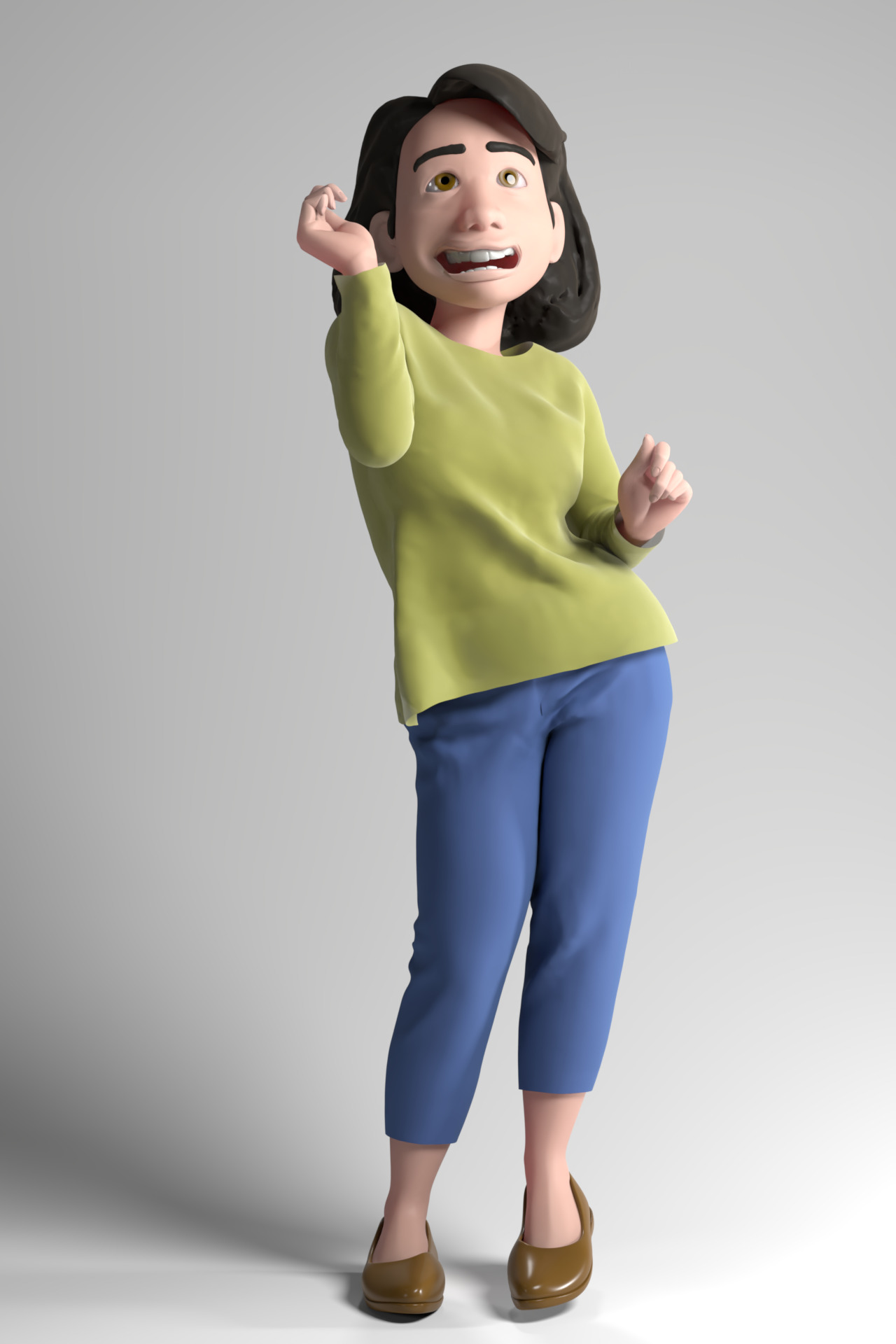} &
        \includegraphics[width=0.1405\linewidth]{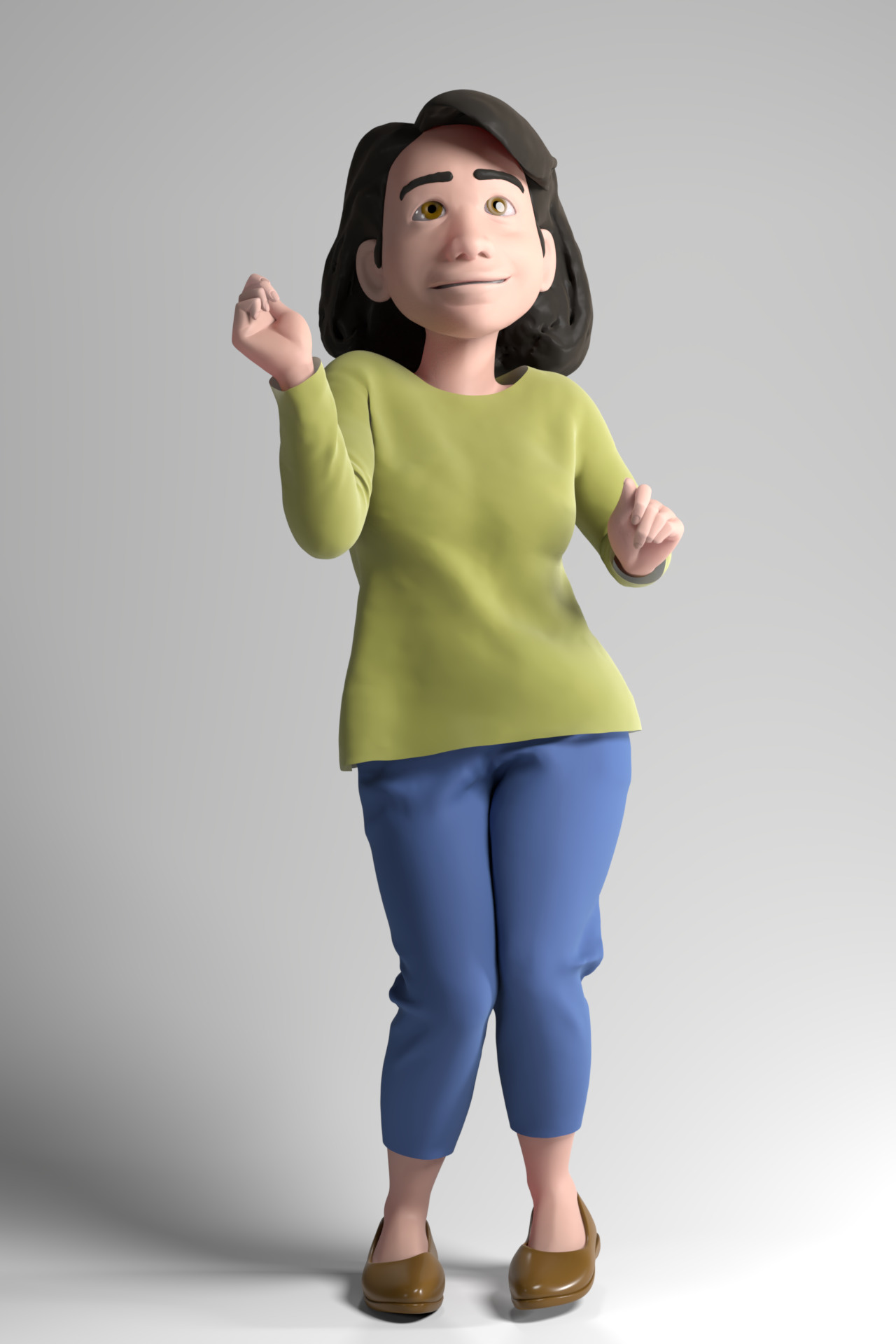} \\
	\end{tabular}
	\caption{Representative predictions on the validation sequences from \jenShirt and \jenPants. Each row shows a different dancing animation.}
	\label{fig:GalleryImage}
\end{figure*}

\paragraph*{Validation sequences.}
\Fig{GalleryImage} shows representative predictions produced by our model on the validation split of \jenShirt and \jenPants datasets. Each row corresponds to a different dancing animation that is fully excluded from training, and columns illustrate different poses sampled along the sequence. Despite complex full-body motions and challenging wrinkling patterns, the model produces stable, high-quality garment deformations with consistent wrinkle structure across poses. These results highlight the ability of the proposed approach to generalize beyond the training distribution and to handle long-range pose variations without temporal information.

\begin{figure*}[t]
	\centering
	\setlength{\tabcolsep}{0.00125\linewidth}
    {\footnotesize
    \begin{tabular}{
          >{\centering\arraybackslash}m{0.06\linewidth}
          m{0.15\linewidth}
          m{0.15\linewidth}
          m{0.15\linewidth}
          m{0.15\linewidth}
          m{0.15\linewidth}
          m{0.15\linewidth}
        }
        Ours &
        \includegraphics[width=\linewidth]{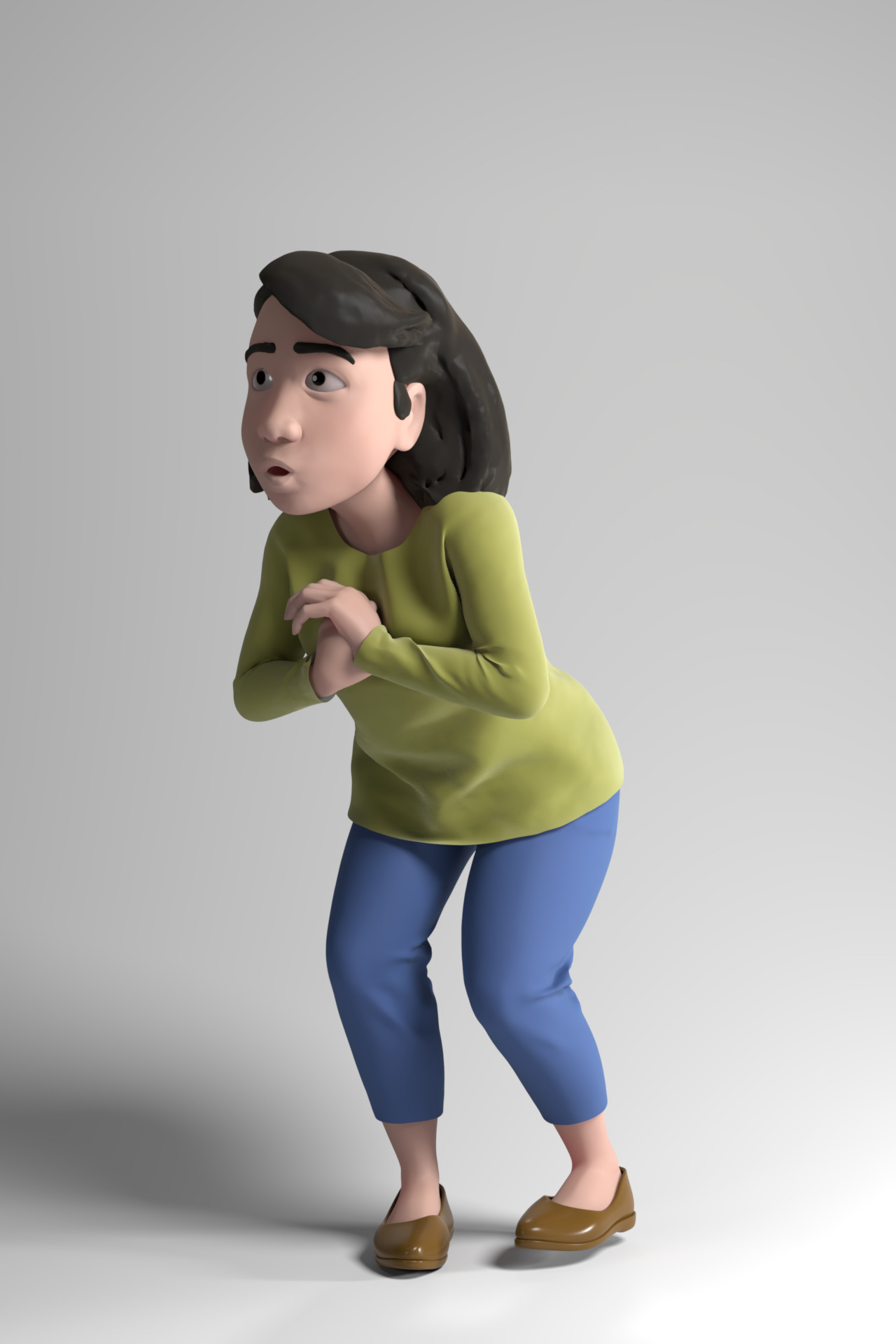} &
        \includegraphics[width=\linewidth]{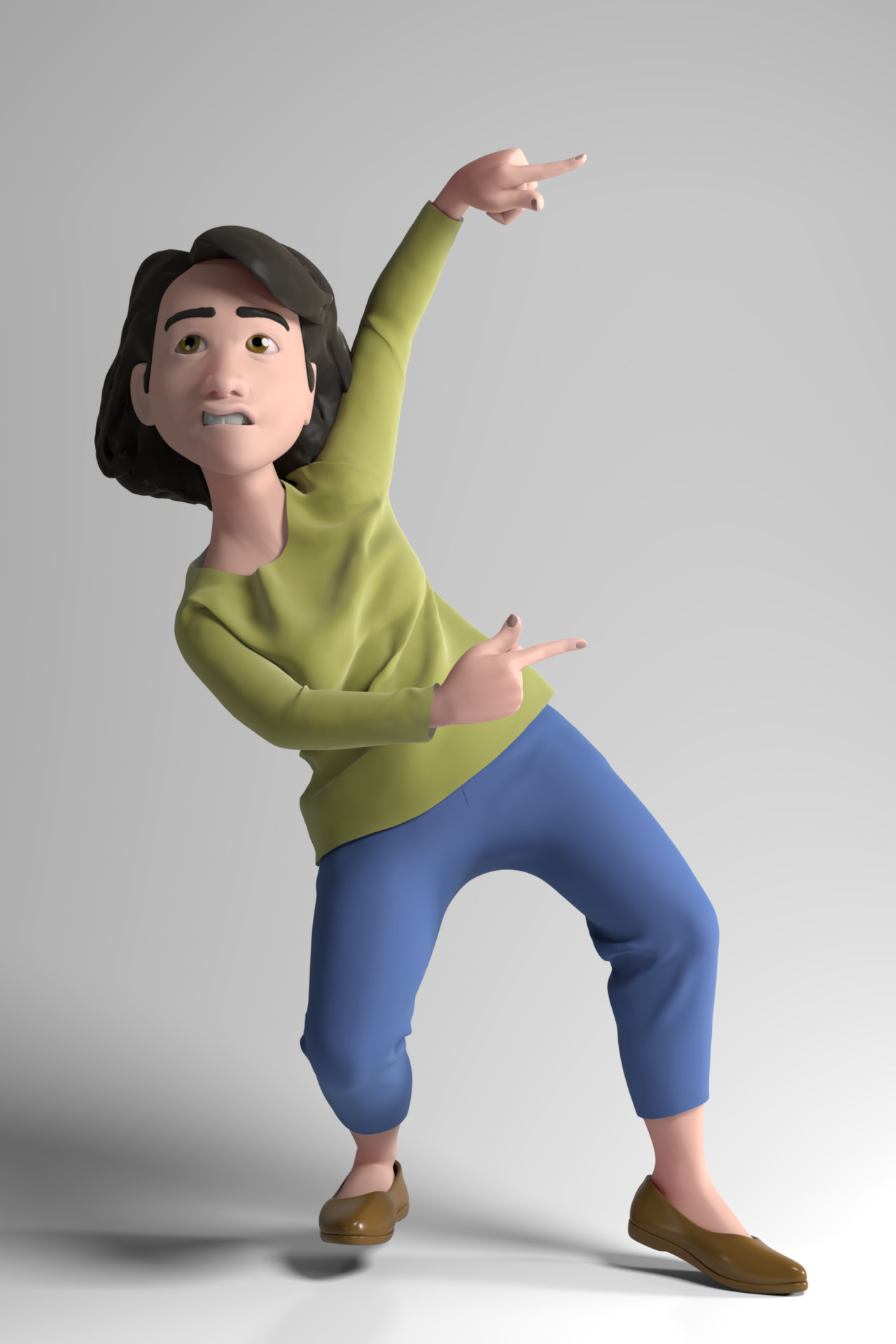} &
        \includegraphics[width=\linewidth]{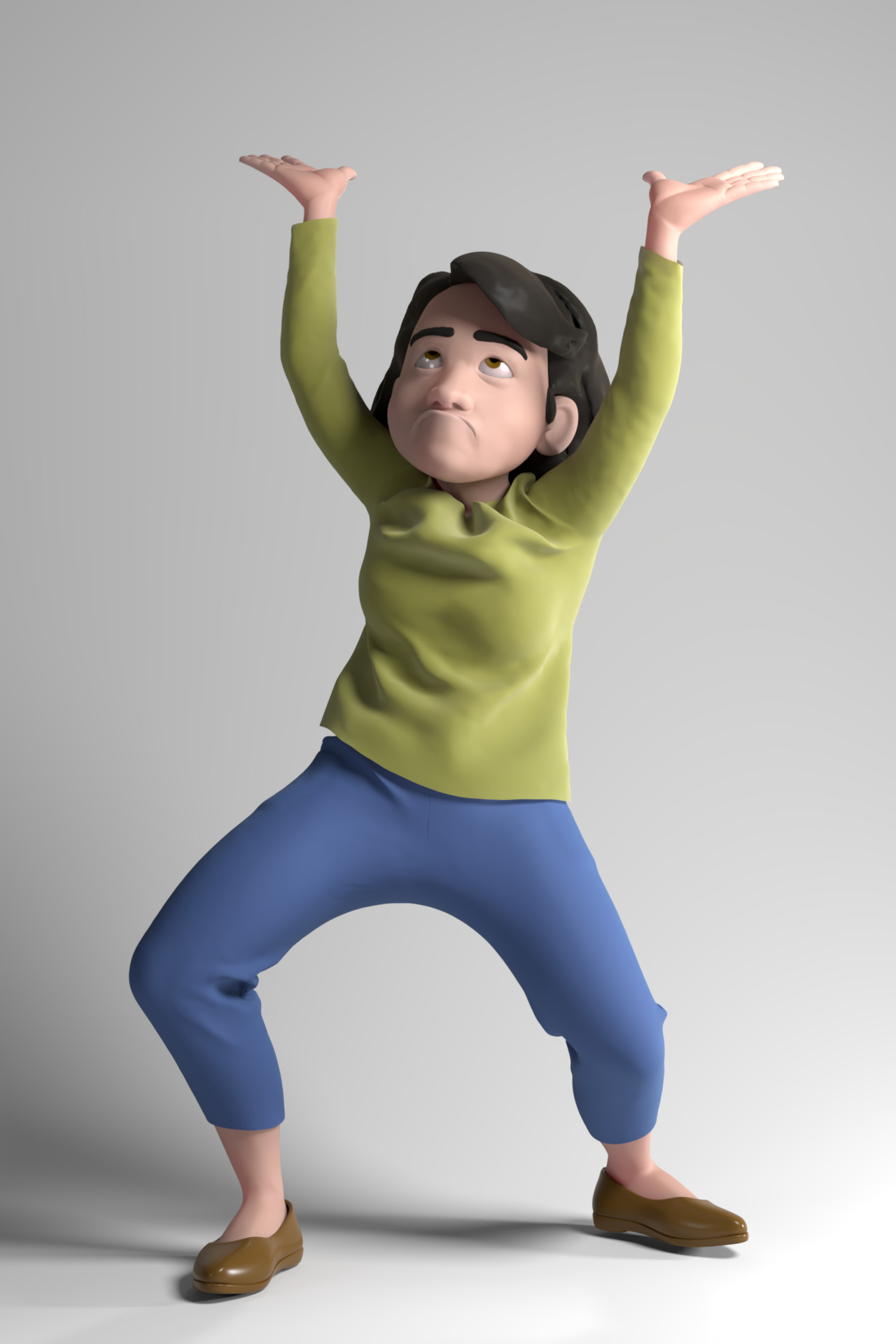} &
        \includegraphics[width=\linewidth]{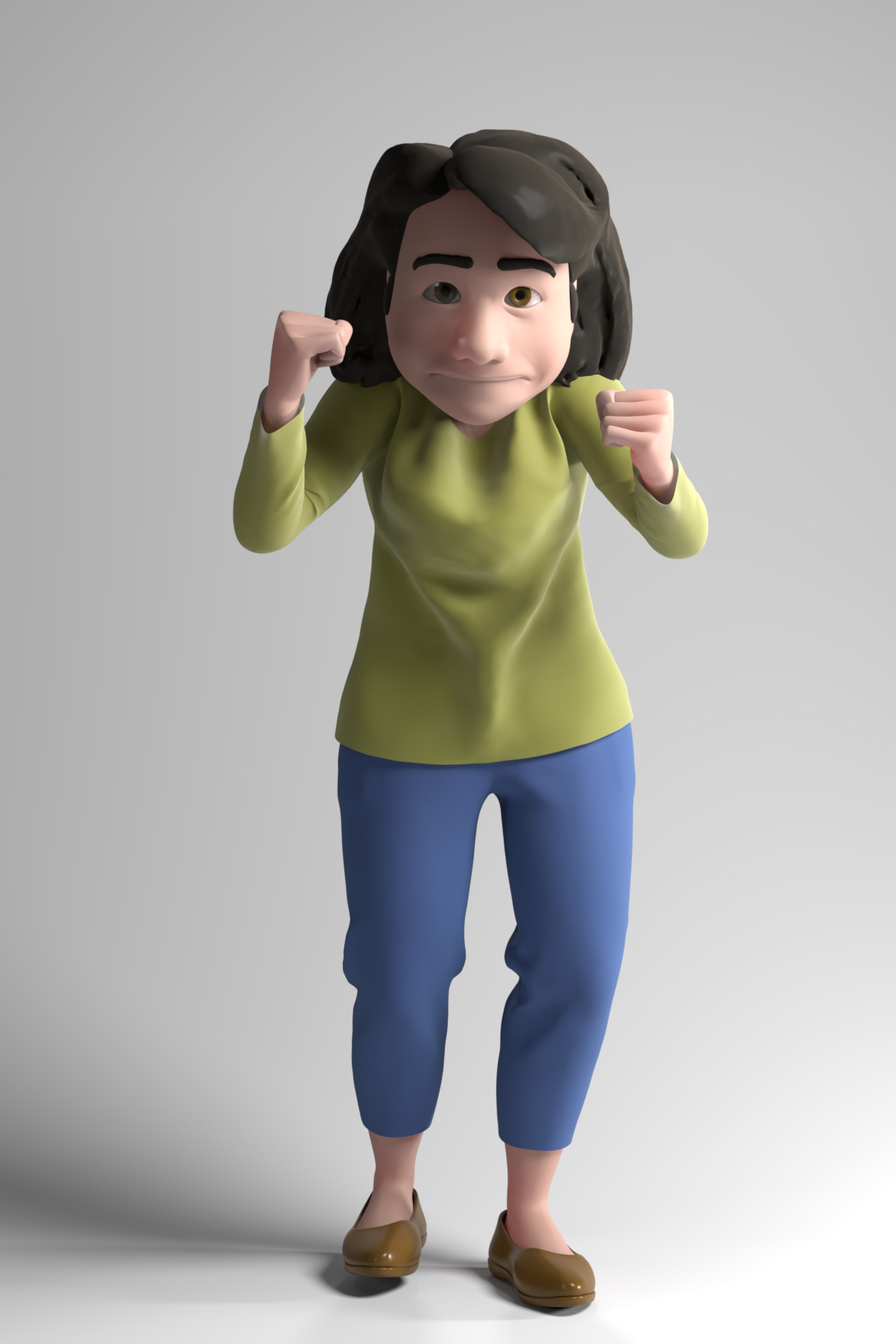} &
        \includegraphics[width=\linewidth]{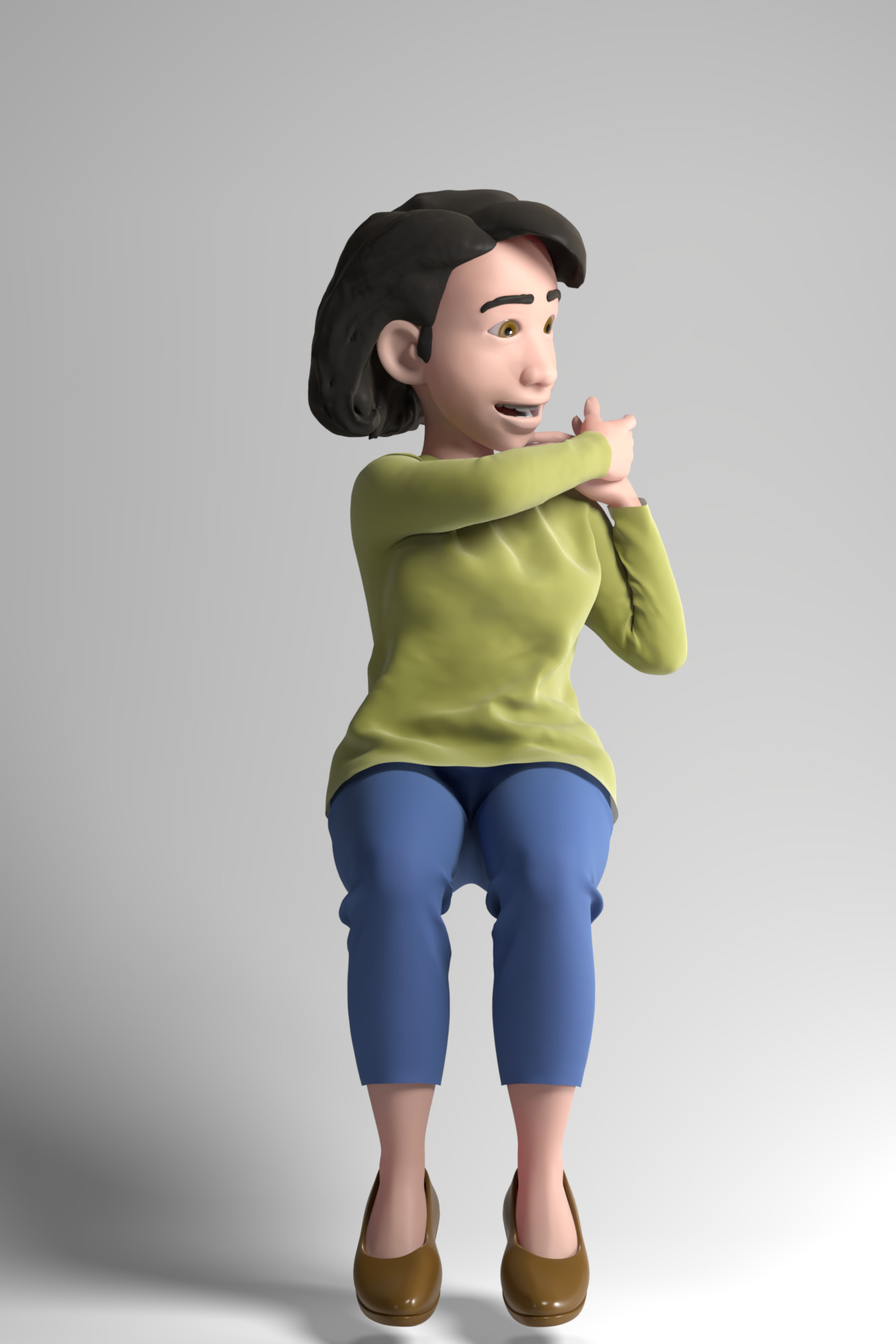} &
        \includegraphics[width=\linewidth]{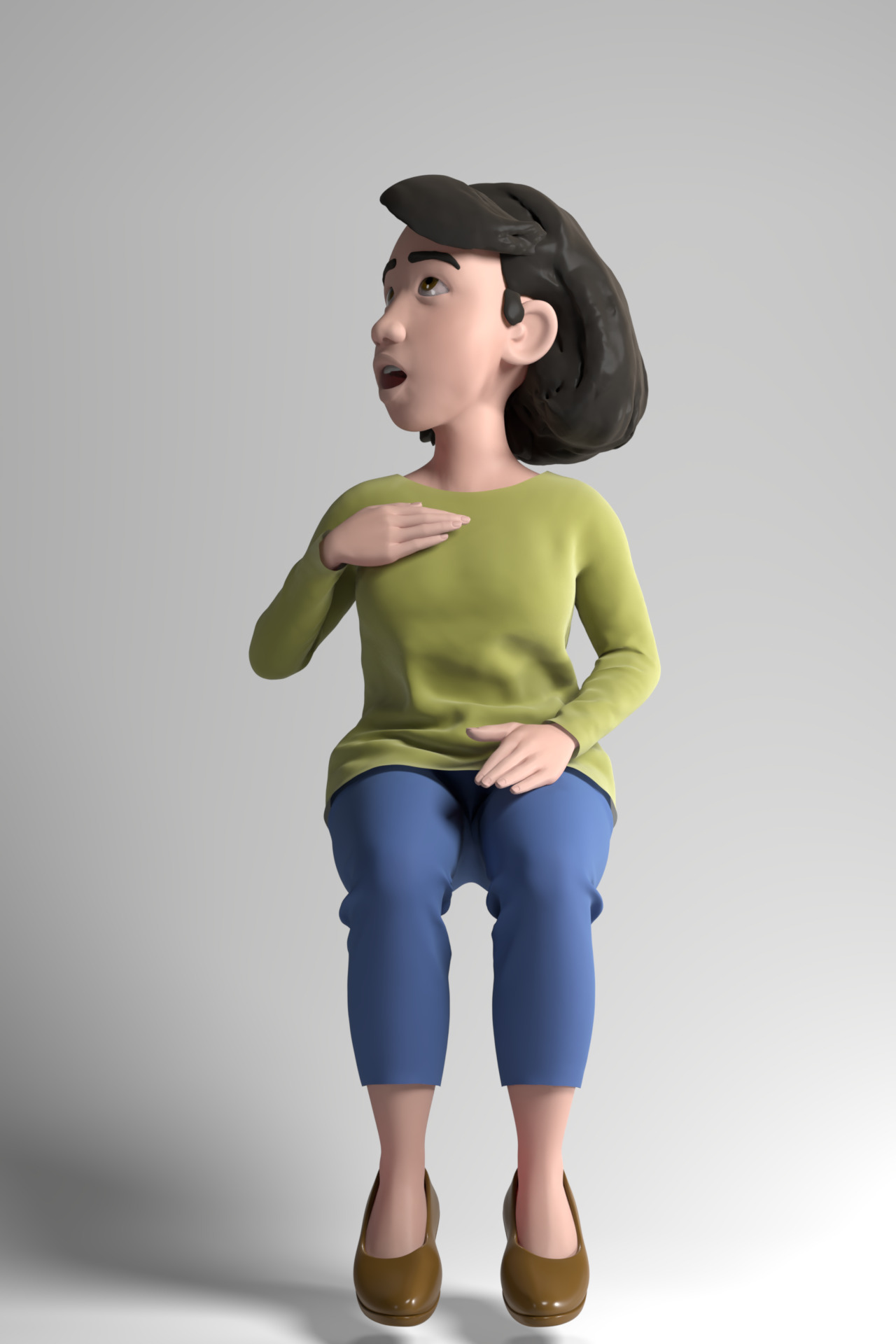} \\

        \shortstack{Ground\\Truth} &
        \includegraphics[width=\linewidth]{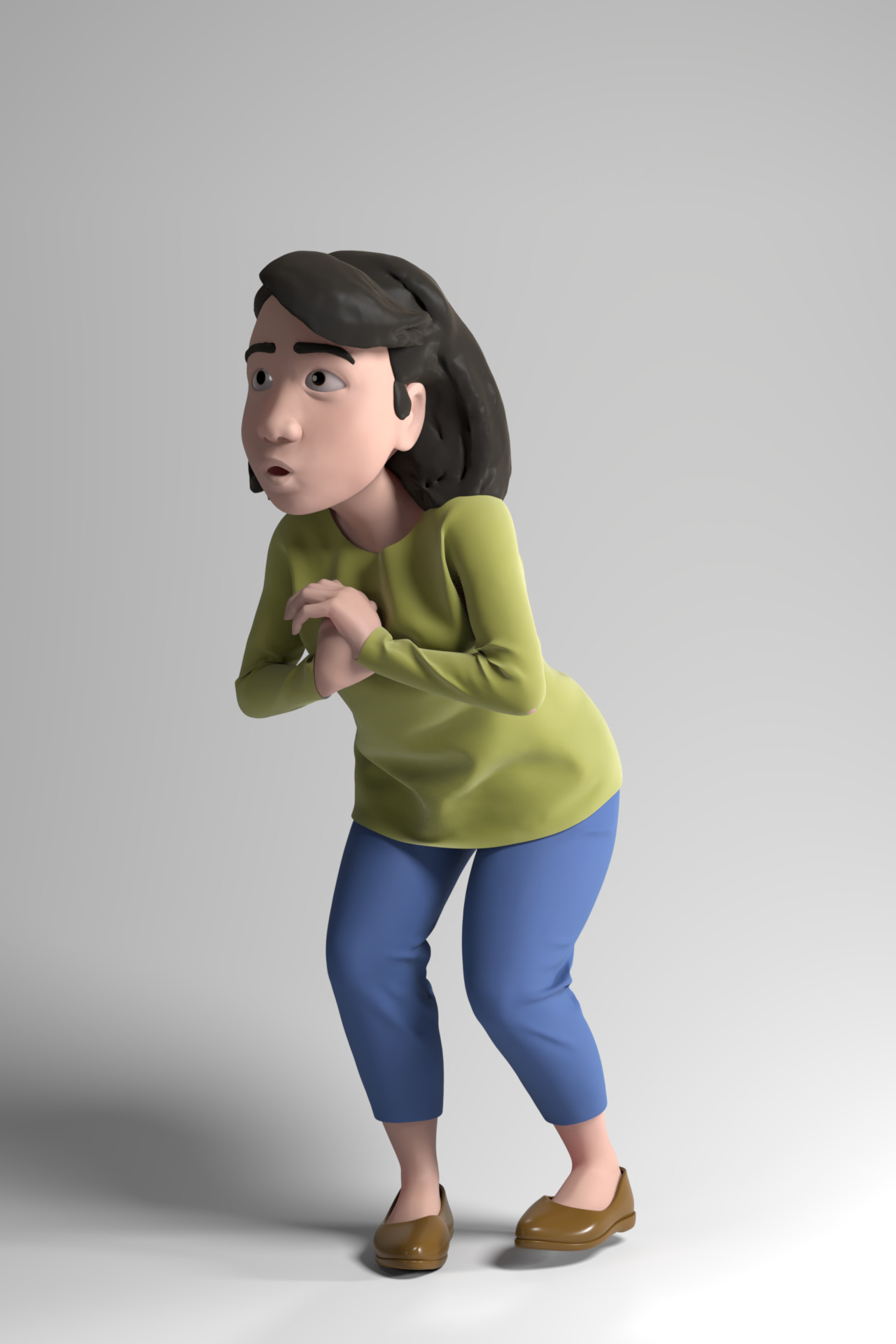} &
        \includegraphics[width=\linewidth]{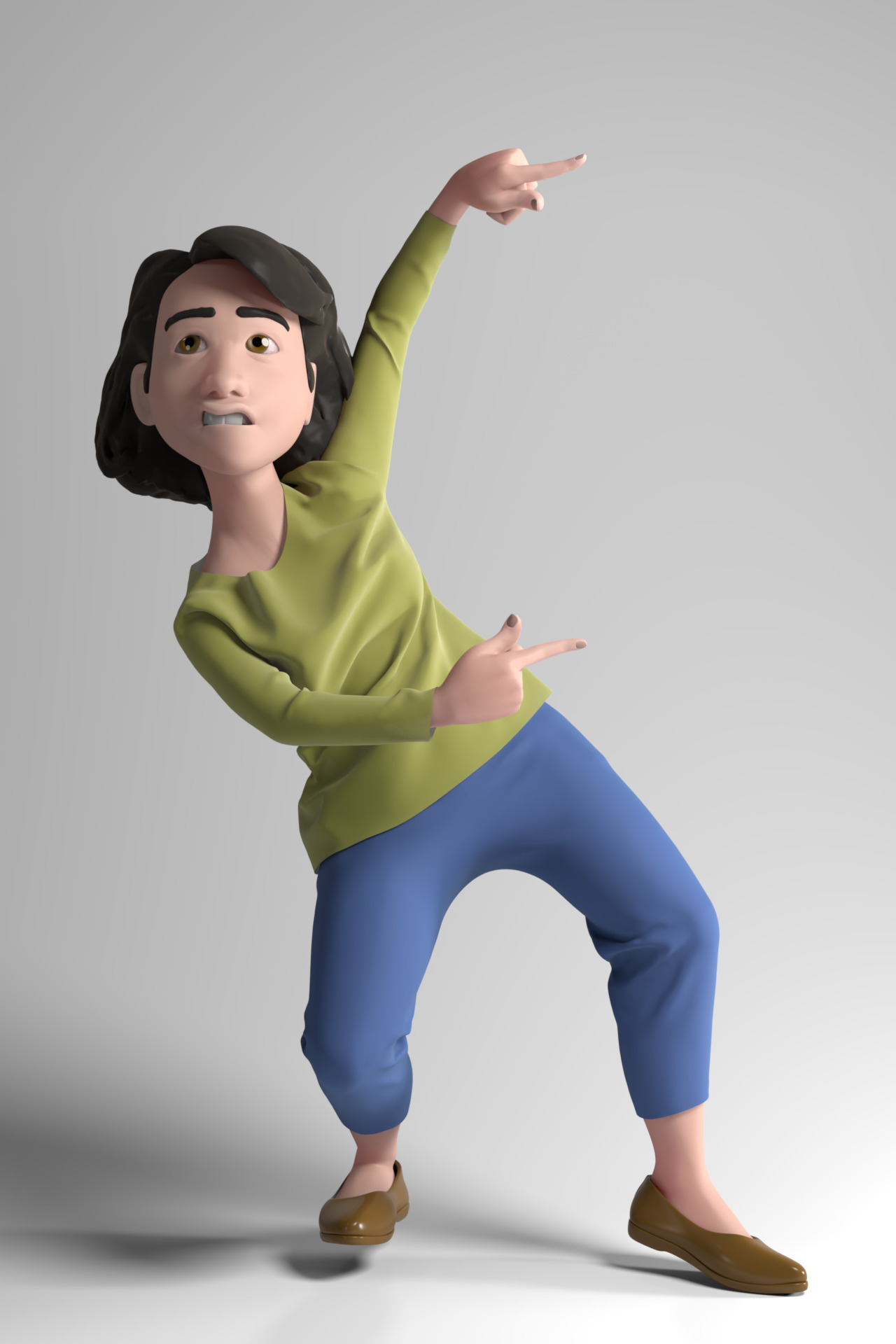} &
        \includegraphics[width=\linewidth]{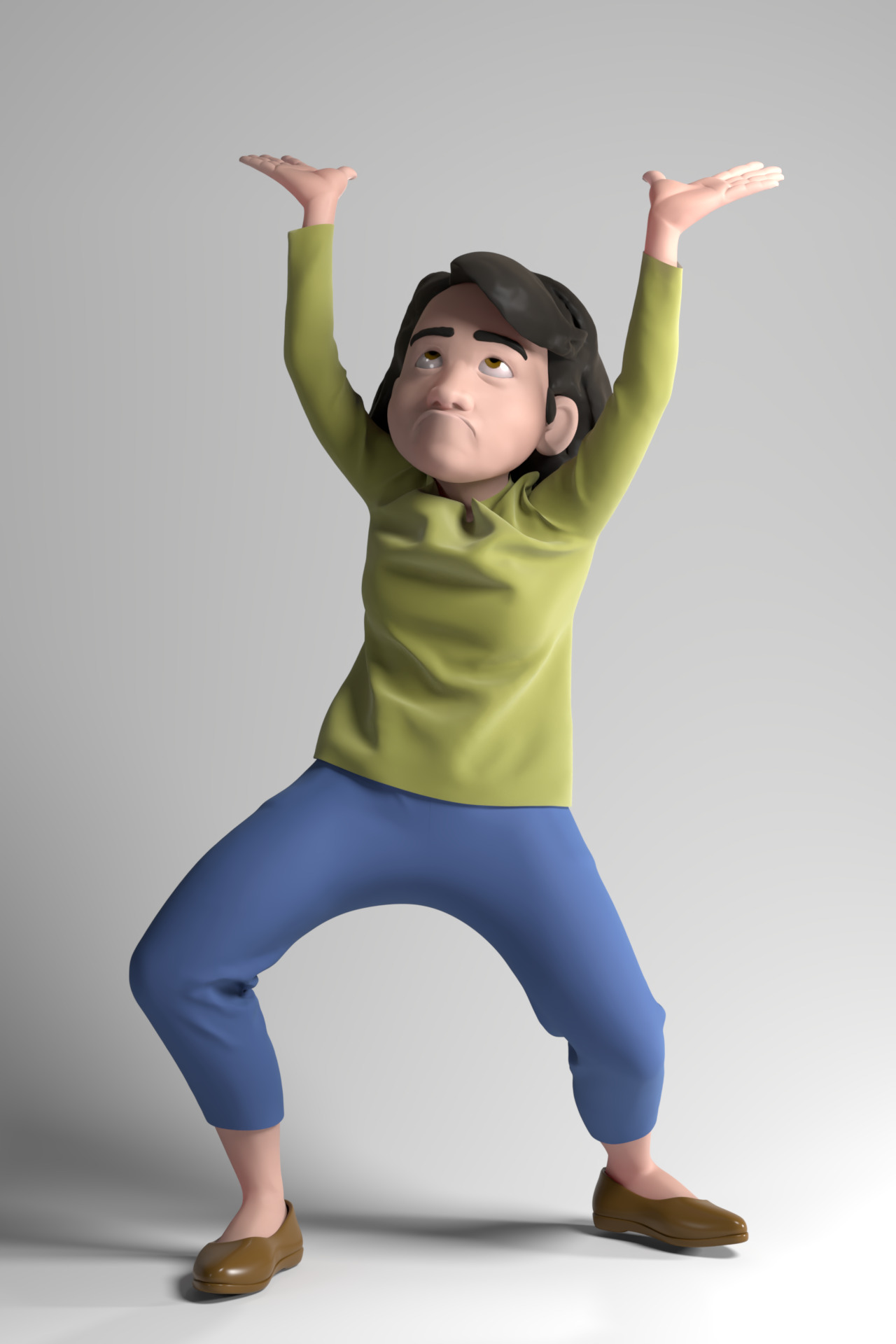} &
        \includegraphics[width=\linewidth]{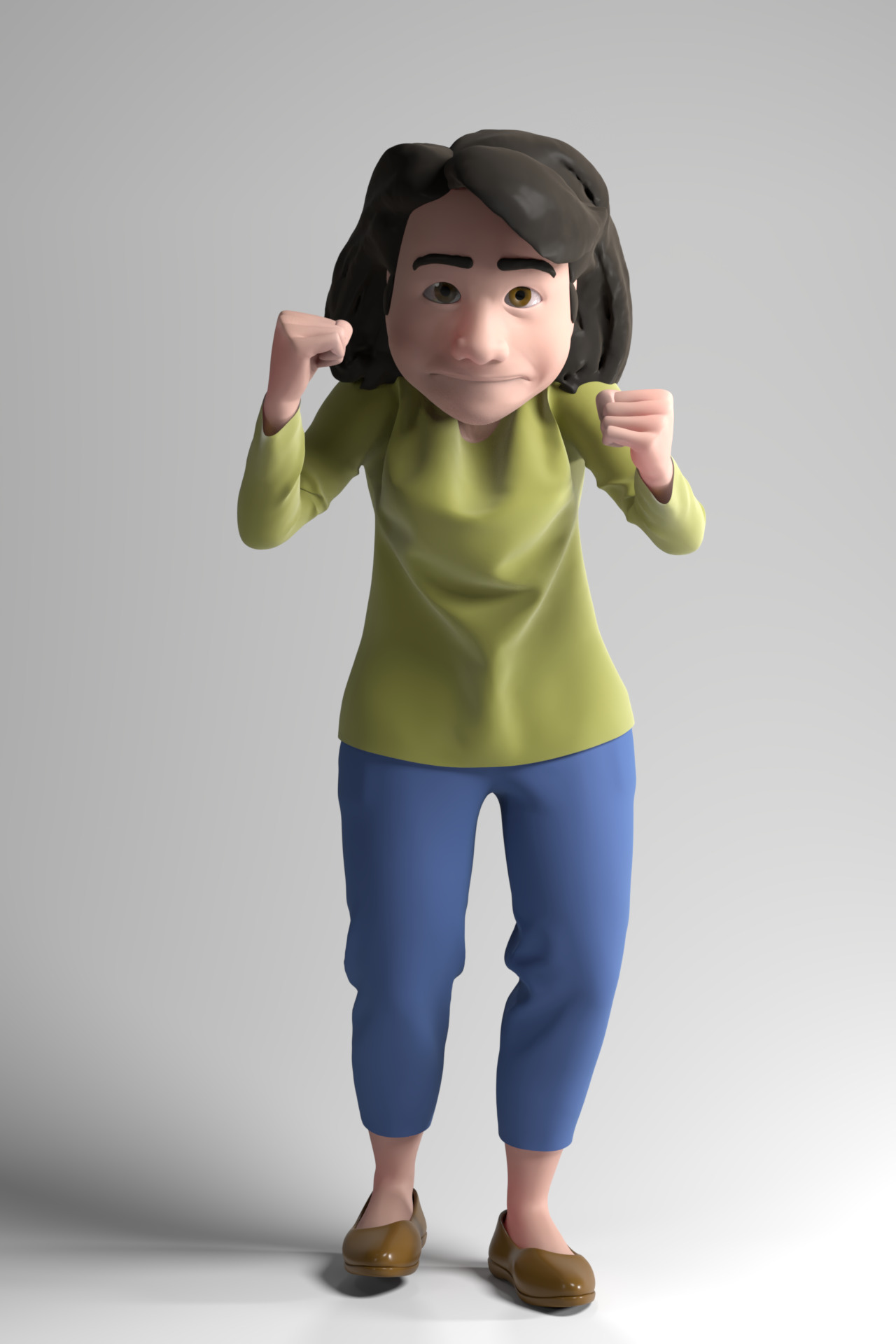} &
        \includegraphics[width=\linewidth]{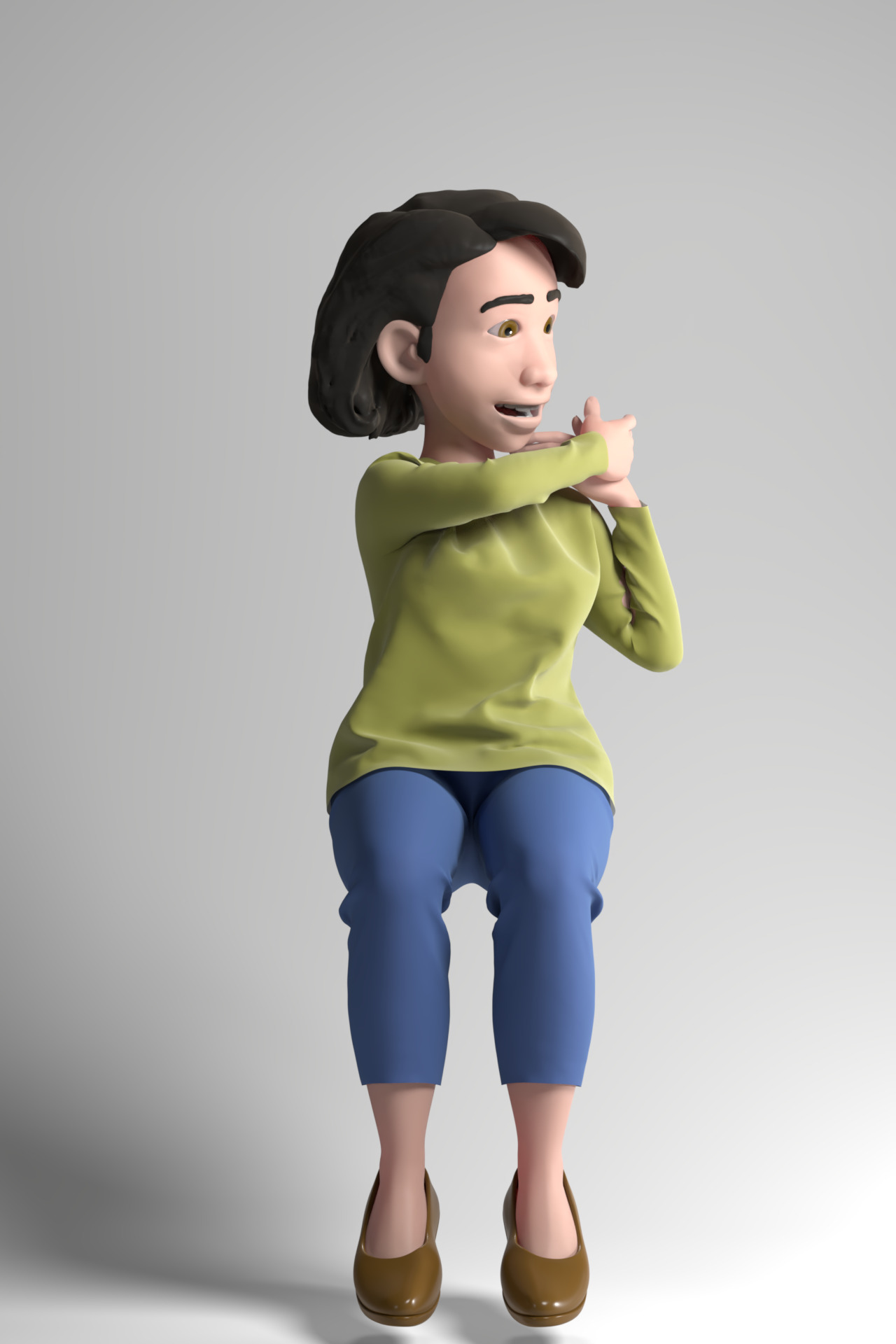} &
        \includegraphics[width=\linewidth]{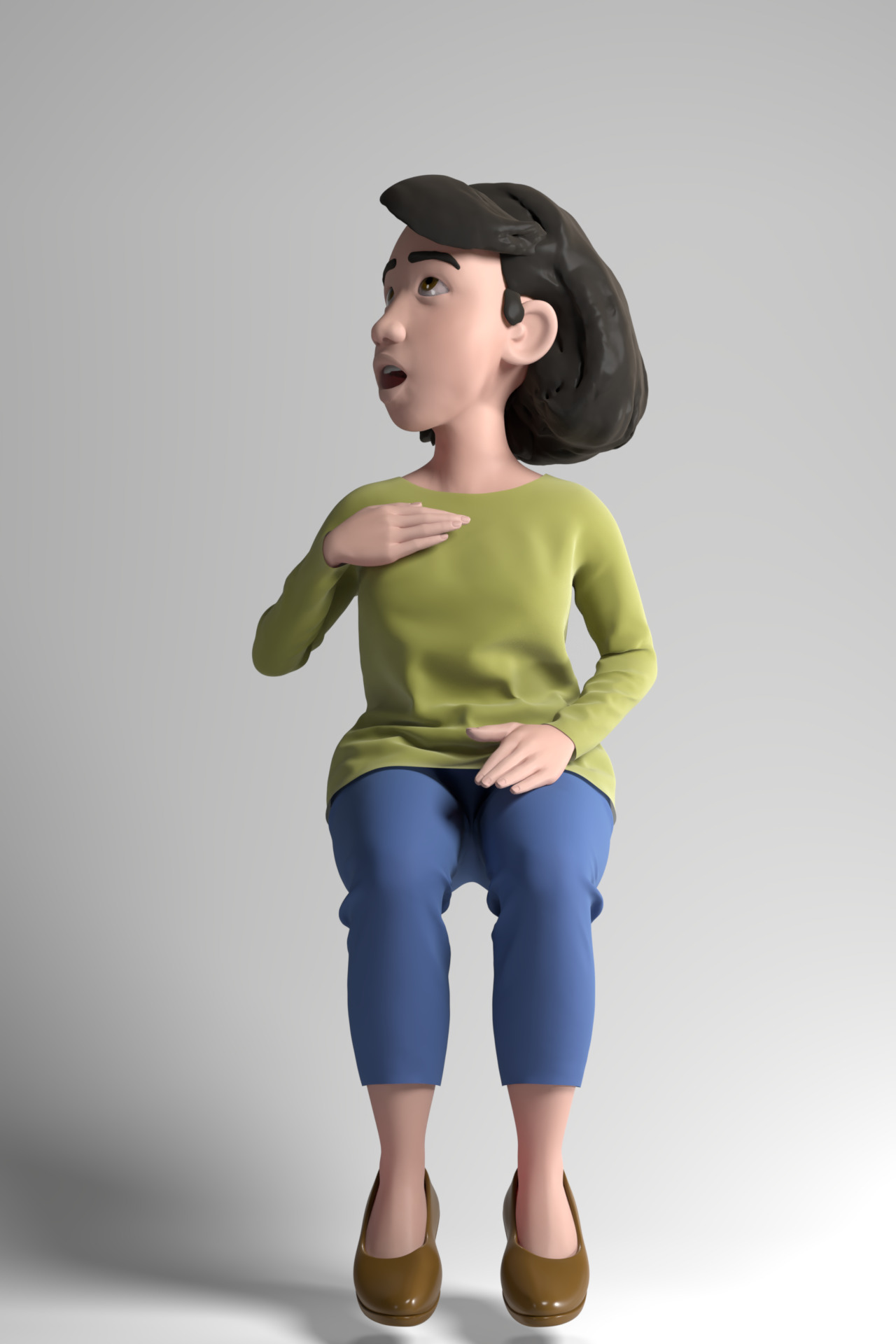} \\

        Ours &
        \includegraphics[width=\linewidth]{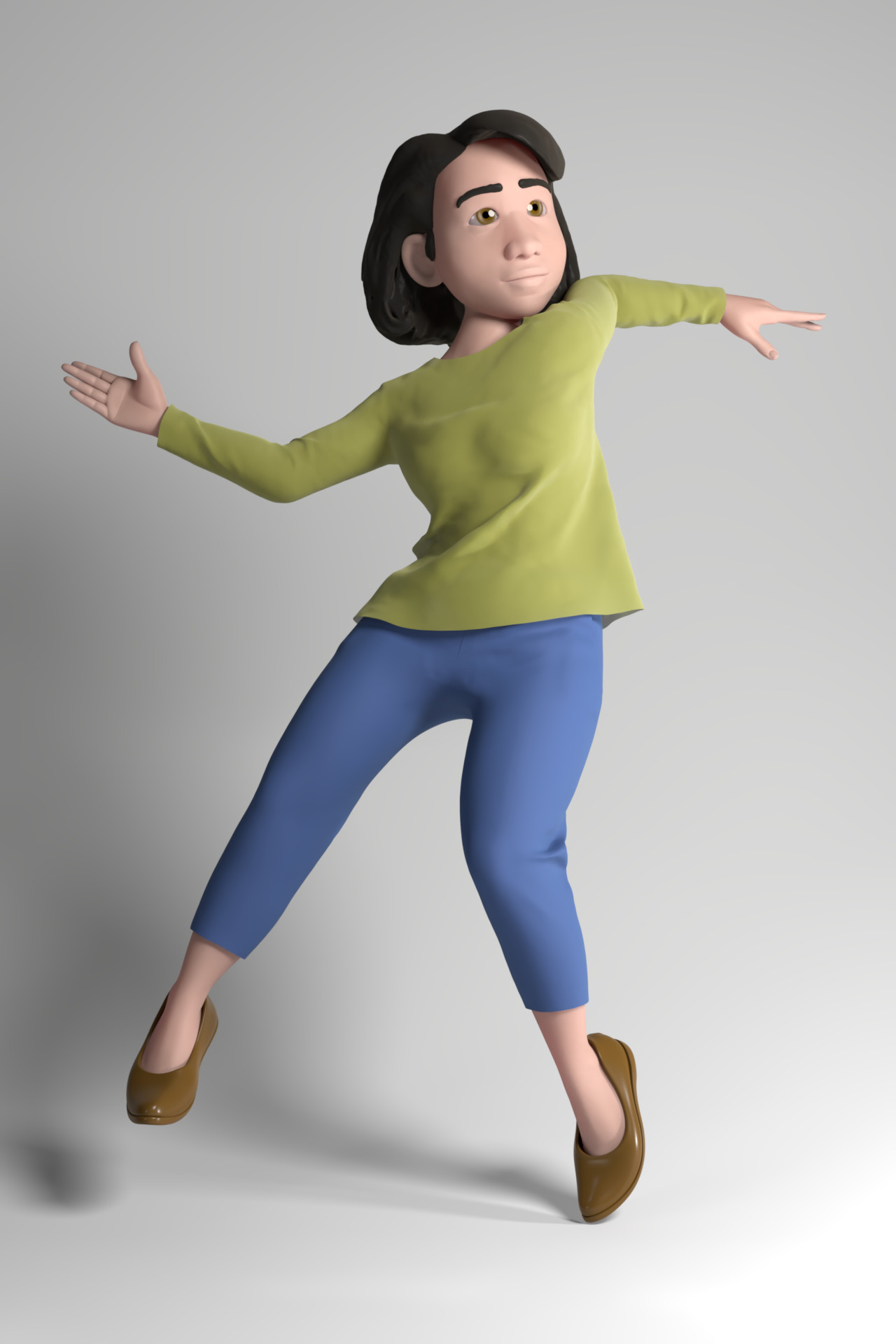} &
        \includegraphics[width=\linewidth]{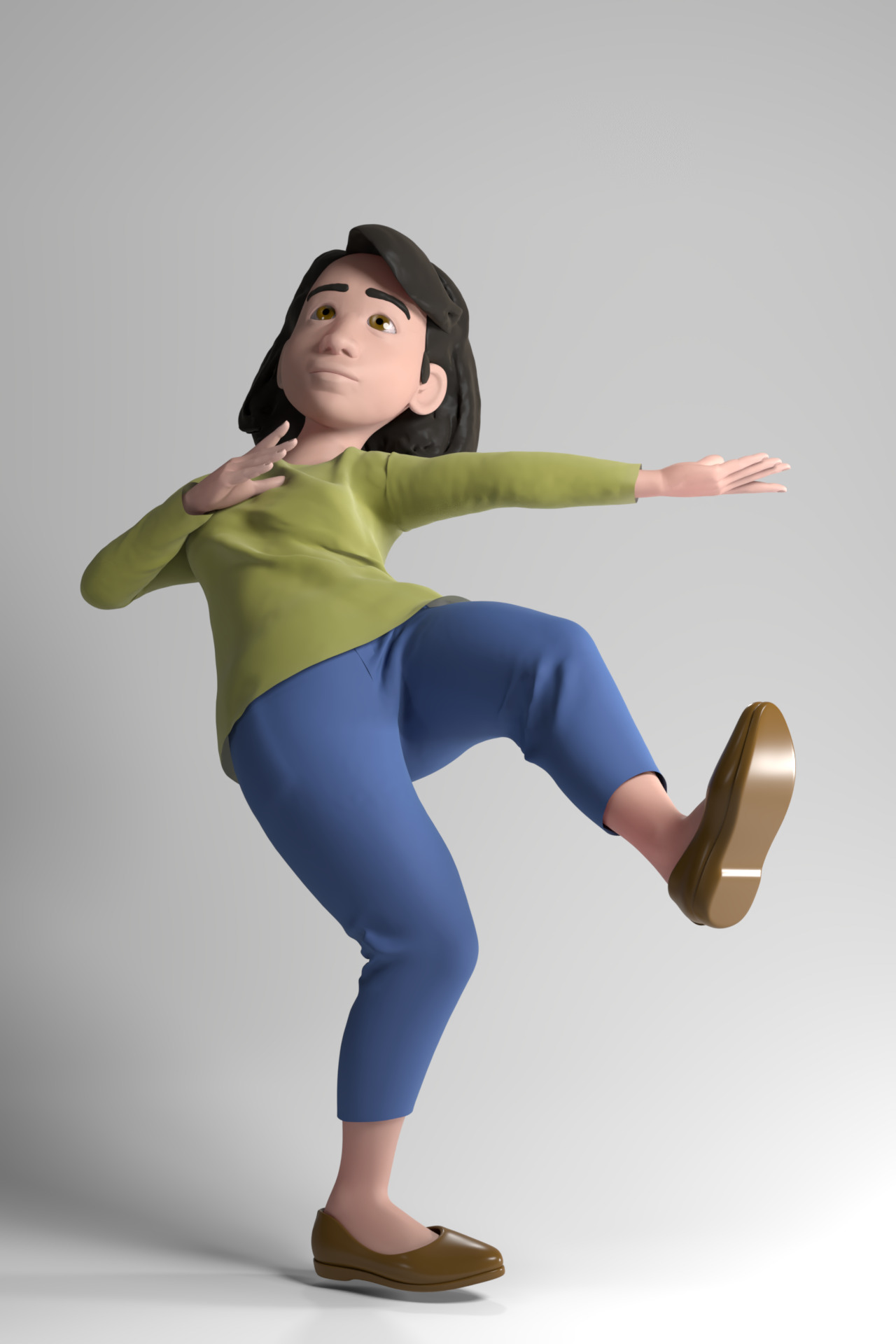} &
        \includegraphics[width=\linewidth]{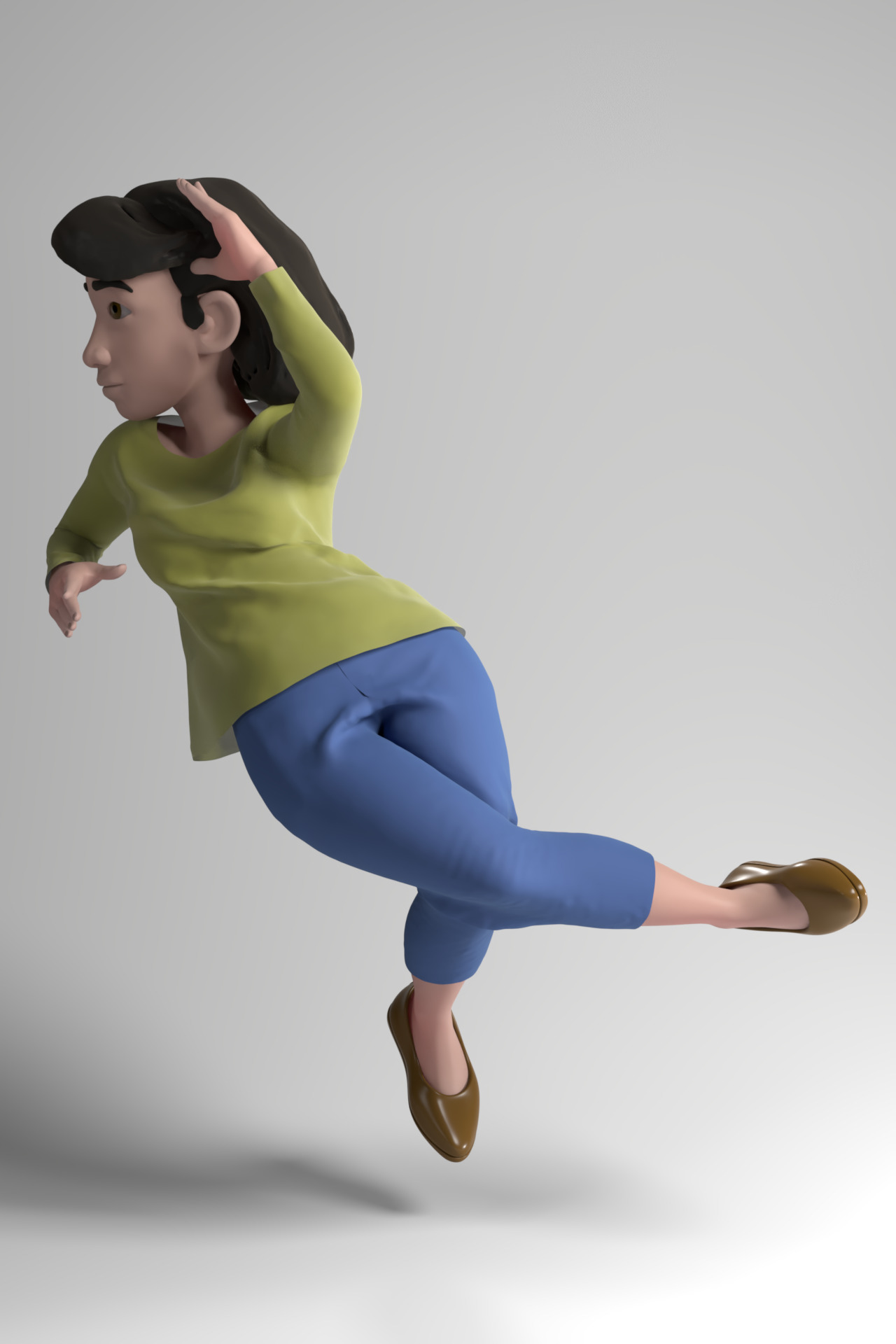} &
        \includegraphics[width=\linewidth]{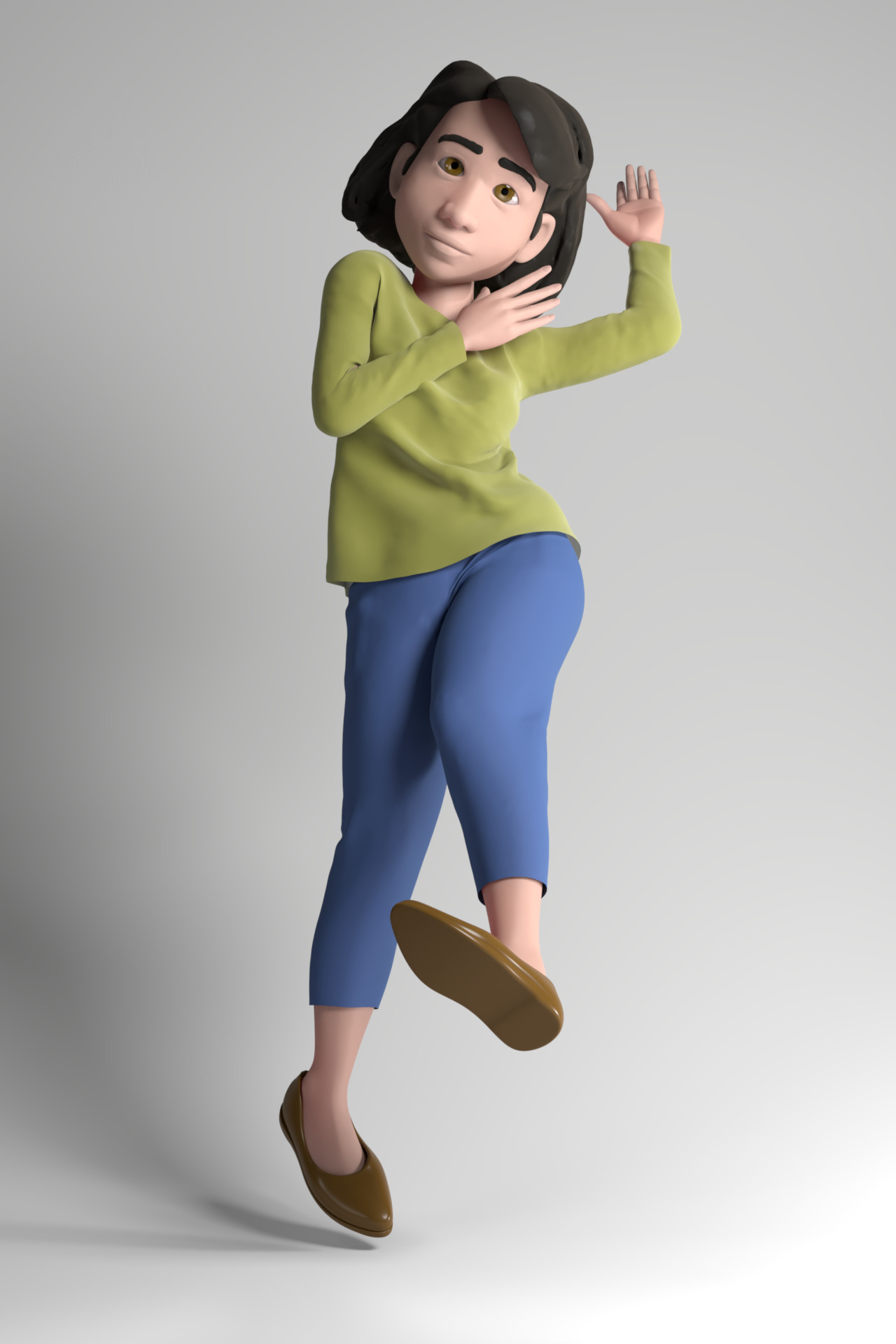} &
        \includegraphics[width=\linewidth]{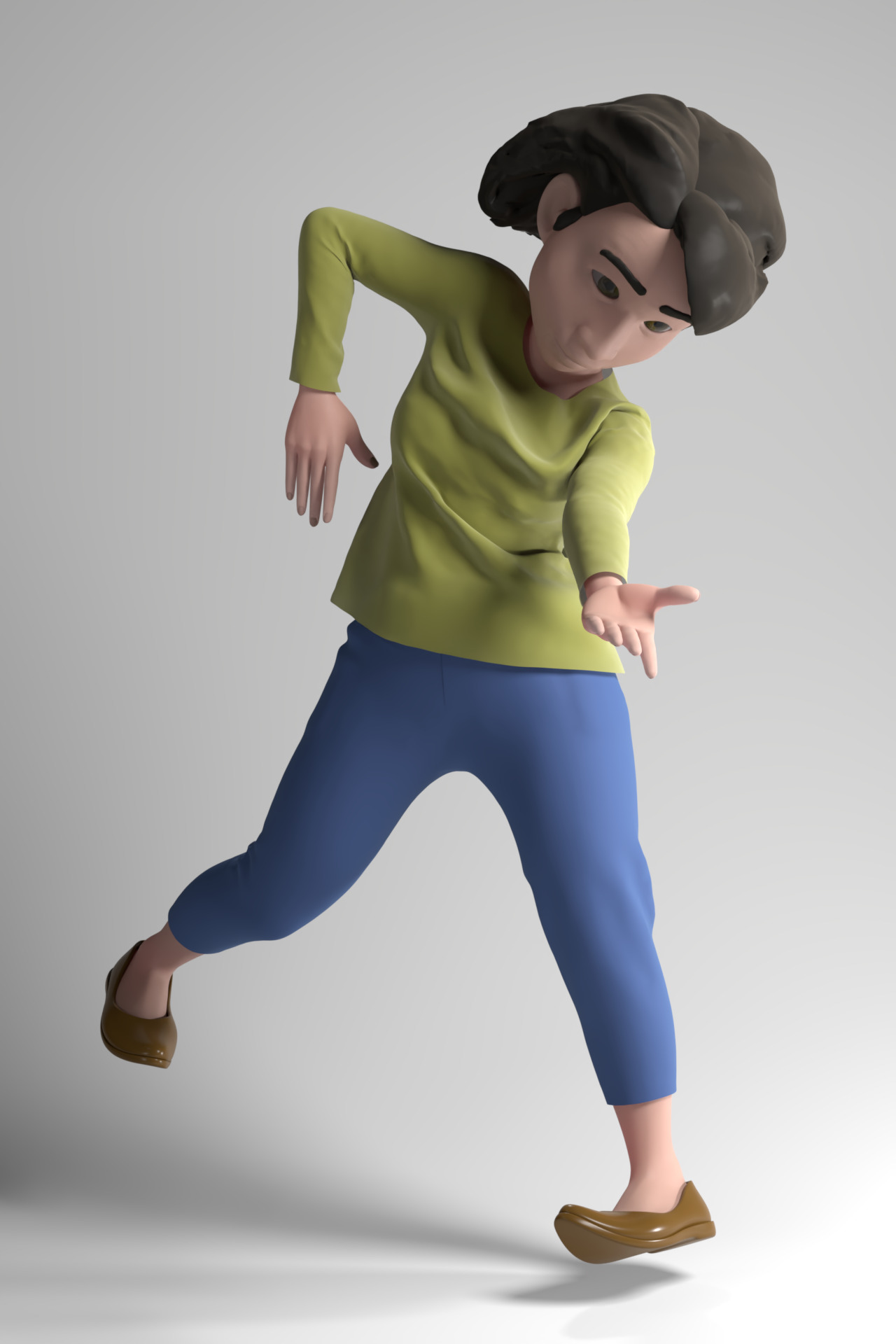} &
        \includegraphics[width=\linewidth]{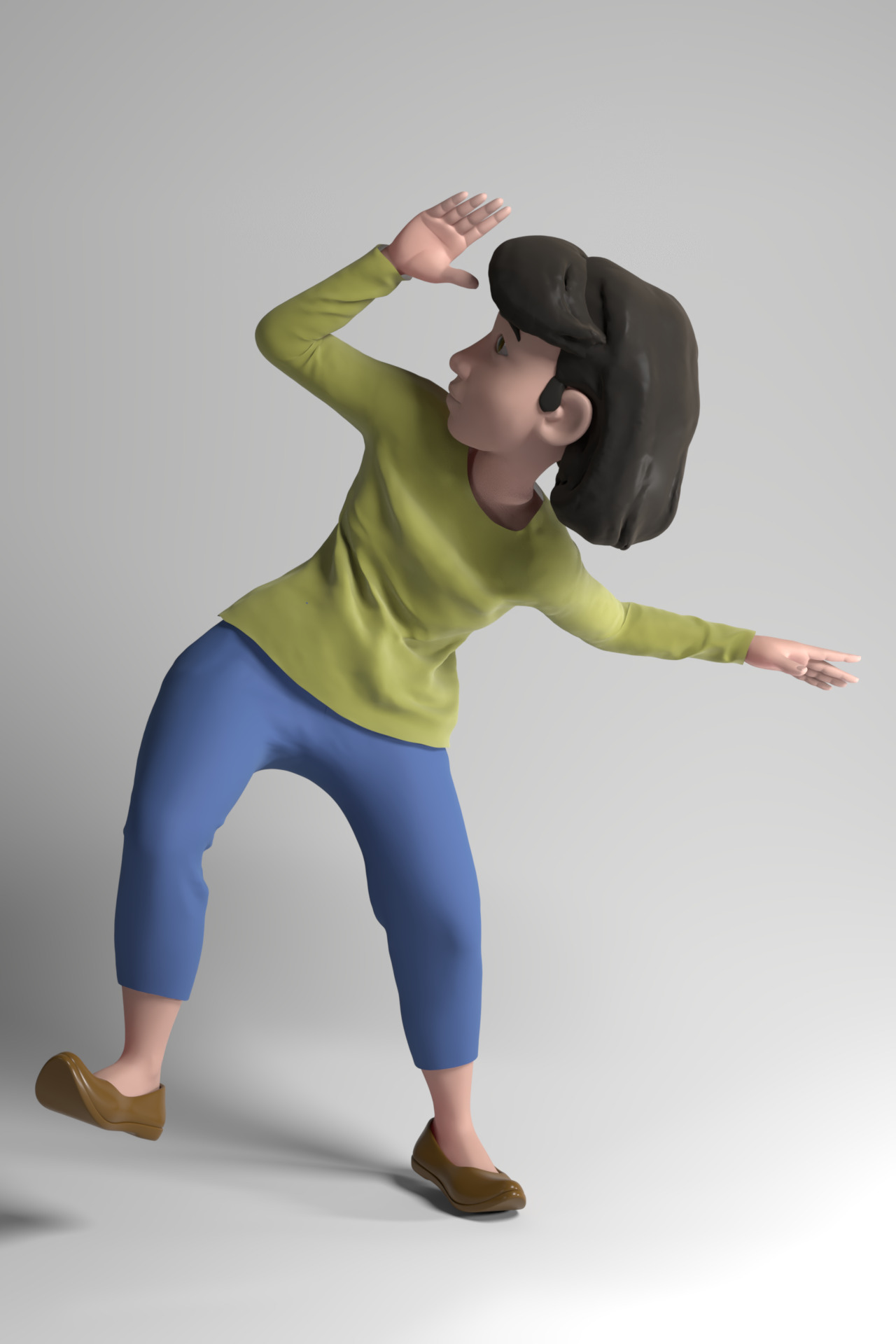} \\

        \shortstack{Ground\\Truth} &
        \includegraphics[width=\linewidth]{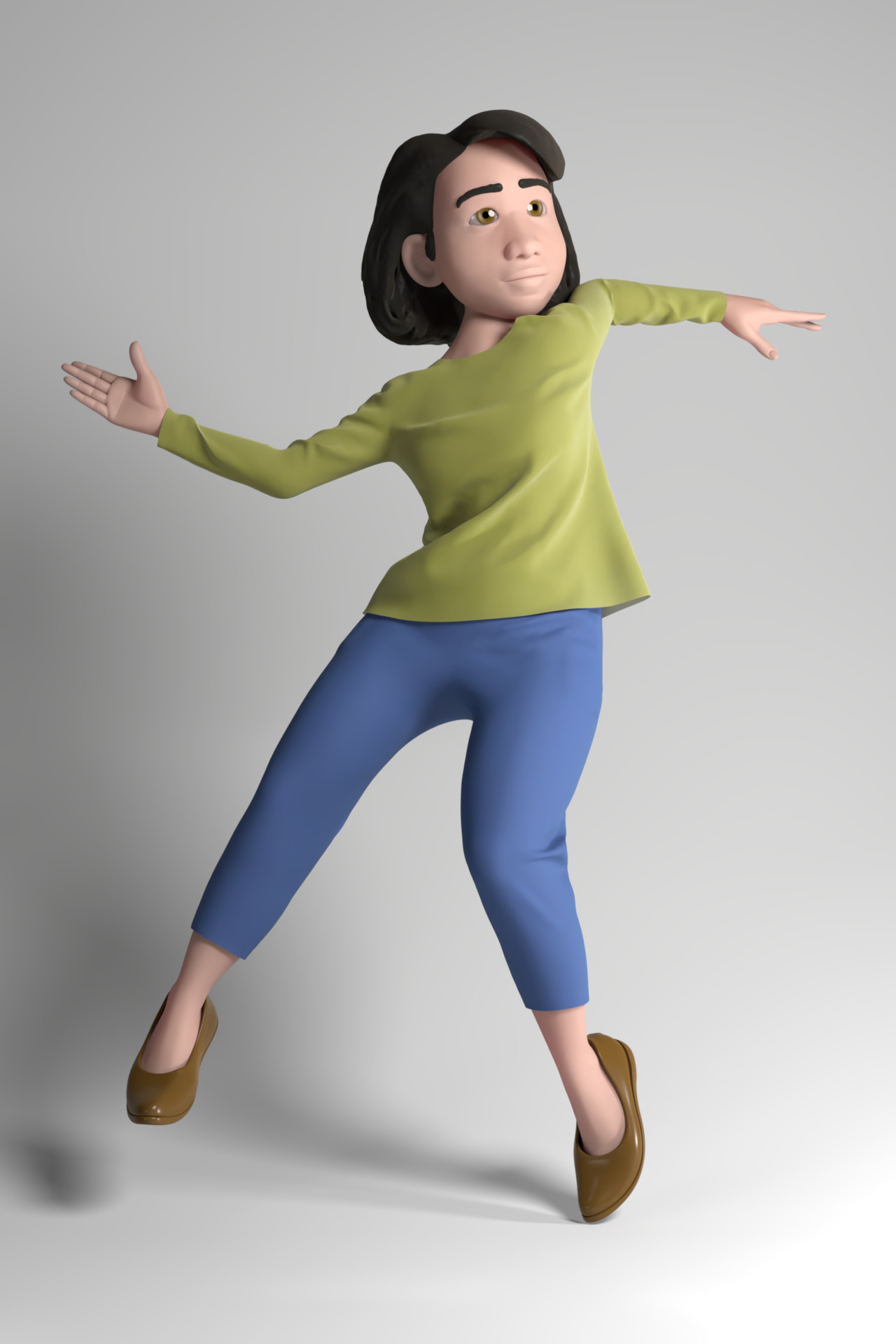} &
        \includegraphics[width=\linewidth]{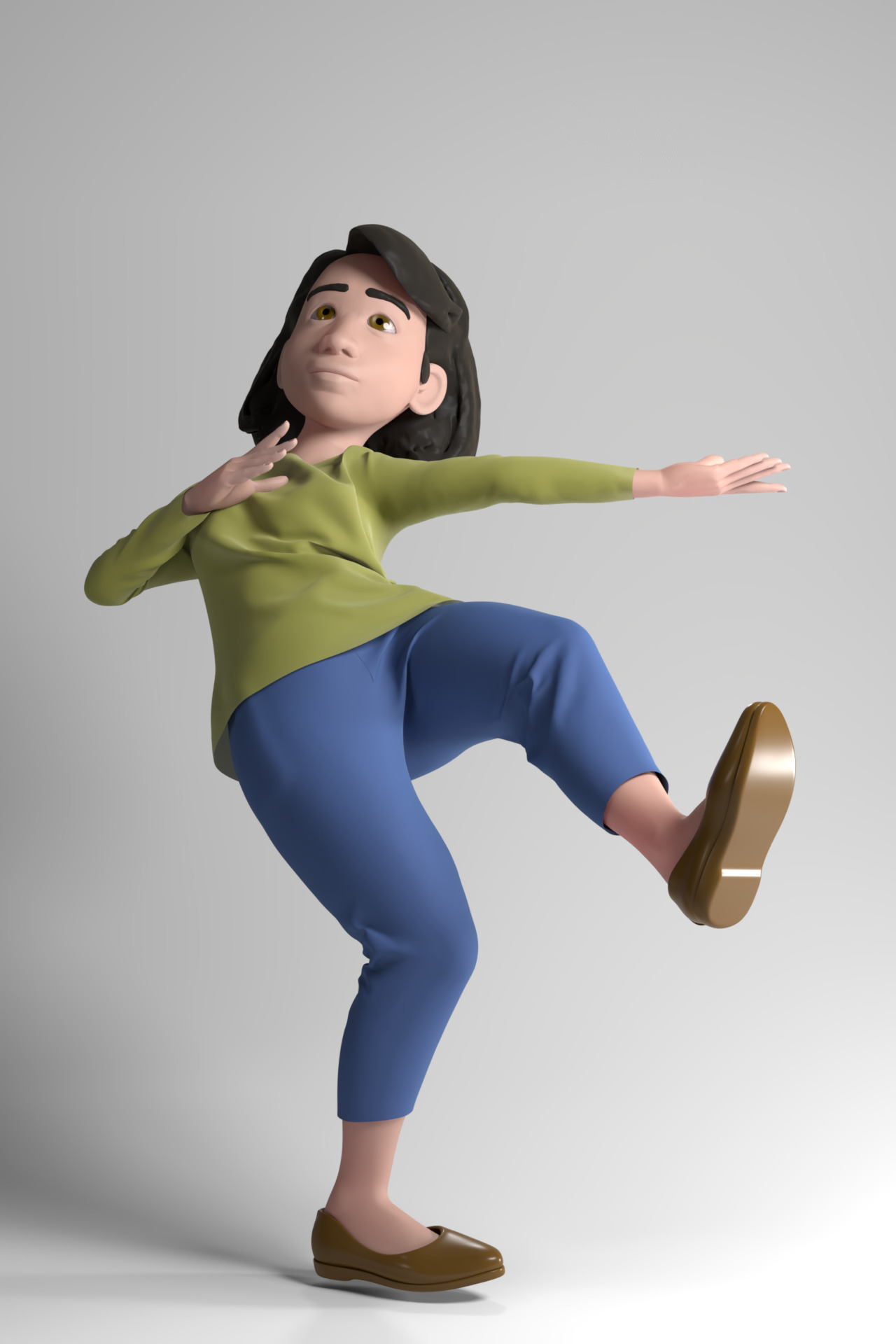} &
        \includegraphics[width=\linewidth]{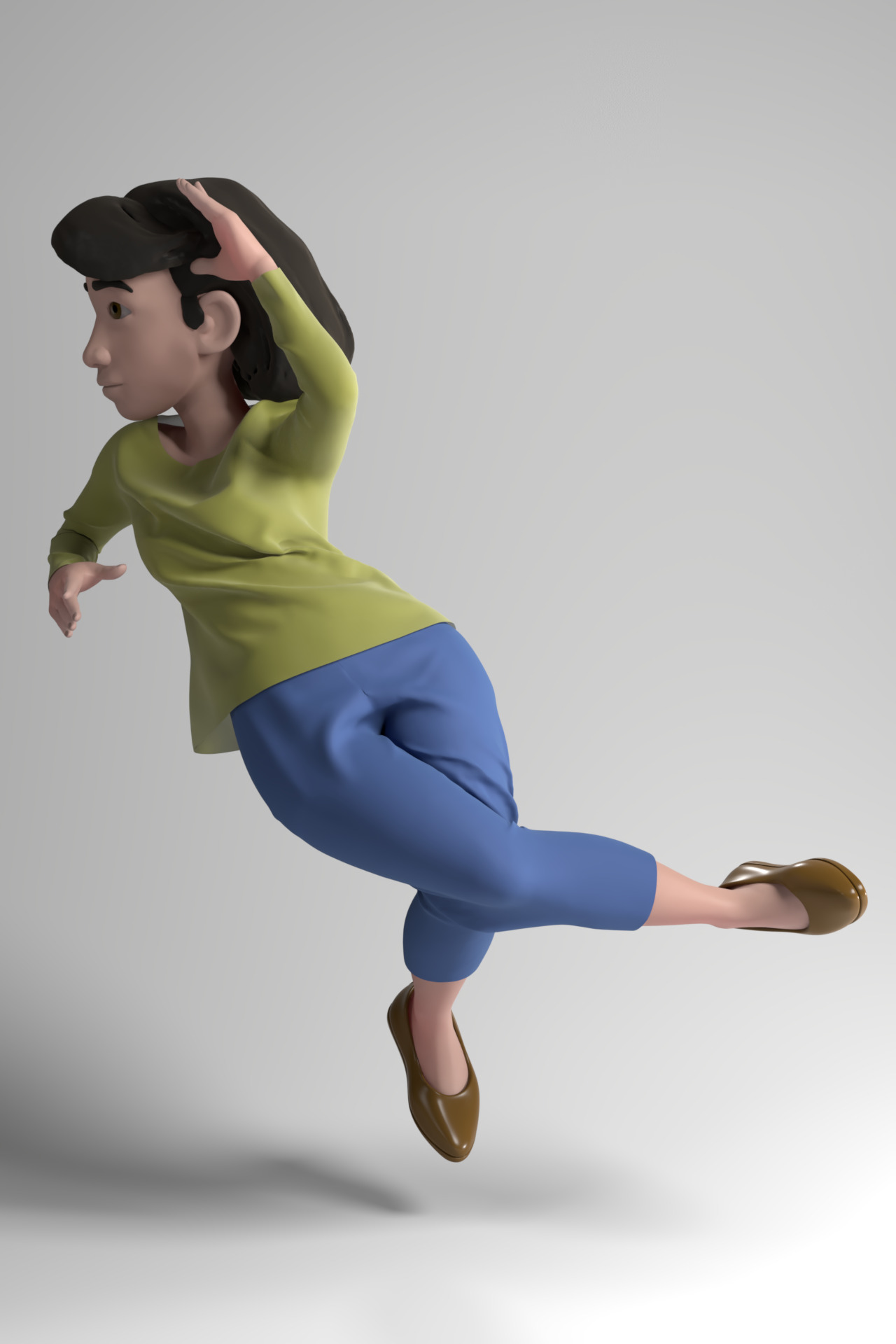} &
        \includegraphics[width=\linewidth]{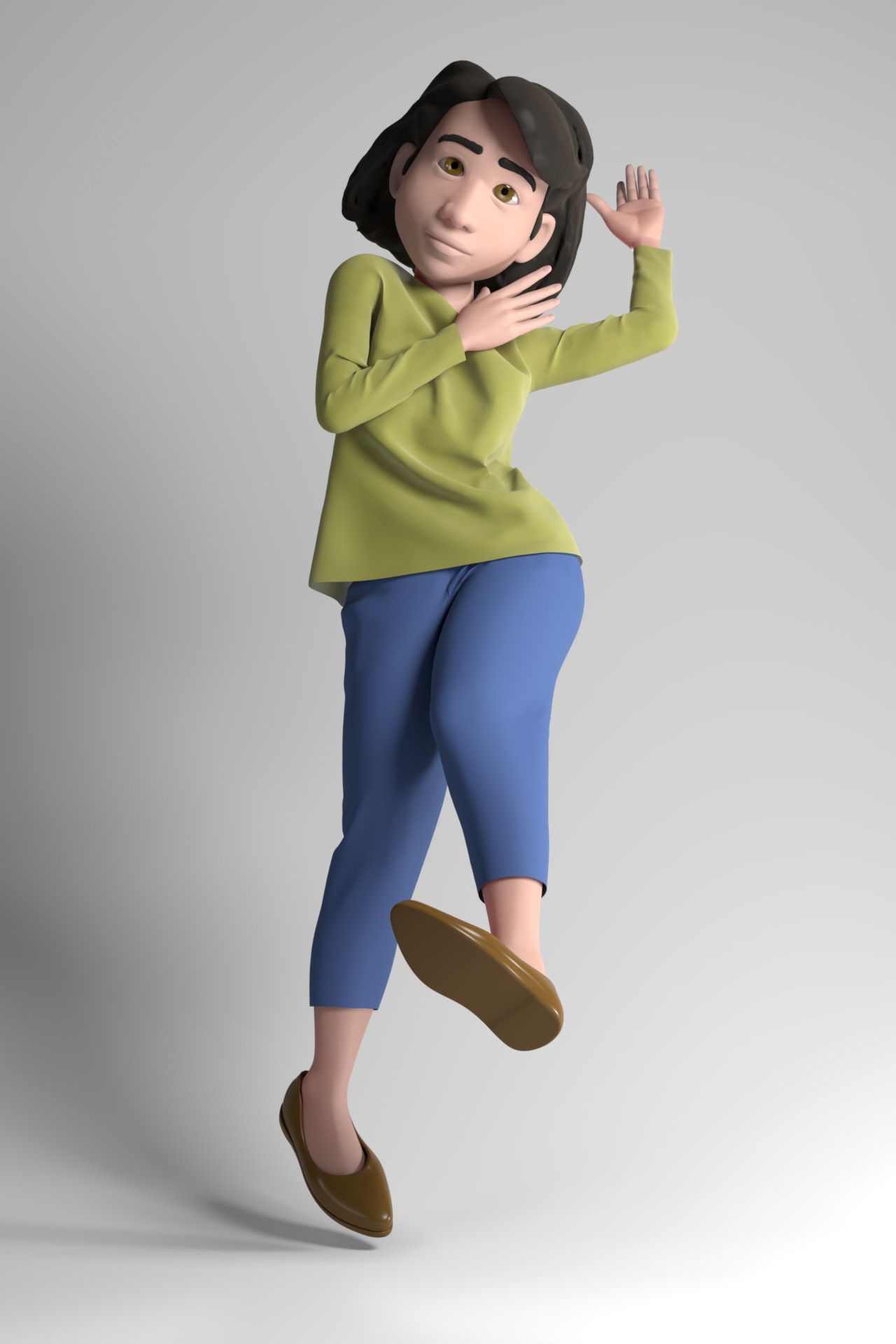} &
        \includegraphics[width=\linewidth]{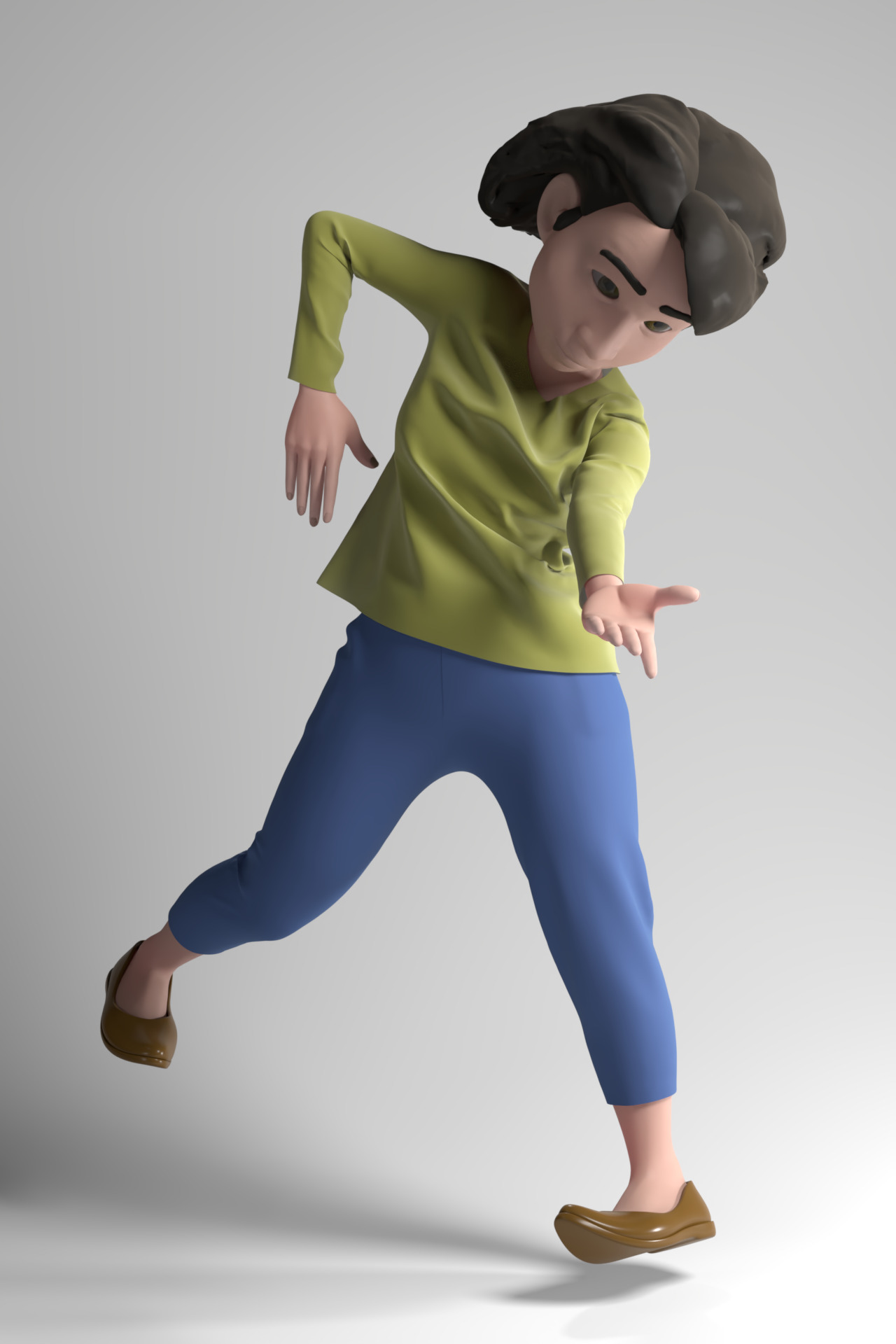} &
        \includegraphics[width=\linewidth]{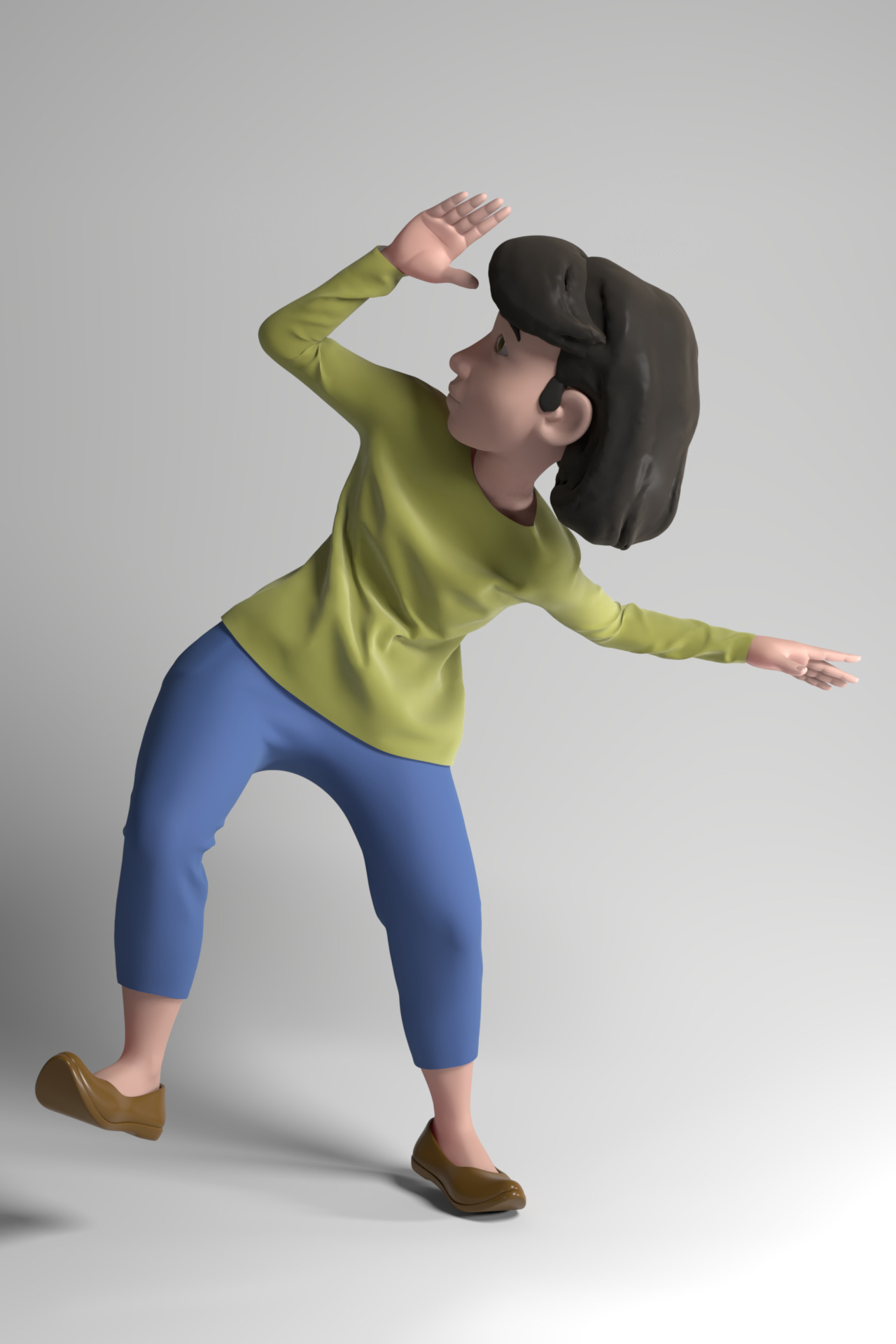}

	\end{tabular}}
  \caption{Predictions of our model in \jenShirt and \jenPants on the training set. The prediction quality is very close to the ground truth.}
  \label{fig:TrainingSamples}
\end{figure*}

\paragraph*{Training set samples.}
\Fig{TrainingSamples} presents qualitative results on samples drawn from the training set. Both our model's predictions and ground-truth are shown in corresponding rows, with columns depicting different poses. These examples confirm that the model accurately reconstructs the supervised training data while preserving fine-scale wrinkle detail and overall garment shape.

\begin{figure*}[t]
	\centering
	\setlength{\tabcolsep}{0.00125\linewidth}
    {\footnotesize
    \begin{tabular}{cccccccc}
        \includegraphics[width=0.123\linewidth]{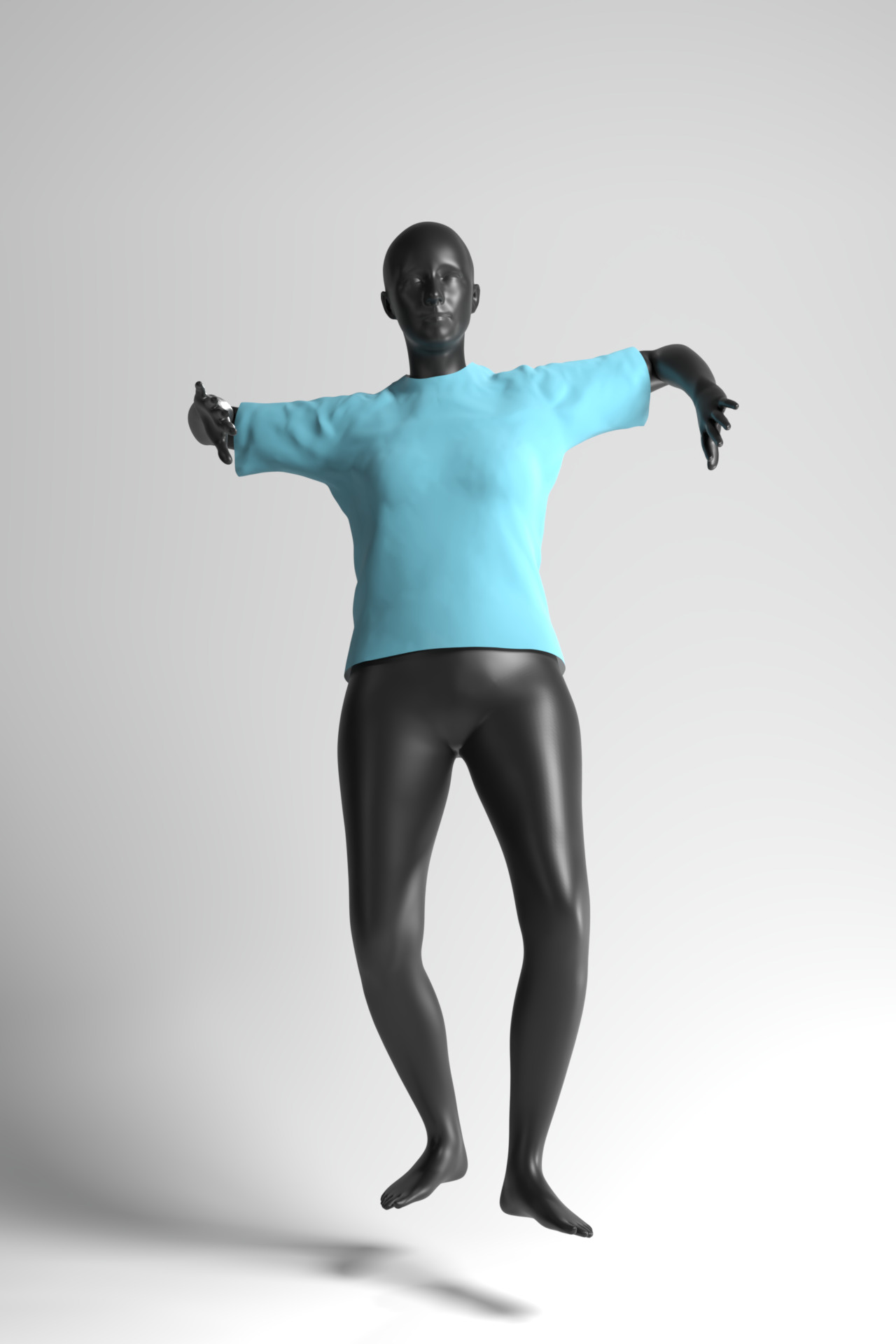} &
        \includegraphics[width=0.123\linewidth]{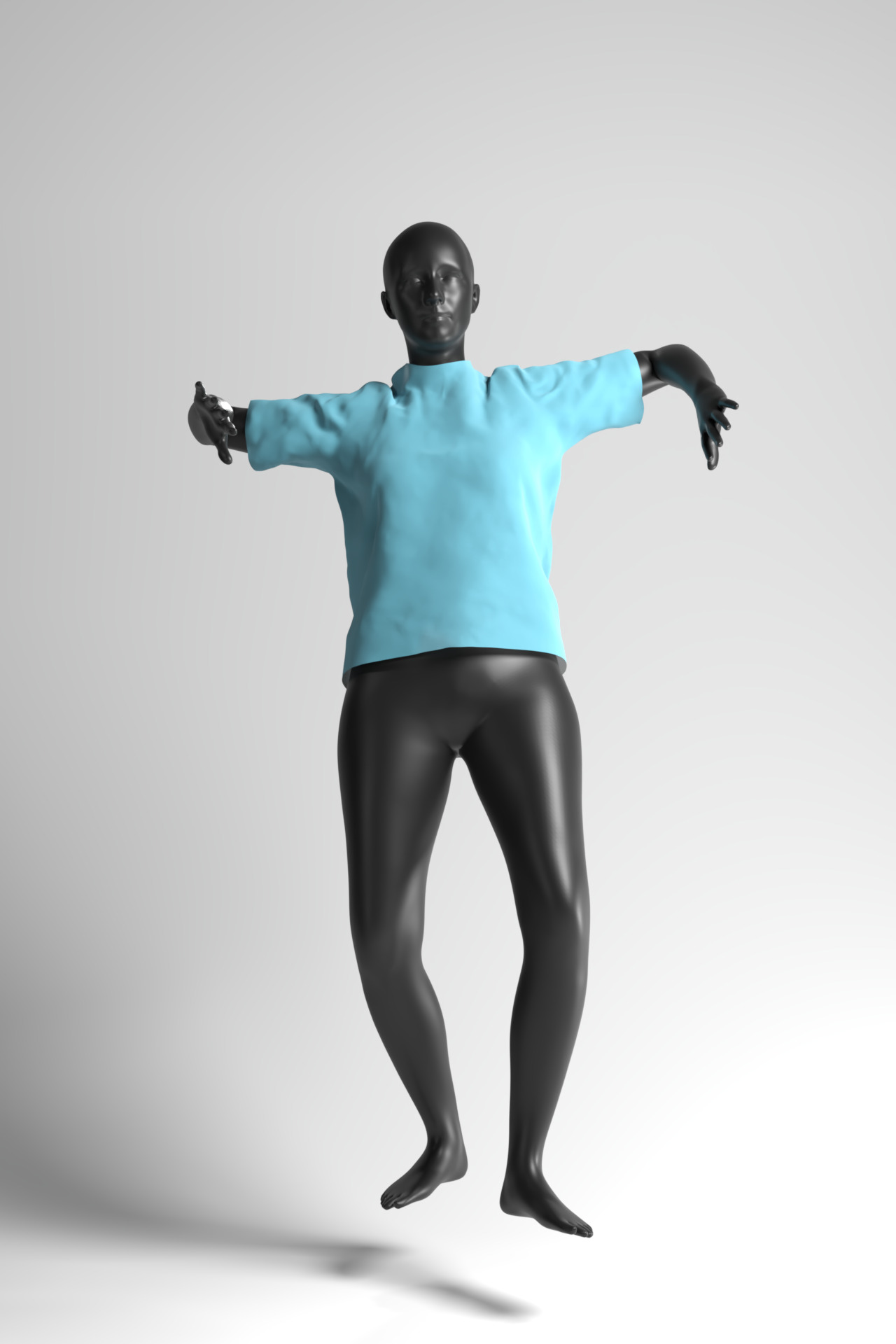} &
        \includegraphics[width=0.123\linewidth]{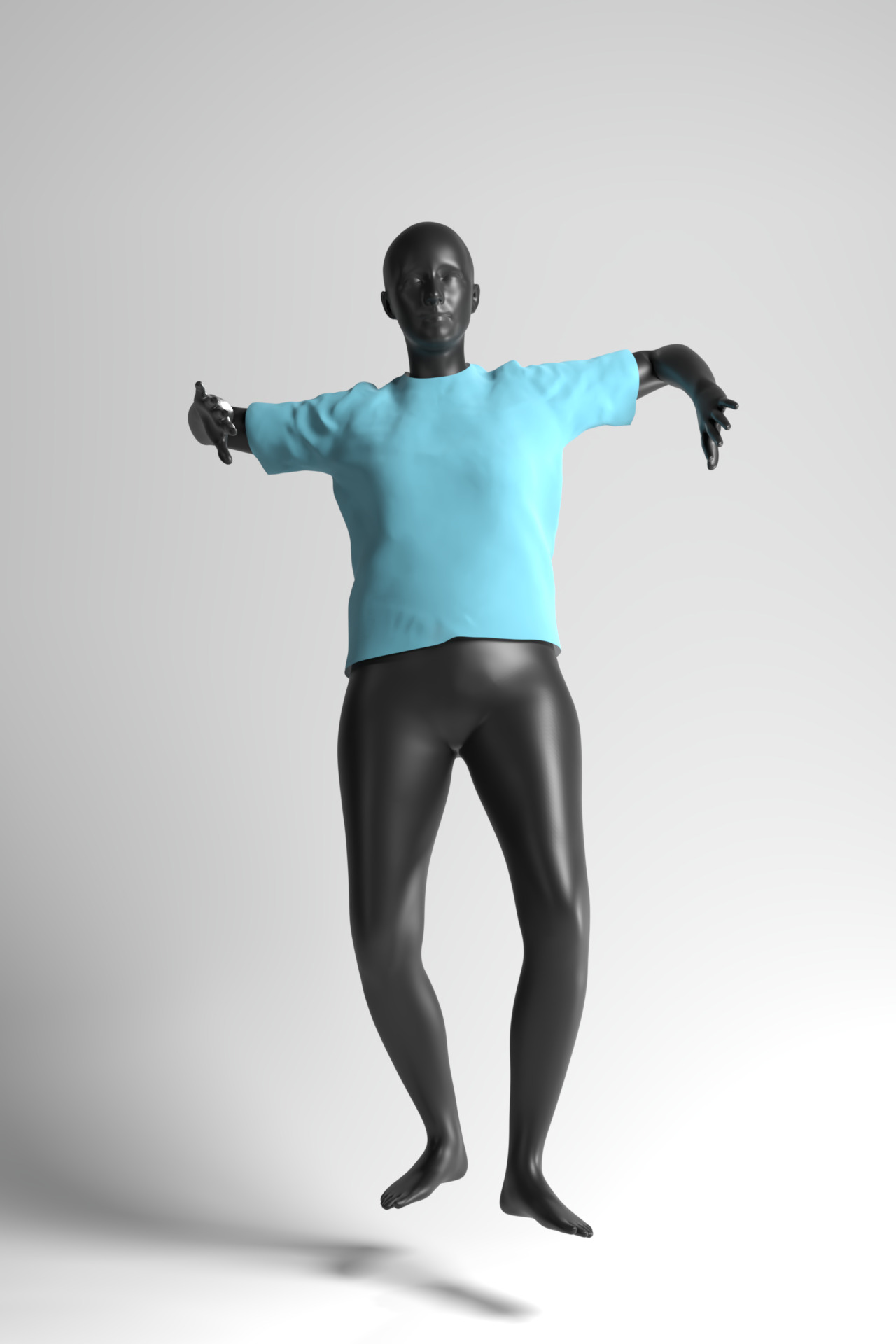} &
        \includegraphics[width=0.123\linewidth]{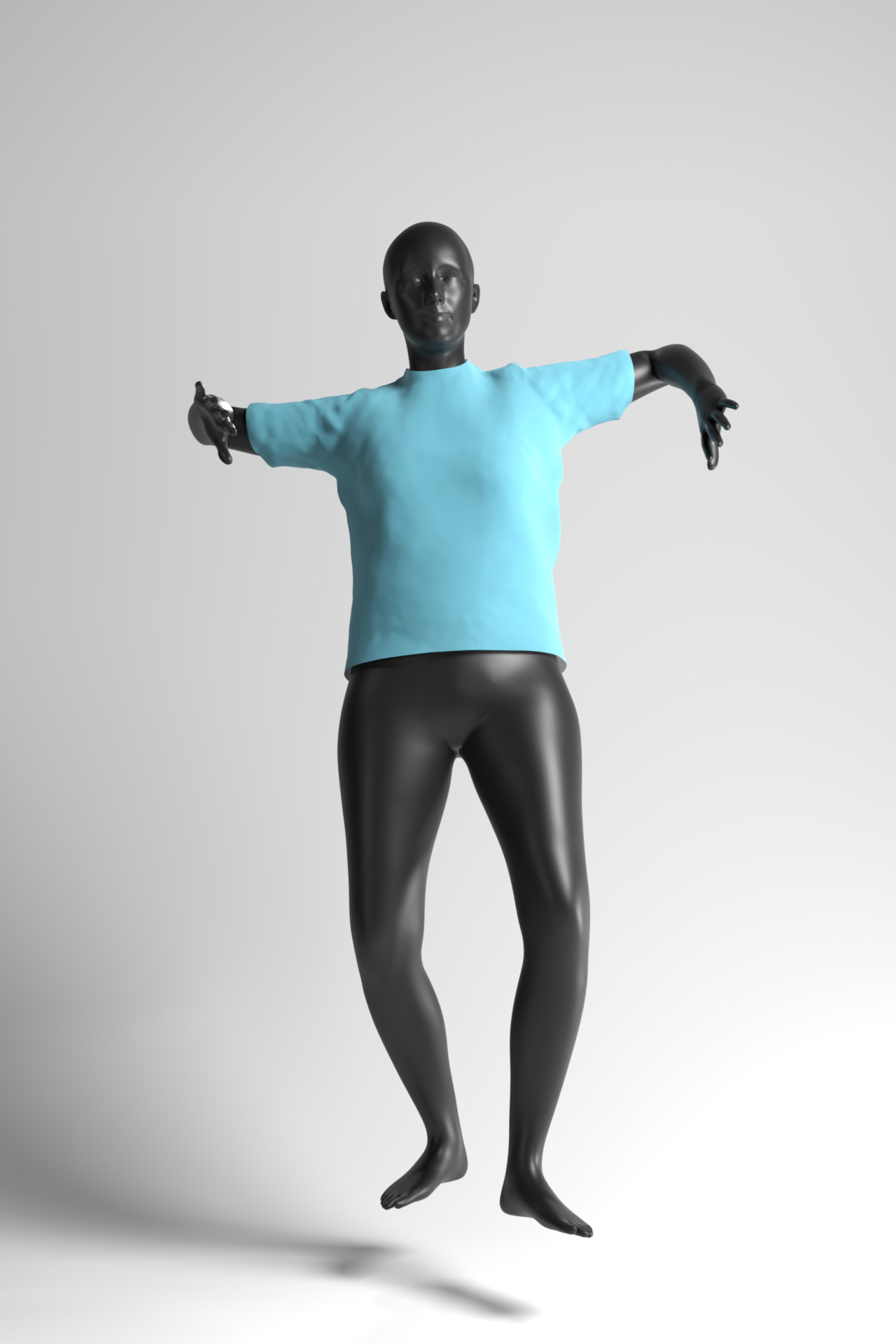} &
        \includegraphics[width=0.123\linewidth]{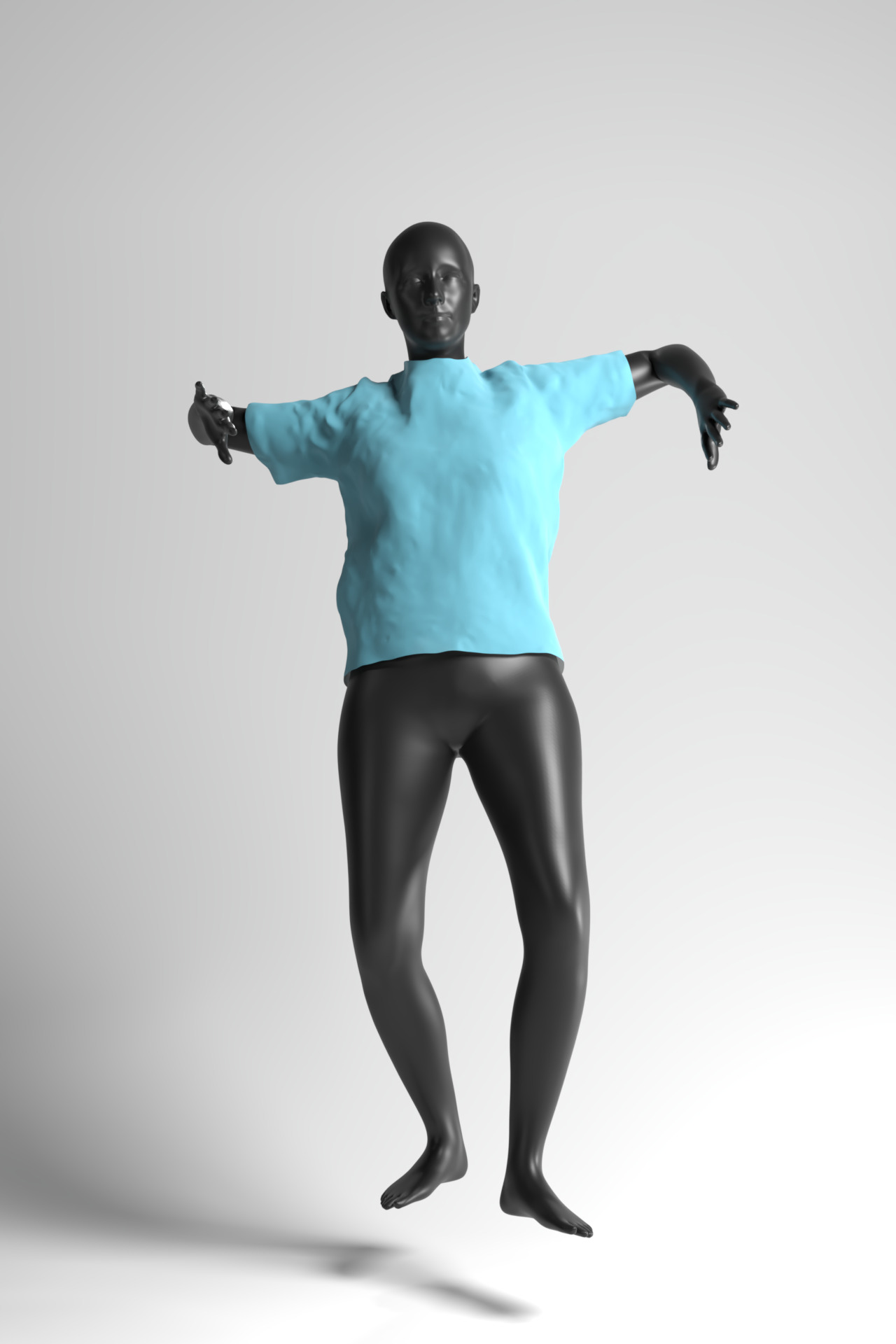} &
        \includegraphics[width=0.123\linewidth]{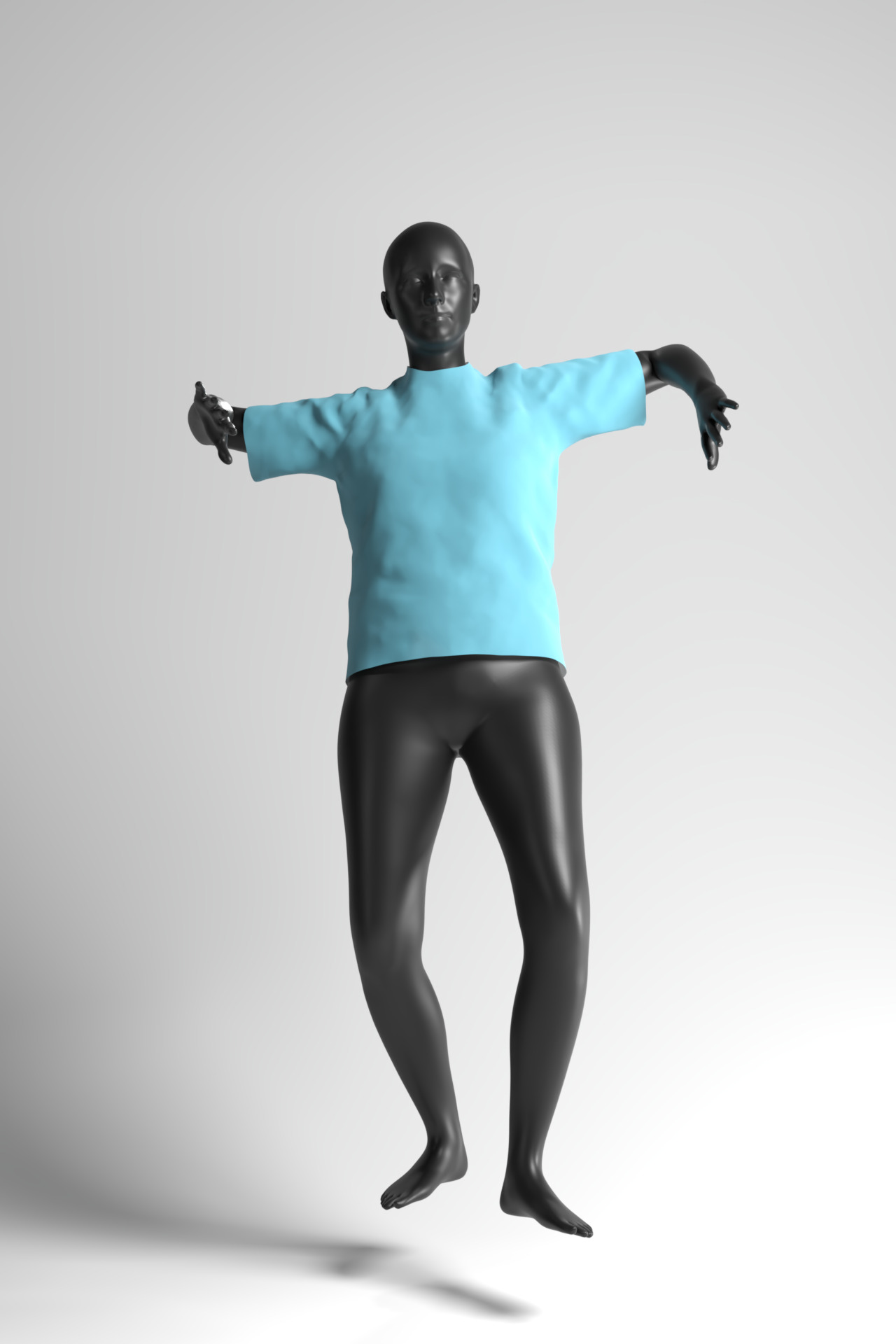} &
        \includegraphics[width=0.123\linewidth]{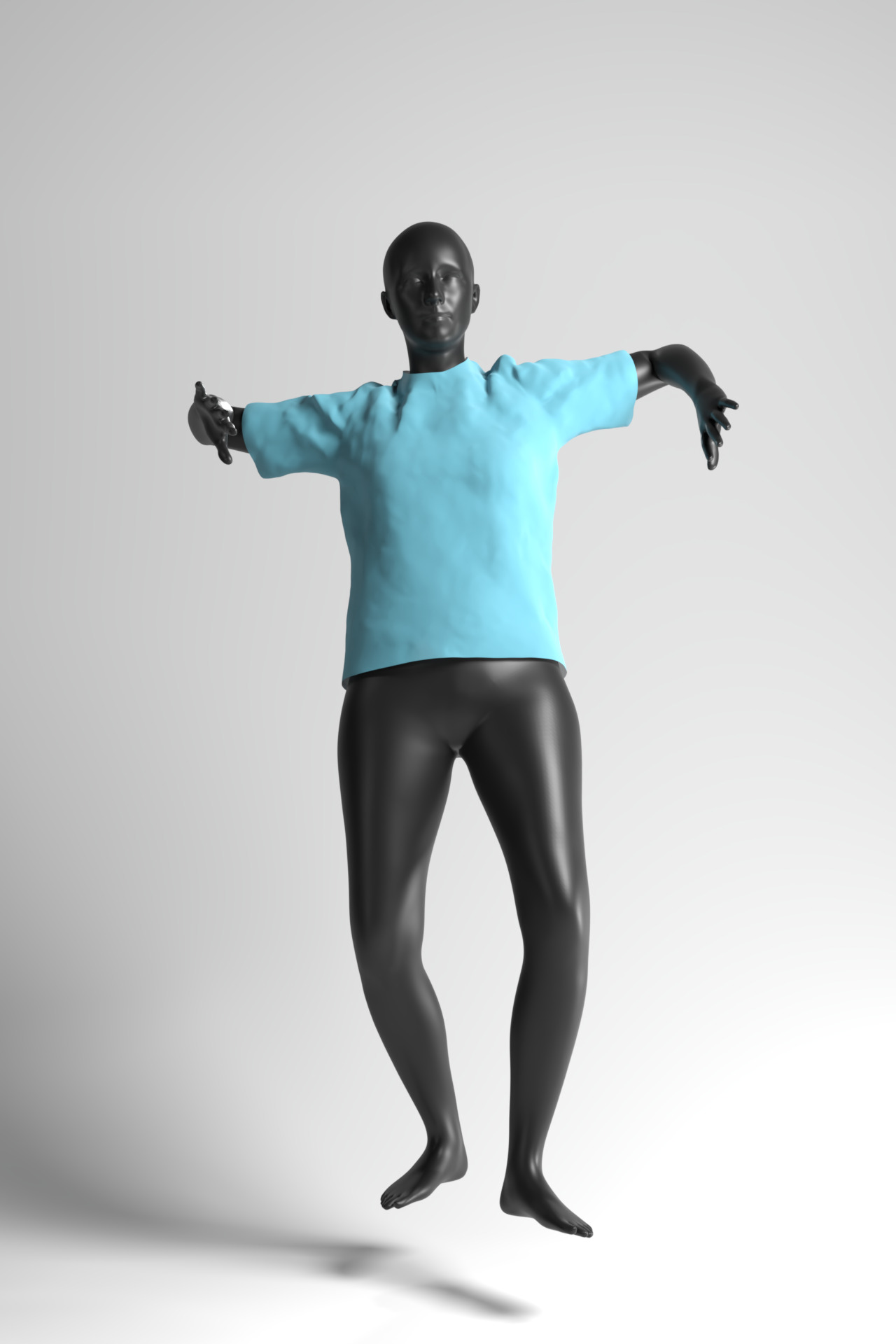} &
        \includegraphics[width=0.123\linewidth]{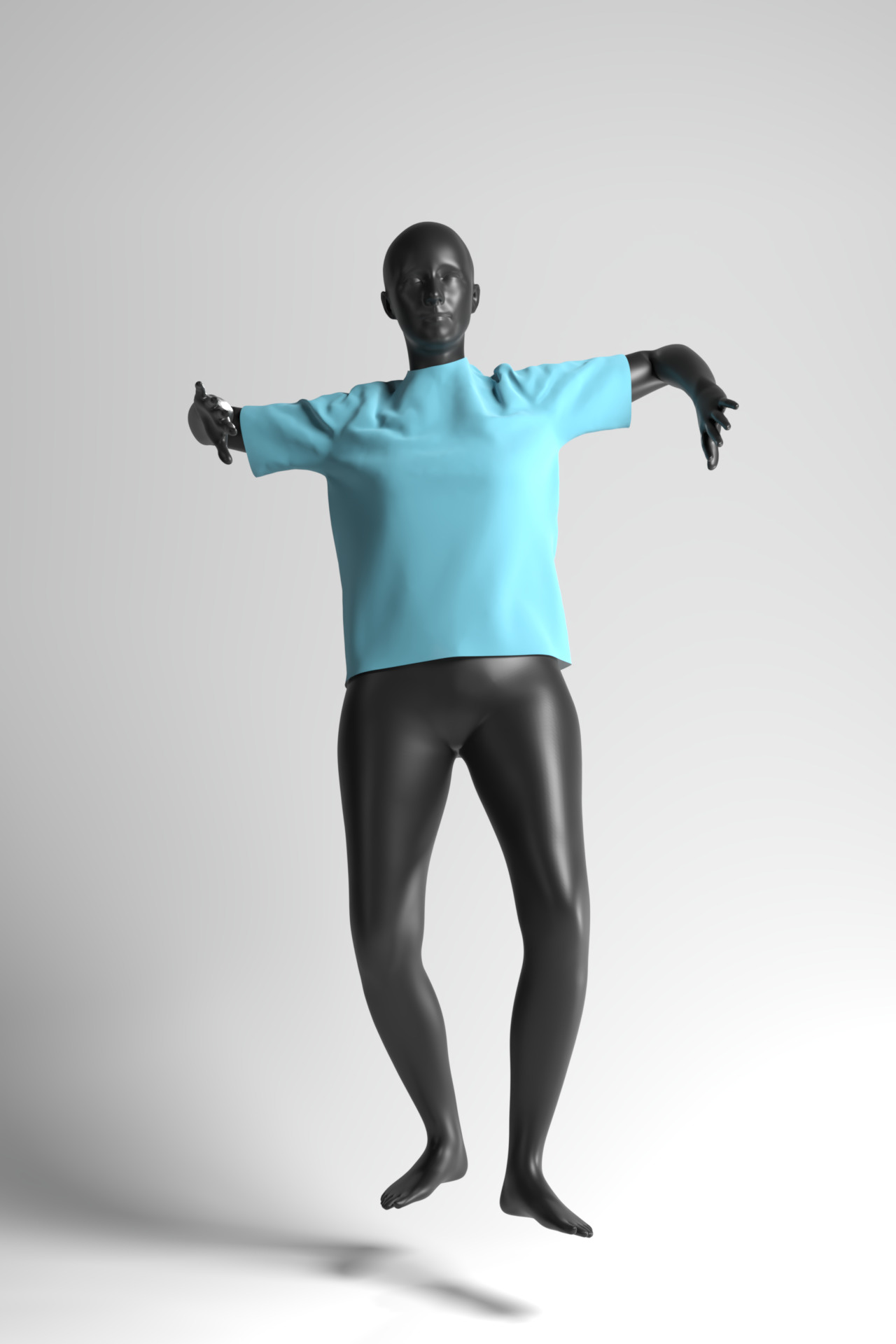}\\

        \includegraphics[width=0.123\linewidth]{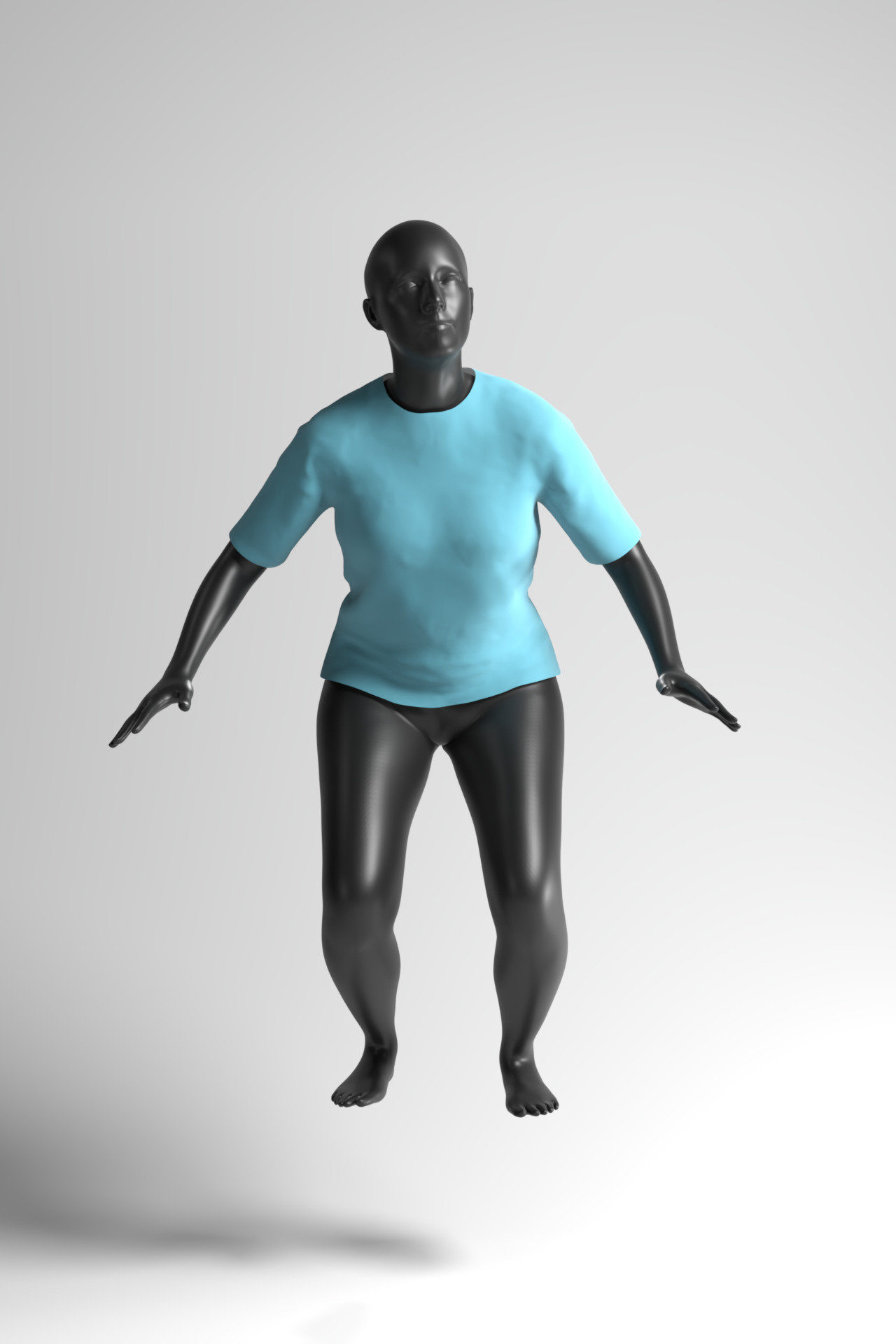} &
        \includegraphics[width=0.123\linewidth]{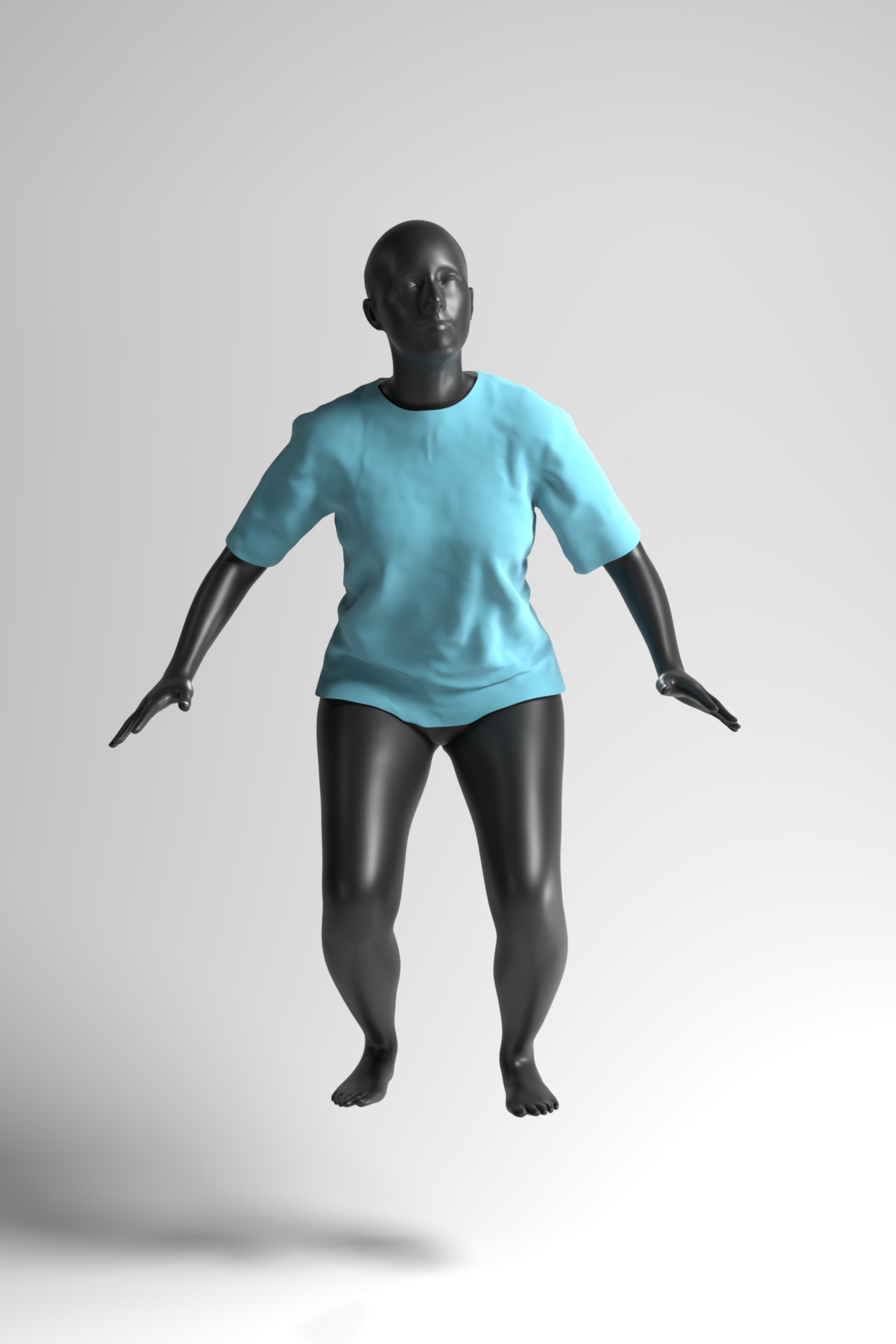} &
        \includegraphics[width=0.123\linewidth]{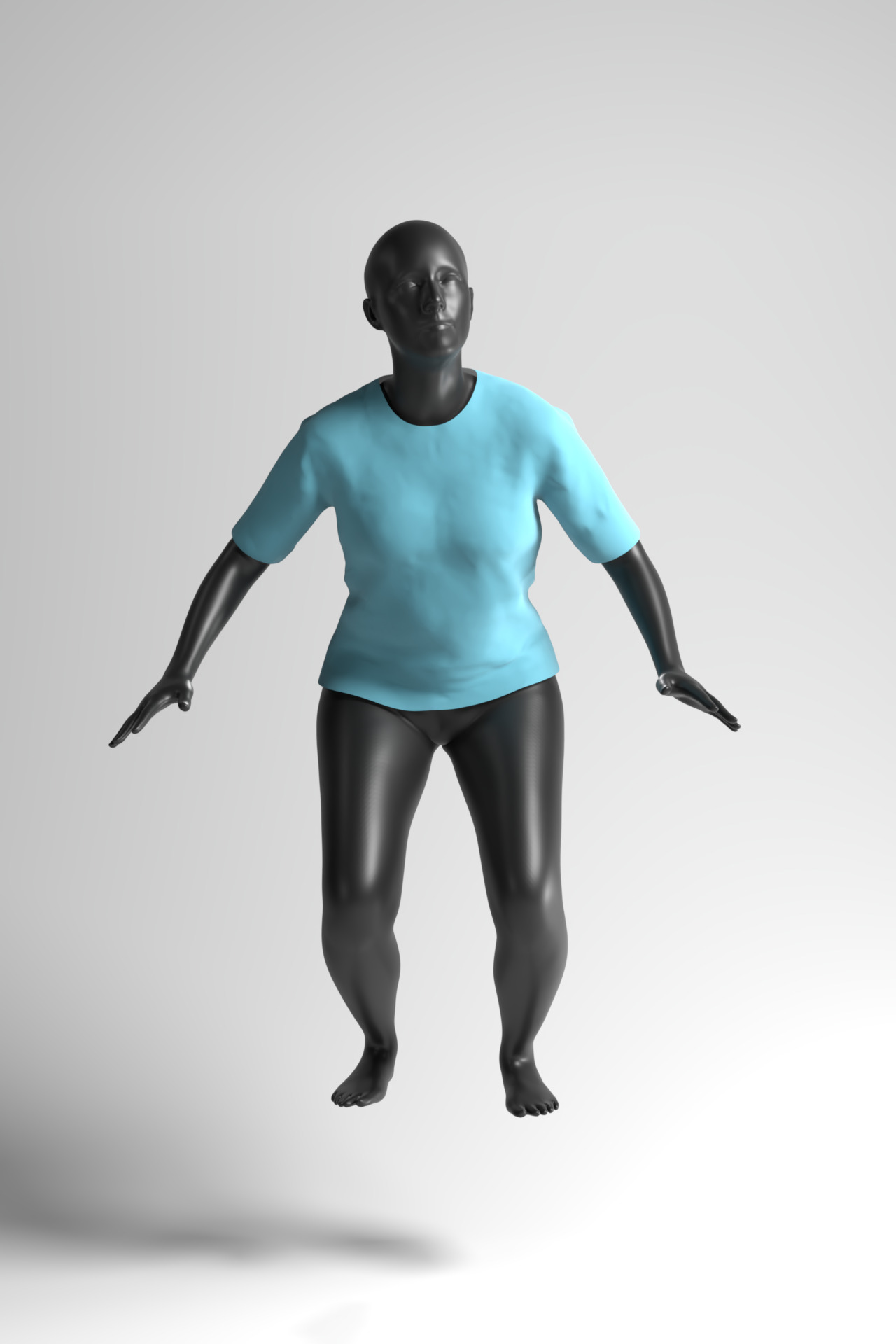} &
        \includegraphics[width=0.123\linewidth]{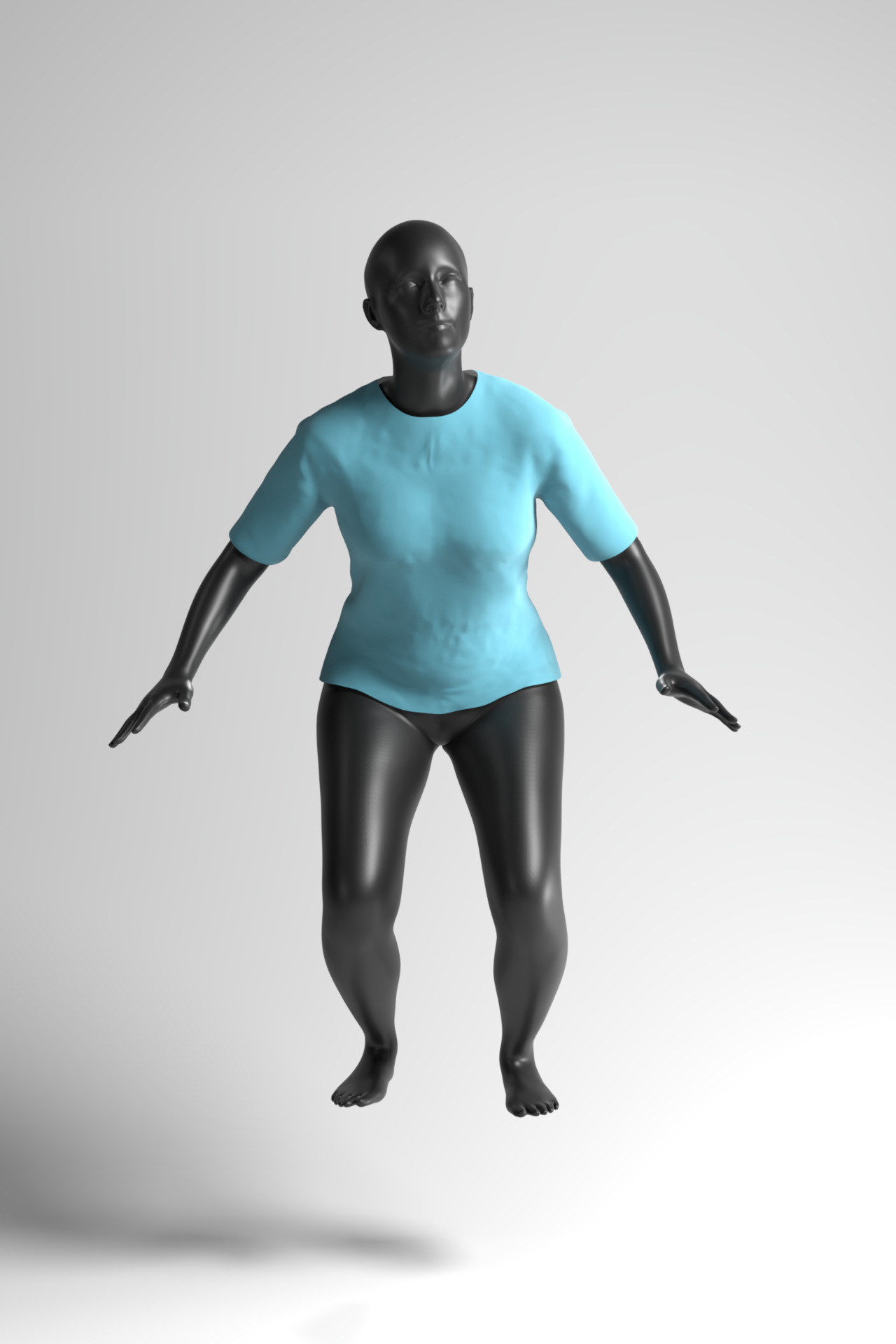} &
        \includegraphics[width=0.123\linewidth]{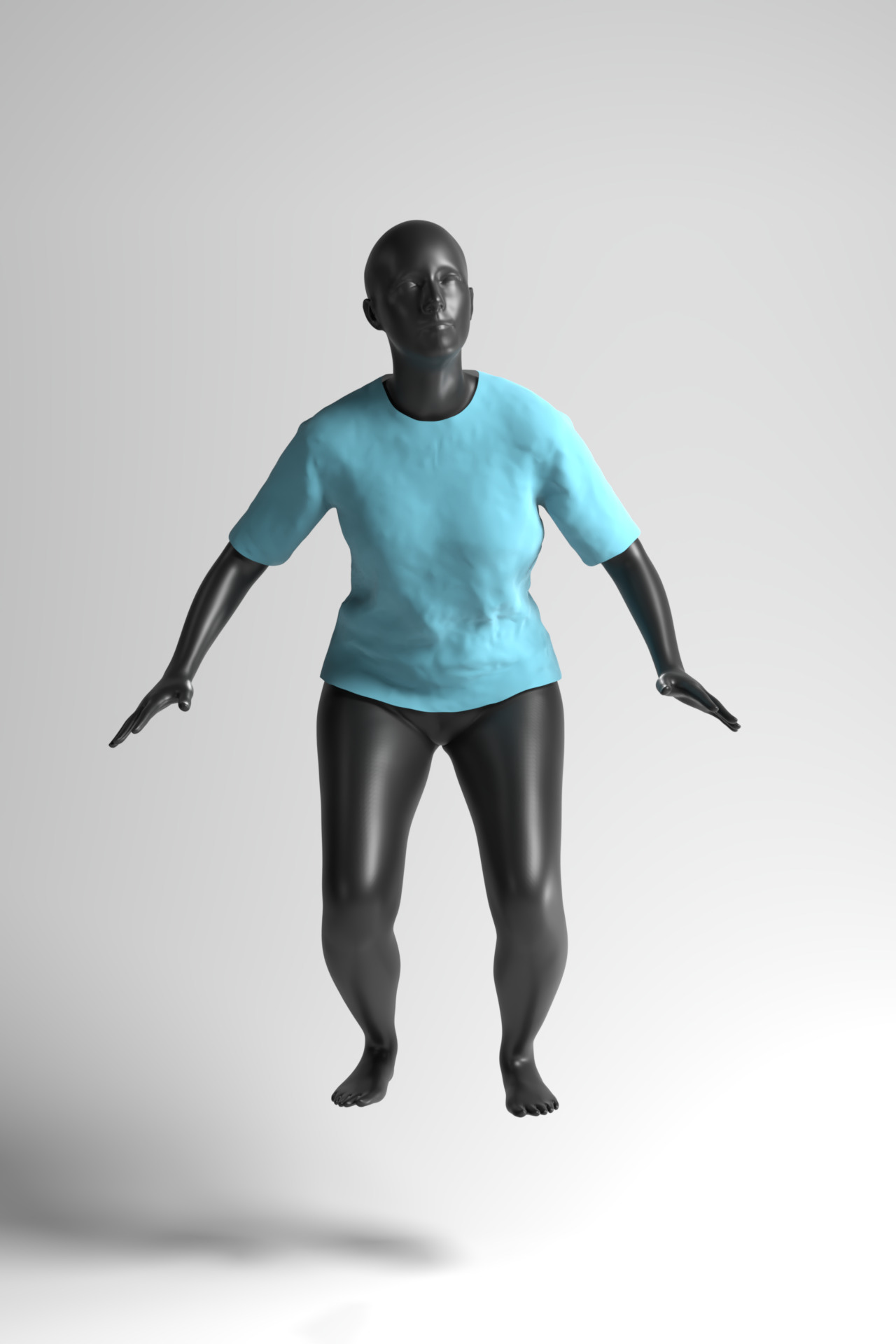} &
        \includegraphics[width=0.123\linewidth]{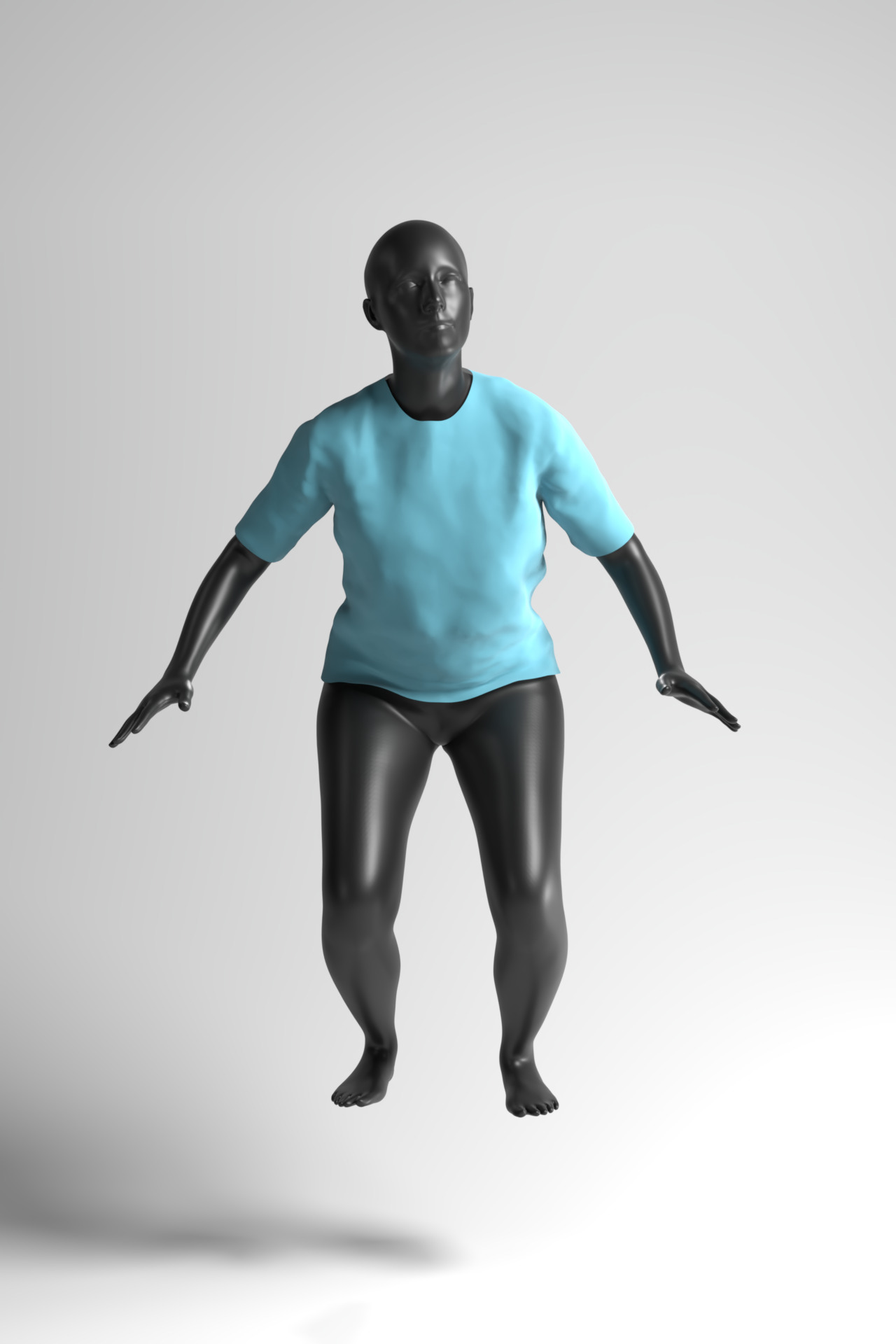} &
        \includegraphics[width=0.123\linewidth]{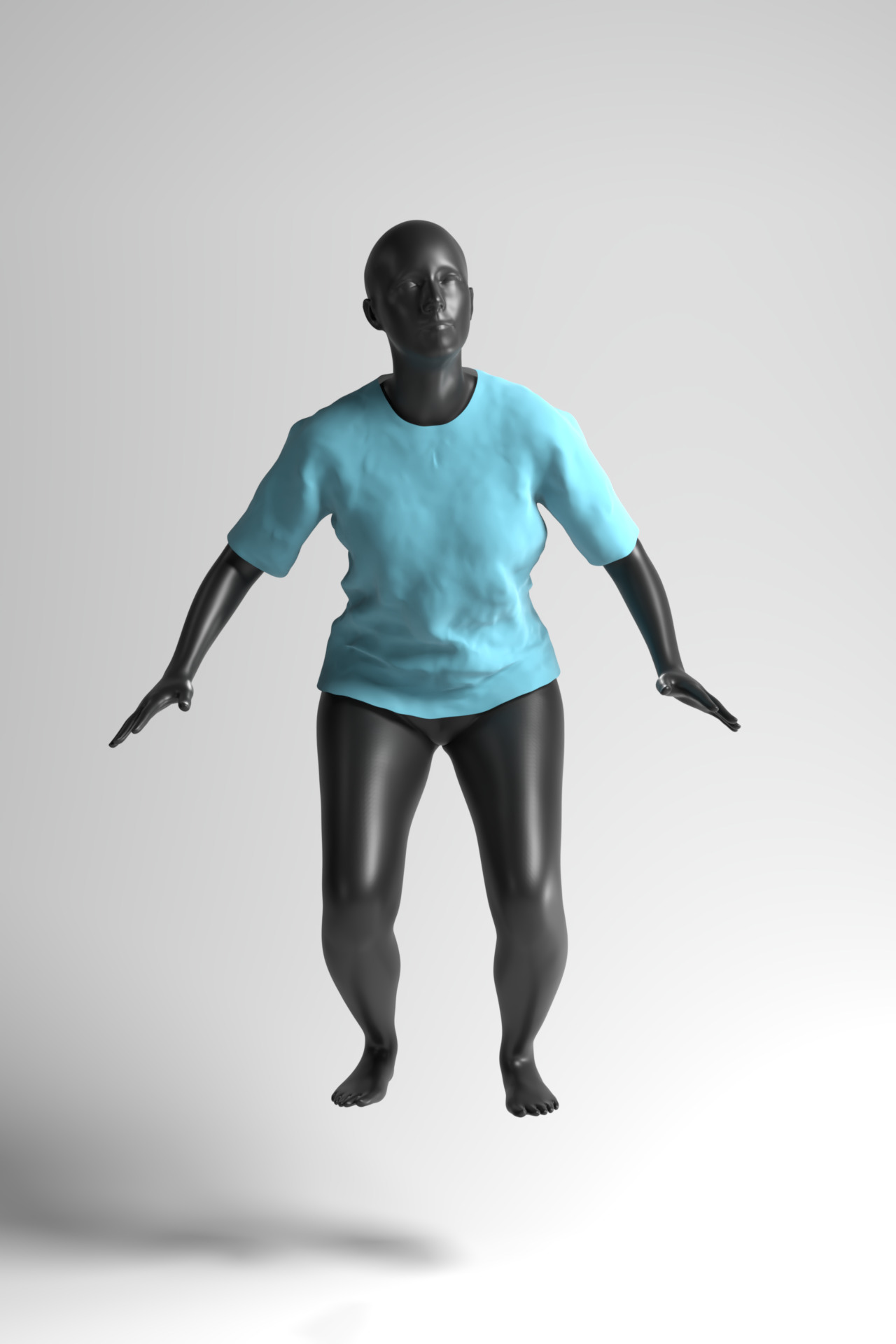} &
        \includegraphics[width=0.123\linewidth]{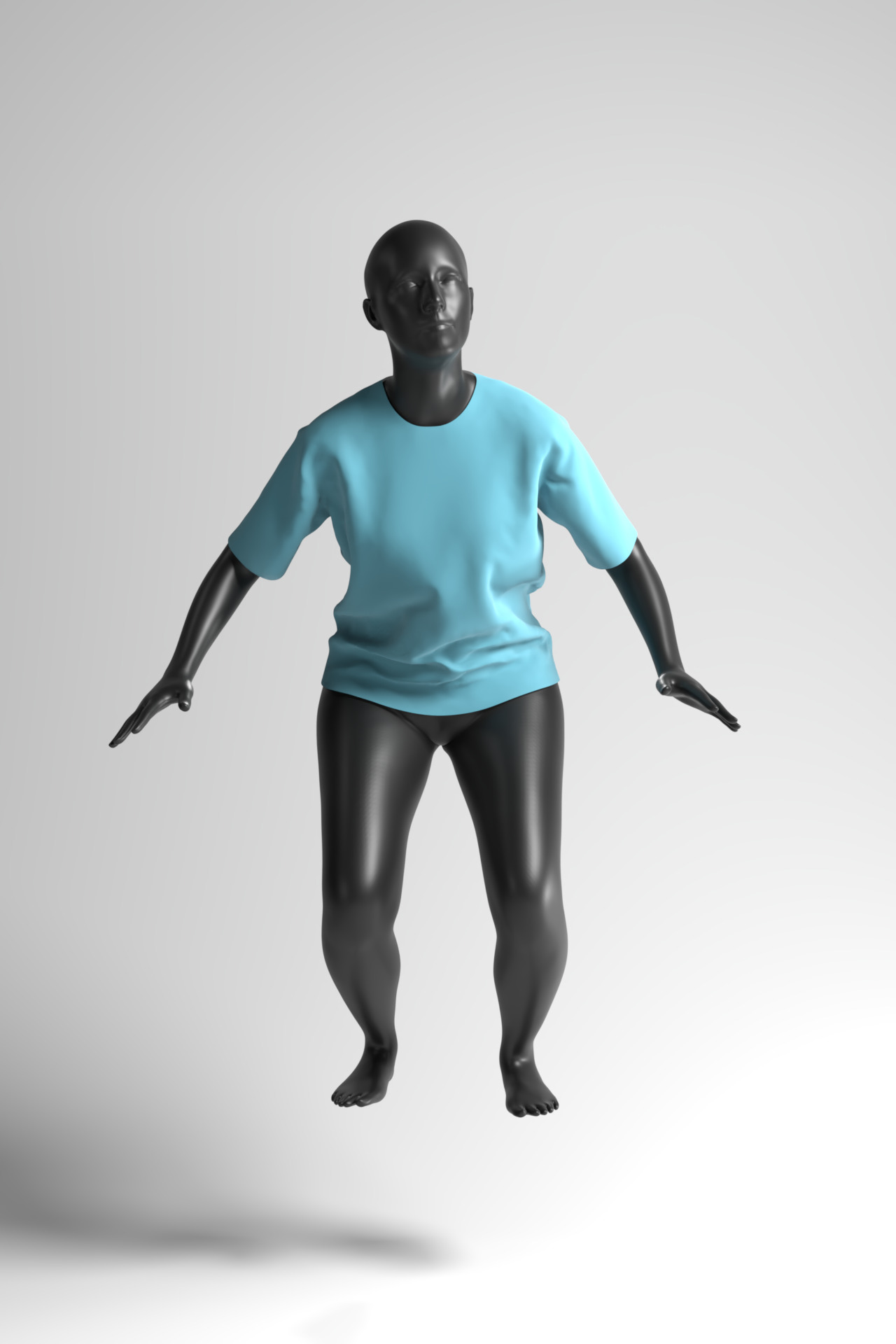}\\

        \includegraphics[width=0.123\linewidth]{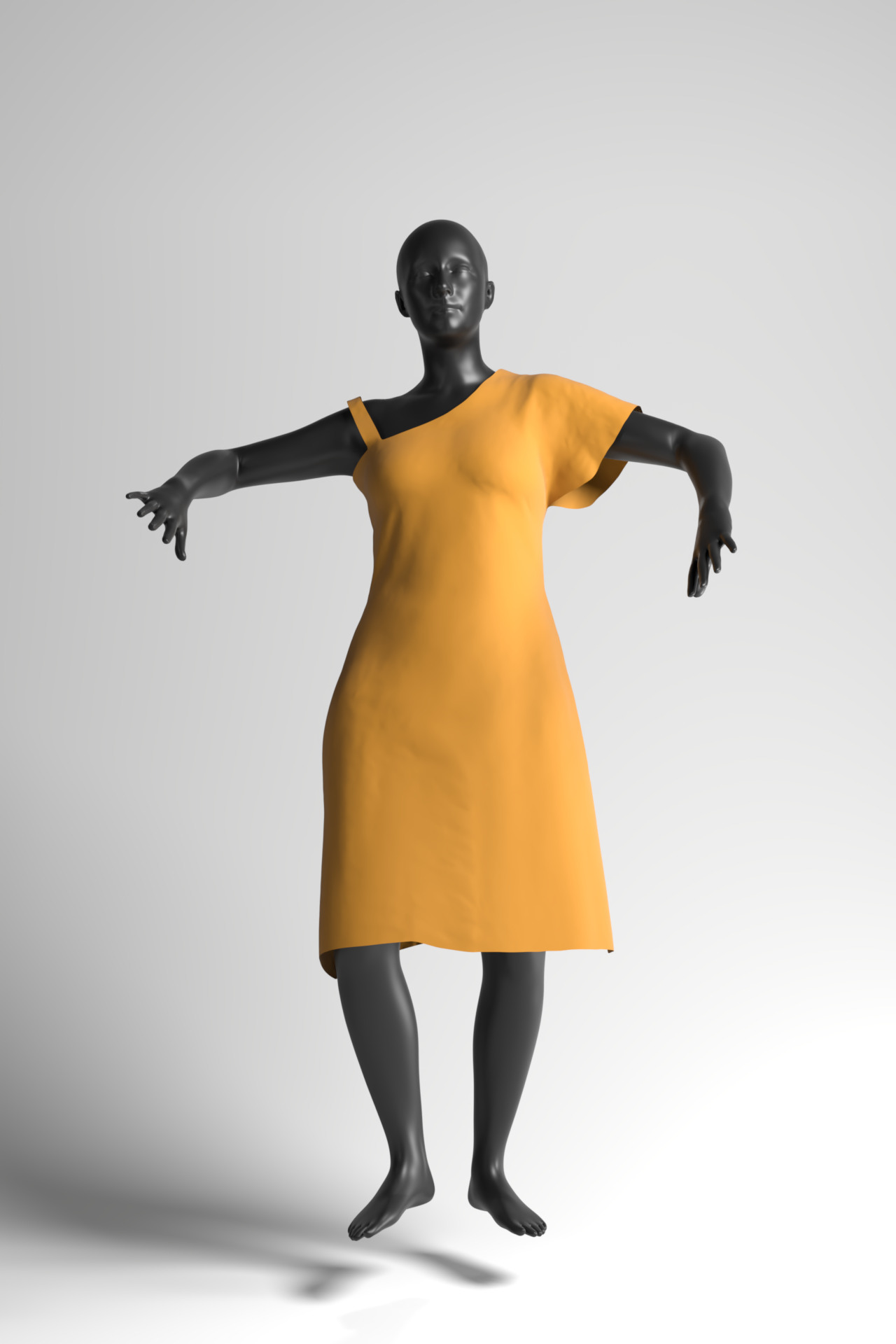} &
        \includegraphics[width=0.123\linewidth]{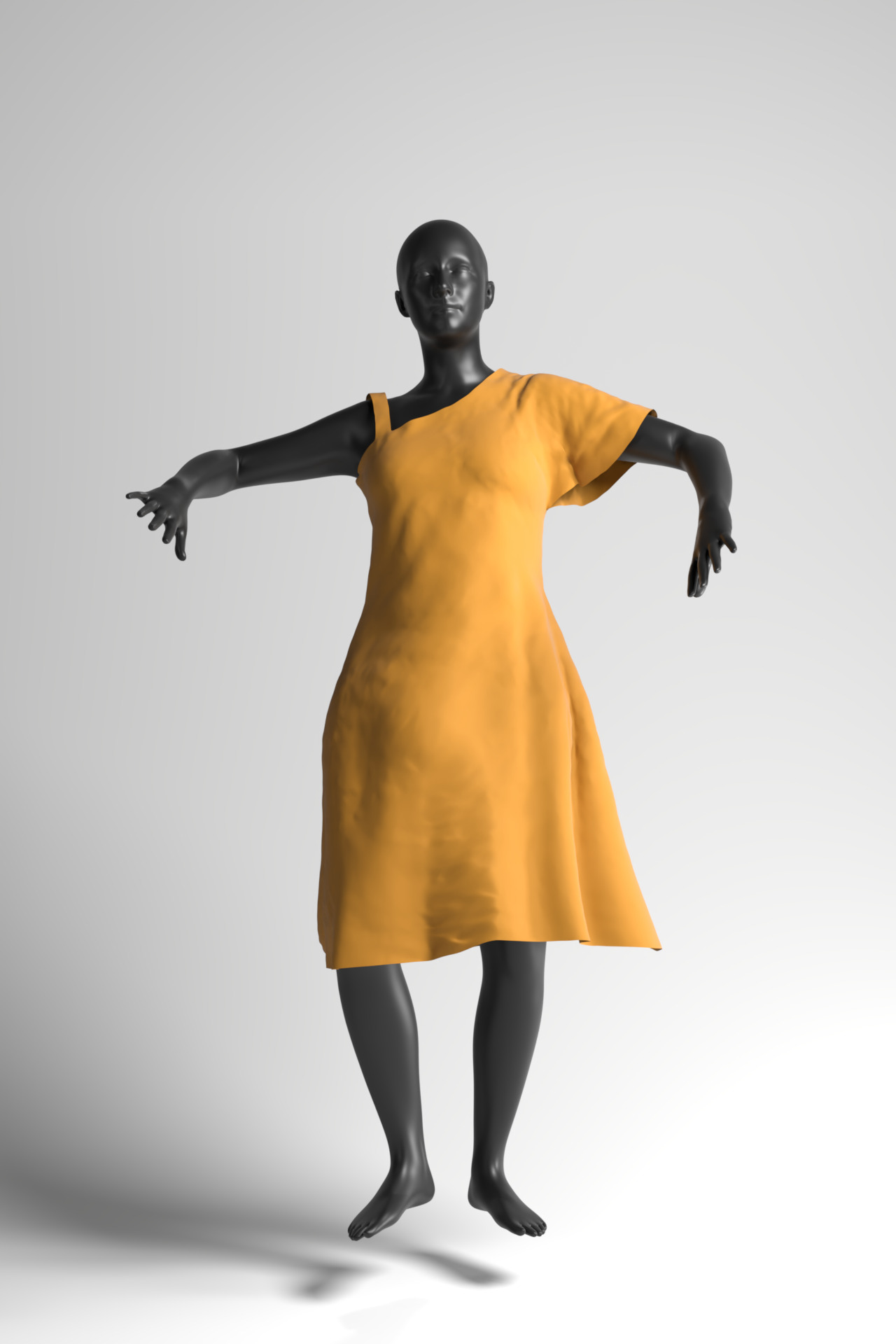} &
        \includegraphics[width=0.123\linewidth]{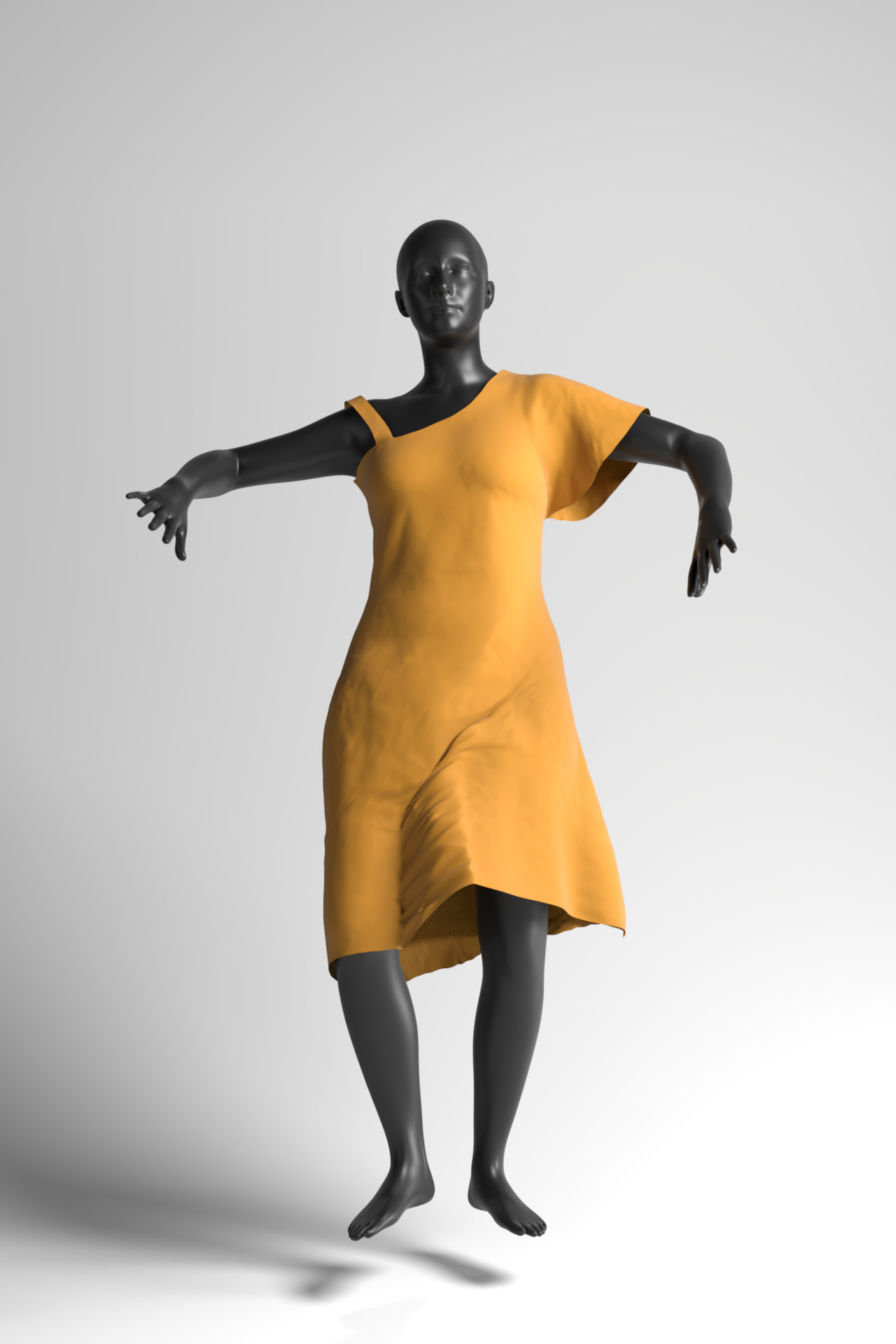} &
        \includegraphics[width=0.123\linewidth]{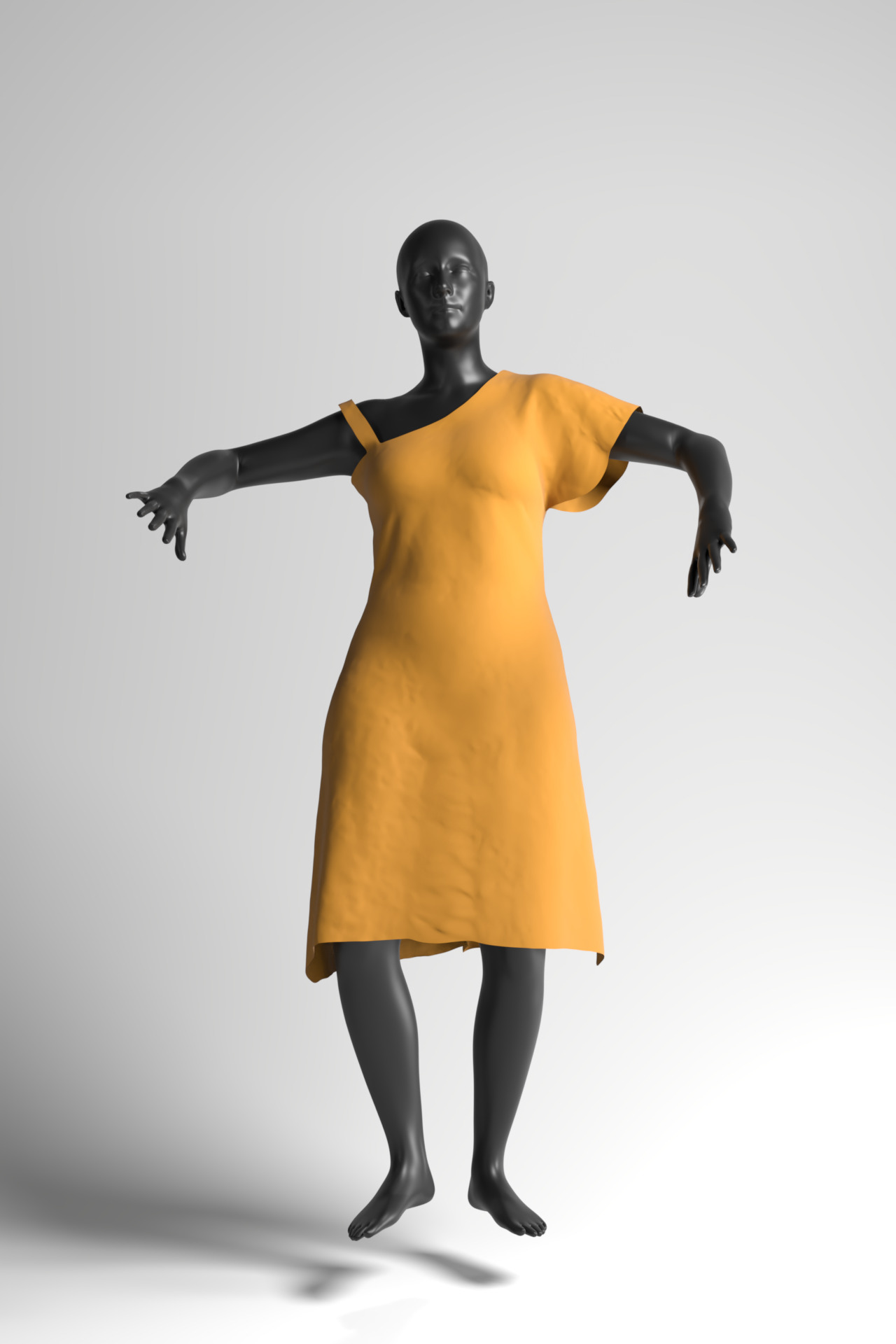} &
        \includegraphics[width=0.123\linewidth]{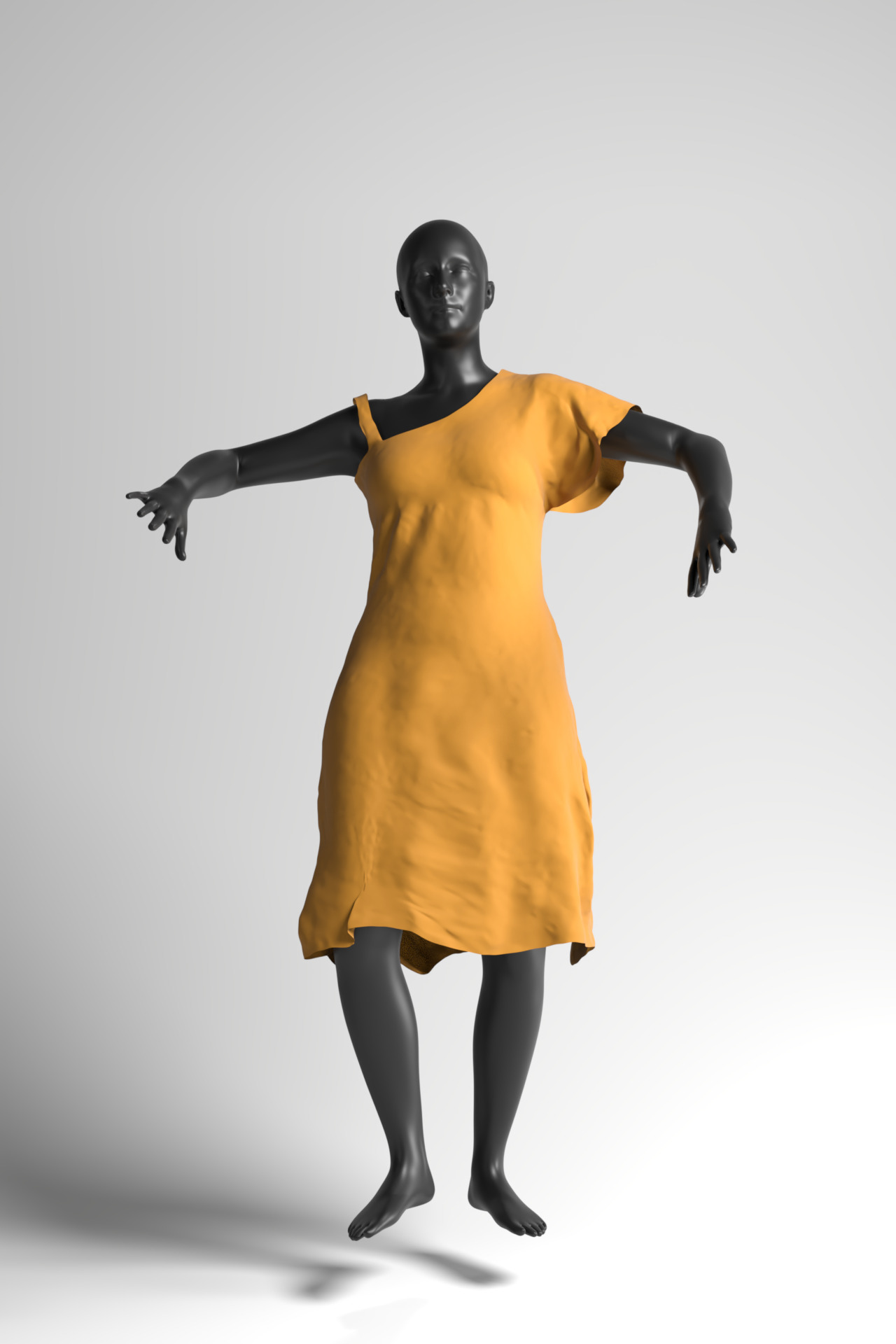} &
        \includegraphics[width=0.123\linewidth]{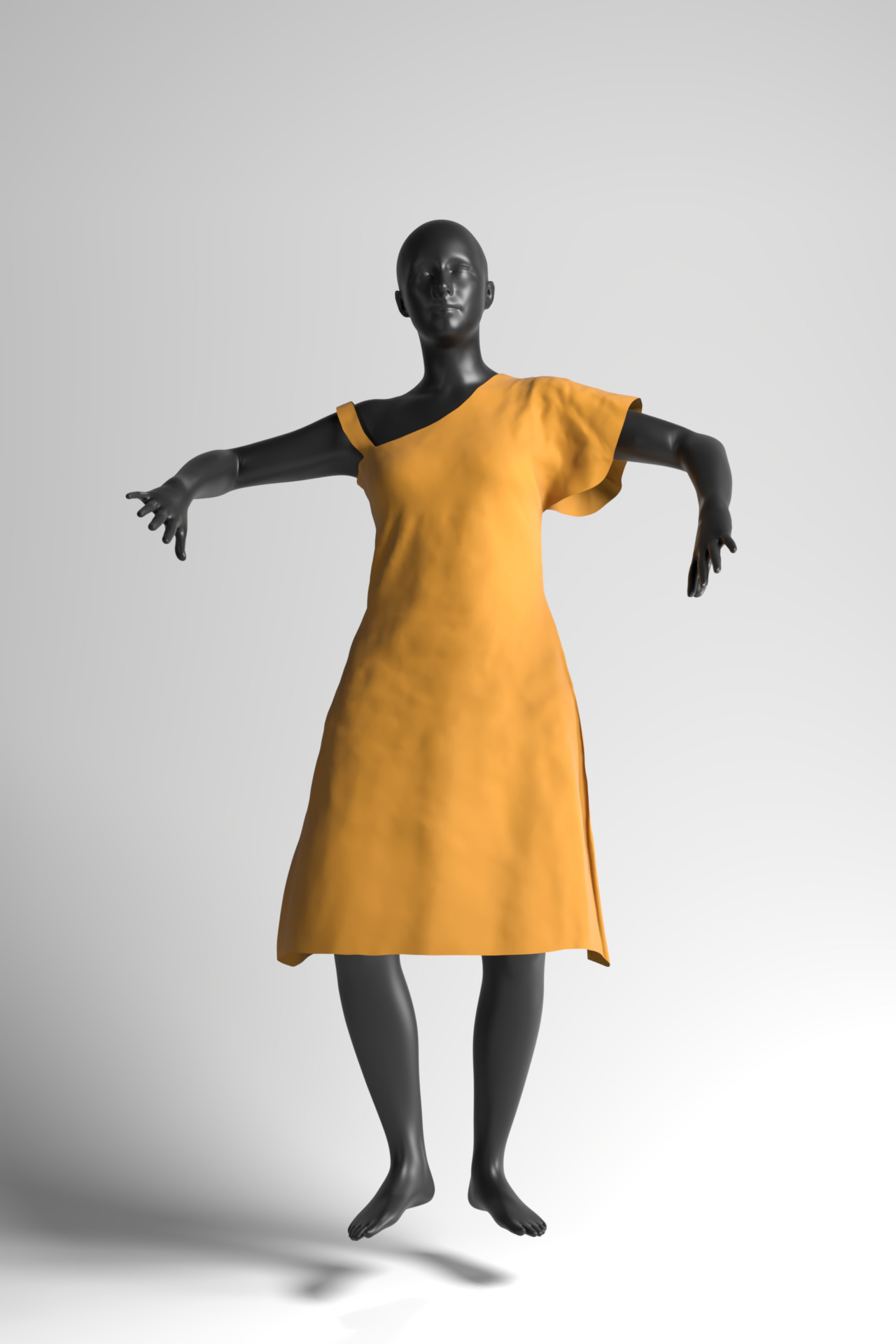} &
        \includegraphics[width=0.123\linewidth]{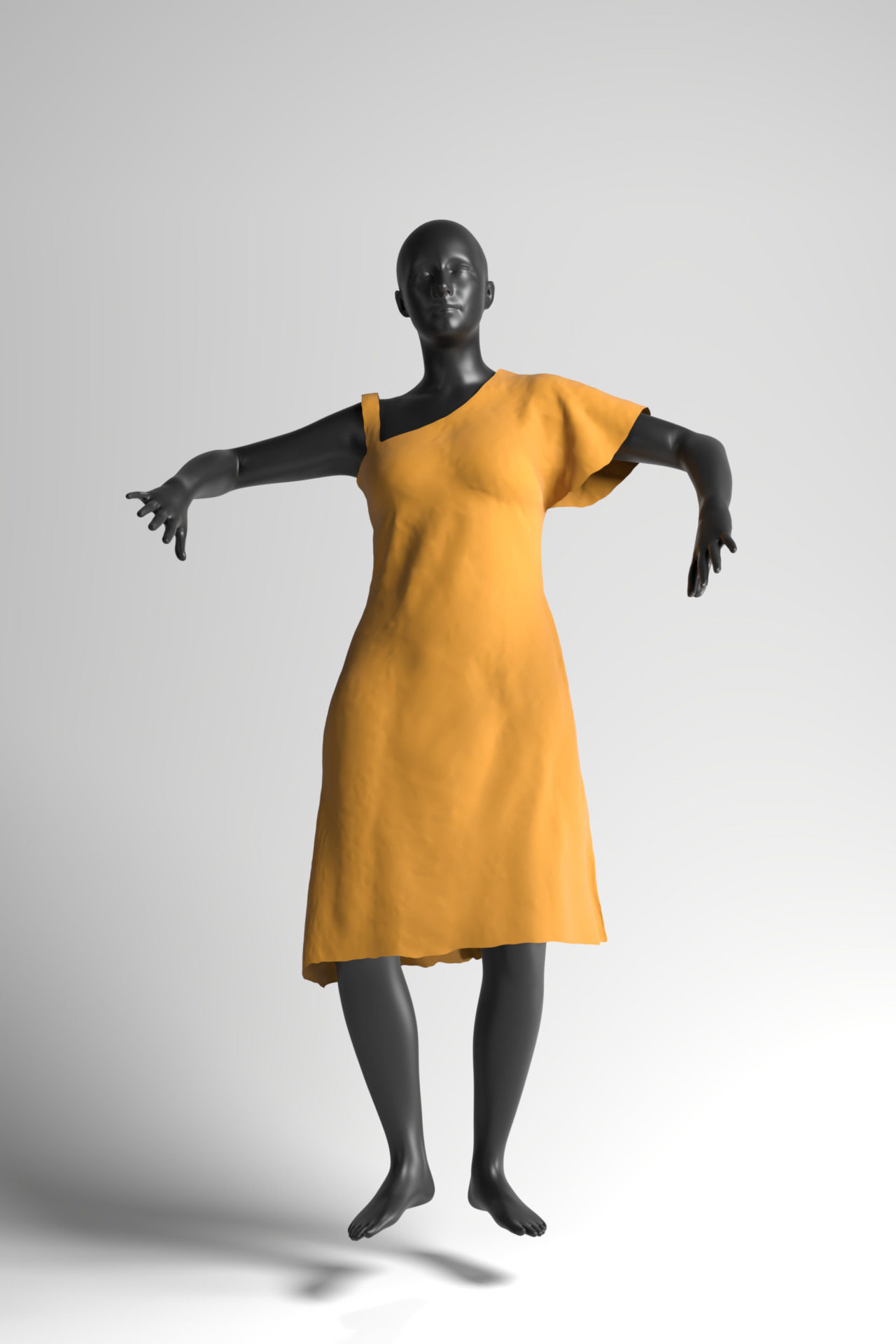} &
        \includegraphics[width=0.123\linewidth]{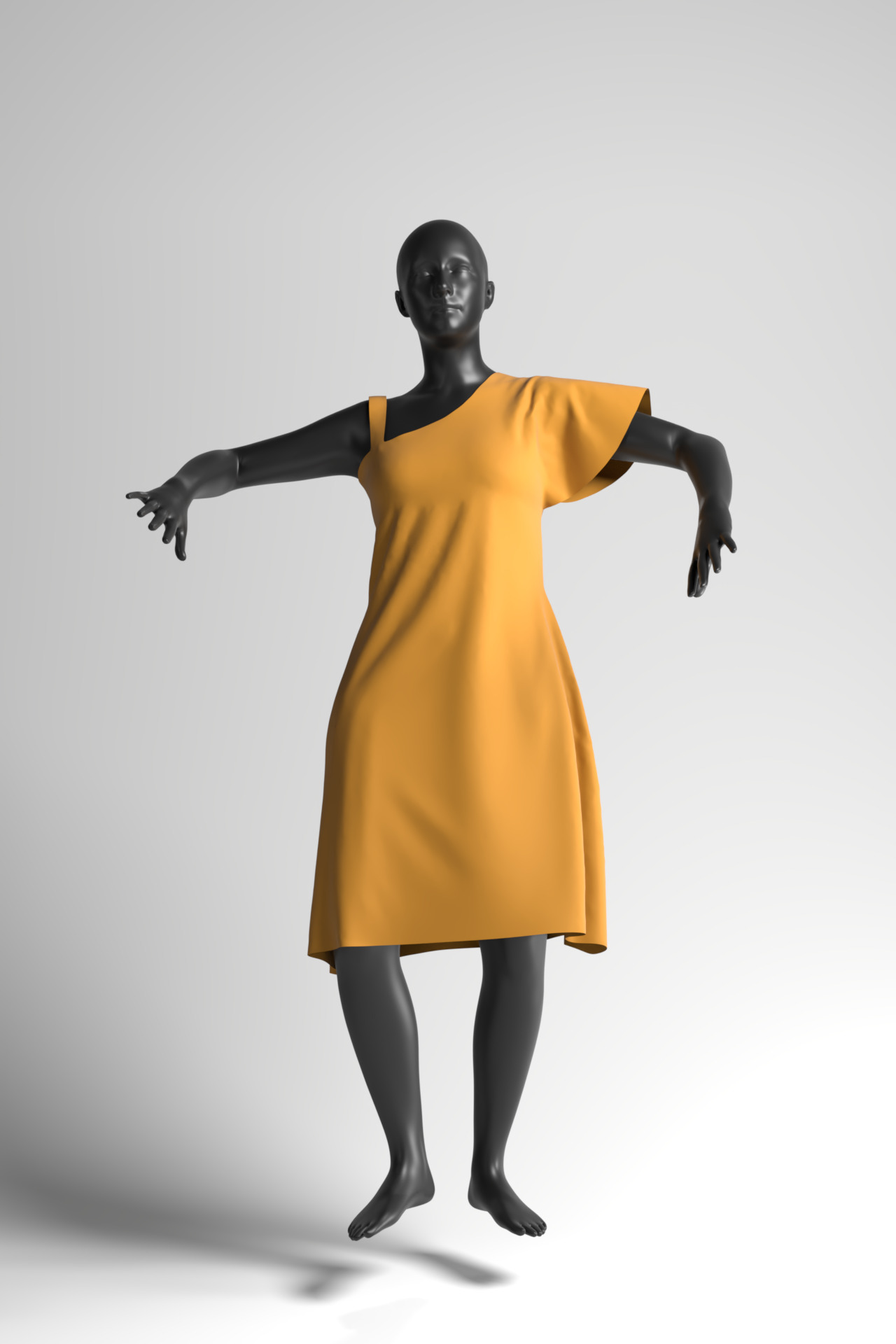}\\

        \includegraphics[width=0.123\linewidth]{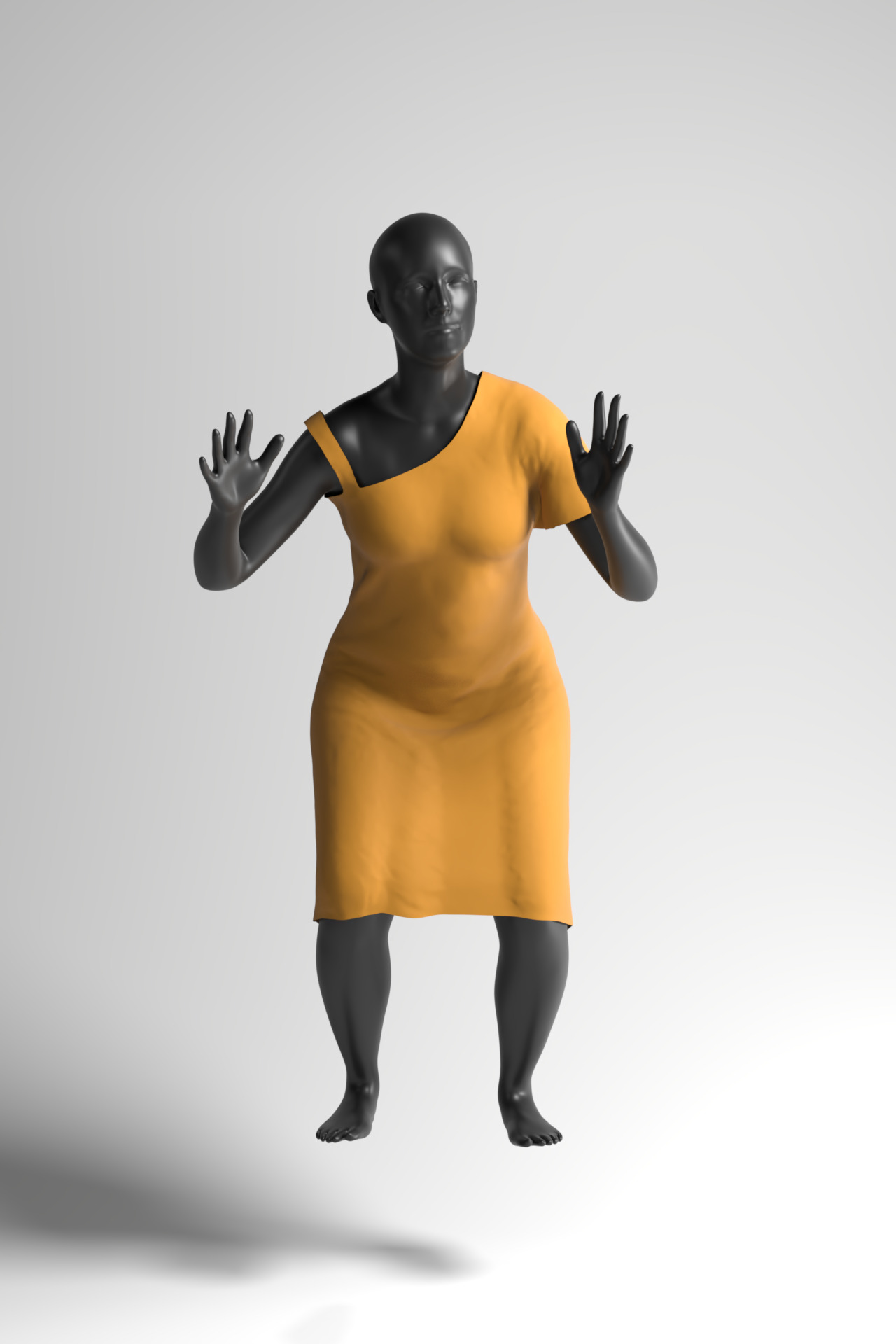} &
        \includegraphics[width=0.123\linewidth]{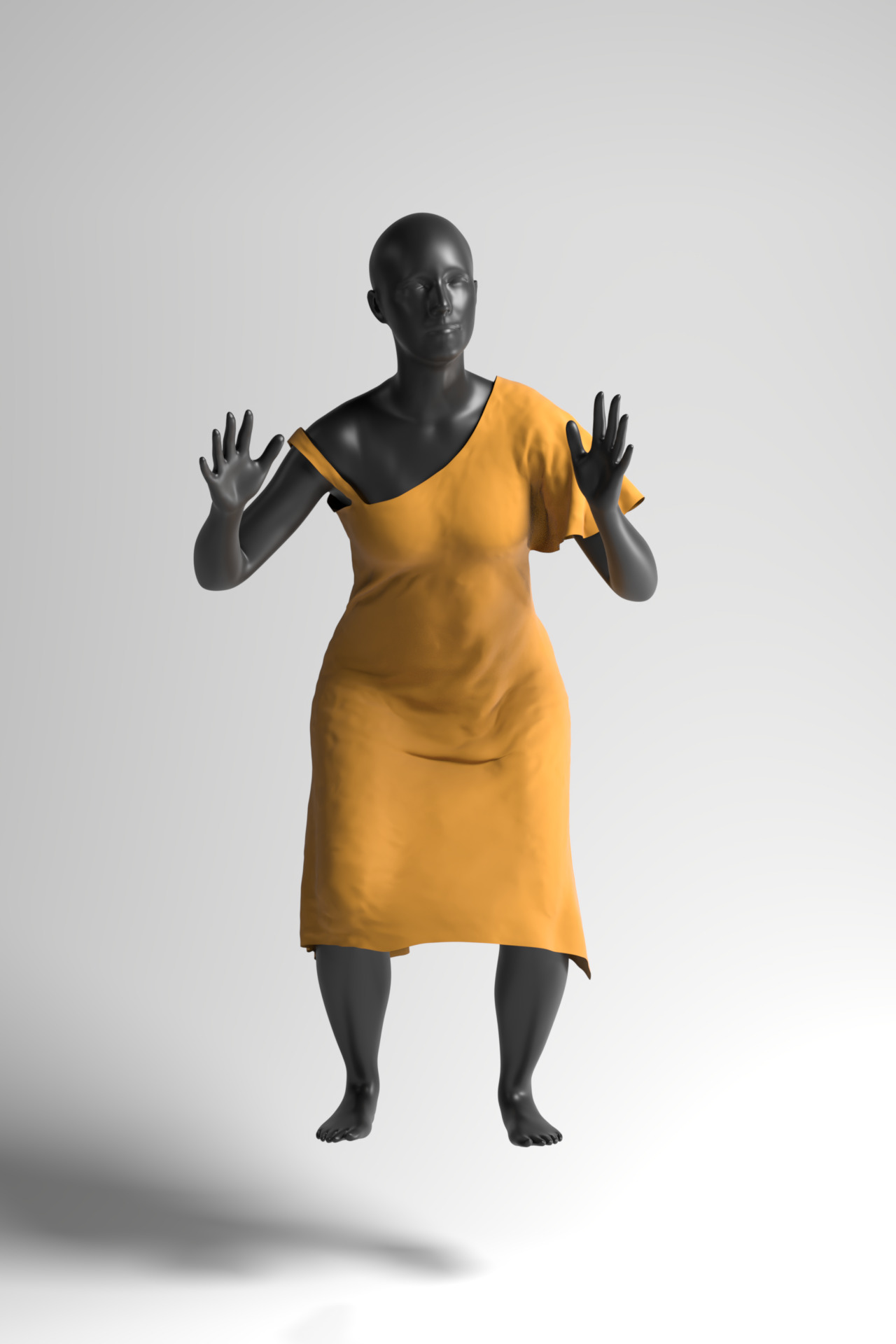} &
        \includegraphics[width=0.123\linewidth]{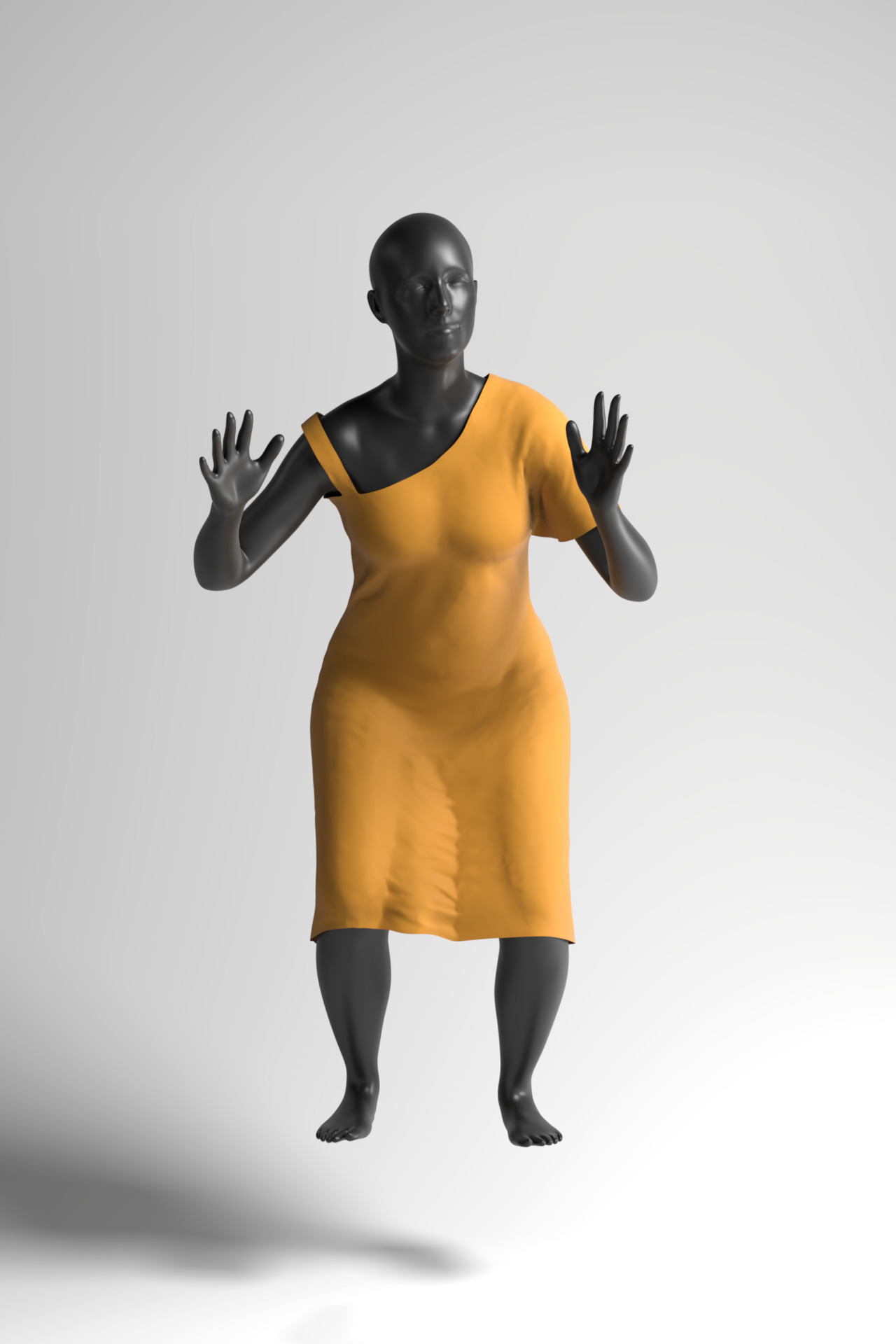} &
        \includegraphics[width=0.123\linewidth]{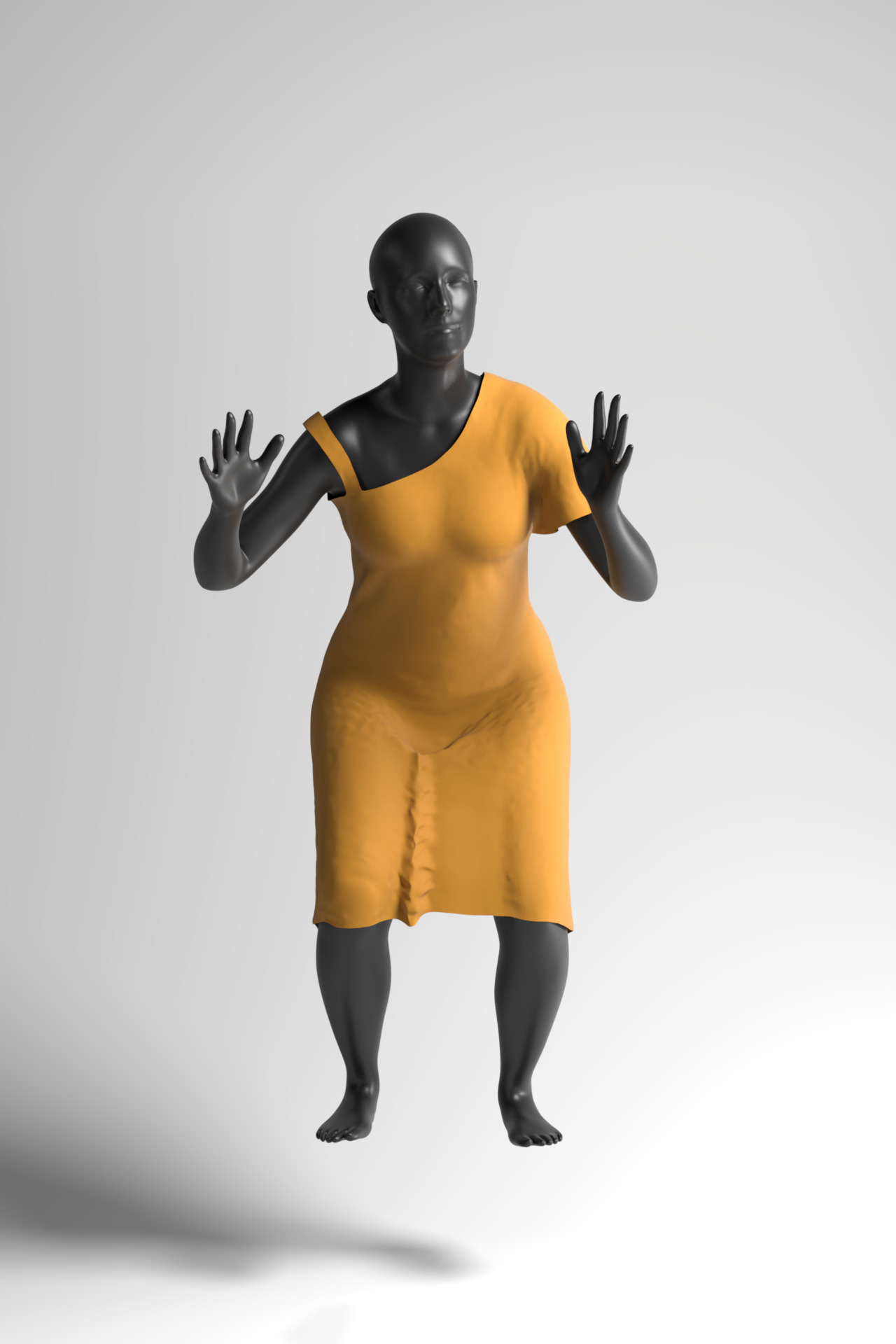} &
        \includegraphics[width=0.123\linewidth]{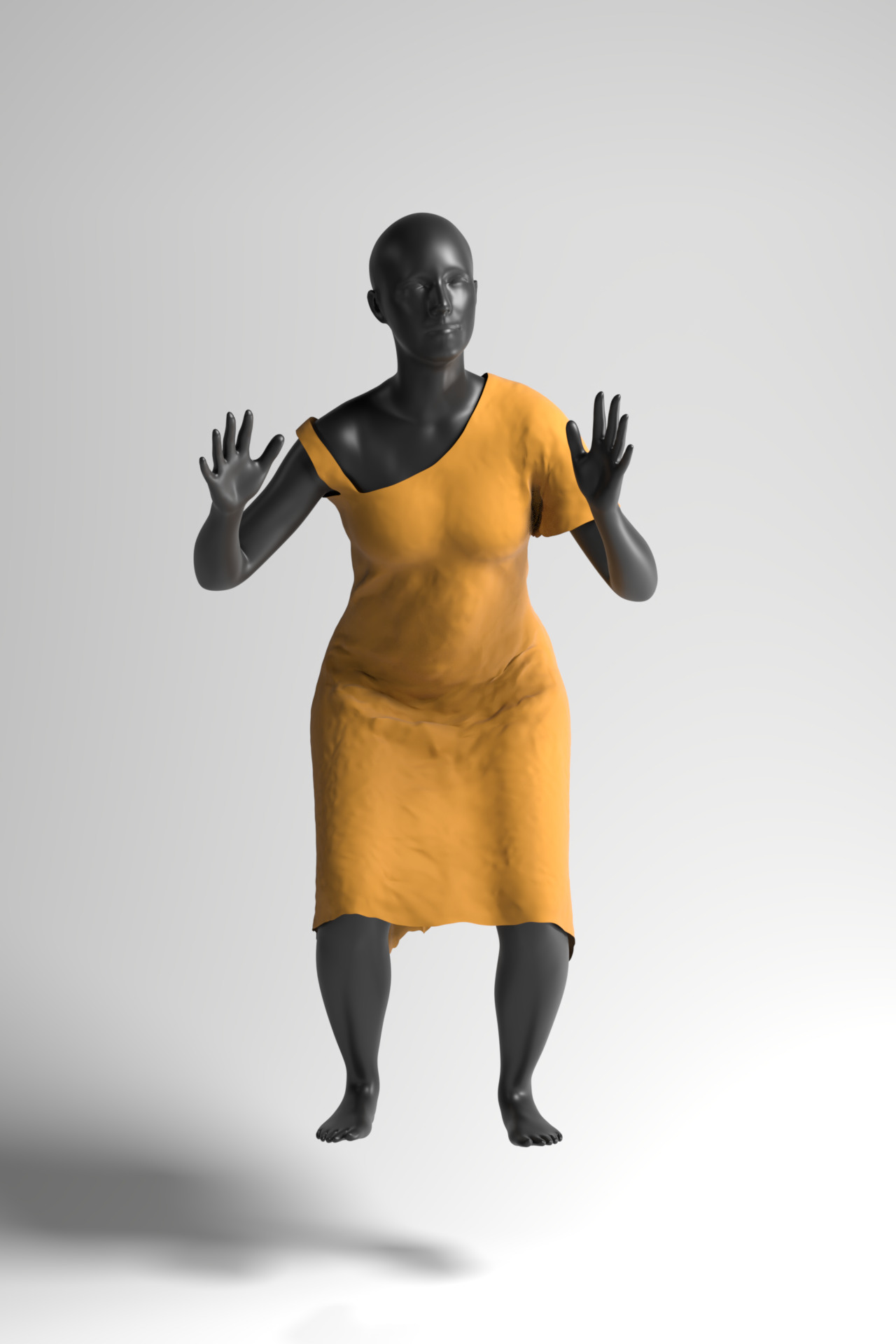} &
        \includegraphics[width=0.123\linewidth]{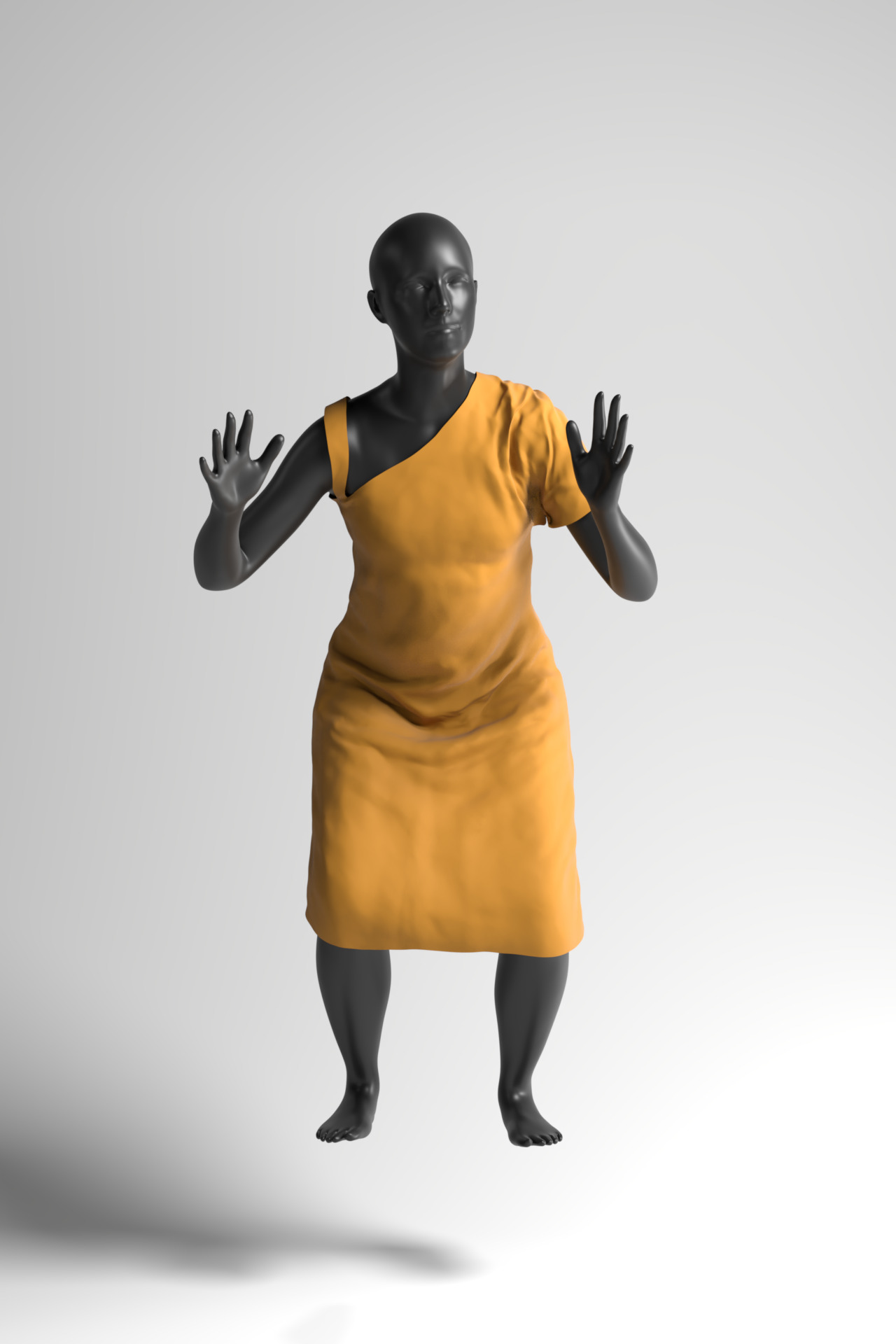} &
        \includegraphics[width=0.123\linewidth]{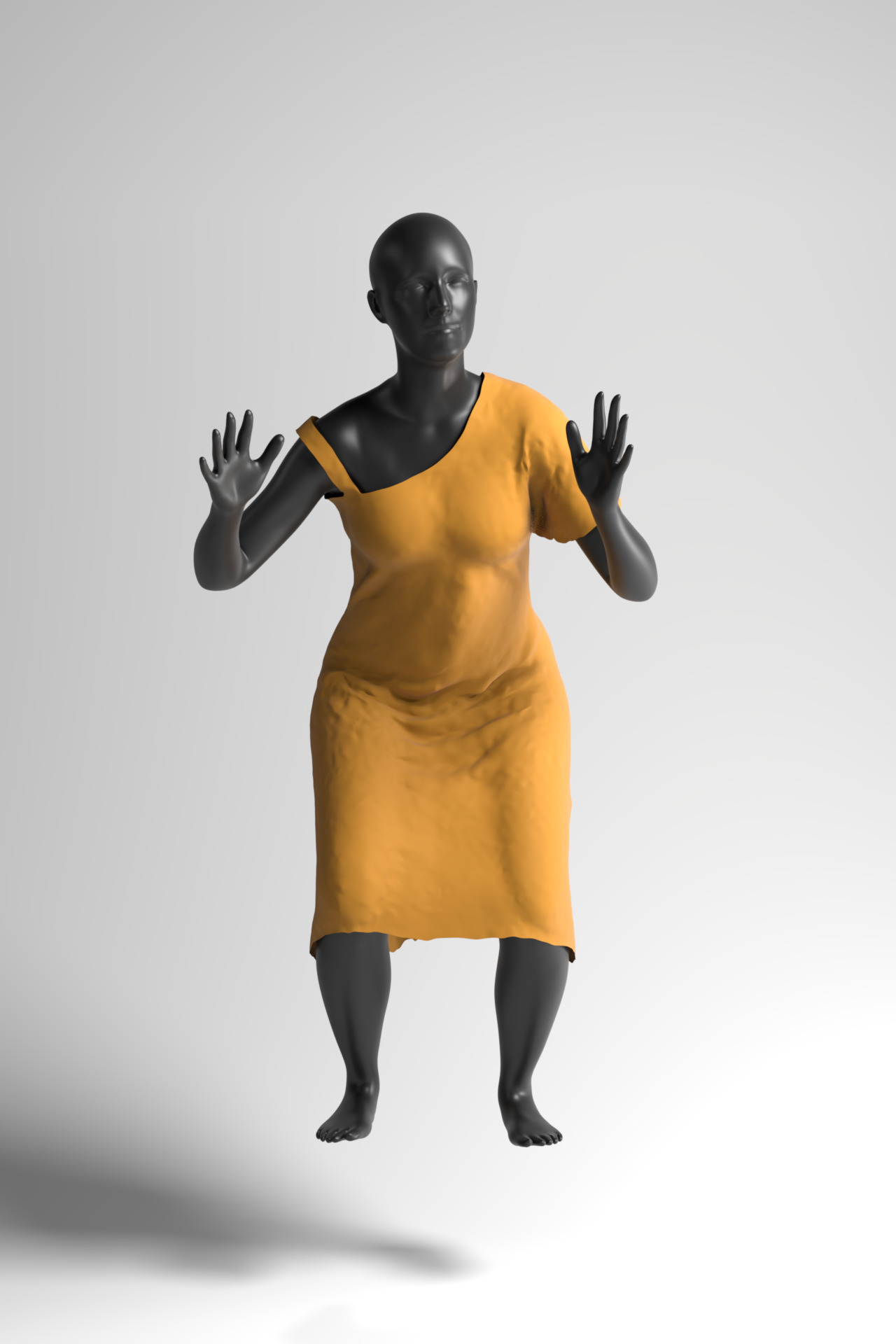} &
        \includegraphics[width=0.123\linewidth]{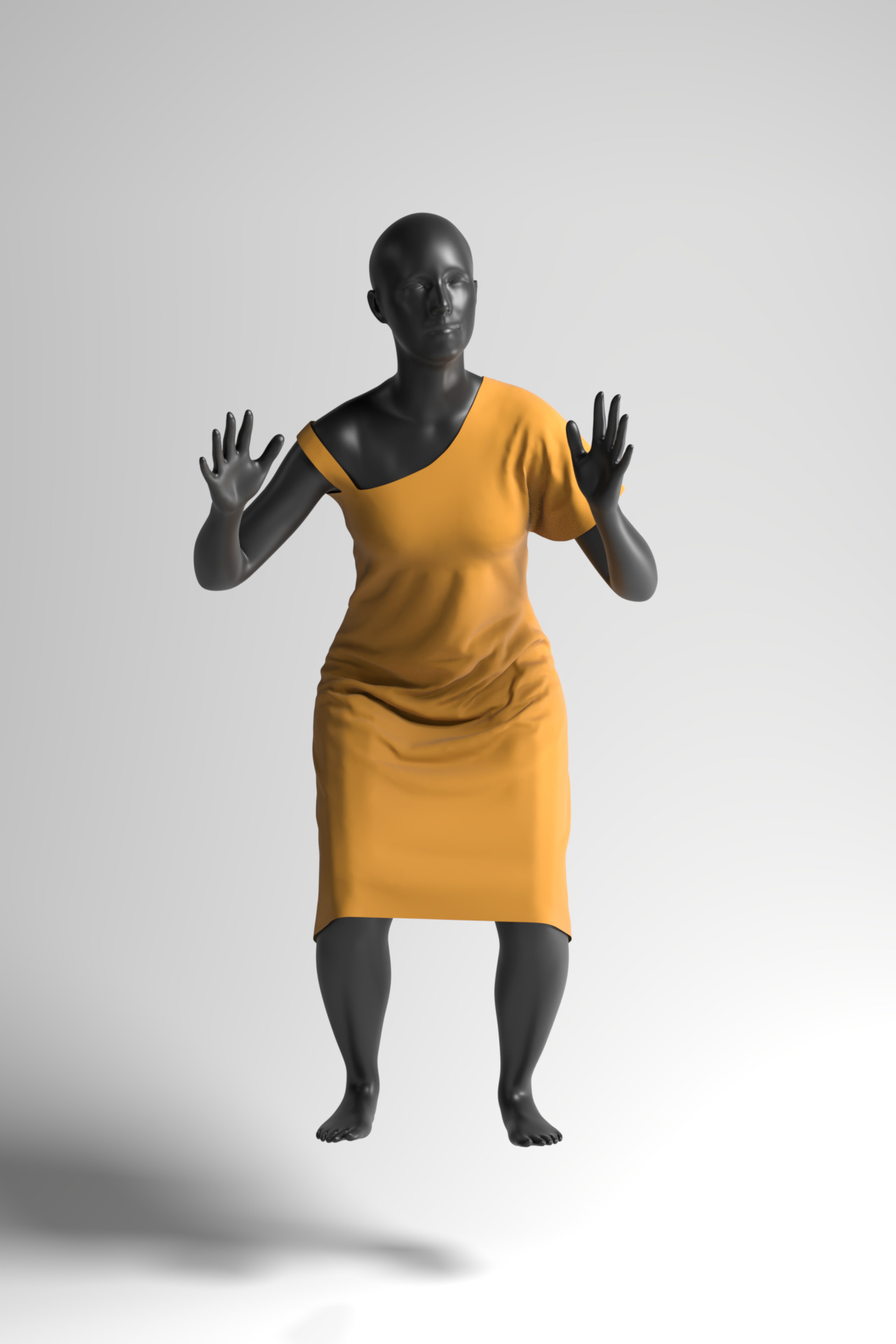}\\

		Maya 2025.2 & Unreal Engine 5.6 & \cite{bailey_fast_2018} & \cite{patel_tailornet_2020} & \cite{chentanez_cloth_2020} & \cite{santesteban_self-supervised_2021} & Ours & Ground Truth
	\end{tabular}}
    \caption{Additional qualitative comparisons on the \vto dataset. Columns correspond to different methods, while rows show examples for T-shirts (top) and dresses (bottom). The loose garments and highly dynamic motions in this dataset pose significant challenges for quasi-static predictors, often resulting in overly smoothed deformations or high-frequency artifacts across methods.}
  \label{fig:ComparisonBaselineModelsVTOSupp}
\end{figure*}

\paragraph*{Additional comparisons on VTO dataset.}
\Fig{ComparisonBaselineModelsVTOSupp} provides additional qualitative comparisons on the \vto dataset, including examples for both the T-shirt and the dress. Columns correspond to different methods, while rows show different garment types and poses. Across these examples, all methods struggle with the highly dynamic motions and loose garments present in the dataset, exhibiting either overly smoothed results or high-frequency artifacts. These examples further illustrate the challenges posed by loose garments and highly dynamic motions in the \vto dataset for quasi-static predictors.

\begin{table}
    \centering
    {\footnotesize
    \rowcolors{3}{grayband}{white}
    \resizebox{\linewidth}{!}{
	\begin{tabular}{ccccc}
		\toprule
        \multirow[c]{2}{*}{Model}
        & \multicolumn{2}{c}{Preference for ours}
        & \multicolumn{2}{c}{Preference for others} \\
        & Strong & Weak & Weak & Strong \\
		\midrule
		LBS & 76.50 & 22.95 & 0.55 & 0.00 \\
		Maya 2025.2 & 61.29 & 34.41 & 4.30 & 0.00 \\
		Unreal Engine 5.6 & 68.31 & 25.68 & 2.19 & 3.83 \\
		\cite{bailey_fast_2018} & 54.95 & 37.36 & 7.14 & 0.55 \\
		\cite{patel_tailornet_2020} & 69.44 & 27.78 & 2.22 & 0.56 \\
		\cite{chentanez_cloth_2020} & 70.33 & 29.12 & 0.55 & 0.00 \\
		\bottomrule
	\end{tabular}
    }
    }
	\caption{User study results for pairwise comparisons against each baseline, reported as percentages of participant responses. Across all baselines, participants more frequently preferred our method, with the majority of votes falling into the strong or weak preference for ours categories.}
	\label{tab:UserStudy}
\end{table}

\begin{figure}[t]
	\includegraphics[width=\linewidth]{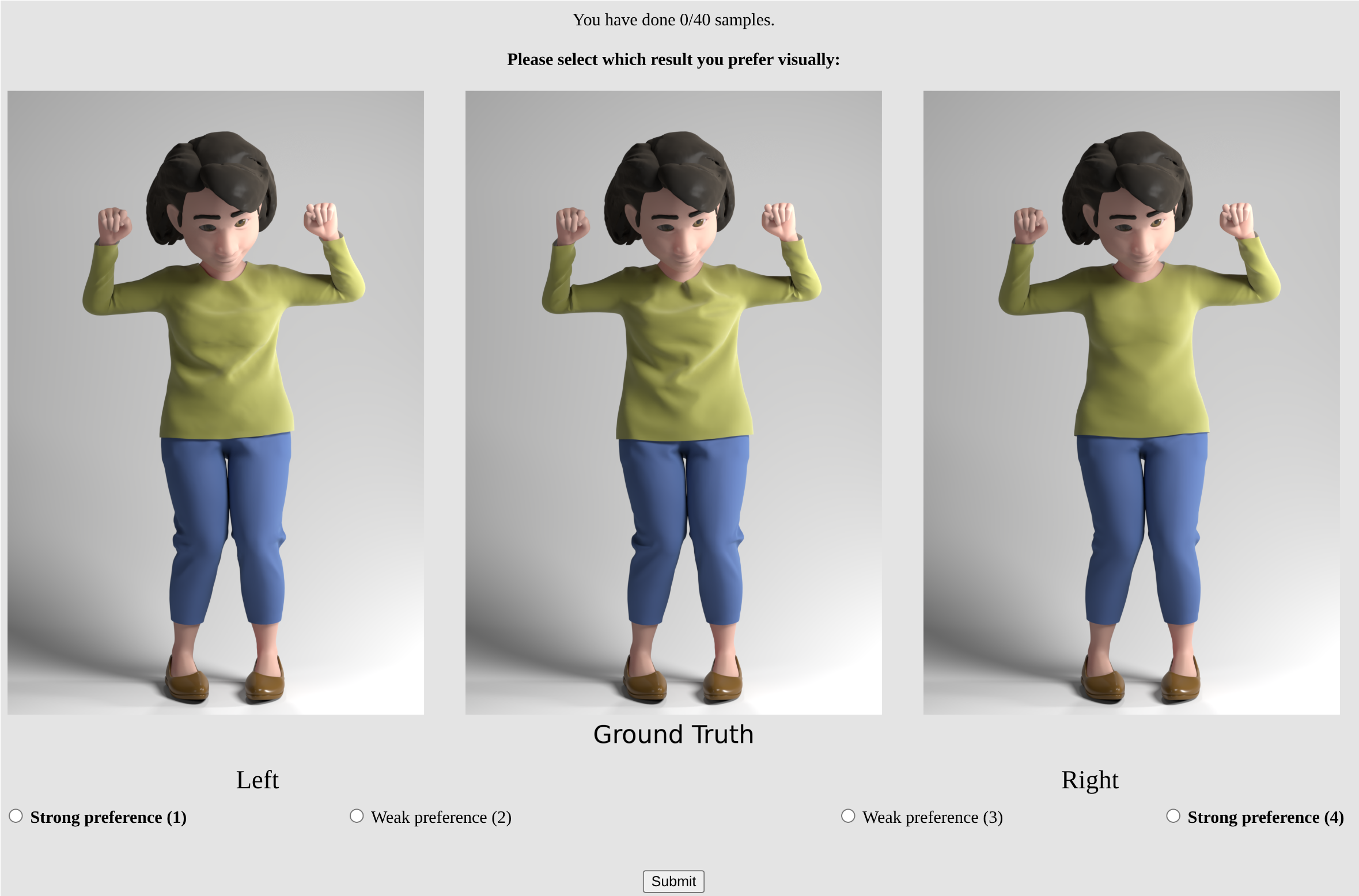}
	\caption{User study setup. The user chooses their preference for each comparison.}
    \label{fig:UserStudySetup}
\end{figure}

\paragraph*{User study setup and results.}
We conducted a user study with 28 participants and collected 1096 votes in total. In each trial, participants were shown a side-by-side comparison between a rendered frame produced by our method and one produced by a baseline method. The corresponding ground-truth frame was always displayed in the center as a visual reference. Participants were asked to indicate which of the two compared results they preferred in terms of visual quality with respect to the ground-truth reference, without being informed of the source of either image. The response options were strongly prefer left, weakly prefer left, weakly prefer right, and strongly prefer right. \Fig{UserStudySetup} shows the interface used in the study.

To reduce presentation bias, the left-right placement of our method and the baseline was randomized for each trial, and the trial order was shuffled independently for each participant. Each participant viewed a randomized subset of all possible trials. Baseline methods were sampled uniformly at random, so each method was represented by approximately the same number of examples overall. Participants could take as much time as needed for each comparison.

For analysis, we mapped the four response options into preference for ours and preference for others, each split into strong and weak preference. \Tab{UserStudy} reports the percentage of responses in each category for every baseline. Across all baselines, participants preferred our method more frequently than the alternatives, with most responses falling into the strong or weak preference for ours categories.

\end{document}